\documentclass{article}
\usepackage{iclr2027_conference,times}

\usepackage{amsmath,amsfonts,bm}

\def\eqref#1{equation~\ref{#1}}

\def\1{\bm{1}}

\DeclareMathAlphabet{\mathsfit}{\encodingdefault}{\sfdefault}{m}{sl}
\SetMathAlphabet{\mathsfit}{bold}{\encodingdefault}{\sfdefault}{bx}{n}

\usepackage{amssymb,amsmath,booktabs,graphicx,xcolor,colortbl}
\usepackage{pifont}
\usepackage{float}
\usepackage{wrapfig}
\usepackage{hyperref}
\hypersetup{
  colorlinks=true,
  linkcolor=black,
  citecolor=blue!55!black,
  urlcolor=blue!55!black
}
\newcommand{\ourmethod}{our method}
\newcommand{\ours}{\textbf{Ours}}
\newcommand{\cmark}{\textcolor{green!55!black}{\ding{51}}}
\newcommand{\xmark}{\textcolor{red!70!black}{\ding{55}}}
\newcommand{\gbf}[1]{\textcolor{green!55!black}{\textbf{#1}}}
\newcommand{\rbf}[1]{\textcolor{red!70!black}{#1}}
\newcommand{\gray}[1]{\textcolor{gray}{#1}}
\newcommand{\gcell}[1]{\cellcolor{gray!12}{#1}}
\newcommand{\bcell}[1]{\cellcolor{blue!9}{#1}}
\newcommand{\brow}{\rowcolor{blue!12}}
\newcommand{\grow}{\rowcolor{gray!10}}
\newcommand{\dE}{\Delta E}
\newif\ifoursappendixcore

\title{Variational Relevance-Informed Sparsification of Term Measurements for Variational Quantum Algorithms}
\author{Yifeng Peng, Bin Yan, Harshal Shah, \\
\textbf{Srinivasan Raghuraman, Kim Wagner \& Sunpreet Singh Arora} \\
Visa \\
Foster City, CA, USA
}

\iclrfinalcopy

\begin{document}
\maketitle
\lhead{Preprint.}

\begin{abstract}
Repeated measurements are a major bottleneck in variational quantum algorithms (VQAs). We uncover and characterize parameter-specific sparsity in the matrix of qubit-wise commuting (QWC) group contributions to parameter-shift gradients: most absolute contribution mass concentrates in a subset of groups whose relevance varies across parameters and optimization states. Building on this structure, we introduce a learned controller that predicts group relevance from Hamiltonian, circuit, light-cone, and current-state features, then selects measurements through a tunable coverage threshold. Exact contributions provide offline supervision without being required at deployment. Across 15 molecular VQE settings, \ourmethod{} saves 49.6\% of nominal gradient-measurement shots with a mean absolute energy gap of 5.67 mHa to Full-QWC. We then ask whether large language models (LLMs) can solve the same selection problem without fine-tuning. The results reveal both promise and a clear limitation: LLMs can identify useful measurement groups, yet without guidance, local models do not consistently prioritize the physical information that matters most. Our learned controller uses these cues more effectively. This contrast points toward a broader direction for AI-assisted science: combining the flexibility of general-purpose models with explicit physical representations, task-specific objectives, and domain-grounded validation.

\end{abstract}
\section{Introduction}

The variational quantum eigensolver (VQE) is the flagship variational quantum algorithm (VQA) for estimating molecular ground-state energies~\citep{peruzzo2014variational,mcclean2016theory}. Much of the research on VQAs has focused on trainability, particularly on mitigating barren plateaus (BPs)~\citep{mcclean2018barren}. Measurement overhead presents an equally fundamental challenge, yet has received comparatively less attention. Even with hardware-efficient circuits, each VQE iteration requires repeated estimation of Hamiltonian observables and gradients. For quantum chemistry, this measurement burden often dominates the practical cost~\citep{mcardle2020quantum,gonthier2022measurements}.

To identify where this measurement burden can be reduced, we first decompose the cost of a single gradient update. Because unknown quantum states cannot be cloned and quantum measurements are destructive~\citep{wootters1982single}, expectation values must be estimated through repeated circuit preparations. On this sampling-based interface, parameter-shift rules are widely
used because they provide hardware-compatible analytic gradients ~\citep{schuld2019evaluating}. Let a qubit Hamiltonian contain $G$ QWC measurement groups and the ansatz $P$ trainable parameters. A parameter-shift gradient evaluates two shifted circuits per parameter and, under the standard Full-QWC estimator, measures all $G$ groups with $S$ shots, so one optimization step costs $2PGS$ circuit shots. Existing methods shrink one factor at a time: grouping compatible Pauli operators reduces $G$~\citep{gokhale2019minimizing,crawford2021efficient}, adaptive shot allocation reduces $S$~\citep{kubler2020adaptive,gu2021adaptive,zhu2024optimizing}, operator sampling subsamples the Hamiltonian~\citep{arrasmith2020operator}, and parameter freezing reduces $P$~\citep{peng2026titan}. None of these asks the finer question: \emph{when parameter $j$ is evaluated, which QWC groups are actually needed for its gradient?}

The answer is strongly parameter dependent: every parameter sees the same Hamiltonian groups, yet the shifted expectation differences depend on gate location, ansatz light cone, and current circuit state, so the contribution matrix over parameter--group pairs is concentrated but not captured by any global coefficient ranking. Exploiting this structure faces a circularity: computing the matrix exactly at each iteration would require precisely the Full-QWC measurements we seek to avoid.

\begin{figure}[H]
\centering
\includegraphics[width=\linewidth]{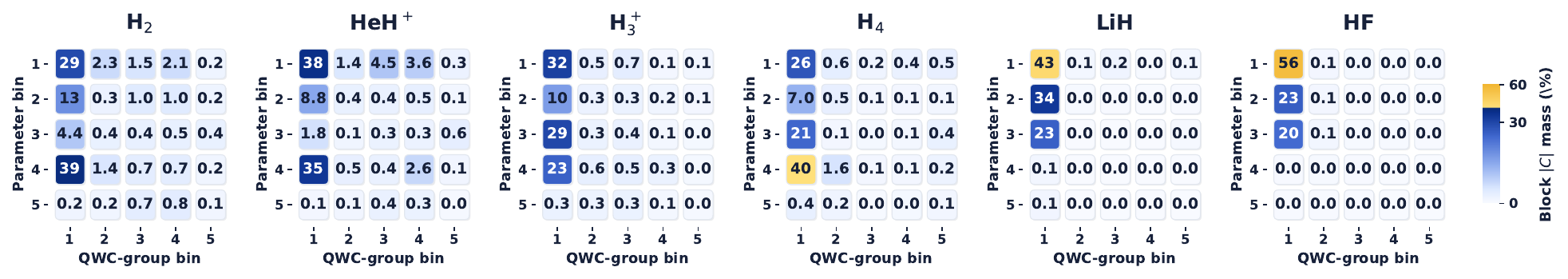}
\caption{\textbf{Parameter-specific QWC sparsity.} Cells show contribution-mass percentages in a $5\times5$ visualization of the exact parameter--group matrix; training and evaluation use the unaggregated matrix. The corresponding result under an IBM \texttt{FakeGuadalupeV2} calibration-derived noise model~\citep{ibm2026fakeprovider,ibm2026noisemodels} is reported in Appendix~\ref{sec:app_noise}, Figure~\ref{fig:vista_qwc_sparsity_ibm_noise}.}
\label{fig:vista_qwc_sparsity}
\vspace{-0.5em}
\end{figure}

Our method resolves the circularity by learning to predict the latent matrix. Separate Transformer encoders contextualize parameter and QWC-group tokens; a pairwise scorer estimates a normalized relevance distribution for every parameter; a row-wise top-mass rule converts the prediction into a measurement mask. $\tau$ thereby becomes a direct post-training accuracy-cost control; exact contributions are used only offline, never at deployment. Our contributions are threefold:
\begin{itemize}
    \item We uncover parameter-specific sparsity in the QWC contribution
    matrix of variational quantum gradients in Figure~\ref{fig:vista_qwc_sparsity}: most absolute contribution
    mass concentrates in a subset of measurement groups whose relevance
    varies across parameters and optimization states.

    \item We develop a size-flexible learned controller that directly
    predicts parameter--group relevance from physical and current-state
    features, using exact contributions only as offline supervision.
    A single threshold converts these predictions into measurement
    subsets, enabling adjustable accuracy--cost trade-offs without
    retraining.

    \item We investigate whether LLMs can perform this scientific
    selection task without fine-tuning. Frontier models achieve
    millihartree-scale gaps with retrieved training examples, while
    local-model audits reveal uneven utility of the supplied physical
    information. These findings demonstrate the potential of LLMs for
    scientific decision making and motivate domain-specific guidance
    as a promising route toward more reliable AI-assisted research.
\end{itemize}

Dominant groups vary by parameter, yet each row concentrates on a small fraction of the measurement set; this is the sparse structure \ourmethod{} exploits, and it persists under the calibrated-noise audit in Appendix~\ref{sec:app_noise}, indicating that it is not merely an artifact of idealized, noiseless execution.

\section{Problem Formulation and Method}
A molecular electronic-structure Hamiltonian mapped to qubits can be written as
\begin{equation}
    H = h_0 I + \sum_{\ell=1}^{L} h_\ell P_\ell,
    \qquad
    E(\boldsymbol\theta)
    = \langle \psi(\boldsymbol\theta)|H|\psi(\boldsymbol\theta)\rangle,
\end{equation}
where $P_\ell$ is a Pauli string and $h_\ell$ is fixed. Qubit-wise commuting strings share a tensor-product basis and form groups $\{\mathcal G_k\}_{k=1}^{G}$~\citep{verteletskyi2020measurement,crawford2021efficient}.

For RY/RZ gates, the parameter-shift rule gives~\citep{schuld2019evaluating,wierichs2022general}
\begin{equation}
    g_j(\boldsymbol\theta)
    = \frac{\partial E}{\partial\theta_j}
    = \frac{1}{2}\left[
        E\!\left(\boldsymbol\theta+\frac{\pi}{2}\boldsymbol e_j\right)
        -E\!\left(\boldsymbol\theta-\frac{\pi}{2}\boldsymbol e_j\right)
      \right].
\end{equation}

\begin{figure}[!t]
\centering
\includegraphics[width=\linewidth]{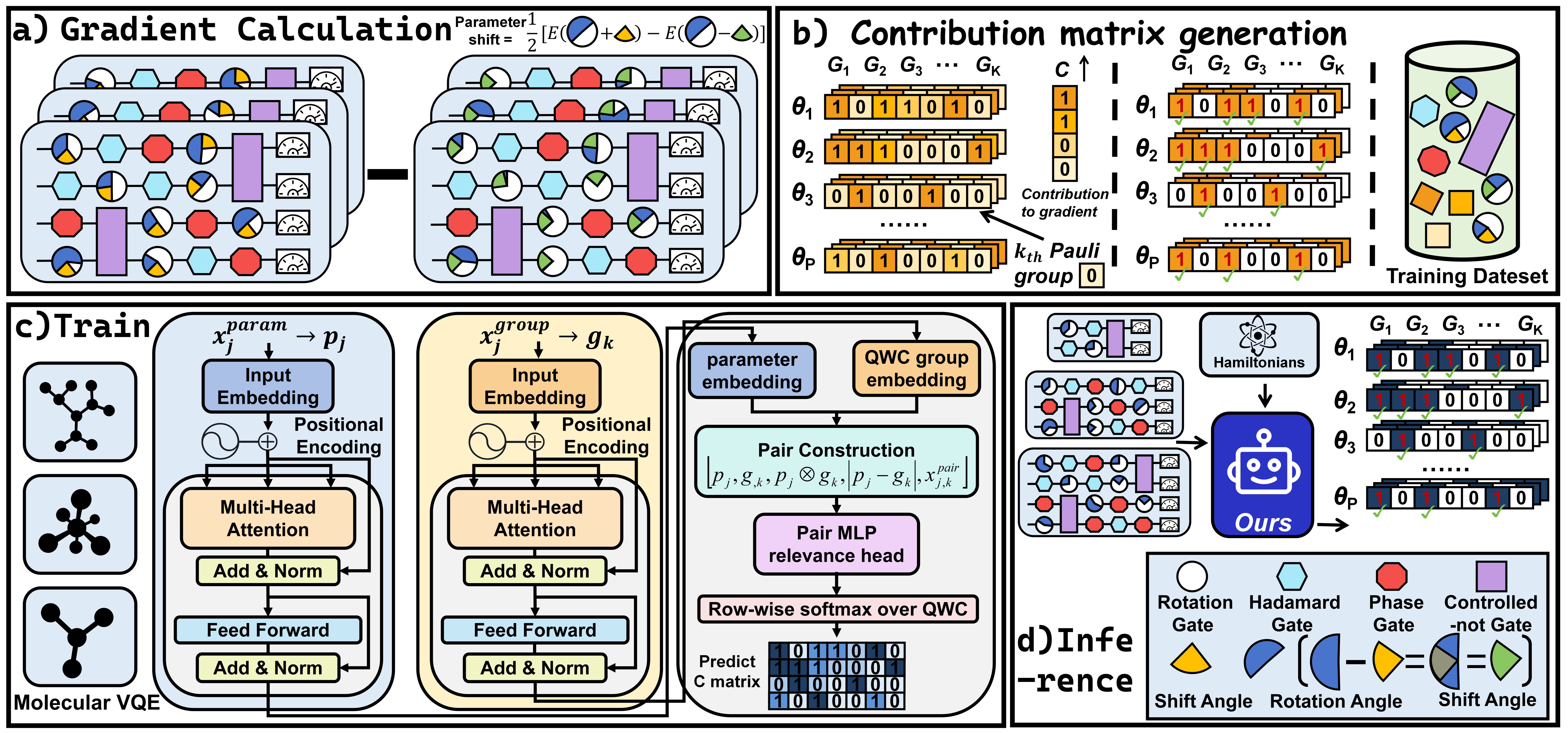}
\caption{\textbf{Method overview in four stages.} (a) A Hamiltonian is partitioned into QWC groups and evaluated with a parameterized ansatz. (b) Full-QWC parameter shift measures every group for every parameter at cost $2PGS$. (c) The learned controller predicts the parameter--group relevance matrix and retains the smallest per-parameter subset covering top-$\tau$ mass. (d) Closed-loop VQE measures only those groups, forms sparse gradients, and updates all parameters at reduced cost.}
\label{fig:vista_method_overview}
\vspace{-1.5em}
\end{figure}

Expanding each shifted energy over QWC groups exposes the signed contribution of group $k$ to $j$:
\begin{equation}
\label{eq:contribution}
    C_{j,k}(\boldsymbol\theta)
    = \frac{1}{2}\sum_{\ell\in\mathcal G_k} h_\ell
      \left[
        \langle P_\ell\rangle_{\boldsymbol\theta+\frac{\pi}{2}\boldsymbol e_j}
        -
        \langle P_\ell\rangle_{\boldsymbol\theta-\frac{\pi}{2}\boldsymbol e_j}
      \right],
    \qquad
    g_j = \sum_{k=1}^{G} C_{j,k}.
\end{equation}
Figure~\ref{fig:vista_method_overview} summarizes our framework containing 4 stages. The coefficients and QWC partition stay fixed, but $C_{j,k}$ changes with the ansatz state. At an offline training snapshot, we evaluate Eq.~\ref{eq:contribution} and normalize its absolute contribution,
\begin{equation}
\label{eq:relevance-target}
    w_{j,k}(\boldsymbol\theta)
    = \frac{|C_{j,k}(\boldsymbol\theta)|}
           {\sum_{r=1}^{G}|C_{j,r}(\boldsymbol\theta)|}.
\end{equation}
Exact $C$ supplies offline labels and simulator diagnostics; selectors never receive current oracle contributions. Separate Transformer encoders~\citep{vaswani2017attention} contextualize parameter and group tokens (position, angle/time, coefficient, Pauli support, light cones). A pair scorer receives
\begin{equation}
    z_{j,k} = [p_j, q_k, p_j\odot q_k, |p_j-q_k|, x^{\mathrm{pair}}_{j,k}],
\end{equation}
and predicts row-normalized relevance $\widehat w_{j,k}$. Training combines distribution, top-set, ranking, budget, and entropy losses:
\begin{equation}
\label{eq:training-loss}
    \mathcal L = \mathcal L_{\mathrm{CE}}
      + \lambda_{\mathrm m}\mathcal L_{\mathrm{mask}}
      + \lambda_{\mathrm r}\mathcal L_{\mathrm{rank}}
      + \lambda_{\mathrm b}\mathcal L_{\mathrm{budget}}
      + \lambda_{\mathrm e}\mathcal L_{\mathrm{entropy}}.
\end{equation}
For each parameter, \ourmethod{} chooses the smallest set $S_j(\tau)$ satisfying $\sum_{k\in S_j(\tau)} \widehat w_{j,k} \geq \tau$.

Only groups in $S_j(\tau)$ are measured on the two shifted circuits; with $S$ shots per setting,
\begin{equation}
    B_{\mathrm{Full}}=2SPG,
    \qquad
    B_{\mathrm{Ours}}=2S\sum_{j=1}^{P}|S_j(\tau)|,
    \qquad
    \mathrm{gain}=1-\frac{B_{\mathrm{Ours}}}{B_{\mathrm{Full}}}.
\end{equation}
For fixed predictions, increasing $\tau$ nests the selected masks and directly controls measurement coverage. In closed-loop optimization, each mask also shapes subsequent circuit states. The omitted signed contributions determine the truncation bias; Appendix~\ref{sec:app_reproducibility} gives its bound, the complete losses, and implementation details. Appendix~\ref{sec:app_checkpoint_provenance} identifies the checkpoint used in each sweep. The formulation covers Pauli-observable VQAs; related work is in Appendix~\ref{sec:app_related_work}.

\section{Experimental Setup}
PySCF/OpenFermion STO-3G Hamiltonians~\citep{sun2018pyscf,mcclean2020openfermion,hehre1969self} for H2, HeH+, H3+, H4, and LiH map under Jordan--Wigner~\citep{seeley2012bravyi} to 4--12 qubits and 5--179 QWC groups; the main ansatz is layered RY/RZ with linear entanglers~\citep{kandala2017hardware}, and structural-OOD tests use named ansatze and spin Hamiltonians.

Training uses depths 1-3 and 8 seeds. Four held-out seeds run 28 Adam steps with exact masked expectations and nominal cost at 1024 shots/setting. We report $\dE=E_{\rm method}-E_{\rm Full}$ (Ha) and shot gain; \emph{no worse} means $\dE\leq1.6$ mHa. Metric and estimator definitions are in Appendix~\ref{sec:app_metrics}; 400-step shot-noise checks are in Appendix~\ref{sec:app_v82_converged_fci}.

\section{Main Results}
\begin{table*}[t]
\centering
\vspace{-0.5em}
\caption{\textbf{Cost-matched online VQE at $\tau=0.995$.} Exact-expectation results; shot gains are nominal. Entries are mean $\pm$ standard deviation over four seeds; $\dE$ is in Ha. Random and coefficient controls use comparable budgets; TITAN$^{\dagger}$ is cell-wise shot matched. \textcolor{green!50!black}{Green}/\textcolor{red}{red} shading indicates the one-sided 1.6 mHa criterion. \textbf{Bold} marks the lowest $|\dE|$ or largest gain; \underline{underlining} marks the second-largest gain. Metrics, full controls, and noise extensions are in Appendices~\ref{sec:app_metrics}, \ref{sec:app_id_results}, and~\ref{sec:app_noise}.}
\vspace{0.25em}
\setlength{\tabcolsep}{0.55mm}
\renewcommand{\arraystretch}{0.88}
\scriptsize
\begingroup
\definecolor{vistatextgreen}{HTML}{278A57}
\definecolor{vistatextred}{HTML}{D84D68}
\newcommand{\tgood}[2]{\textcolor{vistatextgreen}{#2}}
\newcommand{\tbad}[2]{\textcolor{vistatextred}{#2}}
\newcommand{\epos}[1]{+#1}
\newcommand{\eneg}[1]{\textminus#1}
\newcommand{\estat}[2]{#1\,\(\pm\)\,#2}
\newcommand{\pstat}[2]{#1\,\(\pm\)\,#2\%}
\newcommand{\best}[1]{\textbf{#1}}
\newcommand{\runnerup}[1]{\underline{#1}}
\newcommand{\methodrule}{\arrayrulecolor{black!18}\specialrule{0.18pt}{0.13mm}{0.13mm}}
\begin{tabular*}{\textwidth}{@{\extracolsep{\fill}}l|c|c|c|c|c|c|c|@{}}
\arrayrulecolor{black}\toprule
Setting & Metric & Method & H2 (4q) & HeH+ (4q) & H3+ (6q) & H4 chain (8q) & LiH (12q) \\
\midrule\arrayrulecolor{black!18}
\gcell{Depth 1} & \(\dE\) vs Full \(\downarrow\) & \ours{} & \tgood{15}{\best{\estat{\epos{0.0011}}{0.0008}}} & \tgood{20}{\best{\estat{\epos{0.0015}}{0.0012}}} & \tgood{10}{\best{\estat{\epos{0.0005}}{0.0006}}} & \tgood{11}{\best{\estat{\eneg{0.0006}}{0.0024}}} & \tbad{14}{\estat{\epos{0.0023}}{0.0016}} \\
\methodrule
\gcell{Depth 1} & \(\dE\) vs Full \(\downarrow\) & Random & \tbad{45}{\estat{\epos{0.2359}}{0.0507}} & \tbad{65}{\estat{\epos{1.2220}}{0.2247}} & \tbad{58}{\estat{\epos{0.8916}}{0.1887}} & \tbad{68}{\estat{\epos{1.4117}}{0.2745}} & \tbad{78}{\estat{\epos{3.4982}}{0.0793}} \\
\methodrule
\gcell{Depth 1} & \(\dE\) vs Full \(\downarrow\) & Global coeff. & \tbad{17}{\estat{\epos{0.0038}}{0.0053}} & \tbad{12}{\estat{\epos{0.0019}}{0.0041}} & \tbad{25}{\estat{\epos{0.0122}}{0.0093}} & \tbad{22}{\estat{\epos{0.0089}}{0.0074}} & \tbad{10}{\best{\estat{\epos{0.0016}}{0.0026}}} \\
\methodrule
\gcell{Depth 1} & \(\dE\) vs Full \(\downarrow\) & TITAN$^{\dagger}$ & \tbad{25}{\estat{\epos{0.0128}}{0.0170}} & \tbad{40}{\estat{\epos{0.1299}}{0.0870}} & \tbad{35}{\estat{\epos{0.0772}}{0.0483}} & \tgood{58}{\estat{\eneg{0.0642}}{0.0758}} & \tgood{59}{\estat{\eneg{0.0648}}{0.1287}} \\
\methodrule
\gcell{Depth 1} & \(\dE\) vs Full \(\downarrow\) & TITAN$^{\dagger}$+\ours{} & \tbad{25}{\estat{\epos{0.0127}}{0.0160}} & \tbad{40}{\estat{\epos{0.1297}}{0.0845}} & \tbad{35}{\estat{\epos{0.0776}}{0.0483}} & \tgood{59}{\estat{\eneg{0.0649}}{0.0752}} & \tgood{59}{\estat{\eneg{0.0646}}{0.1286}} \\
\cmidrule(lr){2-8}
\bcell{Depth 1} & Shot gain \(\uparrow\) & \ours{} & \pstat{40.4}{6.3} & \pstat{58.9}{3.5} & \pstat{63.0}{2.8} & \pstat{74.7}{1.1} & \pstat{86.9}{0.5} \\
\methodrule
\bcell{Depth 1} & Shot gain \(\uparrow\) & Random & \pstat{42.0}{4.6} & \runnerup{\pstat{60.7}{1.5}} & \runnerup{\pstat{66.5}{2.2}} & \runnerup{\pstat{79.2}{1.3}} & \runnerup{\pstat{90.2}{0.3}} \\
\methodrule
\bcell{Depth 1} & Shot gain \(\uparrow\) & Global coeff. & \pstat{41.7}{6.5} & \pstat{60.0}{3.3} & \pstat{62.9}{2.9} & \pstat{75.0}{0.3} & \pstat{87.0}{0.6} \\
\methodrule
\bcell{Depth 1} & Shot gain \(\uparrow\) & TITAN$^{\dagger}$ & \runnerup{\pstat{43.8}{6.2}} & \pstat{59.4}{6.3} & \pstat{62.7}{6.4} & \pstat{70.3}{5.7} & \pstat{81.8}{5.2} \\
\methodrule
\bcell{Depth 1} & Shot gain \(\uparrow\) & TITAN$^{\dagger}$+\ours{} & \best{\pstat{66.8}{3.7}} & \best{\pstat{91.7}{2.0}} & \best{\pstat{95.0}{2.7}} & \best{\pstat{97.9}{0.9}} & \best{\pstat{99.9}{0.1}} \\
\arrayrulecolor{black}\midrule\arrayrulecolor{black!18}
\gcell{Depth 2} & \(\dE\) vs Full \(\downarrow\) & \ours{} & \tgood{10}{\best{\estat{\epos{0.0003}}{0.0007}}} & \tgood{15}{\estat{\eneg{0.0011}}{0.0026}} & \tgood{20}{\best{\estat{\eneg{0.0015}}{0.0041}}} & \tbad{27}{\best{\estat{\epos{0.0158}}{0.0443}}} & \tgood{28}{\best{\estat{\eneg{0.0049}}{0.0097}}} \\
\methodrule
\gcell{Depth 2} & \(\dE\) vs Full \(\downarrow\) & Random & \tbad{42}{\estat{\epos{0.1656}}{0.1105}} & \tbad{52}{\estat{\epos{0.4704}}{0.1737}} & \tbad{55}{\estat{\epos{0.6527}}{0.0767}} & \tbad{59}{\estat{\epos{0.9343}}{0.1414}} & \tbad{80}{\estat{\epos{4.0043}}{0.3648}} \\
\methodrule
\gcell{Depth 2} & \(\dE\) vs Full \(\downarrow\) & Global coeff. & \tgood{20}{\estat{\epos{0.0015}}{0.0091}} & \tgood{10}{\best{\estat{\eneg{0.0004}}{0.0088}}} & \tbad{29}{\estat{\epos{0.0233}}{0.0429}} & \tbad{31}{\estat{\epos{0.0379}}{0.0540}} & \tgood{28}{\estat{\eneg{0.0049}}{0.0123}} \\
\methodrule
\gcell{Depth 2} & \(\dE\) vs Full \(\downarrow\) & TITAN$^{\dagger}$ & \tbad{21}{\estat{\epos{0.0075}}{0.0093}} & \tbad{24}{\estat{\epos{0.0116}}{0.0467}} & \tbad{27}{\estat{\epos{0.0153}}{0.0618}} & \tbad{33}{\estat{\epos{0.0472}}{0.0558}} & \tbad{47}{\estat{\epos{0.2903}}{0.1308}} \\
\methodrule
\gcell{Depth 2} & \(\dE\) vs Full \(\downarrow\) & TITAN$^{\dagger}$+\ours{} & \tbad{21}{\estat{\epos{0.0072}}{0.0082}} & \tbad{25}{\estat{\epos{0.0129}}{0.0483}} & \tbad{26}{\estat{\epos{0.0140}}{0.0616}} & \tbad{33}{\estat{\epos{0.0493}}{0.0560}} & \tbad{47}{\estat{\epos{0.2920}}{0.1356}} \\
\cmidrule(lr){2-8}
\bcell{Depth 2} & Shot gain \(\uparrow\) & \ours{} & \pstat{25.6}{2.9} & \pstat{31.6}{2.6} & \pstat{44.9}{1.3} & \pstat{51.8}{0.5} & \pstat{82.6}{0.4} \\
\methodrule
\bcell{Depth 2} & Shot gain \(\uparrow\) & Random & \runnerup{\pstat{26.3}{2.5}} & \pstat{33.2}{1.5} & \runnerup{\pstat{47.7}{2.1}} & \runnerup{\pstat{53.7}{0.4}} & \runnerup{\pstat{83.4}{1.3}} \\
\methodrule
\bcell{Depth 2} & Shot gain \(\uparrow\) & Global coeff. & \pstat{25.7}{2.8} & \pstat{31.9}{2.4} & \pstat{44.7}{0.8} & \pstat{51.7}{0.5} & \pstat{82.5}{0.3} \\
\methodrule
\bcell{Depth 2} & Shot gain \(\uparrow\) & TITAN$^{\dagger}$ & \pstat{23.8}{4.7} & \runnerup{\pstat{39.0}{14.5}} & \pstat{41.9}{9.4} & \pstat{46.8}{9.9} & \pstat{82.7}{5.3} \\
\methodrule
\bcell{Depth 2} & Shot gain \(\uparrow\) & TITAN$^{\dagger}$+\ours{} & \best{\pstat{40.2}{5.9}} & \best{\pstat{64.0}{10.6}} & \best{\pstat{74.5}{6.6}} & \best{\pstat{79.0}{6.3}} & \best{\pstat{99.3}{0.3}} \\
\arrayrulecolor{black}\bottomrule
\end{tabular*}
\endgroup
\label{tab:v39_tau0995_main}
\end{table*}

At $\tau=0.995$, \ourmethod{} saves 49.6\% of parameter-shift shots at a 5.67 mHa seed-level mean absolute gap, attains the smallest $|\dE|$ in eight of ten molecule--depth cells of Table~\ref{tab:v39_tau0995_main}, and satisfies the one-sided no-worse criterion in 8/10 cells, more than any alternative (Appendix~\ref{sec:app_id_results}, Table~\ref{tab:v47_table1_full_appendix}). Tightening to $\tau=0.997$ reduces the mean gap to 3.52 mHa while still saving 45.5\%; the complete $\tau$ frontier, including its behavior under depth shift, is in Appendix~\ref{sec:app_threshold_depth} (Figure~\ref{fig:v42_depth_ood_pareto_bar}). 

Each control isolates one candidate explanation of the savings. Count-matched Random inherits \ourmethod{}'s per-parameter budgets and attains slightly \emph{larger} shot gains in every cell, yet drifts by 0.17-4.00 Ha: the savings come from the budget size, while the accuracy comes from the learned ranking. Static coefficient ranking (natural training-free ordering) performs no worse in only 3/10 cells. Shot-matched TITAN$^{\dagger}$ freezing~\citep{peng2026titan} drifts by tens to hundreds of mHa, so the gain is not explained by evaluating fewer parameters; conversely, TITAN$^{\dagger}$+\ours{} achieves the largest gain, showing that relevance selection composes with, but does not repair, aggressive freezing.

The endpoint and trajectory views resolve complementary aspects of the trade-off. LiH depth 1 lies 0.7 mHa above the no-worse threshold, while H4 depth 2 ($+15.8\pm44.3$ mHa) exhibits strong seed dependence. Exact-expectation trajectories (Figure~\ref{fig:vista_shots_to_energy}) reach Full-QWC parity at 74.4\% (H2) and 55.1\% (H3+) of its nominal budget. Separate 400-step runs with simulated shot noise (Appendix~\ref{sec:app_v82_converged_fci}) keep every molecule's mean final gap within 0.18 mHa of Full-QWC while saving 24.7--92.1\%. 

The learned ranking is also not tied to one architecture: swapping the Transformer for message-passing GNN~\citep{gilmer2017neural}, MLP, or ResNet-18~\citep{he2016deep} predictors (Appendix~\ref{sec:app_predictor_tables}) preserves the qualitative picture, while the Transformer has the lowest tail error on the measurement-intensive molecules and the strongest molecule-OOD transfer. Architecture details are in Appendix~\ref{sec:app_baseline_implementations}. The consistency across model families supports parameter-group
relevance as a useful supervised learning target.
The Transformer's lower tail error further motivates its use as
our default predictor, with robustness across molecular systems
guiding the architectural choice.

\newcommand{\appVistaShotsToEnergy}{%
\begin{figure}[!t]
\centering
\includegraphics[width=\linewidth]{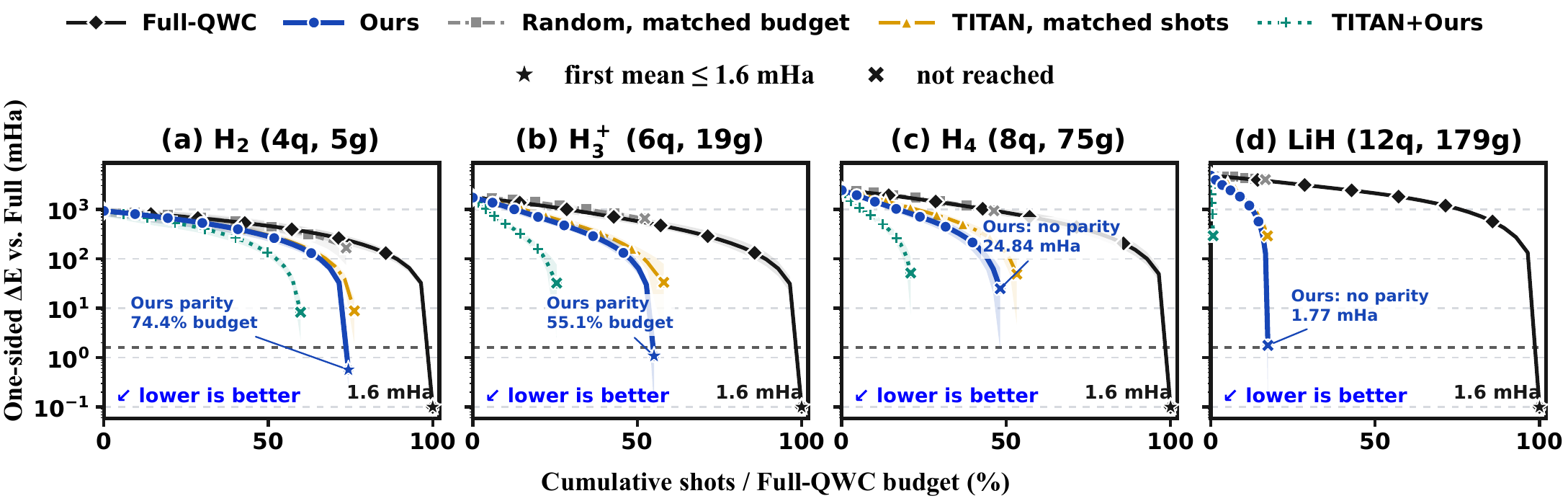}
\caption{\textbf{Shots-to-chemical-parity dynamics at depth 2.} Curves report the mean across four paired seeds for H2, H3+, an H4 chain, and LiH; shaded regions are 95\% bootstrap confidence intervals. The horizontal axis is cumulative parameter-shift shots normalized by each seed's Full-QWC budget. The vertical axis is the one-sided energy excess above the paired Full-QWC final energy, so lower energies incur no penalty; the dashed line marks 1.6 mHa chemical parity. Stars mark the first optimizer checkpoint where the four-seed mean reaches parity, while crosses mark methods that never reach it. Our method first reaches parity at 74.4\% of the Full-QWC budget on H2 and 55.1\% on H3+; it does not reach parity on H4 and finishes narrowly above it on LiH (1.77 mHa at 17.4\% budget). Unlike Table~\ref{tab:v39_tau0995_main}, which compares final endpoints, this figure isolates convergence efficiency and shots-to-target.}
\label{fig:vista_shots_to_energy}
\vspace{-0.5em}
\end{figure}
}

\clearpage
\section{Allocation and Structural Generalization}
\begin{figure}[!t]
\centering
\includegraphics[width=\linewidth]{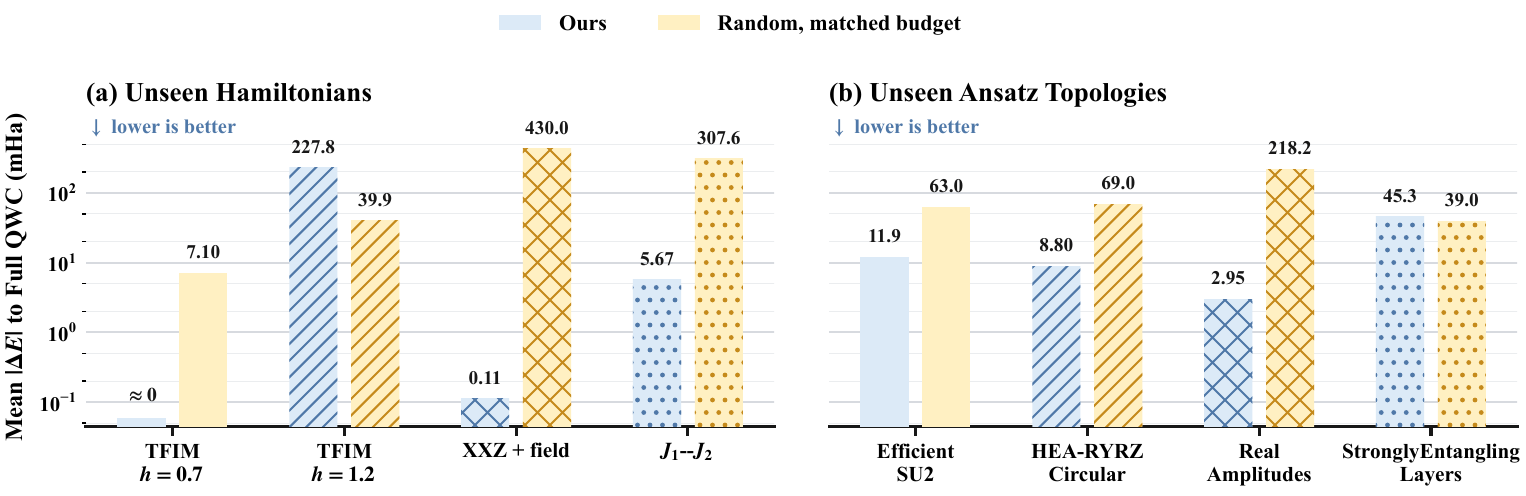}
\caption{\textbf{Matched-budget structural OOD.} Mean absolute gaps for unseen Hamiltonian families (a) and ansatz topologies (b); Random receives the same parameter-specific group counts as \ourmethod{}. Exact values and gains are in Appendix~\ref{sec:app_structural_ood_summary}.}
\label{fig:ood_transfer_bars}
\vspace{-0.5em}
\end{figure}

\begin{table}[H]
\centering
\vspace{-0.6em}
\caption{\textbf{Shot-allocation Pareto benchmark.} Five molecules, depths 1--2, and four seeds under common Adam with simulated shot noise. $\dE$ is in Ha; gain is relative to uniform Full-QWC at budget $b$; no-worse means $\dE\leq1.6$ mHa. \textbf{Bold}/\underline{underline} mark best/second-best non-Full values. Adapted-policy settings and detailed results are in Appendices~\ref{sec:app_allocation_protocol} and~\ref{sec:app_shot_allocation}.}
\vspace{0.2em}
\setlength{\tabcolsep}{0.75mm}
\renewcommand{\arraystretch}{0.82}
\scriptsize
\resizebox{\linewidth}{!}{
\begin{tabular}{l|ccc|ccc|ccc}
\toprule
& \multicolumn{3}{c|}{$b=0.25$} & \multicolumn{3}{c|}{$b=0.50$} & \multicolumn{3}{c}{$b=1.00$} \\
Method & Mean $|\dE|\downarrow$ & Shot gain $\uparrow$ & No-worse $\uparrow$
       & Mean $|\dE|\downarrow$ & Shot gain $\uparrow$ & No-worse $\uparrow$
       & Mean $|\dE|\downarrow$ & Shot gain $\uparrow$ & No-worse $\uparrow$ \\
\midrule
\grow Full-QWC uniform
  & \gray{0.00000} & \gray{0.0\%} & \gray{100.0\%}
  & \gray{0.00000} & \gray{0.0\%} & \gray{100.0\%}
  & \gray{0.00000} & \gray{0.0\%} & \gray{100.0\%} \\
\midrule
iCANS~\citep{kubler2020adaptive}
  & 0.17283 & $-35.0\%$ & 10.0\%
  & 0.11977 & $-49.8\%$ & 37.5\%
  & 0.09657 & $-76.6\%$ & 17.5\% \\
gCANS~\citep{gu2021adaptive}
  & 0.20449 & \underline{92.9\%} & 7.5\%
  & 0.14059 & \underline{92.2\%} & 25.0\%
  & 0.10789 & \underline{91.8\%} & 27.5\% \\
Rosalin~\citep{arrasmith2020operator}
  & \underline{0.06008} & $-78.6\%$ & 45.0\%
  & 0.09137 & $-90.9\%$ & \underline{72.5\%}
  & 0.06912 & $-97.8\%$ & 67.5\% \\
Variance-aware~\citep{zhu2024optimizing}
  & 0.06694 & 0.0\% & \underline{65.0\%}
  & \underline{0.06521} & 0.0\% & \textbf{80.0\%}
  & \underline{0.06278} & 0.0\% & \underline{72.5\%} \\
\midrule
\brow \ours{} + uniform
  & \textbf{0.05472} & 57.5\% & \textbf{72.5\%}
  & \textbf{0.06064} & 57.1\% & 65.0\%
  & 0.06376 & 57.1\% & 65.0\% \\
\brow \ours{} + variance-aware
  & 0.07211 & 57.1\% & \underline{65.0\%}
  & 0.06876 & 57.1\% & \textbf{80.0\%}
  & \textbf{0.06031} & 56.8\% & \textbf{82.5\%} \\
\brow \ours{} + gCANS
  & 0.12300 & \textbf{97.3\%} & 17.5\%
  & 0.16134 & \textbf{97.2\%} & 27.5\%
  & 0.07148 & \textbf{97.2\%} & 30.0\% \\
\bottomrule
\end{tabular}}
\label{tab:v71_shot_allocation_pareto}
\vspace{-0.5em}
\end{table}

Separate transfer studies test four shift axes: Figure~\ref{fig:ood_transfer_bars} shows \ourmethod{} beats count-matched Random on three of four unseen Hamiltonian families and on 44/45 transferable named-ansatz cells. TFIM at $h{=}1.2$ and StronglyEntanglingLayers identify challenging structural shifts (Appendix~\ref{sec:app_ansatz_ood}). Table~\ref{tab:v71_shot_allocation_pareto} demonstrates composition with allocation policies: \ourmethod{}+variance-aware saves 56.8\% and raises the no-worse rate from 72.5\% to 82.5\%. The common Adam protocol controls optimizer dynamics while comparing allocation and selection; exact policy settings are in Appendix~\ref{sec:app_allocation_protocol}. Complete results are in Appendices~\ref{sec:app_structural_ood_summary}, \ref{sec:app_depth_ood}--\ref{sec:app_ansatz_ood}, \ref{sec:app_shot_allocation}, and~\ref{sec:app_noise}; predictor diagnostics are in Appendix~\ref{sec:app_predictor_tables}.

\newcommand{\appMLPredictorComparison}{%
\begin{figure}[!t]
\centering
\includegraphics[width=\linewidth]{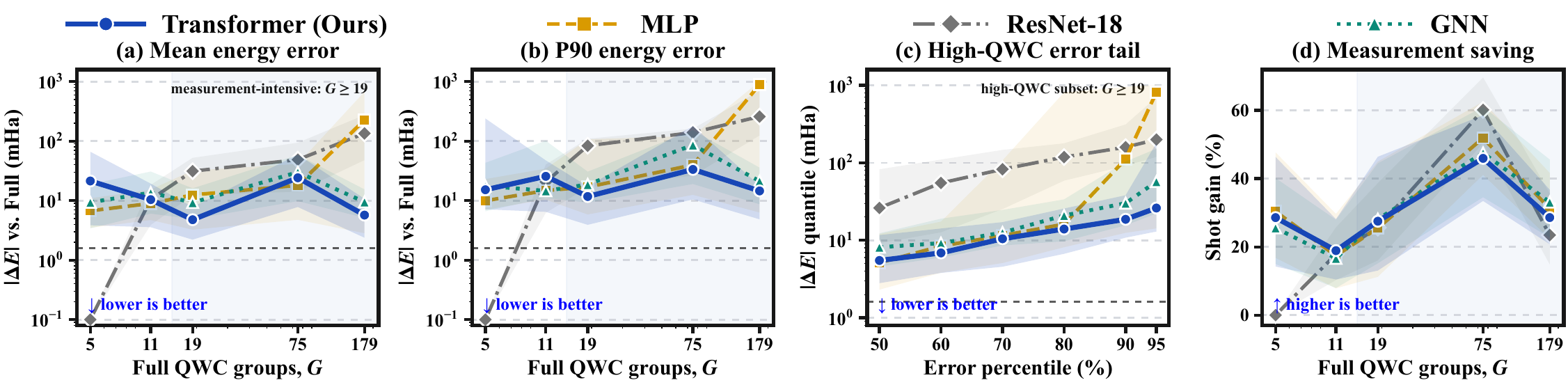}
\caption{\textbf{Predictor robustness as QWC measurement complexity grows.} We use leave-one-molecule-out transfer at $\tau=0.997$ and order the five held-out molecular Hamiltonians by their Full-QWC group counts $G\in\{5,11,19,75,179\}$. Panels (a,b) report the mean and 90th-percentile absolute final-energy gap to Full QWC; panel (c) shows the error-quantile curve over the three measurement-intensive targets with $G\geq19$; and panel (d) reports mean shot gain. Lines are computed from four depths and four paired seeds, and bands are hierarchical-bootstrap 95\% confidence intervals. The Transformer has the lowest P90 error at all three measurement-intensive group counts while retaining a measurement budget comparable to the other learned predictors. The dashed line marks 1.6 mHa. All features, labels, and online-VQE settings are fixed while only the predictor architecture changes. A cell-level mask diagnostic for the same predictors appears in Appendix~\ref{sec:app_predictor_tables}, Figure~\ref{fig:ml_qwc_mask_comparison}; underlying molecule-OOD values and matched held-out-seed/Hamiltonian controls appear in Tables~\ref{tab:v62_lomo_model_zoo}, \ref{tab:v60_model_zoo_table1}, and~\ref{tab:v64_hamiltonian_model_zoo}.}
\label{fig:ml_predictor_comparison}
\vspace{-0.5em}
\end{figure}
}

\section{LLM Controls and Mechanism Audit}
\newcommand{\appVSeventySixPromptProtocol}{%
\begin{figure}[H]
\begin{minipage}[t]{0.44\linewidth}
\vspace{0pt}
\textbf{Controlled rank-only intervention.}
We retain \ourmethod{} only as the cardinality controller: at each query it supplies $k_j$, while a frozen local LLM must infer \emph{which} QWC groups to select from structured parameter, group, and light-cone features.  The oracle contribution matrix is never included in the prompt.

\textbf{No model adaptation.}
Qwen3-8B, Llama-3.1-8B, DeepSeek-R1-8B, and GPT-OSS-20B are evaluated zero-shot with deterministic decoding and no fine-tuning.  Schema validation and coefficient-based repair prevent malformed output from changing the requested cardinality.

\textbf{Sparse refresh.}
The LLM is queried at $t\in\{0,7,14,21\}$ and each returned mask is reused for seven online-VQE steps.  The complete prompt and settings are in Appendix~\ref{sec:app_llm_prompt_protocol}.
\end{minipage}
\hfill
\begin{minipage}[t]{0.52\linewidth}
\vspace{0pt}
\centering
\includegraphics[width=0.88\linewidth]{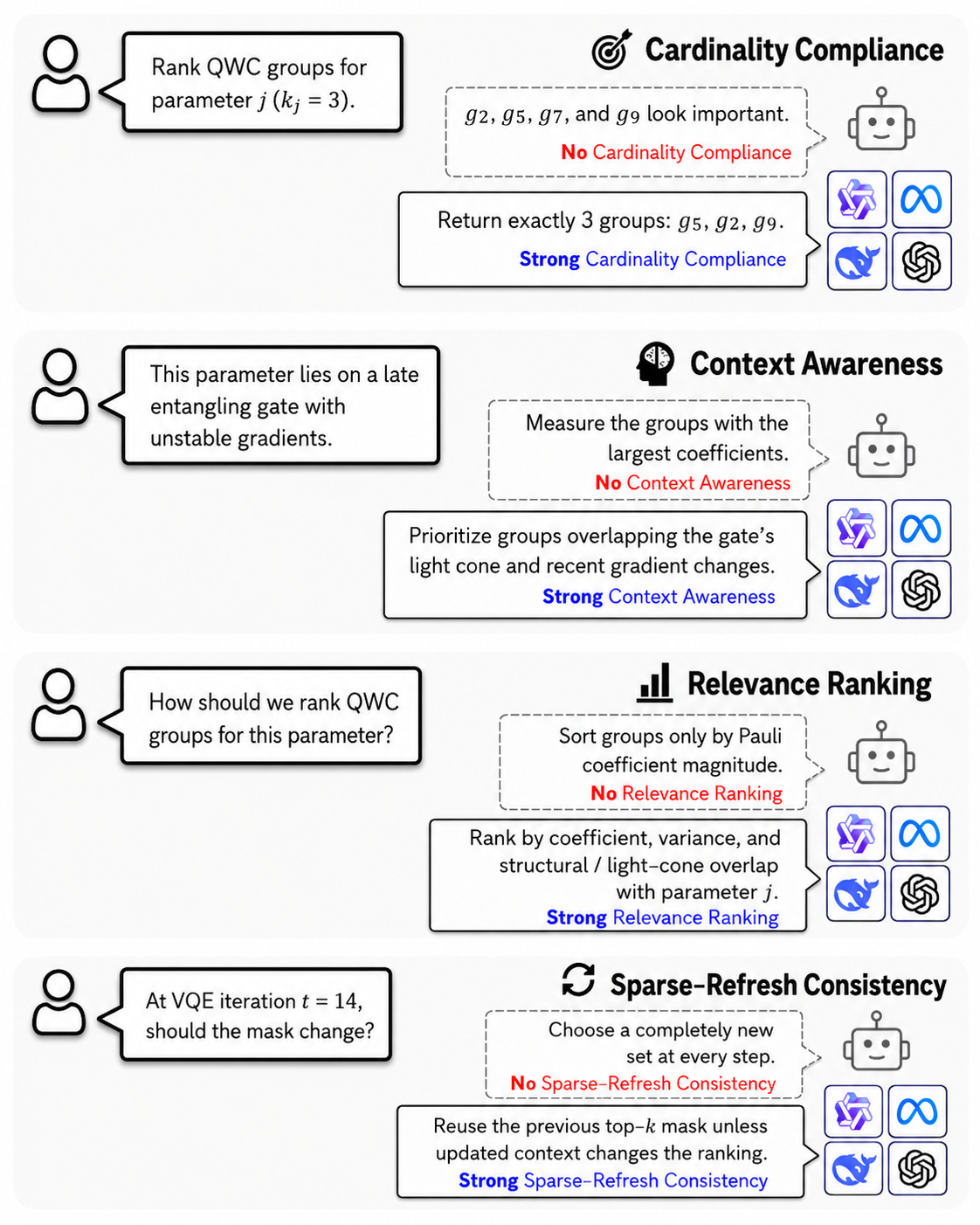}
\caption{\textbf{Controlled diagnostics for frozen local LLMs.} The schematic contrasts failure modes (red) and desired behavior (blue) for cardinality compliance, context awareness, parameter-specific QWC relevance ranking, and sparse-refresh consistency.}
\label{fig:v76b_prompt_protocol}
\end{minipage}
\vspace{-0.5em}
\end{figure}
}

\begin{table*}[t]
\centering
\vspace{-0.5em}
\caption{\textbf{Frozen local-LLM QWC ranking.} Four zero-shot LLMs receive predicted counts and refresh every $K{=}7$ steps. Entries are mean $\pm$ s.d. over four seeds ($\dE$: Ha; gain: \%). \textcolor{green!70!black}{Green}/\textcolor{red}{red} shading indicates the 1.6 mHa no-worse split; \ding{51}/\ding{55} compares $|\dE|$ with Random. \textbf{Bold}/\underline{underline} mark the smallest/second-smallest $|\dE|$ per cell. Protocols are in Appendices~\ref{sec:app_llm_prompt_protocol} and~\ref{sec:app_trace_provenance}.}
\vspace{0.25em}
\setlength{\tabcolsep}{0.4mm}
\renewcommand{\arraystretch}{0.90}
\scriptsize
\begingroup
\definecolor{vistaheatgreen}{HTML}{A8D8A8}
\definecolor{vistaheatred}{HTML}{F3A6B7}
\newcommand{\llmlogo}[1]{\raisebox{-0.28ex}{\includegraphics[height=1.15em]{llm_logos/#1.png}}}
\newcommand{\llmmethod}[2]{\llmlogo{#1}\,#2}
\newcommand{\vpass}[2]{\cellcolor{vistaheatgreen!#1!white}\textcolor{black}{#2}}
\newcommand{\vfail}[2]{\cellcolor{vistaheatred!#1!white}\textcolor{black}{#2}}
\newcommand{\epos}[1]{+#1}
\newcommand{\eneg}[1]{\textminus#1}
\newcommand{\estat}[2]{#1\,\(\pm\)\,#2}
\newcommand{\pstat}[2]{#1\,\(\pm\)\,#2\%}
\newcommand{\best}[1]{\textbf{#1}}
\newcommand{\runnerup}[1]{\underline{#1}}
\newcommand{\randwin}{\textcolor{black}{\ding{51}}}
\newcommand{\randloss}{\textcolor{black}{\ding{55}}}
\newcommand{\grayrow}{\specialrule{0.20pt}{0.35mm}{0.35mm}}
\arrayrulecolor{black!18}
\begin{tabular}{llccccc}
\arrayrulecolor{black}\toprule
Setting & Ranking method & H2 (4q) & HeH+ (4q) & H3+ (6q) & H4 chain (8q) & LiH (12q) \\
\midrule\arrayrulecolor{black!18}
\bcell{Depth 1} & \ours{} gain, $K{=}1$ $\uparrow$ & \pstat{40.4}{6.3} & \pstat{58.9}{3.5} & \pstat{63.0}{2.8} & \pstat{74.7}{1.1} & \pstat{86.9}{0.5} \\
\grayrow
\bcell{Depth 1} & \ours{} gain, $K{=}7$ $\uparrow$ & \pstat{41.2}{5.3} & \pstat{60.7}{3.1} & \pstat{65.1}{2.6} & \pstat{77.3}{0.7} & \pstat{88.2}{0.4} \\
\arrayrulecolor{black}\cmidrule(lr){1-7}\arrayrulecolor{black!18}
\gcell{Depth 1} & Random ($K{=}7$) & \estat{\epos{0.1847}}{0.0868} & \estat{\epos{1.2191}}{0.2759} & \estat{\epos{0.7247}}{0.2856} & \estat{\epos{0.9736}}{0.3101} & \estat{\epos{2.9030}}{0.6058} \\
\grayrow
\gcell{Depth 1} & \llmmethod{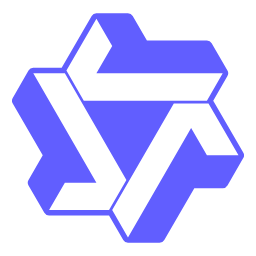}{Qwen3-8B} & \vfail{30}{\estat{\epos{0.1321}}{0.0938}\,\randwin} & \vfail{57}{\estat{\epos{1.3928}}{0.9006}\,\randloss} & \vfail{51}{\estat{\epos{0.9972}}{0.1834}\,\randloss} & \vfail{41}{\estat{\epos{0.4765}}{0.2133}\,\randwin} & \vfail{66}{\estat{\epos{2.0656}}{0.7546}\,\randwin} \\
\grayrow
\gcell{Depth 1} & \llmmethod{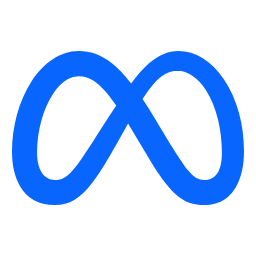}{Llama-3.1-8B} & \vfail{36}{\estat{\epos{0.2840}}{0.1350}\,\randloss} & \vfail{55}{\estat{\epos{1.2347}}{0.3492}\,\randloss} & \vfail{44}{\estat{\epos{0.6195}}{0.3195}\,\randwin} & \vfail{40}{\estat{\epos{0.4219}}{0.3064}\,\randwin} & \vfail{57}{\estat{\epos{1.3266}}{0.8740}\,\randwin} \\
\grayrow
\gcell{Depth 1} & \llmmethod{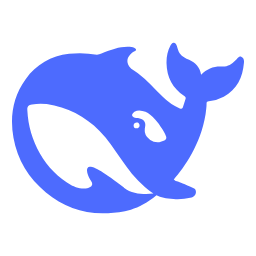}{DeepSeek-R1-8B} & \vfail{34}{\estat{\epos{0.2405}}{0.1463}\,\randloss} & \vfail{53}{\estat{\epos{1.0993}}{0.5373}\,\randwin} & \vfail{36}{\estat{\epos{0.2766}}{0.1157}\,\randwin} & \vfail{26}{\estat{\epos{0.0522}}{0.0653}\,\randwin} & \vfail{25}{\estat{\epos{0.0504}}{0.0576}\,\randwin} \\
\grayrow
\gcell{Depth 1} & \llmmethod{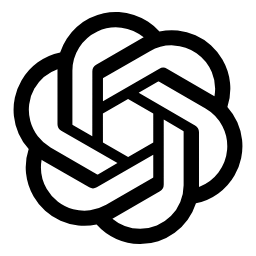}{GPT-OSS-20B} & \vfail{25}{\estat{\epos{0.0471}}{0.0463}\,\randwin} & \vfail{36}{\estat{\epos{0.2783}}{0.2758}\,\randwin} & \vfail{26}{\estat{\epos{0.0568}}{0.0867}\,\randwin} & \vfail{24}{\estat{\epos{0.0379}}{0.0641}\,\randwin} & \vfail{18}{\best{\estat{\epos{0.0018}}{0.0032}}\,\randwin} \\
\grayrow
\gcell{Depth 1} & \ours{} ($K{=}7$) & \vpass{18}{\runnerup{\estat{\epos{0.0013}}{0.0003}}\,\randwin} & \vfail{19}{\runnerup{\estat{\epos{0.0035}}{0.0015}}\,\randwin} & \vpass{18}{\best{\estat{\epos{0.0004}}{0.0006}}\,\randwin} & \vpass{18}{\best{\estat{\epos{0.0001}}{0.0060}}\,\randwin} & \vfail{19}{\estat{\epos{0.0030}}{0.0016}\,\randwin} \\
\grayrow
\gcell{Depth 1} & \ours{} ($K{=}1$) & \vpass{18}{\best{\estat{\epos{0.0011}}{0.0008}}\,\randwin} & \vpass{18}{\best{\estat{\epos{0.0015}}{0.0012}}\,\randwin} & \vpass{18}{\runnerup{\estat{\epos{0.0005}}{0.0006}}\,\randwin} & \vpass{24}{\runnerup{\estat{\eneg{0.0006}}{0.0024}}\,\randwin} & \vfail{19}{\runnerup{\estat{\epos{0.0023}}{0.0016}}\,\randwin} \\
\arrayrulecolor{black}\midrule\arrayrulecolor{black!18}
\bcell{Depth 2} & \ours{} gain, $K{=}1$ $\uparrow$ & \pstat{25.6}{2.9} & \pstat{31.6}{2.6} & \pstat{44.9}{1.3} & \pstat{51.8}{0.5} & \pstat{82.6}{0.4} \\
\grayrow
\bcell{Depth 2} & \ours{} gain, $K{=}7$ $\uparrow$ & \pstat{27.3}{2.5} & \pstat{32.5}{2.4} & \pstat{47.9}{1.4} & \pstat{55.1}{0.5} & \pstat{83.7}{0.4} \\
\arrayrulecolor{black}\cmidrule(lr){1-7}\arrayrulecolor{black!18}
\gcell{Depth 2} & Random ($K{=}7$) & \estat{\epos{0.0930}}{0.0297} & \estat{\epos{0.5668}}{0.4463} & \estat{\epos{0.3054}}{0.1633} & \estat{\epos{0.9327}}{0.3273} & \estat{\epos{2.6969}}{0.2816} \\
\grayrow
\gcell{Depth 2} & \llmmethod{qwen}{Qwen3-8B} & \vfail{28}{\estat{\epos{0.0992}}{0.0689}\,\randloss} & \vfail{49}{\estat{\epos{0.8415}}{0.1479}\,\randloss} & \vfail{45}{\estat{\epos{0.6751}}{0.2741}\,\randloss} & \vfail{32}{\estat{\epos{0.1842}}{0.0985}\,\randwin} & \vfail{73}{\estat{\epos{2.6722}}{0.9356}\,\randwin} \\
\grayrow
\gcell{Depth 2} & \llmmethod{meta}{Llama-3.1-8B} & \vpass{75}{\estat{\eneg{0.0595}}{0.1790}\,\randwin} & \vfail{35}{\estat{\epos{0.2644}}{0.2592}\,\randwin} & \vfail{46}{\estat{\epos{0.6780}}{0.0990}\,\randloss} & \vpass{72}{\estat{\eneg{0.0543}}{0.0502}\,\randwin} & \vfail{41}{\estat{\epos{0.4541}}{0.2847}\,\randwin} \\
\grayrow
\gcell{Depth 2} & \llmmethod{deepseek}{DeepSeek-R1-8B} & \vfail{20}{\estat{\epos{0.0051}}{0.1945}\,\randwin} & \vfail{28}{\estat{\epos{0.0991}}{0.0653}\,\randwin} & \vfail{37}{\estat{\epos{0.3239}}{0.1195}\,\randloss} & \vfail{22}{\estat{\epos{0.0167}}{0.0328}\,\randwin} & \vfail{24}{\estat{\epos{0.0332}}{0.0342}\,\randwin} \\
\grayrow
\gcell{Depth 2} & \llmmethod{openai}{GPT-OSS-20B} & \vpass{18}{\estat{\epos{0.0011}}{0.0098}\,\randwin} & \vfail{20}{\runnerup{\estat{\epos{0.0049}}{0.0044}}\,\randwin} & \vfail{24}{\estat{\epos{0.0322}}{0.0484}\,\randwin} & \vfail{22}{\estat{\epos{0.0167}}{0.0328}\,\randwin} & \vfail{26}{\estat{\epos{0.0603}}{0.1773}\,\randwin} \\
\grayrow
\gcell{Depth 2} & \ours{} ($K{=}7$) & \vpass{24}{\runnerup{\estat{\eneg{0.0007}}{0.0015}}\,\randwin} & \vfail{21}{\estat{\epos{0.0076}}{0.0176}\,\randwin} & \vpass{18}{\best{\estat{\epos{0.0006}}{0.0031}}\,\randwin} & \vfail{22}{\runnerup{\estat{\epos{0.0162}}{0.0438}}\,\randwin} & \vpass{33}{\best{\estat{\eneg{0.0040}}{0.0118}}\,\randwin} \\
\grayrow
\gcell{Depth 2} & \ours{} ($K{=}1$) & \vpass{18}{\best{\estat{\epos{0.0003}}{0.0007}}\,\randwin} & \vpass{26}{\best{\estat{\eneg{0.0011}}{0.0026}}\,\randwin} & \vpass{27}{\runnerup{\estat{\eneg{0.0015}}{0.0041}}\,\randwin} & \vfail{22}{\best{\estat{\epos{0.0158}}{0.0443}}\,\randwin} & \vpass{34}{\runnerup{\estat{\eneg{0.0049}}{0.0097}}\,\randwin} \\
\arrayrulecolor{black}\midrule\arrayrulecolor{black!18}
\bcell{Depth 3} & \ours{} gain, $K{=}1$ $\uparrow$ & \pstat{18.1}{2.7} & \pstat{24.9}{1.3} & \pstat{29.9}{1.0} & \pstat{35.2}{0.9} & \pstat{75.4}{1.8} \\
\grayrow
\bcell{Depth 3} & \ours{} gain, $K{=}7$ $\uparrow$ & \pstat{19.0}{2.4} & \pstat{26.1}{1.2} & \pstat{32.1}{1.2} & \pstat{38.6}{0.7} & \pstat{76.6}{1.3} \\
\arrayrulecolor{black}\cmidrule(lr){1-7}\arrayrulecolor{black!18}
\gcell{Depth 3} & Random ($K{=}7$) & \estat{\epos{0.0730}}{0.0859} & \estat{\epos{0.2617}}{0.2130} & \estat{\epos{0.2750}}{0.0765} & \estat{\epos{0.2773}}{0.2665} & \estat{\epos{2.3928}}{0.4840} \\
\grayrow
\gcell{Depth 3} & \llmmethod{qwen}{Qwen3-8B} & \vfail{27}{\estat{\epos{0.0742}}{0.0922}\,\randloss} & \vfail{48}{\estat{\epos{0.8192}}{0.3035}\,\randloss} & \vfail{37}{\estat{\epos{0.3073}}{0.1583}\,\randloss} & \vfail{20}{\estat{\epos{0.0046}}{0.1338}\,\randwin} & \vfail{50}{\estat{\epos{0.9203}}{0.3562}\,\randwin} \\
\grayrow
\gcell{Depth 3} & \llmmethod{meta}{Llama-3.1-8B} & \vfail{26}{\estat{\epos{0.0631}}{0.0983}\,\randwin} & \vfail{36}{\estat{\epos{0.3042}}{0.1041}\,\randloss} & \vfail{31}{\estat{\epos{0.1583}}{0.0876}\,\randwin} & \vpass{53}{\estat{\eneg{0.0230}}{0.0343}\,\randwin} & \vpass{70}{\estat{\eneg{0.0493}}{0.3344}\,\randwin} \\
\grayrow
\gcell{Depth 3} & \llmmethod{deepseek}{DeepSeek-R1-8B} & \vpass{57}{\estat{\eneg{0.0279}}{0.0752}\,\randwin} & \vfail{24}{\estat{\epos{0.0302}}{0.0339}\,\randwin} & \vfail{28}{\estat{\epos{0.0823}}{0.0558}\,\randwin} & \vpass{41}{\estat{\eneg{0.0093}}{0.0205}\,\randwin} & \vfail{24}{\estat{\epos{0.0306}}{0.0379}\,\randwin} \\
\grayrow
\gcell{Depth 3} & \llmmethod{openai}{GPT-OSS-20B} & \vfail{25}{\estat{\epos{0.0436}}{0.0577}\,\randwin} & \vfail{31}{\estat{\epos{0.1632}}{0.1611}\,\randwin} & \vfail{21}{\estat{\epos{0.0091}}{0.0196}\,\randwin} & \vpass{31}{\estat{\eneg{0.0033}}{0.0142}\,\randwin} & \vpass{64}{\estat{\eneg{0.0392}}{0.1522}\,\randwin} \\
\grayrow
\gcell{Depth 3} & \ours{} ($K{=}7$) & \vpass{18}{\runnerup{\estat{\epos{0.0010}}{0.0027}}\,\randwin} & \vfail{21}{\runnerup{\estat{\epos{0.0108}}{0.0118}}\,\randwin} & \vfail{19}{\runnerup{\estat{\epos{0.0029}}{0.0095}}\,\randwin} & \vfail{19}{\runnerup{\estat{\epos{0.0029}}{0.0113}}\,\randwin} & \vpass{34}{\runnerup{\estat{\eneg{0.0046}}{0.0040}}\,\randwin} \\
\grayrow
\gcell{Depth 3} & \ours{} ($K{=}1$) & \vpass{18}{\best{\estat{\epos{0.0005}}{0.0015}}\,\randwin} & \vfail{21}{\best{\estat{\epos{0.0092}}{0.0131}}\,\randwin} & \vpass{18}{\best{\estat{\epos{0.0005}}{0.0077}}\,\randwin} & \vpass{27}{\best{\estat{\eneg{0.0014}}{0.0097}}\,\randwin} & \vpass{27}{\best{\estat{\eneg{0.0017}}{0.0036}}\,\randwin} \\
\arrayrulecolor{black}\bottomrule
\end{tabular}
\endgroup
\label{tab:v76_ollama_table2}
\vspace{-0.5em}
\end{table*}

\begin{wrapfigure}{r}{0.56\linewidth}
\vspace{-0.9em}
\centering
\includegraphics[width=0.90\linewidth]{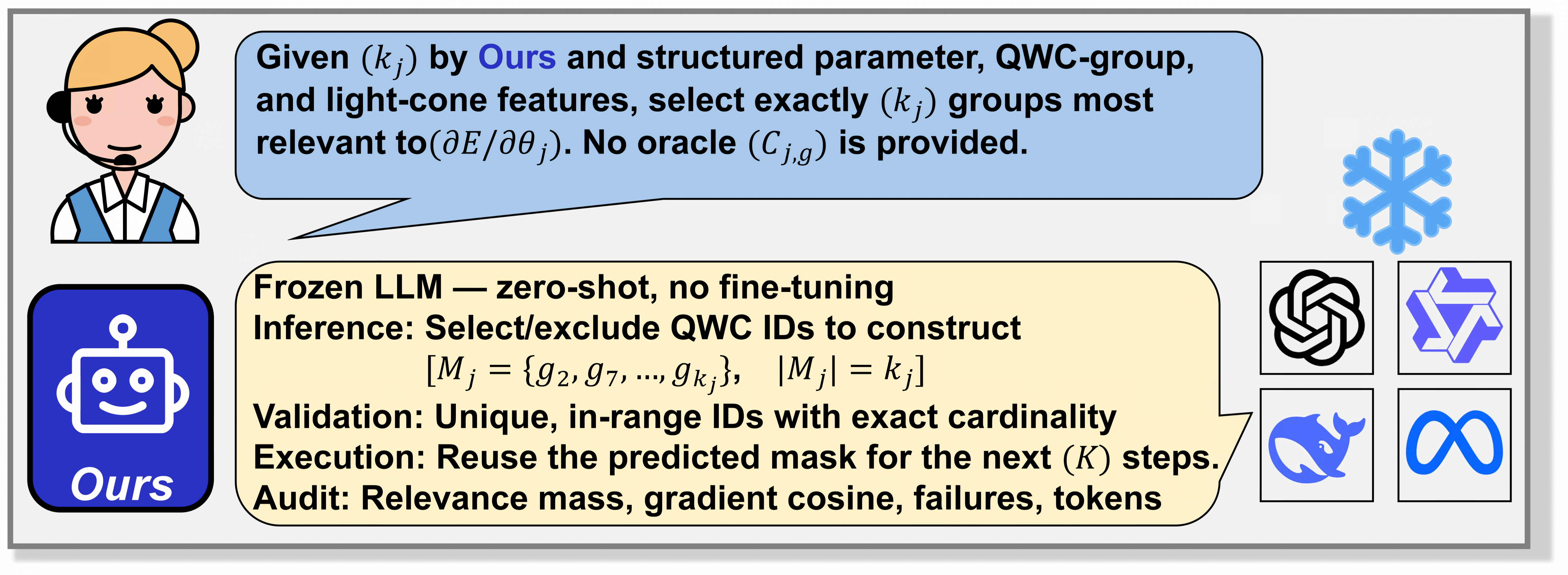}
\caption{\textbf{Clean frozen-LLM protocol.} Our learned controller supplies $k_j$ and features; the LLM returns select/exclude QWC IDs; the mask is reused for $K{=}7$.}
\label{fig:v79_clean_llm_protocol}
\vspace{-0.9em}
\end{wrapfigure}

\emph{Do the gains require task-specific learning, or can a capable general model be prompted into the same decisions?} Figure~\ref{fig:v79_clean_llm_protocol} shows the shared interface: \ourmethod{} supplies each parameter's cardinality $k_j$ and structured physical features, a frozen zero-shot LLM selects the group identities, and the validated mask is reused for $K{=}7$ steps. Matching refresh cadence controls invocation frequency. We evaluate four frozen local LLMs: Qwen3-8B~\citep{yang2025qwen3},
Llama-3.1-8B~\citep{grattafiori2024llama},
DeepSeek-R1-Distill-Llama-8B~\citep{guo2025deepseek},
and GPT-OSS-20B~\citep{agarwal2025gpt}. In Table~\ref{tab:v76_ollama_table2}, relative to per-step \ours{}-$K{=}1$, the four-refresh \ours{}-$K{=}7$ control increases the cross-cell mean absolute gap from 0.00567 to 0.00715 Ha while changing mean shot saving from 49.6\% to 51.4\%; stale masks do cost reliability (the no-worse count falls from 12/15 to 8/15 cells). Even so, \ours{}-$K{=}7$ remains closer to Full-QWC than Qwen, Llama, and DeepSeek in all 15 cells, and closer than GPT-OSS in 13/15, at nearly identical aggregate savings. GPT-OSS is the strongest frozen LLM (0.0694 Ha mean absolute gap), followed by DeepSeek (0.172), Llama (0.421), and Qwen (0.786). Against the protocol-matched Random control, Qwen, Llama, DeepSeek, and GPT-OSS win 7/15, 11/15, 13/15, and 15/15 cells, respectively, while both proposed-method controls win 15/15: physics-conditioned prompting beats random choice, yet the aggregate LLM gap stays almost an order of magnitude above that of our refresh-matched method; the advantage comes from learned parameter-specific ranking, not from invocation frequency.

The repair-free audit (Appendix~\ref{sec:app_llm_details}, Table~\ref{tab:v79_clean_llm_audit}) jointly changes the output contract and replaces coefficient completion with Full-QWC fallback after failed retries. Models retain 52.0--52.4\% mean shot saving and beat Random in 13/15, 10/15, 11/15, and 7/15 cells (Qwen, Llama, DeepSeek, GPT-OSS), with mean absolute gaps of 0.4482--0.6986 Ha. Both proposed-method controls win 15/15. This second interface tests the same scientific selection task under a distinct validation policy. Protocol versions and direct-JSON inference settings are detailed in Appendix~\ref{sec:app_trace_provenance}; Figure~\ref{fig:v76b_prompt_protocol} retains the earlier schematic.

\paragraph{Data-exposed frontier models.}
\label{sec:v87_frontier_models}
We next test whether scale plus data access closes the gap: five GPT and four
Claude variants receive two train-only same-molecule/depth/parameter neighbors
and an eight-neighbor summary (retrieval details in Appendix~\ref{sec:app_frontier_retrieval}). They choose both group IDs and counts, yielding model-specific accuracy--cost operating points. The held-out
contribution matrix is never shown, tools are disabled, and an invalid
response is charged as Full-QWC; all methods share the 28-step, $K{=}7$
protocol and held-out seeds. Table~\ref{tab:v87_frontier_depth1} shows
depth 1 with five GPT variants~\citep{openai2026models} and
four Claude variants~\citep{anthropic2026transparency}; Appendix~\ref{sec:app_llm_details},
Tables~\ref{tab:v87_frontier_depth2}--\ref{tab:v87_frontier_depth3}, give
depths 2--3. Data-exposed frontier models do reach the mHa regime; the
relevance structure is learnable in context, but the outcome is a
cost--accuracy frontier rather than uniform dominance: Claude Fable 5 attains the
best overall mean $|\dE|$ (0.00302 Ha) at 35.8\% saving, \ours{}-$K{=}7$
operates at a more aggressive 51.4\% saving with 0.00715 Ha, and no model
dominates every cell.

\begingroup
\def\OursFrontierMain{1}
\ifdefined\OursFrontierMain
\begin{table*}[t]
\centering
\caption{\textbf{Data-exposed frontier models at depth 1.} Train-only retrieval informs group IDs and counts under four-refresh $K{=}7$. Entries are mean $\pm$ s.d. over four seeds. \textcolor{green!70!black}{Green}/\textcolor{red}{red} shading indicates the 1.6 mHa no-worse split; \ding{51}/\ding{55} compares $|\dE|$ with Random. Among non-Random methods, \textbf{bold}/\underline{underline} mark best/second-best $|\dE|\downarrow$ and gain $\uparrow$; gray marks largest gain. Protocols and full results: Appendices~\ref{sec:app_frontier_retrieval} and~\ref{sec:app_llm_details}.}
\vspace{0.25em}
\setlength{\tabcolsep}{0.55mm}
\renewcommand{\arraystretch}{1.04}
\fontsize{7.6}{9.1}\selectfont
\begingroup
\definecolor{vistaheatgreen}{HTML}{A8D8A8}
\definecolor{vistaheatred}{HTML}{F3A6B7}
\newcommand{\frontierlogo}[1]{\raisebox{-0.28ex}{\includegraphics[height=1.10em]{llm_logos/#1.png}}}
\newcommand{\frontiermethod}[2]{\frontierlogo{#1}\,#2}
\newcommand{\vpass}[2]{\cellcolor{vistaheatgreen!#1!white}\textcolor{black}{#2}}
\newcommand{\vfail}[2]{\cellcolor{vistaheatred!#1!white}\textcolor{black}{#2}}
\newcommand{\epos}[1]{+#1}
\newcommand{\eneg}[1]{\textminus#1}
\newcommand{\estat}[2]{#1\,\(\pm\)\,#2}
\newcommand{\pstat}[2]{#1\,\(\pm\)\,#2\%}
\newcommand{\best}[1]{\textbf{#1}}
\newcommand{\runnerup}[1]{\underline{#1}}
\newcommand{\shotbest}[1]{\cellcolor{black!12}\best{#1}}
\newcommand{\randwin}{\textcolor{black}{\ding{51}}}
\newcommand{\randloss}{\textcolor{black}{\ding{55}}}
\newcommand{\grayrow}{\specialrule{0.20pt}{0.30mm}{0.30mm}}
\newcommand{\shotcell}[1]{\multicolumn{1}{|c}{#1}}
\newcommand{\shotlast}[1]{\multicolumn{1}{|c|}{#1}}
\arrayrulecolor{black!18}
\begin{tabular*}{\textwidth}{@{\extracolsep{\fill}}lccccc@{}}
\arrayrulecolor{black}\toprule
Method & H2 (4q) & HeH+ (4q) & H3+ (6q) & H4 chain (8q) & LiH (12q) \\
\midrule\arrayrulecolor{black!18}
\multicolumn{6}{c}{\textbf{Energy fidelity $\downarrow$: $\dE$ vs. Full (Ha)}} \\
\arrayrulecolor{black!30}\hline\arrayrulecolor{black!18}
\addlinespace[0.10em]
Random ($K{=}7$) & \estat{\epos{0.1847}}{0.0868} & \estat{\epos{1.2191}}{0.2759} & \estat{\epos{0.7247}}{0.2856} & \estat{\epos{0.9736}}{0.3101} & \estat{\epos{2.9030}}{0.6058} \\
\grayrow
\arrayrulecolor{black!18}\cmidrule(lr){1-6}
\frontiermethod{openai}{GPT-5.4} & \vpass{18}{\estat{\epos{0.0002}}{0.0003}\,\randwin} & \vpass{55}{\estat{\eneg{0.0007}}{0.0017}\,\randwin} & \vpass{18}{\estat{\epos{0.0010}}{0.0011}\,\randwin} & \vfail{31}{\estat{\epos{0.0021}}{0.0015}\,\randwin} & \vpass{18}{\best{\estat{\epos{0.0010}}{0.0005}}\,\randwin} \\
\grayrow
\frontiermethod{openai}{GPT-5.5} & \vpass{18}{\estat{\epos{0.0001}}{0.0007}\,\randwin} & \vpass{18}{\estat{\epos{0.0006}}{0.0011}\,\randwin} & \vpass{18}{\estat{\epos{0.0008}}{0.0009}\,\randwin} & \vfail{45}{\estat{\epos{0.0026}}{0.0022}\,\randwin} & \vfail{22}{\estat{\epos{0.0017}}{0.0012}\,\randwin} \\
\grayrow
\frontiermethod{openai}{GPT-5.6 Sol} & \vpass{18}{\estat{\epos{0.0002}}{0.0005}\,\randwin} & \vpass{54}{\estat{\eneg{0.0007}}{0.0023}\,\randwin} & \vpass{18}{\estat{\epos{0.0015}}{0.0015}\,\randwin} & \vpass{18}{\runnerup{\estat{\epos{0.0002}}{0.0023}}\,\randwin} & \vfail{27}{\estat{\epos{0.0018}}{0.0007}\,\randwin} \\
\grayrow
\frontiermethod{openai}{GPT-5.6 Terra} & \vpass{27}{\best{\estat{\eneg{0.0000}}{0.0007}}\,\randwin} & \vpass{18}{\estat{\epos{0.0009}}{0.0015}\,\randwin} & \vfail{65}{\estat{\epos{0.0041}}{0.0050}\,\randwin} & \vfail{23}{\estat{\epos{0.0017}}{0.0033}\,\randwin} & \vpass{18}{\estat{\epos{0.0016}}{0.0005}\,\randwin} \\
\grayrow
\frontiermethod{openai}{GPT-5.6 Luna} & \vpass{18}{\estat{\epos{0.0003}}{0.0004}\,\randwin} & \vpass{18}{\runnerup{\estat{\epos{0.0001}}{0.0010}}\,\randwin} & \vpass{18}{\estat{\epos{0.0011}}{0.0013}\,\randwin} & \vfail{65}{\estat{\epos{0.0035}}{0.0024}\,\randwin} & \vfail{27}{\estat{\epos{0.0018}}{0.0009}\,\randwin} \\
\grayrow
\arrayrulecolor{black!18}\cmidrule(lr){1-6}
\frontiermethod{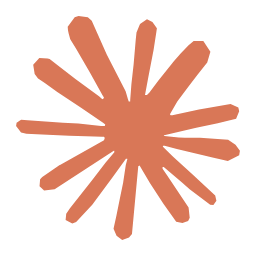}{Claude Haiku 4.5} & \vpass{55}{\estat{\eneg{0.0001}}{0.0004}\,\randwin} & \vpass{18}{\estat{\epos{0.0002}}{0.0004}\,\randwin} & \vpass{18}{\estat{\epos{0.0006}}{0.0010}\,\randwin} & \vfail{28}{\estat{\epos{0.0020}}{0.0033}\,\randwin} & \vpass{18}{\estat{\epos{0.0011}}{0.0007}\,\randwin} \\
\grayrow
\frontiermethod{claude}{Claude Sonnet 5} & \vpass{18}{\runnerup{\estat{\epos{0.0000}}{0.0004}}\,\randwin} & \vpass{18}{\estat{\epos{0.0004}}{0.0010}\,\randwin} & \vpass{18}{\best{\estat{\epos{0.0003}}{0.0010}}\,\randwin} & \vfail{22}{\estat{\epos{0.0017}}{0.0022}\,\randwin} & \vfail{20}{\estat{\epos{0.0016}}{0.0005}\,\randwin} \\
\grayrow
\frontiermethod{claude}{Claude Opus 5} & \vpass{18}{\estat{\epos{0.0002}}{0.0004}\,\randwin} & \vpass{18}{\estat{\epos{0.0005}}{0.0009}\,\randwin} & \vpass{18}{\estat{\epos{0.0004}}{0.0009}\,\randwin} & \vfail{39}{\estat{\epos{0.0024}}{0.0040}\,\randwin} & \vpass{18}{\estat{\epos{0.0015}}{0.0008}\,\randwin} \\
\grayrow
\frontiermethod{claude}{Claude Fable 5} & \vpass{18}{\estat{\epos{0.0003}}{0.0007}\,\randwin} & \vpass{22}{\best{\estat{\eneg{0.0000}}{0.0006}}\,\randwin} & \vpass{18}{\estat{\epos{0.0004}}{0.0010}\,\randwin} & \vpass{18}{\estat{\epos{0.0010}}{0.0044}\,\randwin} & \vpass{18}{\runnerup{\estat{\epos{0.0011}}{0.0005}}\,\randwin} \\
\grayrow
\arrayrulecolor{black!18}\cmidrule(lr){1-6}
\textbf{\ours{}} ($K{=}7$) & \vpass{18}{\estat{\epos{0.0013}}{0.0003}\,\randwin} & \vfail{65}{\estat{\epos{0.0035}}{0.0015}\,\randwin} & \vpass{18}{\runnerup{\estat{\epos{0.0004}}{0.0006}}\,\randwin} & \vpass{18}{\best{\estat{\epos{0.0001}}{0.0060}}\,\randwin} & \vfail{65}{\estat{\epos{0.0030}}{0.0016}\,\randwin} \\
\arrayrulecolor{black}\midrule
\multicolumn{6}{c}{\textbf{Measurement efficiency: shot gain (\%; higher is better $\uparrow$)}} \\
\arrayrulecolor{black!30}\hline\arrayrulecolor{black!18}
\addlinespace[0.10em]\arrayrulecolor{black!18}
Random ($K{=}7$) & \shotcell{\pstat{42.7}{3.8}} & \shotcell{\pstat{61.3}{2.6}} & \shotcell{\pstat{68.2}{2.5}} & \shotcell{\pstat{80.2}{1.2}} & \shotlast{\pstat{90.8}{0.1}} \\
\grayrow
\arrayrulecolor{black!18}\cmidrule(lr){1-6}
\frontiermethod{openai}{GPT-5.4} & \shotcell{\pstat{17.2}{1.0}} & \shotcell{\pstat{46.7}{4.2}} & \shotcell{\pstat{52.2}{3.7}} & \shotcell{\pstat{64.6}{1.9}} & \shotlast{\pstat{80.4}{0.3}} \\
\grayrow
\frontiermethod{openai}{GPT-5.5} & \shotcell{\pstat{18.6}{1.6}} & \shotcell{\pstat{45.4}{3.8}} & \shotcell{\pstat{51.8}{3.3}} & \shotcell{\runnerup{\pstat{70.4}{2.6}}} & \shotlast{\pstat{82.9}{2.2}} \\
\grayrow
\frontiermethod{openai}{GPT-5.6 Sol} & \shotcell{\pstat{18.1}{1.4}} & \shotcell{\runnerup{\pstat{47.3}{4.5}}} & \shotcell{\pstat{53.3}{3.7}} & \shotcell{\pstat{64.3}{1.2}} & \shotlast{\pstat{80.7}{0.3}} \\
\grayrow
\frontiermethod{openai}{GPT-5.6 Terra} & \shotcell{\pstat{18.1}{1.0}} & \shotcell{\pstat{46.7}{2.3}} & \shotcell{\runnerup{\pstat{53.6}{2.5}}} & \shotcell{\pstat{64.1}{1.1}} & \shotlast{\pstat{80.2}{0.4}} \\
\grayrow
\frontiermethod{openai}{GPT-5.6 Luna} & \shotcell{\pstat{19.4}{2.5}} & \shotcell{\pstat{46.8}{4.1}} & \shotcell{\pstat{52.3}{3.6}} & \shotcell{\pstat{64.7}{1.0}} & \shotlast{\pstat{80.8}{0.3}} \\
\grayrow
\arrayrulecolor{black!18}\cmidrule(lr){1-6}
\frontiermethod{claude}{Claude Haiku 4.5} & \shotcell{\pstat{18.1}{1.6}} & \shotcell{\pstat{46.9}{3.3}} & \shotcell{\pstat{52.0}{3.1}} & \shotcell{\pstat{65.1}{1.8}} & \shotlast{\pstat{81.3}{1.1}} \\
\grayrow
\frontiermethod{claude}{Claude Sonnet 5} & \shotcell{\runnerup{\pstat{19.7}{1.4}}} & \shotcell{\pstat{43.8}{3.3}} & \shotcell{\pstat{50.3}{3.1}} & \shotcell{\pstat{69.6}{0.9}} & \shotlast{\runnerup{\pstat{85.7}{0.6}}} \\
\grayrow
\frontiermethod{claude}{Claude Opus 5} & \shotcell{\pstat{17.2}{1.1}} & \shotcell{\pstat{42.3}{3.1}} & \shotcell{\pstat{44.7}{3.0}} & \shotcell{\pstat{68.9}{2.2}} & \shotlast{\pstat{81.9}{1.0}} \\
\grayrow
\frontiermethod{claude}{Claude Fable 5} & \shotcell{\pstat{18.3}{1.5}} & \shotcell{\pstat{44.3}{3.4}} & \shotcell{\pstat{50.4}{3.0}} & \shotcell{\pstat{67.2}{2.1}} & \shotlast{\pstat{80.1}{0.5}} \\
\grayrow
\arrayrulecolor{black!18}\cmidrule(lr){1-6}
\textbf{\ours{}} ($K{=}7$) & \shotcell{\shotbest{\pstat{41.2}{5.3}}} & \shotcell{\shotbest{\pstat{60.7}{3.1}}} & \shotcell{\shotbest{\pstat{65.1}{2.6}}} & \shotcell{\shotbest{\pstat{77.3}{0.7}}} & \shotlast{\shotbest{\pstat{88.2}{0.4}}} \\
\arrayrulecolor{black}\bottomrule
\end{tabular*}
\endgroup
\label{tab:v87_frontier_depth1}
\end{table*}
\fi

\ifdefined\OursFrontierAppendix
\begin{table*}[t]
\centering
\caption{\textbf{Data-exposed frontier-model online VQE at depth 2.} The train/test split, $\tau=0.995$, four-refresh $K{=}7$ protocol, and reporting conventions match Table~\ref{tab:v87_frontier_depth1}.}
\vspace{0.25em}
\setlength{\tabcolsep}{0.55mm}
\renewcommand{\arraystretch}{1.04}
\fontsize{7.6}{9.1}\selectfont
\begingroup
\definecolor{vistaheatgreen}{HTML}{A8D8A8}
\definecolor{vistaheatred}{HTML}{F3A6B7}
\newcommand{\frontierlogo}[1]{\raisebox{-0.28ex}{\includegraphics[height=1.10em]{llm_logos/#1.png}}}
\newcommand{\frontiermethod}[2]{\frontierlogo{#1}\,#2}
\newcommand{\vpass}[2]{\cellcolor{vistaheatgreen!#1!white}\textcolor{black}{#2}}
\newcommand{\vfail}[2]{\cellcolor{vistaheatred!#1!white}\textcolor{black}{#2}}
\newcommand{\epos}[1]{+#1}
\newcommand{\eneg}[1]{\textminus#1}
\newcommand{\estat}[2]{#1\,\(\pm\)\,#2}
\newcommand{\pstat}[2]{#1\,\(\pm\)\,#2\%}
\newcommand{\best}[1]{\textbf{#1}}
\newcommand{\runnerup}[1]{\underline{#1}}
\newcommand{\shotbest}[1]{\cellcolor{black!12}\best{#1}}
\newcommand{\randwin}{\textcolor{black}{\ding{51}}}
\newcommand{\randloss}{\textcolor{black}{\ding{55}}}
\newcommand{\grayrow}{\specialrule{0.20pt}{0.30mm}{0.30mm}}
\arrayrulecolor{black!18}
\begin{tabular*}{\textwidth}{@{\extracolsep{\fill}}lccccc@{}}
\arrayrulecolor{black}\toprule
Method & H2 (4q) & HeH+ (4q) & H3+ (6q) & H4 chain (8q) & LiH (12q) \\
\midrule\arrayrulecolor{black!18}
\multicolumn{6}{c}{\textbf{Energy fidelity $\downarrow$: $\dE$ vs. Full (Ha)}} \\
\arrayrulecolor{black!30}\hline\arrayrulecolor{black!18}
\addlinespace[0.10em]
Random ($K{=}7$) & \estat{\epos{0.0930}}{0.0297} & \estat{\epos{0.5668}}{0.4463} & \estat{\epos{0.3054}}{0.1633} & \estat{\epos{0.9327}}{0.3273} & \estat{\epos{2.6969}}{0.2816} \\
\grayrow
\arrayrulecolor{black!18}\cmidrule(lr){1-6}
\frontiermethod{openai}{GPT-5.4} & \vfail{27}{\estat{\epos{0.0032}}{0.0064}\,\randwin} & \vfail{36}{\estat{\epos{0.0090}}{0.0164}\,\randwin} & \vpass{18}{\estat{\epos{0.0011}}{0.0056}\,\randwin} & \vpass{18}{\best{\estat{\epos{0.0000}}{0.0098}}\,\randwin} & \vfail{26}{\estat{\epos{0.0020}}{0.0041}\,\randwin} \\
\grayrow
\frontiermethod{openai}{GPT-5.5} & \vpass{32}{\runnerup{\estat{\eneg{0.0002}}{0.0021}}\,\randwin} & \vfail{22}{\estat{\epos{0.0026}}{0.0057}\,\randwin} & \vfail{39}{\estat{\epos{0.0051}}{0.0077}\,\randwin} & \vfail{23}{\estat{\epos{0.0031}}{0.0178}\,\randwin} & \vfail{20}{\estat{\epos{0.0016}}{0.0042}\,\randwin} \\
\grayrow
\frontiermethod{openai}{GPT-5.6 Sol} & \vfail{65}{\estat{\epos{0.0117}}{0.0135}\,\randwin} & \vfail{62}{\estat{\epos{0.0205}}{0.0326}\,\randwin} & \vpass{18}{\runnerup{\estat{\epos{0.0009}}{0.0074}}\,\randwin} & \vfail{65}{\estat{\epos{0.0231}}{0.0383}\,\randwin} & \vpass{18}{\estat{\epos{0.0015}}{0.0046}\,\randwin} \\
\grayrow
\frontiermethod{openai}{GPT-5.6 Terra} & \vfail{52}{\estat{\epos{0.0088}}{0.0150}\,\randwin} & \vfail{65}{\estat{\epos{0.0219}}{0.0367}\,\randwin} & \vfail{29}{\estat{\epos{0.0033}}{0.0062}\,\randwin} & \vfail{21}{\runnerup{\estat{\epos{0.0022}}{0.0063}}\,\randwin} & \vfail{36}{\estat{\epos{0.0026}}{0.0055}\,\randwin} \\
\grayrow
\frontiermethod{openai}{GPT-5.6 Luna} & \vfail{25}{\estat{\epos{0.0027}}{0.0057}\,\randwin} & \vfail{48}{\estat{\epos{0.0141}}{0.0188}\,\randwin} & \vfail{65}{\estat{\epos{0.0100}}{0.0099}\,\randwin} & \vfail{61}{\estat{\epos{0.0211}}{0.0385}\,\randwin} & \vfail{34}{\estat{\epos{0.0025}}{0.0057}\,\randwin} \\
\grayrow
\arrayrulecolor{black!18}\cmidrule(lr){1-6}
\frontiermethod{claude}{Claude Haiku 4.5} & \vfail{26}{\estat{\epos{0.0030}}{0.0056}\,\randwin} & \vpass{52}{\best{\estat{\eneg{0.0001}}{0.0013}}\,\randwin} & \vfail{28}{\estat{\epos{0.0030}}{0.0070}\,\randwin} & \vfail{47}{\estat{\epos{0.0146}}{0.0514}\,\randwin} & \vpass{18}{\best{\estat{\epos{0.0012}}{0.0028}}\,\randwin} \\
\grayrow
\frontiermethod{claude}{Claude Sonnet 5} & \vpass{18}{\estat{\epos{0.0003}}{0.0003}\,\randwin} & \vpass{55}{\estat{\eneg{0.0001}}{0.0009}\,\randwin} & \vpass{18}{\estat{\epos{0.0016}}{0.0062}\,\randwin} & \vpass{55}{\estat{\eneg{0.0039}}{0.0115}\,\randwin} & \vfail{65}{\estat{\epos{0.0043}}{0.0092}\,\randwin} \\
\grayrow
\frontiermethod{claude}{Claude Opus 5} & \vpass{18}{\best{\estat{\epos{0.0000}}{0.0005}}\,\randwin} & \vpass{18}{\estat{\epos{0.0001}}{0.0003}\,\randwin} & \vfail{40}{\estat{\epos{0.0054}}{0.0073}\,\randwin} & \vpass{47}{\estat{\eneg{0.0031}}{0.0186}\,\randwin} & \vfail{55}{\estat{\epos{0.0037}}{0.0082}\,\randwin} \\
\grayrow
\frontiermethod{claude}{Claude Fable 5} & \vfail{51}{\estat{\epos{0.0086}}{0.0147}\,\randwin} & \vpass{18}{\runnerup{\estat{\epos{0.0001}}{0.0009}}\,\randwin} & \vfail{21}{\estat{\epos{0.0017}}{0.0063}\,\randwin} & \vfail{33}{\estat{\epos{0.0077}}{0.0136}\,\randwin} & \vpass{18}{\runnerup{\estat{\epos{0.0015}}{0.0030}}\,\randwin} \\
\grayrow
\arrayrulecolor{black!18}\cmidrule(lr){1-6}
\textbf{\ours{}} ($K{=}7$) & \vpass{55}{\estat{\eneg{0.0007}}{0.0015}\,\randwin} & \vfail{33}{\estat{\epos{0.0076}}{0.0176}\,\randwin} & \vpass{18}{\best{\estat{\epos{0.0006}}{0.0031}}\,\randwin} & \vfail{51}{\estat{\epos{0.0162}}{0.0438}\,\randwin} & \vpass{55}{\estat{\eneg{0.0040}}{0.0118}\,\randwin} \\
\arrayrulecolor{black}\midrule
\multicolumn{6}{c}{\textbf{Measurement efficiency: shot gain (\%; higher is better $\uparrow$)}} \\
\arrayrulecolor{black!30}\hline\arrayrulecolor{black!18}
\addlinespace[0.10em]\arrayrulecolor{black!18}
Random ($K{=}7$) & \pstat{28.0}{2.2} & \pstat{34.0}{1.4} & \pstat{49.2}{1.4} & \pstat{56.7}{0.8} & \pstat{84.7}{0.6} \\
\grayrow
\arrayrulecolor{black!18}\cmidrule(lr){1-6}
\frontiermethod{openai}{GPT-5.4} & \pstat{13.0}{0.3} & \pstat{21.4}{1.3} & \pstat{29.6}{1.0} & \pstat{39.2}{1.2} & \pstat{72.0}{1.8} \\
\grayrow
\frontiermethod{openai}{GPT-5.5} & \pstat{15.1}{0.8} & \pstat{22.3}{1.6} & \pstat{33.1}{0.3} & \shotbest{\pstat{60.6}{3.7}} & \pstat{73.5}{0.5} \\
\grayrow
\frontiermethod{openai}{GPT-5.6 Sol} & \runnerup{\pstat{16.3}{1.3}} & \pstat{25.3}{1.8} & \pstat{32.7}{2.1} & \pstat{40.5}{0.8} & \pstat{67.6}{0.8} \\
\grayrow
\frontiermethod{openai}{GPT-5.6 Terra} & \pstat{13.7}{1.8} & \pstat{23.6}{2.6} & \pstat{30.6}{0.6} & \pstat{39.7}{1.2} & \pstat{67.1}{1.0} \\
\grayrow
\frontiermethod{openai}{GPT-5.6 Luna} & \pstat{15.0}{0.9} & \runnerup{\pstat{26.2}{1.3}} & \runnerup{\pstat{36.3}{1.5}} & \pstat{41.0}{1.2} & \pstat{68.9}{1.3} \\
\grayrow
\arrayrulecolor{black!18}\cmidrule(lr){1-6}
\frontiermethod{claude}{Claude Haiku 4.5} & \pstat{13.0}{1.6} & \pstat{21.6}{1.0} & \pstat{32.2}{1.2} & \pstat{41.6}{1.9} & \pstat{68.2}{1.0} \\
\grayrow
\frontiermethod{claude}{Claude Sonnet 5} & \pstat{11.6}{0.6} & \pstat{21.3}{1.4} & \pstat{29.5}{0.9} & \runnerup{\pstat{57.6}{4.4}} & \runnerup{\pstat{77.4}{4.4}} \\
\grayrow
\frontiermethod{claude}{Claude Opus 5} & \pstat{12.4}{1.0} & \pstat{20.7}{1.0} & \pstat{28.9}{0.7} & \pstat{54.0}{3.6} & \pstat{76.0}{2.1} \\
\grayrow
\frontiermethod{claude}{Claude Fable 5} & \pstat{14.6}{2.2} & \pstat{21.1}{1.5} & \pstat{29.4}{0.9} & \pstat{29.0}{6.8} & \pstat{66.9}{1.0} \\
\grayrow
\arrayrulecolor{black!18}\cmidrule(lr){1-6}
\textbf{\ours{}} ($K{=}7$) & \shotbest{\pstat{27.3}{2.5}} & \shotbest{\pstat{32.5}{2.4}} & \shotbest{\pstat{47.9}{1.4}} & \pstat{55.1}{0.5} & \shotbest{\pstat{83.7}{0.4}} \\
\arrayrulecolor{black}\bottomrule
\end{tabular*}
\endgroup
\label{tab:v87_frontier_depth2}
\end{table*}

\begin{table*}[t]
\centering
\caption{\textbf{Data-exposed frontier-model online VQE at depth 3.} The train/test split, $\tau=0.995$, four-refresh $K{=}7$ protocol, and reporting conventions match Table~\ref{tab:v87_frontier_depth1}.}
\vspace{0.25em}
\setlength{\tabcolsep}{0.55mm}
\renewcommand{\arraystretch}{1.04}
\fontsize{7.6}{9.1}\selectfont
\begingroup
\definecolor{vistaheatgreen}{HTML}{A8D8A8}
\definecolor{vistaheatred}{HTML}{F3A6B7}
\newcommand{\frontierlogo}[1]{\raisebox{-0.28ex}{\includegraphics[height=1.10em]{llm_logos/#1.png}}}
\newcommand{\frontiermethod}[2]{\frontierlogo{#1}\,#2}
\newcommand{\vpass}[2]{\cellcolor{vistaheatgreen!#1!white}\textcolor{black}{#2}}
\newcommand{\vfail}[2]{\cellcolor{vistaheatred!#1!white}\textcolor{black}{#2}}
\newcommand{\epos}[1]{+#1}
\newcommand{\eneg}[1]{\textminus#1}
\newcommand{\estat}[2]{#1\,\(\pm\)\,#2}
\newcommand{\pstat}[2]{#1\,\(\pm\)\,#2\%}
\newcommand{\best}[1]{\textbf{#1}}
\newcommand{\runnerup}[1]{\underline{#1}}
\newcommand{\shotbest}[1]{\cellcolor{black!12}\best{#1}}
\newcommand{\randwin}{\textcolor{black}{\ding{51}}}
\newcommand{\randloss}{\textcolor{black}{\ding{55}}}
\newcommand{\grayrow}{\specialrule{0.20pt}{0.30mm}{0.30mm}}
\arrayrulecolor{black!18}
\begin{tabular*}{\textwidth}{@{\extracolsep{\fill}}lccccc@{}}
\arrayrulecolor{black}\toprule
Method & H2 (4q) & HeH+ (4q) & H3+ (6q) & H4 chain (8q) & LiH (12q) \\
\midrule\arrayrulecolor{black!18}
\multicolumn{6}{c}{\textbf{Energy fidelity $\downarrow$: $\dE$ vs. Full (Ha)}} \\
\arrayrulecolor{black!30}\hline\arrayrulecolor{black!18}
\addlinespace[0.10em]
Random ($K{=}7$) & \estat{\epos{0.0730}}{0.0859} & \estat{\epos{0.2617}}{0.2130} & \estat{\epos{0.2750}}{0.0765} & \estat{\epos{0.2773}}{0.2665} & \estat{\epos{2.3928}}{0.4840} \\
\grayrow
\arrayrulecolor{black!18}\cmidrule(lr){1-6}
\frontiermethod{openai}{GPT-5.4} & \vfail{65}{\estat{\epos{0.0023}}{0.0032}\,\randwin} & \vpass{18}{\estat{\epos{0.0006}}{0.0016}\,\randwin} & \vpass{55}{\estat{\eneg{0.0018}}{0.0029}\,\randwin} & \vfail{20}{\best{\estat{\epos{0.0019}}{0.0031}}\,\randwin} & \vfail{24}{\estat{\epos{0.0018}}{0.0076}\,\randwin} \\
\grayrow
\frontiermethod{openai}{GPT-5.5} & \vpass{18}{\estat{\epos{0.0010}}{0.0036}\,\randwin} & \vpass{43}{\estat{\eneg{0.0003}}{0.0027}\,\randwin} & \vpass{18}{\estat{\epos{0.0007}}{0.0036}\,\randwin} & \vfail{29}{\estat{\epos{0.0134}}{0.0121}\,\randwin} & \vpass{29}{\runnerup{\estat{\eneg{0.0012}}{0.0122}}\,\randwin} \\
\grayrow
\frontiermethod{openai}{GPT-5.6 Sol} & \vpass{18}{\estat{\epos{0.0014}}{0.0025}\,\randwin} & \vpass{24}{\best{\estat{\eneg{0.0001}}{0.0052}}\,\randwin} & \vfail{21}{\estat{\epos{0.0017}}{0.0049}\,\randwin} & \vfail{25}{\estat{\epos{0.0080}}{0.0056}\,\randwin} & \vpass{18}{\estat{\epos{0.0012}}{0.0099}\,\randwin} \\
\grayrow
\frontiermethod{openai}{GPT-5.6 Terra} & \vpass{18}{\runnerup{\estat{\epos{0.0005}}{0.0018}}\,\randwin} & \vpass{18}{\estat{\epos{0.0004}}{0.0072}\,\randwin} & \vfail{65}{\estat{\epos{0.0079}}{0.0100}\,\randwin} & \vfail{23}{\estat{\epos{0.0057}}{0.0053}\,\randwin} & \vpass{47}{\estat{\eneg{0.0036}}{0.0046}\,\randwin} \\
\grayrow
\frontiermethod{openai}{GPT-5.6 Luna} & \vpass{37}{\estat{\eneg{0.0010}}{0.0046}\,\randwin} & \vfail{21}{\estat{\epos{0.0018}}{0.0044}\,\randwin} & \vfail{29}{\estat{\epos{0.0028}}{0.0121}\,\randwin} & \vfail{26}{\estat{\epos{0.0097}}{0.0088}\,\randwin} & \vfail{65}{\estat{\epos{0.0037}}{0.0125}\,\randwin} \\
\grayrow
\arrayrulecolor{black!18}\cmidrule(lr){1-6}
\frontiermethod{claude}{Claude Haiku 4.5} & \vpass{38}{\estat{\eneg{0.0011}}{0.0050}\,\randwin} & \vpass{18}{\estat{\epos{0.0012}}{0.0015}\,\randwin} & \vpass{18}{\estat{\epos{0.0008}}{0.0039}\,\randwin} & \vfail{65}{\estat{\epos{0.0631}}{0.1113}\,\randwin} & \vfail{65}{\estat{\epos{0.0037}}{0.0101}\,\randwin} \\
\grayrow
\frontiermethod{claude}{Claude Sonnet 5} & \vpass{55}{\estat{\eneg{0.0021}}{0.0020}\,\randwin} & \vpass{18}{\estat{\epos{0.0001}}{0.0014}\,\randwin} & \vpass{18}{\estat{\epos{0.0016}}{0.0011}\,\randwin} & \vfail{28}{\estat{\epos{0.0131}}{0.0133}\,\randwin} & \vpass{40}{\estat{\eneg{0.0026}}{0.0122}\,\randwin} \\
\grayrow
\frontiermethod{claude}{Claude Opus 5} & \vpass{26}{\best{\estat{\eneg{0.0003}}{0.0009}}\,\randwin} & \vpass{55}{\estat{\eneg{0.0005}}{0.0032}\,\randwin} & \vpass{18}{\runnerup{\estat{\epos{0.0004}}{0.0035}}\,\randwin} & \vfail{26}{\estat{\epos{0.0095}}{0.0099}\,\randwin} & \vfail{35}{\estat{\epos{0.0023}}{0.0109}\,\randwin} \\
\grayrow
\frontiermethod{claude}{Claude Fable 5} & \vpass{18}{\estat{\epos{0.0014}}{0.0022}\,\randwin} & \vpass{18}{\runnerup{\estat{\epos{0.0001}}{0.0012}}\,\randwin} & \vpass{18}{\best{\estat{\epos{0.0001}}{0.0009}}\,\randwin} & \vfail{20}{\runnerup{\estat{\epos{0.0019}}{0.0045}}\,\randwin} & \vpass{22}{\best{\estat{\eneg{0.0003}}{0.0039}}\,\randwin} \\
\grayrow
\arrayrulecolor{black!18}\cmidrule(lr){1-6}
\textbf{\ours{}} ($K{=}7$) & \vpass{18}{\estat{\epos{0.0010}}{0.0027}\,\randwin} & \vfail{65}{\estat{\epos{0.0108}}{0.0118}\,\randwin} & \vfail{30}{\estat{\epos{0.0029}}{0.0095}\,\randwin} & \vfail{21}{\estat{\epos{0.0029}}{0.0113}\,\randwin} & \vpass{55}{\estat{\eneg{0.0046}}{0.0040}\,\randwin} \\
\arrayrulecolor{black}\midrule
\multicolumn{6}{c}{\textbf{Measurement efficiency: shot gain (\%; higher is better $\uparrow$)}} \\
\arrayrulecolor{black!30}\hline\arrayrulecolor{black!18}
\addlinespace[0.10em]\arrayrulecolor{black!18}
Random ($K{=}7$) & \pstat{19.3}{1.6} & \pstat{26.3}{1.0} & \pstat{32.1}{1.4} & \pstat{39.0}{0.9} & \pstat{77.5}{0.9} \\
\grayrow
\arrayrulecolor{black!18}\cmidrule(lr){1-6}
\frontiermethod{openai}{GPT-5.4} & \pstat{7.7}{1.5} & \pstat{13.9}{1.3} & \pstat{19.6}{0.8} & \pstat{27.2}{2.3} & \pstat{58.6}{1.9} \\
\grayrow
\frontiermethod{openai}{GPT-5.5} & \pstat{9.7}{0.4} & \pstat{15.9}{0.8} & \pstat{17.6}{0.3} & \shotbest{\pstat{39.8}{2.6}} & \pstat{70.5}{2.7} \\
\grayrow
\frontiermethod{openai}{GPT-5.6 Sol} & \runnerup{\pstat{10.1}{0.6}} & \pstat{18.3}{0.7} & \pstat{19.7}{1.4} & \pstat{28.8}{0.7} & \pstat{56.4}{0.5} \\
\grayrow
\frontiermethod{openai}{GPT-5.6 Terra} & \pstat{7.9}{0.1} & \pstat{17.2}{0.7} & \pstat{20.6}{1.6} & \pstat{28.4}{0.8} & \pstat{55.5}{1.0} \\
\grayrow
\frontiermethod{openai}{GPT-5.6 Luna} & \pstat{10.0}{1.1} & \runnerup{\pstat{19.2}{0.7}} & \runnerup{\pstat{24.4}{0.9}} & \pstat{28.6}{0.5} & \pstat{57.1}{2.5} \\
\grayrow
\arrayrulecolor{black!18}\cmidrule(lr){1-6}
\frontiermethod{claude}{Claude Haiku 4.5} & \pstat{7.6}{1.3} & \pstat{14.1}{1.1} & \pstat{19.1}{0.9} & \pstat{29.0}{0.9} & \pstat{56.6}{0.8} \\
\grayrow
\frontiermethod{claude}{Claude Sonnet 5} & \pstat{7.5}{0.4} & \pstat{12.5}{0.2} & \pstat{18.0}{0.9} & \pstat{28.1}{7.3} & \pstat{70.6}{1.4} \\
\grayrow
\frontiermethod{claude}{Claude Opus 5} & \pstat{7.3}{0.8} & \pstat{14.7}{1.0} & \pstat{21.2}{0.9} & \pstat{36.9}{3.5} & \runnerup{\pstat{71.3}{1.2}} \\
\grayrow
\frontiermethod{claude}{Claude Fable 5} & \pstat{9.8}{0.5} & \pstat{13.5}{0.8} & \pstat{17.2}{0.6} & \pstat{19.1}{1.2} & \pstat{55.8}{1.7} \\
\grayrow
\arrayrulecolor{black!18}\cmidrule(lr){1-6}
\textbf{\ours{}} ($K{=}7$) & \shotbest{\pstat{19.0}{2.4}} & \shotbest{\pstat{26.1}{1.2}} & \shotbest{\pstat{32.1}{1.2}} & \runnerup{\pstat{38.6}{0.7}} & \shotbest{\pstat{76.6}{1.3}} \\
\arrayrulecolor{black}\bottomrule
\end{tabular*}
\endgroup
\label{tab:v87_frontier_depth3}
\end{table*}
\fi

\endgroup

Appendix Table~\ref{tab:v87_frontier_token_cost} records 17.0--32.4M tokens
per frontier model; our controller has 984,454 parameters and a 3.78-MiB
checkpoint. These are distinct resource measures; shot gain excludes offline
training and classical costs (Appendix~\ref{sec:app_estimator_scope}).

\paragraph{Prompt, context, and feature audit.}
\label{sec:v80_llm_ablation}
\emph{Which parts of the interface improve scientific selection?} We freeze 12 Full-QWC
snapshots and test the four local LLMs under 15 prompt, context, output,
feature-deletion, and shuffle conditions (720 cases). The oracle matrix is
audit-only, and cardinality is fixed. Figure~\ref{fig:v80_trace_llama} shows
one observable Llama decision; the corresponding Qwen, DeepSeek, and GPT-OSS
traces are in Appendix Figures~\ref{fig:v80_trace_qwen},
\ref{fig:v80_trace_deepseek}, and~\ref{fig:v80_trace_gptoss}. These are
observable outputs paired with external feature audits. The harness performs
validation and VQE updates; frozen LLMs produce structured selection JSON
through text inference with tool execution disabled.

\newcommand{\methodtracepoint}[1]{%
\par\noindent\hangindent=1.15em\hangafter=1
\textbullet\hspace{0.45em}#1}
\newcommand{\methodtracebox}[4]{%
\par\noindent
\fcolorbox{#1}{#2}{%
\parbox{\dimexpr\linewidth-2\fboxsep-2\fboxrule\relax}{%
\normalsize
\textbf{\textcolor{#1}{#3}}
#4\par
}}%
\par
}
\newcommand{\methodtraceprose}[1]{%
\noindent\begin{minipage}{1\textwidth}
\normalsize #1
\end{minipage}}

\newcommand{\appTraceQwen}{%
\begin{figure*}[p]
\centering
\includegraphics[width=\textwidth]{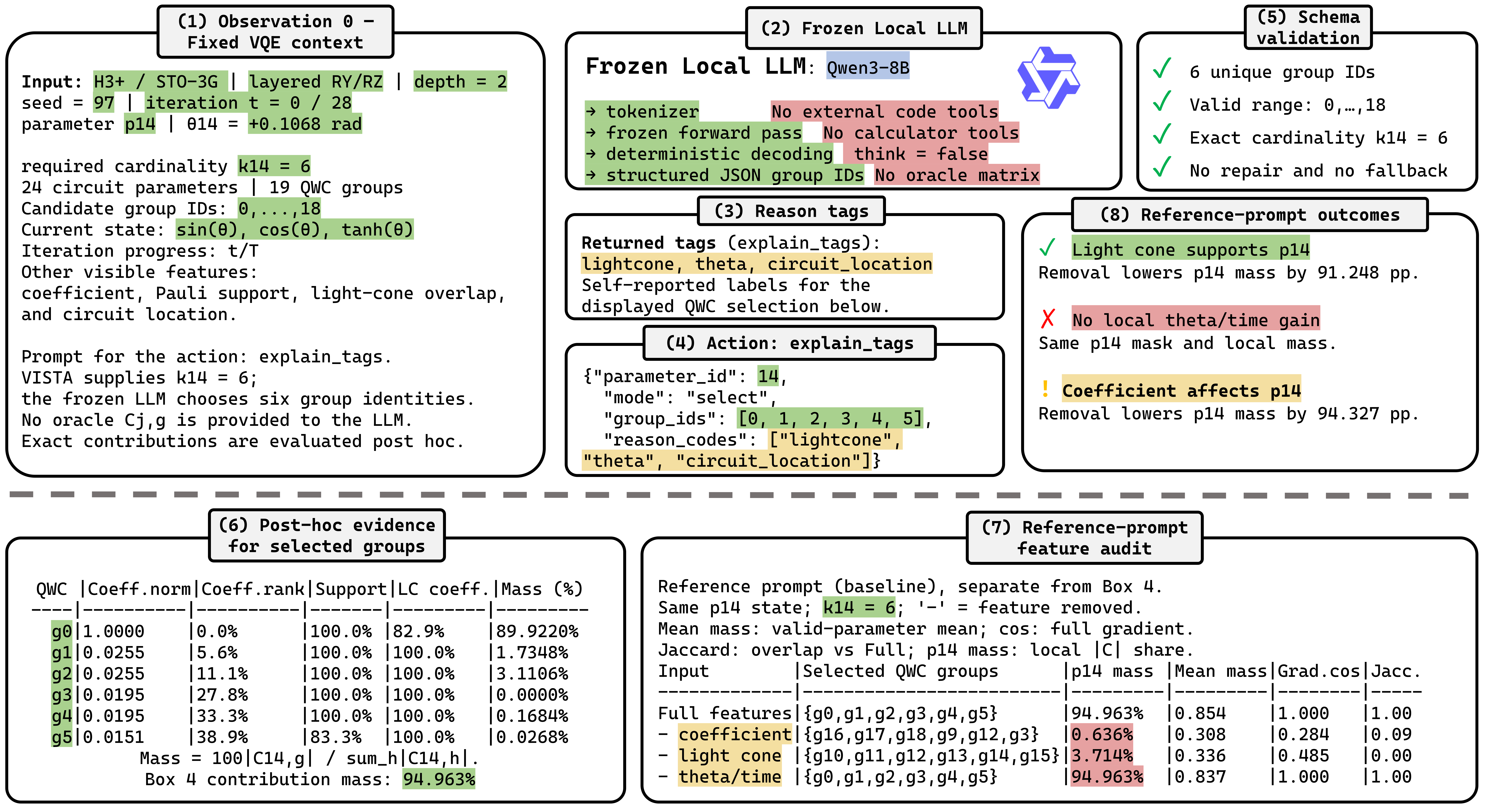}
\caption{\textbf{Qwen3-8B observable decision trace.}
On the fixed H$_3^+$, depth-2, seed-97 snapshot, Qwen selects
$\{g_0,\ldots,g_5\}$ and captures 94.963\% of the exact $p_{14}$ contribution
mass. Removing coefficient or light-cone features reduces that mass to
0.636\% and 3.714\%, respectively, whereas removing $\theta$/time features
leaves the reference mask unchanged. Box (4) uses the explanation prompt;
Box (7) uses the separate reference prompt. Thus deletion measures
reference-prompt selection quality.
Raw reason tags are \texttt{lightcone}, \texttt{theta}, and
\texttt{circuit\_location}, as shown in Box (3). Box (8) reports reference-prompt
utility. Exact contributions are audit-only (Appendix~\ref{sec:app_trace_provenance}).}
\label{fig:v80_trace_qwen}
\vspace{0.35em}
\setlength{\fboxsep}{4pt}
\methodtraceprose{\textbf{How to read the trace.}
Boxes (1)--(5) show the complete model-visible interaction: a fixed VQE
snapshot with 24 circuit parameters and 19 QWC groups is serialized into
numerical descriptors, the frozen LLM returns
reason tags and six group IDs, and a deterministic validator checks only
schema, range, and cardinality. Boxes (6)--(8) are post-hoc scientific
audits. They attach the exact contribution matrix only after the decision
and delete feature families under the reference prompt. Prompt
identity distinguishes the explanation action from its companion feature audit.

\smallskip
\textbf{Observed decision.}
Qwen's executable mask $\{g_0,\ldots,g_5\}$ captures 94.963\% of the local
contribution mass without repair or fallback. The trace therefore separates
two evaluation levels: executable output is assessed by the validator,
and physical relevance is assessed by exact captured mass.}
\par\vspace{0.30em}
\methodtracebox{green!50!black}{green!4}{Reference-prompt feature utility}
{\methodtracepoint{Removing coefficients lowers captured mass from 94.963\% to 0.636\%, a 94.327 percentage-point loss.}
\methodtracepoint{Removing light-cone features lowers it to 3.714\%, a 91.248 percentage-point loss.}}
\par\vspace{0.30em}
\methodtracebox{orange!85!black}{yellow!10}{State sensitivity at this snapshot}
{\methodtracepoint{Removing $\theta$/time leaves the selected $p_{14}$ mask and its 94.963\% local mass unchanged; mean mass across valid parameters changes from 0.854 to 0.837.}
\methodtracepoint{The unchanged local mask coexists with strong coefficient and light-cone effects. The explanation action and reference-prompt sensitivities provide complementary observations.}}
\par\vspace{0.35em}
\methodtraceprose{\textbf{Scientific interpretation.}
Qwen's reference selection responds strongly to physically meaningful
coefficient and light-cone inputs, and its selected groups retain most of the
exact row contribution. The aggregate online experiments extend this local
view across molecules and optimization trajectories.}
\end{figure*}
}
\newcommand{\mainTraceLlama}{%
\begin{figure*}[!t]
\centering
\includegraphics[width=\textwidth]{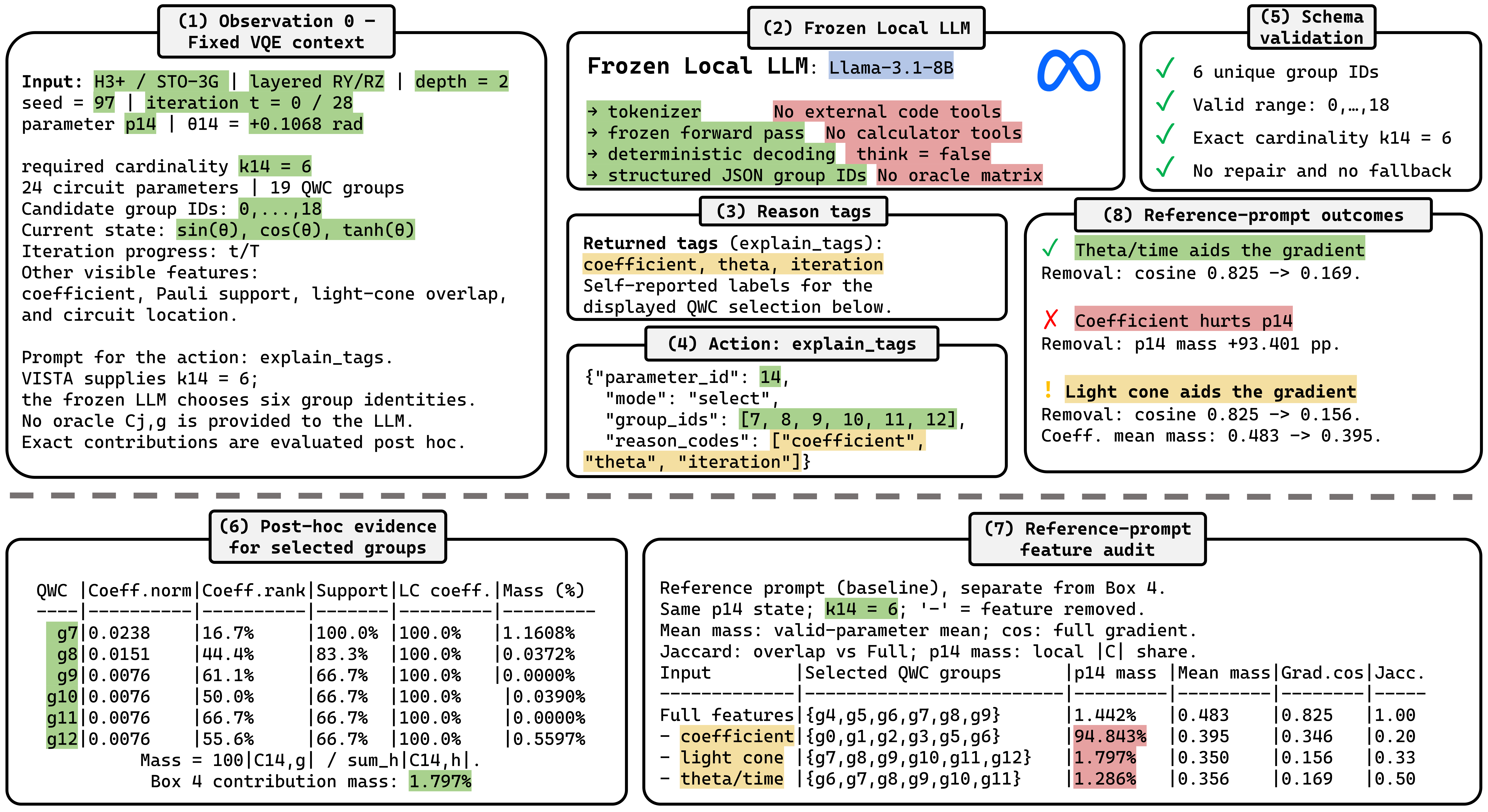}
\caption{\textbf{Llama-3.1-8B observable decision trace.}
On the shared H$_3^+$ snapshot, the explanation-prompt action captures
1.797\% of the exact $p_{14}$ contribution mass. Under the separate
reference prompt, deleting coefficient fields raises local mass from
1.442\% to 94.843\%, identifying a harmful coefficient effect on that
reference decision. Box (8) reports reference-prompt feature utility;
prompt-specific actions and metrics are documented in
Appendix~\ref{sec:app_trace_provenance}.}
\label{fig:v80_trace_llama}
\vspace{2em}
\setlength{\fboxsep}{4pt}
\methodtraceprose{\textbf{How to read the trace.}
The upper half records the model-visible input, frozen inference, returned
IDs, and schema validation under the explanation prompt. The lower half
attaches exact contributions and probes feature utility under the reference
prompt. These views answer complementary questions: \emph{what action the model
returns, how much physical relevance that action retains, and how the
reference selection responds to changes in its inputs.}

\smallskip
\textbf{Observed decision.}
Llama returns six unique, in-range IDs, $\{g_7,\ldots,g_{12}\}$, with the
requested cardinality and a valid schema. The action is directly executable,
yet its captured mass is 1.797\%. Because the measurement count is already
fixed, the scientific distinction lies in the identities of the measured
groups. The example makes the learned ranking task concrete: the controller
must allocate the same number of measurements to the groups carrying the
parameter's actual contribution.}
\par\vspace{0.30em}
\methodtracebox{red!65!black}{red!4}
{Physical information does not guarantee effective selection}
{\methodtracepoint{Physical descriptors can help or mislead the same
LLM: removing coefficients improves Llama's local selection but
reduces its average contribution coverage.}
\methodtracepoint{Removing light-cone or trajectory cues sharply
degrades gradient alignment, showing that their importance extends
beyond the contribution captured for a single parameter.}}

\par\vspace{0.30em}

\methodtracebox{orange!85!black}{yellow!10}
{Task-specific relevance guides scientific decisions}
{\methodtracepoint{Our controller learns which measurement groups
matter for each parameter. Matched-budget comparisons support the
value of this learned ranking for energy--cost performance.}
\methodtracepoint{These findings motivate expert-guided LLM workflows
that prioritize task-relevant physical information and validate
decisions through their scientific outcomes.}}
\par\vspace{0.35em}
\methodtraceprose{\textbf{From one decision to closed-loop performance.}
Captured mass measures absolute contribution coverage for one parameter,
while gradient cosine assesses alignment with the full gradient.
Closed-loop evaluations assess the cumulative impact of repeated
selections on final energy and measurement cost, extending the local
trace analysis to complete optimization trajectories.

\smallskip
\textbf{Across the feature audit.}
Over the paired audit, deleting coefficients or light-cone features reduces
captured mass by 6.4 and 4.9 percentage points
(Figure~\ref{fig:v80_feature_attribution}).
Figure~\ref{fig:v80_reason_consistency} relates these reference-prompt effects
to separately reported tags. Positive and negative deletion effects
distinguish beneficial and harmful feature utility, with model-specific
responses to the same physical descriptors.}
\end{figure*}
}

\newcommand{\appTraceGPTOss}{%
\begin{figure*}[p]
\centering
\includegraphics[width=\textwidth]{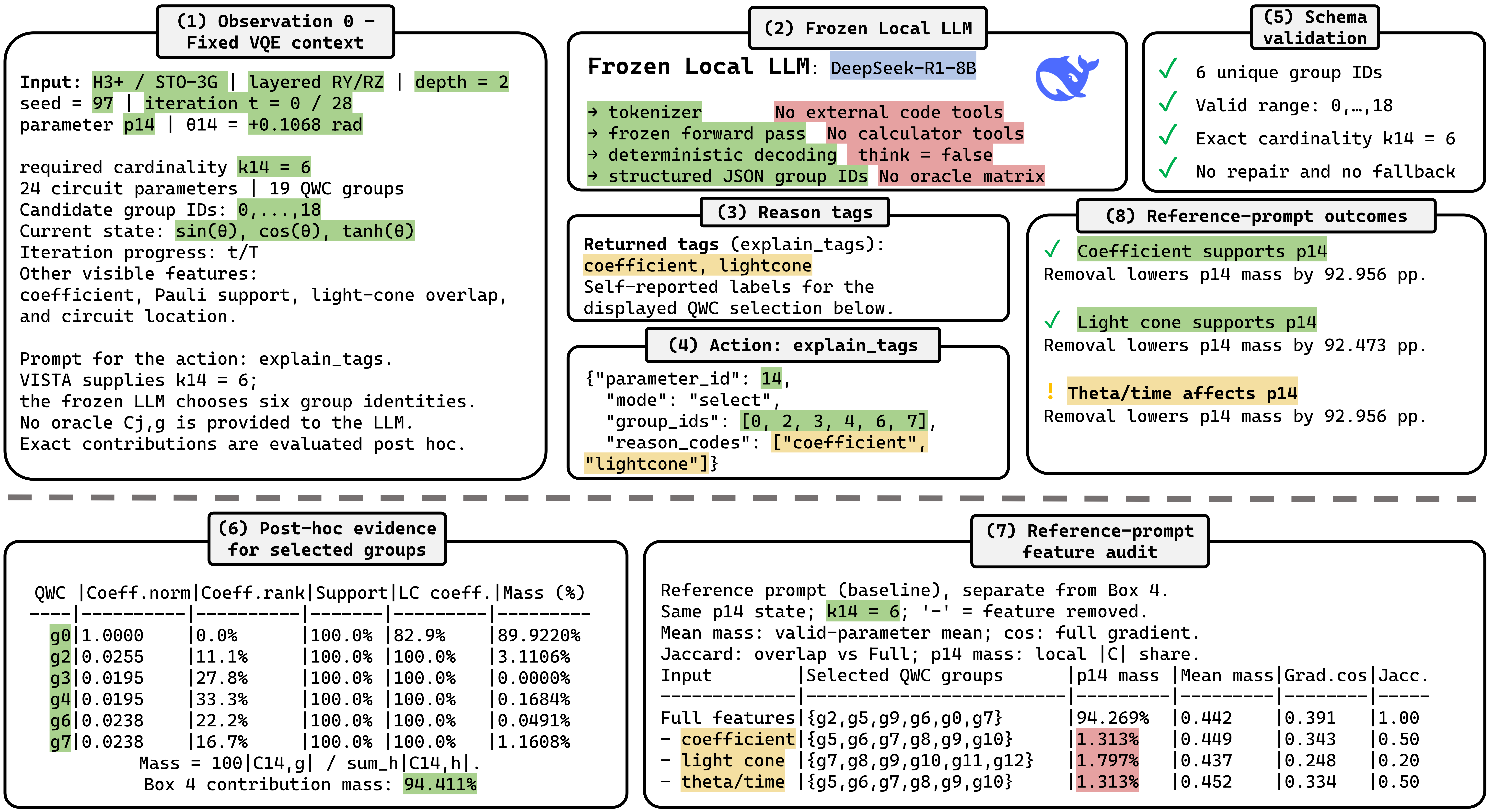}
\caption{\textbf{DeepSeek-R1-8B observable decision trace.}
DeepSeek selects $\{g_0,g_2,g_3,g_4,g_6,g_7\}$ and captures 94.411\% of the
exact $p_{14}$ mass; its full-feature reference mask captures 94.269\%.
Removing coefficient or light-cone features reduces local mass to 1.313\%
and 1.797\% under the separate reference prompt. Removing
$\theta$/time also reduces mass to 1.313\%, identifying current-state
sensitivity in the reference selection. Appendix~\ref{sec:app_trace_provenance}
records the explanation action and each reference-prompt intervention.}
\label{fig:v80_trace_deepseek}
\vspace{0.35em}
\setlength{\fboxsep}{4pt}
\methodtraceprose{\textbf{How to read the trace.}
The action panel contains only the IDs and observable reason tags emitted by
the frozen model. Exact $C_{j,k}$ values in the evidence panel are never
prompt features; they are attached afterward to score the action. The
counterfactual panel tests reference-prompt feature utility alongside the
separately logged explanation-producing action.

\smallskip
\textbf{Observed decision.}
DeepSeek chooses $\{g_0,g_2,g_3,g_4,g_6,g_7\}$ and captures 94.411\% of local
mass, close to the 94.269\% full-feature audit reference. Thus the
model produces both a valid action and a high-quality local ranking on this
snapshot.}
\par\vspace{0.30em}
\methodtracebox{green!50!black}{green!4}{Large reference-prompt quality effects}
{\methodtracepoint{Removing coefficients reduces local mass to 1.313\% (a 92.956-point loss); removing light-cone features reduces it to 1.797\% (a 92.473-point loss).}
\methodtracepoint{Both input families support the high-quality local reference selection by retaining the dominant group $g_0$.}}
\par\vspace{0.30em}
\methodtracebox{orange!85!black}{yellow!10}{Broader feature sensitivity}
{\methodtracepoint{Removing $\theta$/time under the reference prompt also reduces local mass to 1.313\%. All three deletions omit the dominant group $g_0$.}
\methodtracepoint{The reference ranking responds to current-state information as well as coefficients and light cones; the reported tags summarize the separate explanation action.}}
\par\vspace{0.35em}
\methodtraceprose{\textbf{Scientific interpretation.}
DeepSeek produces a high-quality action on this cell, while the companion
audit reveals sensitivity to several physical feature families. Together,
the action score and intervention profile characterize the model's local
selection behavior under the two documented prompt settings.}
\end{figure*}
\clearpage

\begin{figure*}[p]
\centering
\includegraphics[width=\textwidth]{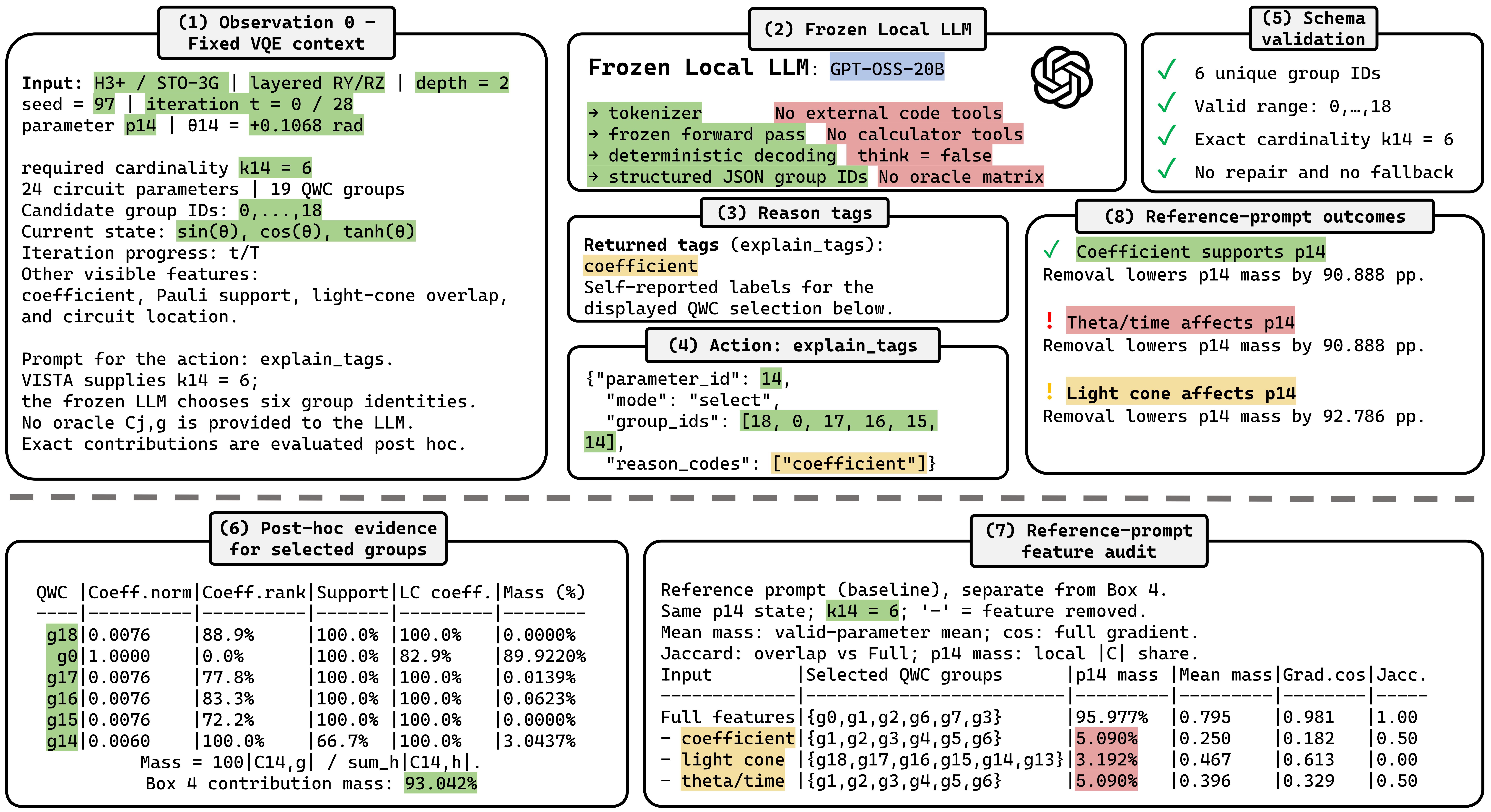}
\caption{\textbf{GPT-OSS-20B observable decision trace.}
GPT-OSS reports coefficient evidence and selects
$\{g_{18},g_0,g_{17},g_{16},g_{15},g_{14}\}$, capturing 93.042\% of the exact
$p_{14}$ mass; its reference mask captures 95.977\%. Removing coefficient,
light-cone, or $\theta$/time features reduces local mass to 5.090\%, 3.192\%,
and 5.090\%, respectively, under the separate reference prompt. This ranking
shows multi-feature quality sensitivity alongside the explanation action's
coefficient tag (Appendix~\ref{sec:app_trace_provenance}).}
\label{fig:v80_trace_gptoss}
\par\vspace{0.8em}
\setlength{\fboxsep}{4pt}
\methodtraceprose{\textbf{How to read the trace.}
This example pairs a high-mass action with reference-prompt interventions.
The frozen LLM generates selection JSON from serialized physical features;
the external Python harness performs validation and all VQE simulation.
The lower audit scores reference-prompt masks using exact contributions
and feature-family deletion, with LLM tool execution disabled.

\smallskip
\textbf{Observed decision.}
GPT-OSS selects $\{g_{18},g_0,g_{17},g_{16},g_{15},g_{14}\}$ and captures
93.042\% of local mass. This is a strong mask, alongside the 95.977\%
full-feature reference. Both actions concentrate measurements on groups
carrying most of the exact parameter-row contribution.}
\par\vspace{0.30em}
\methodtracebox{green!50!black}{green!4}{Reference-prompt feature utility}
{\methodtracepoint{Removing coefficients lowers local captured mass to 5.090\%, a 90.888 percentage-point loss.}
\methodtracepoint{Coefficients improve the reference-prompt selection. The separately prompted explanation action also reports a coefficient tag.}}
\par\vspace{0.30em}
\methodtracebox{orange!85!black}{yellow!10}{Multiple sensitive input families}
{\methodtracepoint{Removing light-cone or $\theta$/time features lowers mass to 3.192\% and 5.090\%, respectively.}
\methodtracepoint{The reference ranking draws useful information from several physical families, complementing the coefficient tag returned under the explanation prompt.}}
\par\vspace{0.35em}
\methodtraceprose{\textbf{Scientific interpretation.}
The deletion pattern is consistent with joint integration of several
serialized physical inputs. The external harness evaluates this behavior
through captured mass and feature sensitivity, with model-generated output
restricted to selection JSON. The successful action and its companion audit
provide two concrete views of GPT-OSS's physical selection capability.}
\end{figure*}
}

\mainTraceLlama

The Qwen, DeepSeek, and GPT-OSS traces (Appendix
Figures~\ref{fig:v80_trace_qwen}, \ref{fig:v80_trace_deepseek},
\ref{fig:v80_trace_gptoss}) capture 95.0\%, 94.4\%, and 93.0\% mass,
versus Llama's 1.8\%. Together, these cases reveal substantial model-specific
differences within the same physical task. Appendix~\ref{sec:app_trace_provenance}
links each action to its prompt and feature audit. Table~\ref{tab:v80_llm_ablation}
reports the 720-case study; Figures~\ref{fig:v83_token_group_attribution}--%
\ref{fig:v85_vista_feature_group_attribution} resolve input-to-output attribution. At the same measurement budget, valid outputs can therefore differ
sharply in their physical relevance.
\clearpage

\begin{figure*}[!t]
\centering
\includegraphics[width=\textwidth]{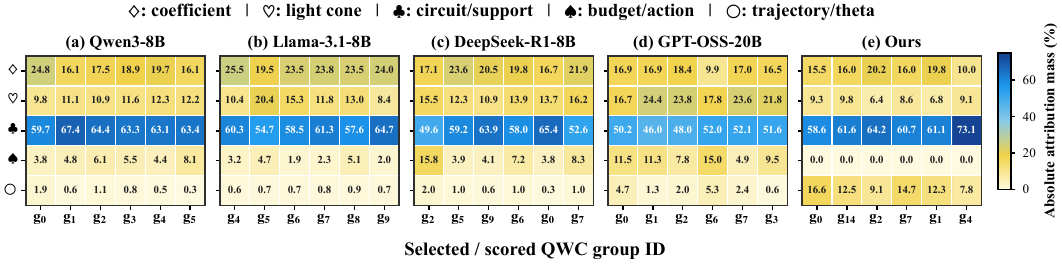}
\caption{\textbf{Feature-family attribution on a shared H$_3^+$ snapshot.}
Panels (a)--(d): frozen LLMs; (e): ours. Cells show within-model
absolute IG mass. Despite receiving the full input feature set, frozen LLMs
do not automatically make substantial use of trajectory information without
targeted intervention: it ranks last throughout the numerical audit
(Appendix~\ref{sec:app_ig_stability}). Our controller assigns 12.16\% to
current-state features. Scores and baselines:
Appendix~\ref{sec:app_ig_quality}; full maps: Appendix
Figures~\ref{fig:v83_token_group_attribution}--%
\ref{fig:v85_vista_feature_group_attribution}.}
\label{fig:feature_family_attribution_strip}
\end{figure*}

\paragraph{The learned controller uses optimization state.}
Our controller combines circuit/support information (63.24\%) with a clear
current-state contribution (12.16\%) in
Figure~\ref{fig:feature_family_attribution_strip}. Trajectory denotes
$[\sin\theta_j,\cos\theta_j,\tanh\theta_j,t/T]$: the angle and normalized
optimization step. Native-checkpoint attribution identifies their explicit
use in QWC ranking, combining observable structure with the circuit's
current state. The fixed-snapshot audit in
Appendix~\ref{sec:app_ig_stability} corroborates consistently low trajectory
attribution in all four frozen LLMs across quadrature and precision settings.

\paragraph{Context capacity and ranking quality play different roles.}
In Table~\ref{tab:v80_prompt_context_cost}, increasing capacity from 8K to
16K and 32K raises protocol success from 60.4\% to 81.2\% and 97.9\%;
valid-case captured mass is 0.562, 0.564, and 0.554. A minimal prompt reaches
100\% success at 140.6k prompt tokens per case, versus 160.2k for the domain
reference. Direct-ID output improves paired mass by 3.5 percentage points
at 95.8\% success (Table~\ref{tab:v80_llm_ablation}). Thus context capacity
improves executable coverage, while physical ranking determines selection quality. The best tested GPT-OSS prompt yields $\Delta E=+0.5056\pm0.6562$ Ha
at 52.4\% saving online
(Appendix Tables~\ref{tab:v80b_best_prompt_online}
and~\ref{tab:v80b_best_prompt_online_full}).

\newcommand{\appQwenAttribution}{%
\begin{figure*}[t]
\centering
\includegraphics[width=0.90\textwidth]{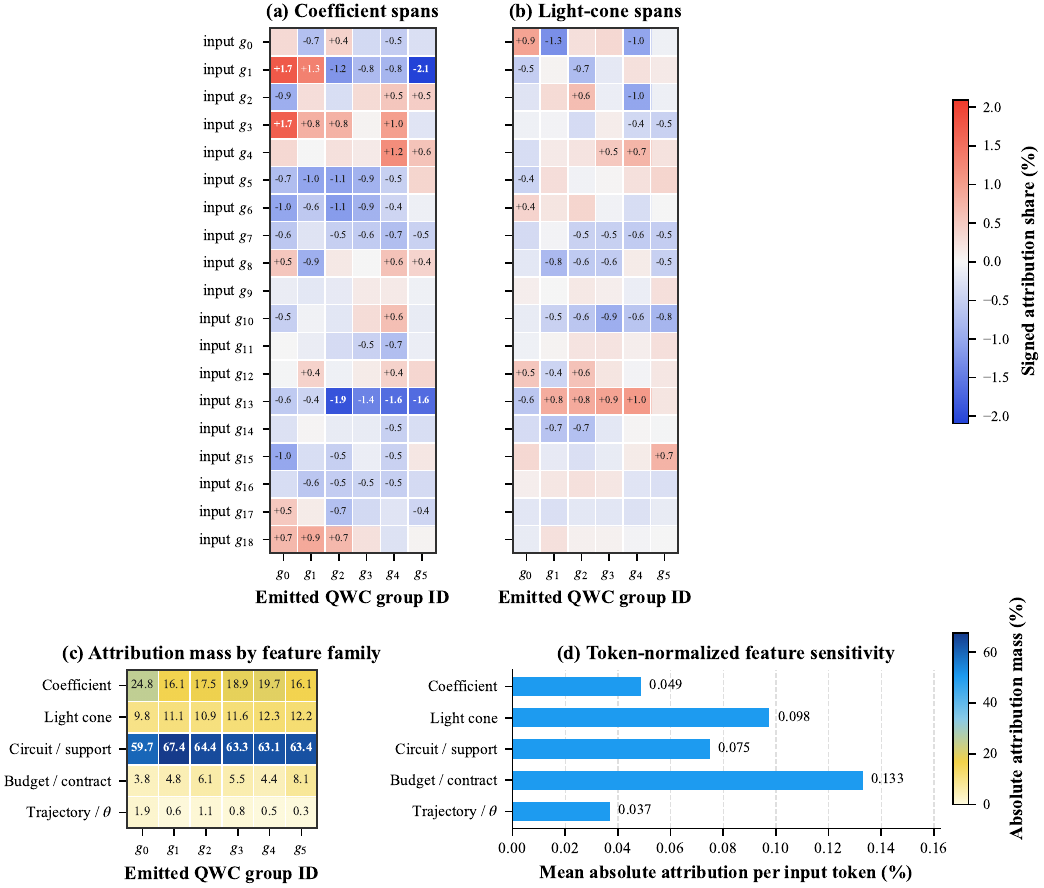}
\caption{\textbf{Token-to-QWC-group attribution for one representative frozen-Qwen decision.}
The audited state is H$_3^+$ with STO-3G, depth 2, seed 97, iteration 0, and
parameter $p_{14}$; \ourmethod{} supplies $k_{14}=6$, and Qwen emits
$[g_0,g_1,g_2,g_3,g_4,g_5]$. Panels (a,b) report the signed share of
integrated-gradient attribution from each input group's coefficient and
parameter--group light-cone spans to each emitted ID: red raises and blue lowers
the grammar-conditioned target score relative to the numeric-zero baseline.
Panel (c) aggregates absolute attribution mass by feature family, while panel
(d) divides that mass by the number of input tokens in each family. The
observable action comes from the audit-panel Ollama \texttt{qwen3:8b} run; gradients are
computed with the official frozen \texttt{Qwen/Qwen3-8B}
checkpoint~\citep{yang2025qwen3} in NF4
4-bit mode because the Ollama/GGUF runtime does not expose input gradients.
This matched white-box surrogate is not bitwise identical to the generation
runtime. The 32-point completeness audit has maximum residual 0.122; for the
informative first decision $g_0$, this is 1.61\% of its score change, but later columns have large relative errors.
The plot attributes the teacher-forced output score; Appendix~\ref{sec:app_ig_quality}
reports the per-column numerical audit.}
\label{fig:v83_token_group_attribution}
\vspace{0.35em}
{\setlength{\fboxsep}{5pt}%
\fcolorbox{red!70!black}{red!3}{%
\parbox{0.87\textwidth}{%
\small\textbf{\textcolor{red!70!black}{Key finding.}}
The saved estimates emphasize circuit/support, coefficient, and light-cone
spans on this snapshot. The first decision has a 1.61\% relative completeness
residual; later continuation scores have larger relative residuals.
The displayed profile describes this fixed-prefix attribution calculation.}}}
\vspace{-0.45em}
\end{figure*}
}

\newcommand{\appLlamaAttribution}{%
\begin{figure*}[t]
\centering
\includegraphics[width=0.90\textwidth]{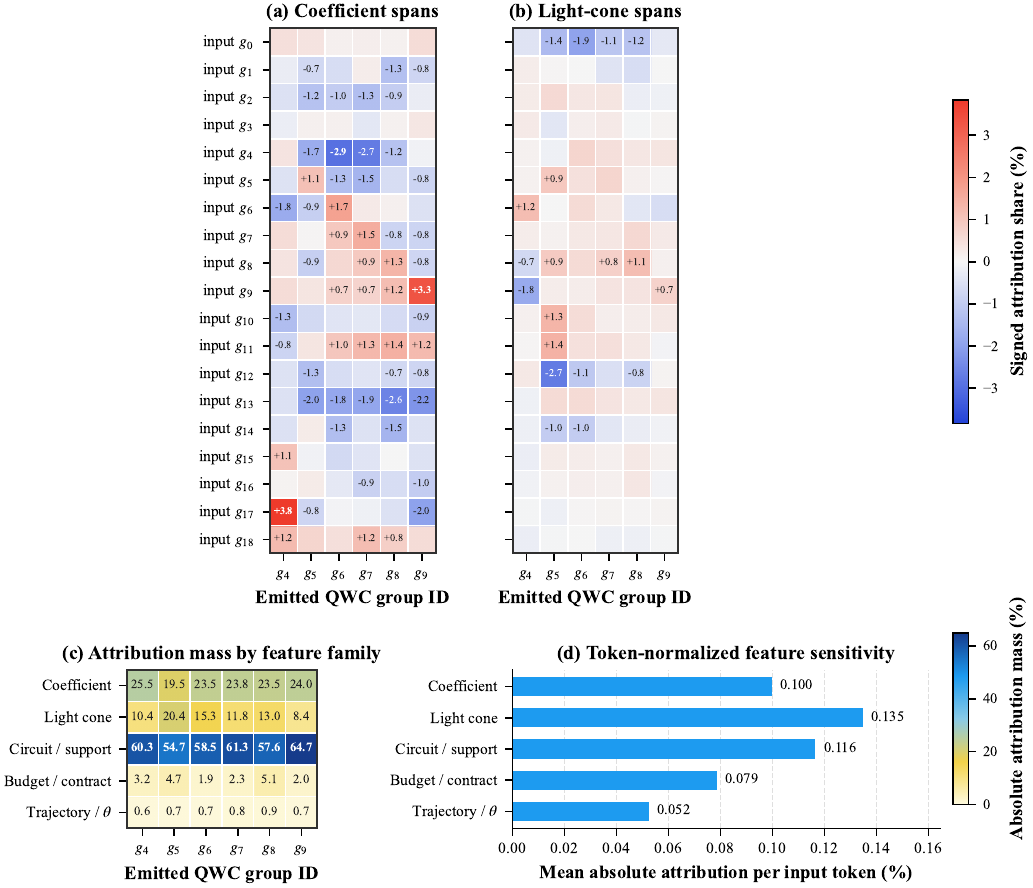}
\caption{\textbf{Token-to-QWC-group attribution for Llama-3.1-8B.}
The state and panel definitions match Figure~\ref{fig:v83_token_group_attribution},
while the accepted reference-prompt action is
$[g_4,g_5,g_6,g_7,g_8,g_9]$.  The observable action comes from Ollama
\texttt{llama3.1:8b}; gradients use a public NF4 conversion of the frozen
Meta-Llama-3.1-8B-Instruct checkpoint~\citep{grattafiori2024llama}, which is not
bitwise identical to the GGUF generation runtime.  The 32-point maximum
completeness residual is 1.149 log-probability units. Appendix~\ref{sec:app_ig_quality}
reports absolute and relative numerical residuals alongside the estimates.}
\label{fig:v84_llama_token_group_attribution}
\vspace{0.35em}
{\setlength{\fboxsep}{5pt}%
\fcolorbox{red!70!black}{red!3}{%
\parbox{0.87\textwidth}{%
\small\textbf{\textcolor{red!70!black}{Key finding.}}
The saved Llama map reports a 0.73\% trajectory share with the numerical
residuals tabulated in Appendix~\ref{sec:app_ig_quality}.
Its first ID $g_4$ has
only 1.06\% conditional probability (rank 15), whereas $g_5$--$g_9$ become
rank-one continuations, exposing a strong autoregressive list-completion
pattern in the white-box runtime.}}}
\vspace{-0.45em}
\end{figure*}
}

\newcommand{\appDeepSeekAttribution}{%
\begin{figure*}[t]
\centering
\includegraphics[width=0.90\textwidth]{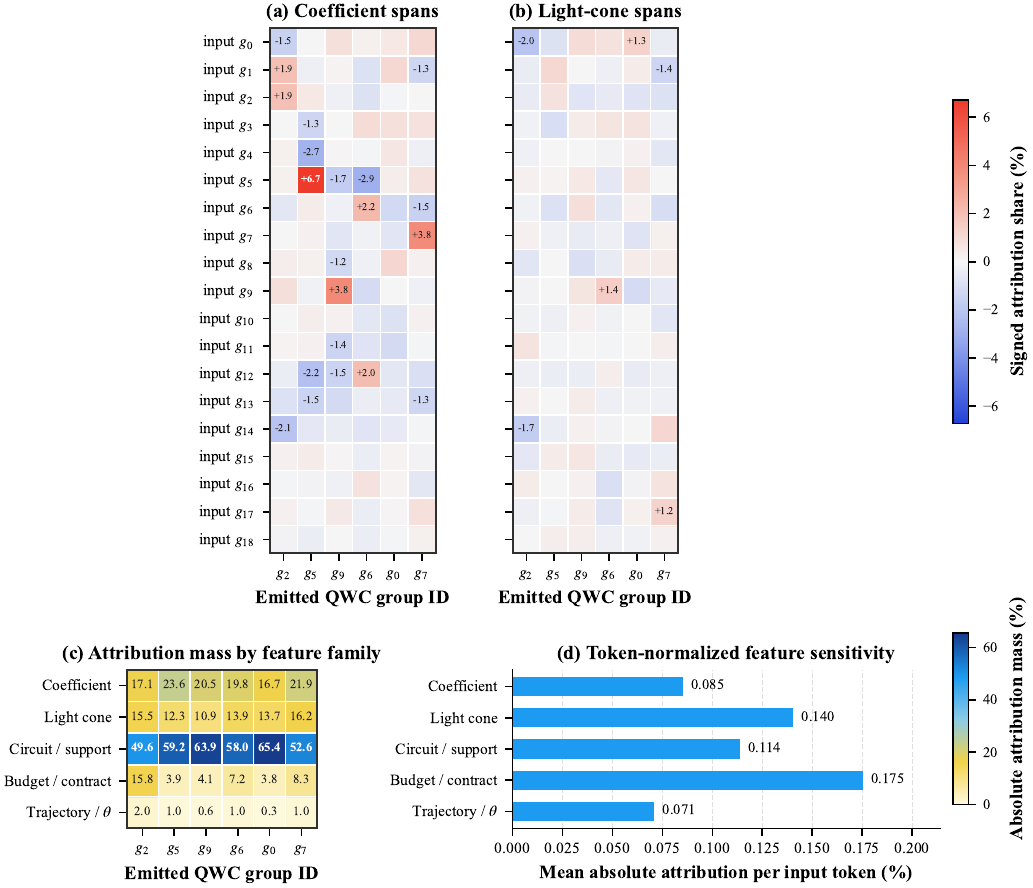}
\caption{\textbf{Token-to-QWC-group attribution for DeepSeek-R1-Distill-Llama-8B.}
The accepted reference-prompt action is $[g_2,g_5,g_9,g_6,g_0,g_7]$ under the same
H$_3^+$ snapshot and attribution protocol.  The action comes from Ollama
\texttt{deepseek-r1:8b}; white-box gradients use the official frozen
DeepSeek-R1-Distill-Llama-8B checkpoint~\citep{guo2025deepseek} with runtime
NF4 quantization.  This surrogate is not bitwise identical to the GGUF
runtime.  The 32-point maximum completeness residual is 0.437
log-probability units; absolute and relative numerical residuals are reported
in Appendix~\ref{sec:app_ig_quality}.}
\label{fig:v84_deepseek_token_group_attribution}
\vspace{0.35em}
{\setlength{\fboxsep}{5pt}%
\fcolorbox{red!70!black}{red!3}{%
\parbox{0.87\textwidth}{%
\small\textbf{\textcolor{red!70!black}{Key finding.}}
The saved DeepSeek map emphasizes static structure, coefficients, and
light-cone spans. The accepted IDs receive only 0.14--30.85\% conditional
probability and ranks 2--15 in the white-box runtime. The panel records
input sensitivity of these fixed-prefix scores, with runtime settings and
integration residuals specified by the accompanying audit.}}}
\vspace{-0.45em}
\end{figure*}
}

\newcommand{\appGPTOssAttribution}{%
\begin{figure*}[t]
\centering
\includegraphics[width=0.90\textwidth]{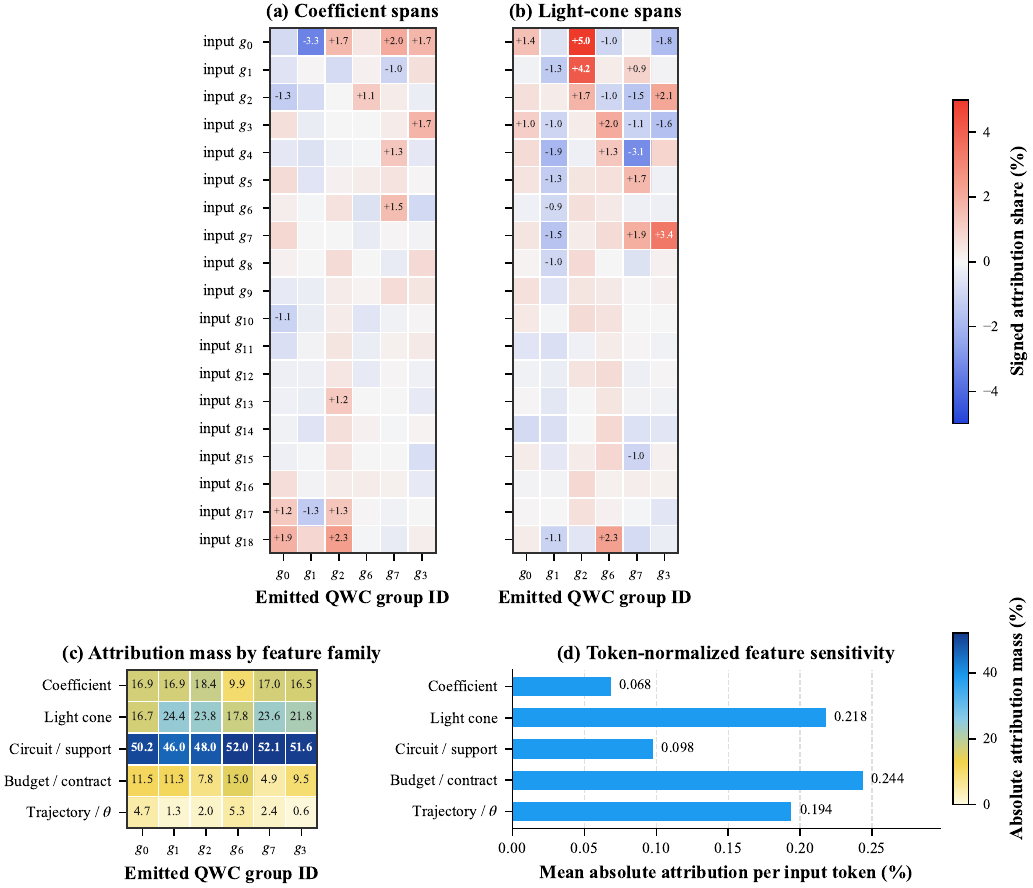}
\caption{\textbf{Token-to-QWC-group attribution for GPT-OSS-20B.}
The accepted reference-prompt action is $[g_0,g_1,g_2,g_6,g_7,g_3]$ under the same
snapshot and 32-point path.  The action comes from Ollama
\texttt{gpt-oss:20b}; gradients use the official frozen OpenAI
GPT-OSS-20B native-MXFP4 checkpoint~\citep{agarwal2025gpt}.  Flex Attention
and one forward graph per output bound memory without changing the scalar
scores or quadrature nodes.  The runtime is not bitwise identical to Ollama,
and the measured maximum completeness residual is 1.414
log-probability units; the full numerical audit appears in Appendix~\ref{sec:app_ig_quality}.}
\label{fig:v84_gptoss_token_group_attribution}
\vspace{0.35em}
{\setlength{\fboxsep}{5pt}%
\fcolorbox{red!70!black}{red!3}{%
\parbox{0.87\textwidth}{%
\small\textbf{\textcolor{red!70!black}{Key finding.}}
GPT-OSS's saved light-cone estimate is 21.36\%. Four emitted
IDs are high-confidence while $g_6$ and $g_3$ remain below 0.4\%, indicating
a heterogeneous fixed-prefix score profile in the white-box runtime.}}}
\vspace{-0.45em}
\end{figure*}
}

\newcommand{\appVistaAttribution}{%
\begin{figure*}[t]
\centering
\includegraphics[width=0.90\textwidth]{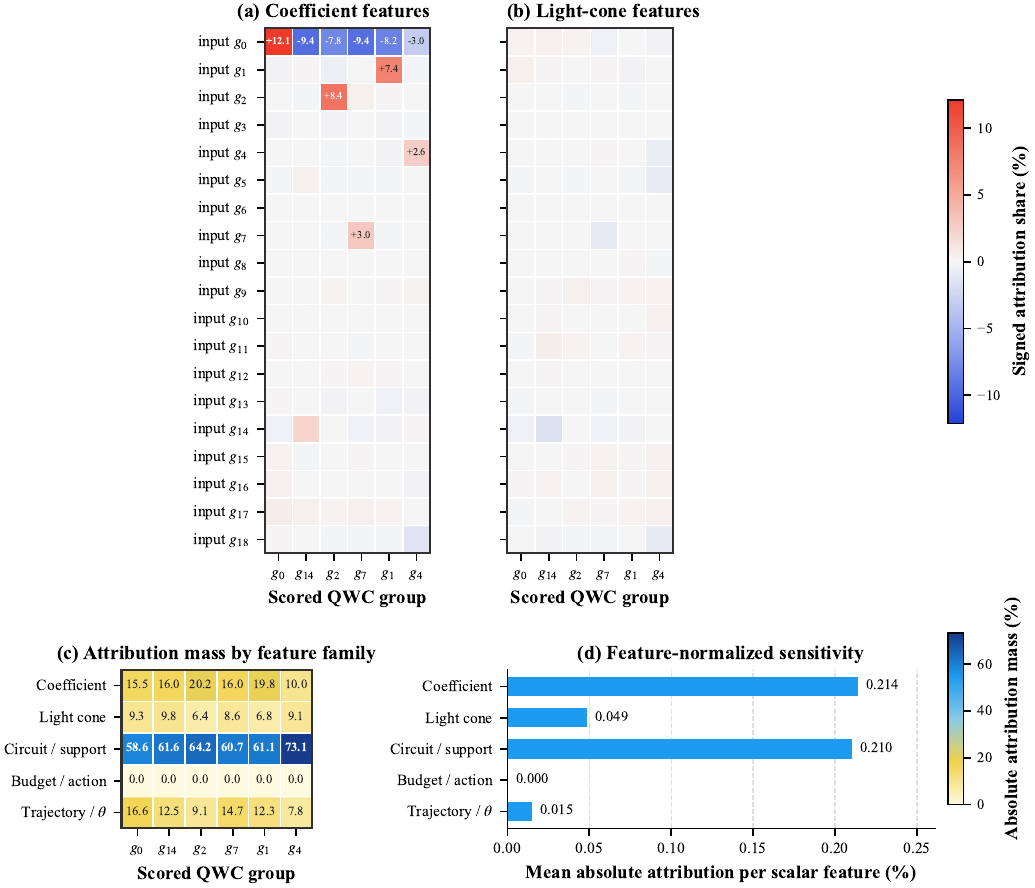}
\caption{\textbf{Model-native feature-to-QWC attribution for the trained
controller.}  We freeze the saved Transformer checkpoint and use
the same H$_3^+$/STO-3G, depth-2, seed-97, iteration-0, parameter-$p_{14}$
snapshot as in Figures~\ref{fig:v83_token_group_attribution}--%
\ref{fig:v84_gptoss_token_group_attribution}.  The six displayed columns are
our method's highest-scoring groups
($g_0,g_{14},g_2,g_7,g_1,g_4$), fixing the LLM comparison cardinality; the
native rounded budget head predicts five groups.  Panels (a) and (b) trace
coefficient and light-cone inputs from every QWC token to each scored group,
(c) aggregates absolute attribution by feature family, and (d) normalizes
that mass by the number of scalar features in each family.  Attributions use
1,024-point Gauss--Legendre integrated gradients from the training-mean
baseline in standardized feature space and include all parameter tokens,
group tokens, and parameter--group pair features.  The maximum completeness
residual is 0.0126 log-probability units, with maximum relative residual
0.46\%. This panel measures native feature-score dependence; the LLM panels
measure token-score dependence under their respective baselines.}
\label{fig:v85_vista_feature_group_attribution}
\vspace{0.35em}
{\setlength{\fboxsep}{5pt}%
\fcolorbox{red!70!black}{red!3}{%
\parbox{0.87\textwidth}{%
\small\textbf{\textcolor{red!70!black}{Key finding.}}
Our method combines strong circuit/support dependence (63.24\%) with
explicit use of trajectory features (12.16\%). Its top-six ranking contains
all five groups in the exact $\tau=0.995$ top-mass mask. The native-checkpoint
attribution connects current optimization state to the scores used for
parameter-specific measurement selection.}}}
\vspace{-0.45em}
\end{figure*}
}

\newcommand{\appReasonConsistency}{%
\begin{figure*}[t]
\centering
\includegraphics[width=0.96\textwidth]{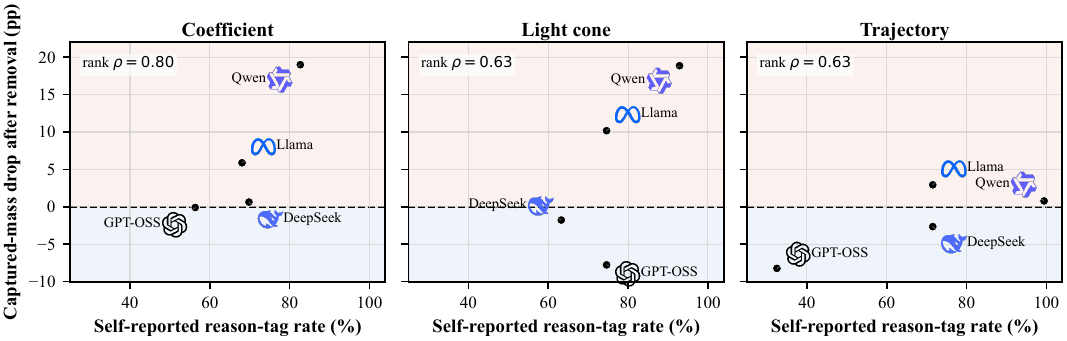}
\caption{\textbf{Self-reported reasons versus measured perturbation sensitivity.}
We audit three input families: coefficient denotes static Hamiltonian-coefficient
summaries for each QWC group; light cone denotes parameter--group support and
coefficient overlap; and trajectory denotes only the current optimization-state
proxy, not a stored history. Specifically, for parameter $j$ it contains
$[\sin\theta_j,\cos\theta_j,\tanh\theta_j,t/T]$, together with prompt-level
iteration $t$ and total iterations $T$. It does not include previous energies,
gradients, Adam moments, measurement outcomes, or earlier masks. For each frozen
LLM and feature family, the horizontal axis is the fraction of explained rows
that name the corresponding reason code (the trajectory rate combines
$\theta$ and iteration tags). The vertical axis is the paired decrease in exact
captured contribution mass when that entire feature family is removed from the
same Full-QWC snapshot, without retraining or changing the requested cardinality.
Positive values indicate beneficial input utility under the reference prompt;
negative values indicate harmful utility. Near-zero effects may reflect
redundancy. Reason tags come from a separate explanation prompt, giving a
cross-prompt association between reported tags and measured feature utility.
Black points are measured model--feature pairs and adjacent labels
carry the model-family logo. The rank correlation is descriptive over only four
models, with prompt provenance in Appendix~\ref{sec:app_trace_provenance}.}
\label{fig:v80_reason_consistency}
\vspace{-0.45em}
\end{figure*}
}

\paragraph{Long-horizon validation.}
\label{sec:converged_fci}
Appendix~\ref{sec:app_v82_converged_fci}, Table~\ref{tab:v82_converged_fci},
extends quantum optimization to 400 Adam steps with simulated shot noise
and frozen predictor weights. All 15 molecule--seed pairs reach within
1.6 mHa of their attainable exact-Adam reference at a checkpoint. Each molecule's mean
final energy stays within 0.18 mHa of Full-QWC, with 24.7--92.1\% shot savings.
FCI anchors the absolute energy scale; the paired Full-QWC comparison
isolates sparse measurement under the same ansatz. Savings persist across
the long-horizon optimization.

\begin{center}
\setlength{\fboxsep}{6pt}
\fcolorbox{red!70!black}{red!3}{%
\parbox{0.94\linewidth}{%
\textbf{\textcolor{red!70!black}{Takeaway.}}
Figure~\ref{fig:feature_family_attribution_strip} suggests that the learned controller better exploits
trajectory information, while the tested frozen LLMs struggle to
prioritize these cues without explicit human guidance.
This motivates expert-guided scientific workflows that help LLMs
recognize and effectively use information critical to the task. This raises an interesting question: \emph{What happens if we explicitly
guide LLMs to use trajectory information?} We will explore this question
across a broader range of quantum tasks in future work.}}
\end{center}

\section{Conclusion}
In this work, we uncover parameter-specific sparsity in QWC contributions to
variational quantum gradients and develop a compact learned controller
that exploits this structure for adjustable accuracy--cost trade-offs.
Our evaluation of LLMs without fine-tuning reveals both capabilities
and limitations: frontier models with retrieved training examples
reach millihartree-scale gaps to Full-QWC, while local-model audits
reveal uneven utility of the supplied physical information.
These findings motivate AI-assisted scientific workflows in which
human domain expertise guides problem representation, task objectives,
and physical validation. By characterizing model capabilities and
failure modes in a concrete scientific task, this work provides
an important first step toward developing and evaluating such
expert-guided workflows.




\clearpage
\subsection*{AI use statement}
In this work, we used generative AI tools for research ideation and feedback on methodology and experimental design, assistance with implementation, literature discovery, and editing the manuscript for clarity. The LLMs evaluated as measurement-selection baselines and their protocols are described separately in the paper. The authors reviewed AI-assisted material, checked reported numerical claims against experimental artifacts, and take full responsibility for the final text, claims, and research artifacts.

\subsection*{Ethics statement}
This study evaluates measurement selection in simulated quantum optimization using computationally generated Hamiltonians and structured physical features. It does not involve human participants or personal data.

\subsection*{Reproducibility statement}
Appendices~\ref{sec:app_reproducibility}--\ref{sec:app_evidence_audit}
specify implementations, metrics, and protocol provenance; preceding
appendices retain all experimental results.

\bibliography{reference}
\bibliographystyle{iclr2027_conference}

\clearpage
\appendix
\raggedbottom
\oursappendixcoretrue
\newcommand{\appendixinput}[1]{%
  \begingroup
  \let\section\subsection
  \let\subsection\subsubsection
  \input{#1}%
  \endgroup
}

\section{Extended Related Work}
\label{sec:app_related_work}

\paragraph{Measurement-frugal VQE.}
Commuting partitioning, unitary partitioning, and compatible-observable
measurement reduce the number of bases used to estimate the
Hamiltonian~\citep{zhao2020measurement,izmaylov2019unitary,yen2020measuring};
classical-shadow variants instead reuse randomized measurements across many
observables~\citep{huang2020predicting,hadfield2022measurements}. A
complementary line allocates a finite shot budget according to gradient or
observable variance~\citep{kubler2020adaptive,gu2021adaptive,zhu2024optimizing},
while operator sampling constructs stochastic gradient estimates from selected
Hamiltonian components~\citep{arrasmith2020operator}. These approaches optimize
the global grouping, the number of shots, or a sampled operator distribution.
Our method operates after a QWC partition is fixed and asks which groups should
be retained for each individual parameter.

\paragraph{Learned measurement adaptation.}
Learning has also been used to adapt generalized measurements and reduce
observation cost~\citep{garcia2021learning,liang2024artificial,anders2025adaptive}.
Our output space is different: it is a variable-size $P\times G$ relevance
matrix tied directly to parameter-shift gradient decomposition. To our
knowledge, prior work does not jointly learn this parameter-by-QWC grid and
charge cost at the physical measurement-setting level.

\paragraph{Ansatz adaptation and parameter freezing.}
Ansatz adaptation and parameter freezing reduce optimization cost along an
orthogonal axis. ADAPT-VQE grows an operator pool
selectively~\citep{grimsley2019adaptive,ikhtiarudin2025shot}, whereas TITAN
predicts parameters that can be frozen along a
trajectory~\citep{peng2026titan}. Meta-VQE learns across related Hamiltonians
to accelerate energy-profile estimation~\citep{cervera2021meta}. Our method
neither changes the ansatz nor requires a parameter to be frozen: for every
parameter that is evaluated, it sparsifies the QWC groups entering that
gradient. We therefore include both TITAN and TITAN+\ours{} to separate and
compose the two cost axes.

\clearpage
\section{Additional In-Distribution Results}
\label{sec:app_id_results}
\subsection{Complete Main-Panel Baselines}
\begin{table}[H]
\centering
\vspace{-0.5em}
\caption{\textbf{Complete version of Table~\ref{tab:v39_tau0995_main}.} This table preserves the default-threshold TITAN~\citep{peng2026titan} and TITAN+\ours{} rows, the cost-matched \(^{\dagger}\) rows, and all one-sided no-worse indicators omitted from the main table for readability. All molecules, seeds, online-VQE settings, and highlighting rules are unchanged.}
\vspace{0.25em}
\setlength{\tabcolsep}{0.45mm}
\renewcommand{\arraystretch}{0.88}
\tiny
\resizebox{1.0\linewidth}{!}{
\begin{tabular}{lll|ccccc}
\toprule
Setting & Metric & Method & H2 (4q) & HeH+ (4q) & H3+ (6q) & H4 chain (8q) & LiH (12q) \\
\midrule
\gcell{Depth 1} & \(\dE\) vs Full \(\downarrow\) & \ours{} & \textbf{\gbf{\(+0.0011\)}} & \textbf{\gbf{\(+0.0015\)}} & \textbf{\gbf{\(+0.0005\)}} & \textbf{\gbf{\(-0.0006\)}} & \rbf{\(+0.0023\)} \\
\gcell{Depth 1} & \(\dE\) vs Full \(\downarrow\) & Random & \rbf{\(+0.2359\)} & \rbf{\(+1.2220\)} & \rbf{\(+0.8916\)} & \rbf{\(+1.4117\)} & \rbf{\(+3.4982\)} \\
\gcell{Depth 1} & \(\dE\) vs Full \(\downarrow\) & Global coeff. & \rbf{\(+0.0038\)} & \rbf{\(+0.0019\)} & \rbf{\(+0.0122\)} & \rbf{\(+0.0089\)} & \textbf{\rbf{\(+0.0016\)}} \\
\gcell{Depth 1} & \(\dE\) vs Full \(\downarrow\) & TITAN~(\citeauthor{peng2026titan}) & \rbf{\(+0.0228\)} & \rbf{\(+0.1040\)} & \rbf{\(+0.0871\)} & \rbf{\(+0.0220\)} & \gbf{\(-0.0881\)} \\
\gcell{Depth 1} & \(\dE\) vs Full \(\downarrow\) & TITAN+\ours{} & \rbf{\(+0.0227\)} & \rbf{\(+0.1040\)} & \rbf{\(+0.0869\)} & \rbf{\(+0.0218\)} & \gbf{\(-0.0875\)} \\
\gcell{Depth 1} & \(\dE\) vs Full \(\downarrow\) & TITAN$^{\dagger}$ & \rbf{\(+0.0128\)} & \rbf{\(+0.1299\)} & \rbf{\(+0.0772\)} & \gbf{\(-0.0642\)} & \gbf{\(-0.0648\)} \\
\gcell{Depth 1} & \(\dE\) vs Full \(\downarrow\) & TITAN$^{\dagger}$+\ours{} & \rbf{\(+0.0127\)} & \rbf{\(+0.1297\)} & \rbf{\(+0.0776\)} & \gbf{\(-0.0649\)} & \gbf{\(-0.0646\)} \\
\cmidrule(lr){2-8}
\bcell{Depth 1} & Shot gain \(\uparrow\) & \ours{} & 40.4\% & 58.9\% & 63.0\% & 74.7\% & 86.9\% \\
\bcell{Depth 1} & Shot gain \(\uparrow\) & Random & 42.0\% & 60.7\% & 66.5\% & 79.2\% & 90.2\% \\
\bcell{Depth 1} & Shot gain \(\uparrow\) & Global coeff. & 41.7\% & 60.0\% & 62.9\% & 75.0\% & 87.0\% \\
\bcell{Depth 1} & Shot gain \(\uparrow\) & TITAN~(\citeauthor{peng2026titan}) & 61.8\% & 56.2\% & 58.2\% & 62.7\% & 71.9\% \\
\bcell{Depth 1} & Shot gain \(\uparrow\) & TITAN+\ours{} & \textbf{\underline{80.0\%}} & 90.5\% & 93.5\% & 96.1\% & 99.7\% \\
\bcell{Depth 1} & Shot gain \(\uparrow\) & TITAN$^{\dagger}$ & 43.8\% & 59.4\% & 62.7\% & 70.3\% & 81.8\% \\
\bcell{Depth 1} & Shot gain \(\uparrow\) & TITAN$^{\dagger}$+\ours{} & 66.8\% & \textbf{\underline{91.7\%}} & \textbf{\underline{95.0\%}} & \textbf{\underline{97.9\%}} & \textbf{\underline{99.9\%}} \\
\cmidrule(lr){2-8}
\gcell{Depth 1} & No-worse vs Full & \ours{} & \cmark & \cmark & \cmark & \cmark & \xmark \\
\gcell{Depth 1} & No-worse vs Full & Random & \xmark & \xmark & \xmark & \xmark & \xmark \\
\gcell{Depth 1} & No-worse vs Full & Global coeff. & \xmark & \xmark & \xmark & \xmark & \xmark \\
\gcell{Depth 1} & No-worse vs Full & TITAN~(\citeauthor{peng2026titan}) & \xmark & \xmark & \xmark & \xmark & \cmark \\
\gcell{Depth 1} & No-worse vs Full & TITAN+\ours{} & \xmark & \xmark & \xmark & \xmark & \cmark \\
\gcell{Depth 1} & No-worse vs Full & TITAN$^{\dagger}$ & \xmark & \xmark & \xmark & \cmark & \cmark \\
\gcell{Depth 1} & No-worse vs Full & TITAN$^{\dagger}$+\ours{} & \xmark & \xmark & \xmark & \cmark & \cmark \\
\midrule
\gcell{Depth 2} & \(\dE\) vs Full \(\downarrow\) & \ours{} & \textbf{\gbf{\(+0.0003\)}} & \gbf{\(-0.0011\)} & \textbf{\gbf{\(-0.0015\)}} & \textbf{\rbf{\(+0.0158\)}} & \textbf{\gbf{\(-0.0049\)}} \\
\gcell{Depth 2} & \(\dE\) vs Full \(\downarrow\) & Random & \rbf{\(+0.1656\)} & \rbf{\(+0.4704\)} & \rbf{\(+0.6527\)} & \rbf{\(+0.9343\)} & \rbf{\(+4.0043\)} \\
\gcell{Depth 2} & \(\dE\) vs Full \(\downarrow\) & Global coeff. & \gbf{\(+0.0015\)} & \textbf{\gbf{\(-0.0004\)}} & \rbf{\(+0.0233\)} & \rbf{\(+0.0379\)} & \gbf{\(-0.0049\)} \\
\gcell{Depth 2} & \(\dE\) vs Full \(\downarrow\) & TITAN~(\citeauthor{peng2026titan}) & \gbf{\(-0.0200\)} & \gbf{\(-0.1194\)} & \gbf{\(-0.0031\)} & \rbf{\(+0.1521\)} & \rbf{\(+0.0806\)} \\
\gcell{Depth 2} & \(\dE\) vs Full \(\downarrow\) & TITAN+\ours{} & \gbf{\(-0.0200\)} & \gbf{\(-0.1200\)} & \gbf{\(-0.0032\)} & \rbf{\(+0.1538\)} & \rbf{\(+0.0784\)} \\
\gcell{Depth 2} & \(\dE\) vs Full \(\downarrow\) & TITAN$^{\dagger}$ & \rbf{\(+0.0075\)} & \rbf{\(+0.0116\)} & \rbf{\(+0.0153\)} & \rbf{\(+0.0472\)} & \rbf{\(+0.2903\)} \\
\gcell{Depth 2} & \(\dE\) vs Full \(\downarrow\) & TITAN$^{\dagger}$+\ours{} & \rbf{\(+0.0072\)} & \rbf{\(+0.0129\)} & \rbf{\(+0.0140\)} & \rbf{\(+0.0493\)} & \rbf{\(+0.2920\)} \\
\cmidrule(lr){2-8}
\bcell{Depth 2} & Shot gain \(\uparrow\) & \ours{} & 25.6\% & 31.6\% & 44.9\% & 51.8\% & 82.6\% \\
\bcell{Depth 2} & Shot gain \(\uparrow\) & Random & 26.3\% & 33.2\% & 47.7\% & 53.7\% & 83.4\% \\
\bcell{Depth 2} & Shot gain \(\uparrow\) & Global coeff. & 25.7\% & 31.9\% & 44.7\% & 51.7\% & 82.5\% \\
\bcell{Depth 2} & Shot gain \(\uparrow\) & TITAN~(\citeauthor{peng2026titan}) & 54.0\% & 54.4\% & 50.0\% & 59.1\% & 77.1\% \\
\bcell{Depth 2} & Shot gain \(\uparrow\) & TITAN+\ours{} & \textbf{\underline{70.9\%}} & \textbf{\underline{75.7\%}} & \textbf{\underline{79.9\%}} & \textbf{\underline{86.1\%}} & 98.7\% \\
\bcell{Depth 2} & Shot gain \(\uparrow\) & TITAN$^{\dagger}$ & 23.8\% & 39.0\% & 41.9\% & 46.8\% & 82.7\% \\
\bcell{Depth 2} & Shot gain \(\uparrow\) & TITAN$^{\dagger}$+\ours{} & 40.2\% & 64.0\% & 74.5\% & 79.0\% & \textbf{\underline{99.3\%}} \\
\cmidrule(lr){2-8}
\gcell{Depth 2} & No-worse vs Full & \ours{} & \cmark & \cmark & \cmark & \xmark & \cmark \\
\gcell{Depth 2} & No-worse vs Full & Random & \xmark & \xmark & \xmark & \xmark & \xmark \\
\gcell{Depth 2} & No-worse vs Full & Global coeff. & \cmark & \cmark & \xmark & \xmark & \cmark \\
\gcell{Depth 2} & No-worse vs Full & TITAN~(\citeauthor{peng2026titan}) & \cmark & \cmark & \cmark & \xmark & \xmark \\
\gcell{Depth 2} & No-worse vs Full & TITAN+\ours{} & \cmark & \cmark & \cmark & \xmark & \xmark \\
\gcell{Depth 2} & No-worse vs Full & TITAN$^{\dagger}$ & \xmark & \xmark & \xmark & \xmark & \xmark \\
\gcell{Depth 2} & No-worse vs Full & TITAN$^{\dagger}$+\ours{} & \xmark & \xmark & \xmark & \xmark & \xmark \\
\bottomrule
\end{tabular}}
\label{tab:v47_table1_full_appendix}
\vspace{-0.5em}
\end{table}

\clearpage
\subsection{In-Distribution Aggregate and Budget Details}
\label{sec:app_id_budget_details}

\begin{table}[H]
\centering
\vspace{-0.5em}
\caption{\textbf{Depth-wise aggregate at $\tau=0.995$.} Mean $|\dE|$ averages seed-level absolute gaps over four seeds and five molecules at each depth. No-worse marks whether the aggregate mean signed $\dE$ is at most $1.6$ mHa above Full.}
\vspace{0.25em}
\setlength{\tabcolsep}{1.2mm}
\small
\begin{tabular}{l|ccc}
\toprule
Setting & Mean \(|\dE|\downarrow\) & No-worse & Mean shot gain \(\uparrow\) \\
\midrule
\grow Full-QWC reference & 0.00000 & \cmark & \gray{0.0\%} \\
\brow \ours{} depth 1 & 0.00150 & \cmark & \gbf{64.8\%} \\
\brow \ours{} depth 2 & 0.00978 & \xmark & \gbf{47.3\%} \\
\brow \ours{} depth 3 & 0.00571 & \cmark & \gbf{36.7\%} \\
\midrule
Random depth 1 & 1.45188 & \xmark & \gbf{67.7\%} \\
Random depth 2 & 1.24548 & \xmark & \gbf{48.9\%} \\
Random depth 3 & 0.82189 & \xmark & \gbf{37.3\%} \\
\midrule
\brow \ours{} all & 0.00567 & \cmark & \gbf{49.6\%} \\
Random all & 1.17308 & \xmark & \gbf{51.3\%} \\
\bottomrule
\end{tabular}
\label{tab:v39_tau0995_agg}
\vspace{-0.5em}
\end{table}

\begin{table}[H]
\centering
\vspace{-0.5em}
\caption{\textbf{Detailed budget usage at $\tau=0.995$.} All molecular Hamiltonians use the STO-3G minimal basis; molecule labels report the resulting qubit counts. Groups/param. reports the mean selected QWC groups per parameter, shot gain is relative to full $2PGS$, and No-worse vs Full uses the one-sided $1.6$ mHa criterion ($\dE \le 1.6$ mHa; lower energy is accepted). Boldface marks the better \ours{}-vs-Random entry for each molecule-depth comparison.}
\vspace{0.25em}
\setlength{\tabcolsep}{0.9mm}
\resizebox{1.0\linewidth}{!}{
\begin{tabular}{ll|c|ccc|ccc|ccc|c}
\toprule
Molecule & Method & Full QWC & \multicolumn{3}{c|}{Groups / param. \(\downarrow\)} & \multicolumn{3}{c|}{Shot gain \(\uparrow\)} & \multicolumn{3}{c|}{No-worse vs Full} & Best \(\dE\downarrow\) \\
\cmidrule(lr){4-6}\cmidrule(lr){7-9}\cmidrule(lr){10-12}
& & groups/param. & D1 & D2 & D3 & D1 & D2 & D3 & D1 & D2 & D3 & \\
\midrule
H2 (4q) & \ours{} & \gcell{5} & \bcell{2.98} & \bcell{3.72} & \bcell{\textbf{4.09}} & \gbf{40.4\%} & \gbf{25.6\%} & \textbf{\gbf{18.1\%}} & \cmark & \cmark & \cmark & \textbf{\gbf{\(+0.0003\)}} \\
H2 (4q) & Random & \gcell{5} & \bcell{\textbf{2.90}} & \bcell{\textbf{3.68}} & \bcell{4.11} & \textbf{\gbf{42.0\%}} & \textbf{\gbf{26.3\%}} & \gbf{17.9\%} & \xmark & \xmark & \cmark & \(-0.0145\) \\
\midrule
HeH+ (4q) & \ours{} & \gcell{11} & \bcell{4.52} & \bcell{7.53} & \bcell{\textbf{8.26}} & \gbf{58.9\%} & \gbf{31.6\%} & \textbf{\gbf{24.9\%}} & \cmark & \cmark & \xmark & \textbf{\gbf{\(-0.0011\)}} \\
HeH+ (4q) & Random & \gcell{11} & \bcell{\textbf{4.32}} & \bcell{\textbf{7.35}} & \bcell{8.30} & \textbf{\gbf{60.7\%}} & \textbf{\gbf{33.2\%}} & \gbf{24.6\%} & \xmark & \xmark & \xmark & \rbf{\(+0.0513\)} \\
\midrule
H3+ (6q) & \ours{} & \gcell{19} & \bcell{7.03} & \bcell{10.47} & \bcell{13.33} & \gbf{63.0\%} & \gbf{44.9\%} & \gbf{29.9\%} & \cmark & \cmark & \cmark & \textbf{\gbf{\(+0.0005\)}} \\
H3+ (6q) & Random & \gcell{19} & \bcell{\textbf{6.37}} & \bcell{\textbf{9.94}} & \bcell{\textbf{13.31}} & \textbf{\gbf{66.5\%}} & \textbf{\gbf{47.7\%}} & \textbf{\gbf{29.9\%}} & \xmark & \xmark & \xmark & \rbf{\(+0.2864\)} \\
\midrule
H4 chain (8q) & \ours{} & \gcell{75} & \bcell{18.96} & \bcell{36.15} & \bcell{48.60} & \gbf{74.7\%} & \gbf{51.8\%} & \gbf{35.2\%} & \cmark & \xmark & \cmark & \textbf{\gbf{\(-0.0006\)}} \\
H4 chain (8q) & Random & \gcell{75} & \bcell{\textbf{15.61}} & \bcell{\textbf{34.71}} & \bcell{\textbf{47.43}} & \textbf{\gbf{79.2\%}} & \textbf{\gbf{53.7\%}} & \textbf{\gbf{36.8\%}} & \xmark & \xmark & \xmark & \rbf{\(+0.2903\)} \\
\midrule
LiH (12q) & \ours{} & \gcell{179} & \bcell{23.40} & \bcell{31.16} & \bcell{44.09} & \gbf{86.9\%} & \gbf{82.6\%} & \gbf{75.4\%} & \xmark & \cmark & \cmark & \textbf{\(-0.0017\)} \\
LiH (12q) & Random & \gcell{179} & \bcell{\textbf{17.57}} & \bcell{\textbf{29.78}} & \bcell{\textbf{40.82}} & \textbf{\gbf{90.2\%}} & \textbf{\gbf{83.4\%}} & \textbf{\gbf{77.2\%}} & \xmark & \xmark & \xmark & \rbf{\(+3.3113\)} \\
\bottomrule
\end{tabular}}
\label{tab:v39_tau0995_budget}
\vspace{-0.5em}
\end{table}

\clearpage
\subsection{Shots-to-Chemical-Parity Dynamics}
Figure~\ref{fig:vista_shots_to_energy} complements the endpoint comparison
with cumulative-shot convergence trajectories.
\appVistaShotsToEnergy

\clearpage
\section{Threshold Control and Depth-OOD Pareto Details}
\label{sec:app_threshold_depth}
\subsection{Depth-OOD Pareto Visualization}
\begin{figure}[!t]
\begin{minipage}[t]{0.43\linewidth}
\vspace{0pt}
\textbf{Post-hoc budget control.}
We vary only the cumulative top-mass threshold $\tau$ after training; model weights, labels, and online oracle access remain unchanged. One selector trained on depths 1--4 is evaluated on unseen depths 5--7. Increasing $\tau$ from 0.98 to 0.997 reduces the proposed method's mean absolute energy error from 22.5 to 7.2 mHa while retaining 16.2\% shot saving. At closely matched budgets, Random remains at 264--606 mHa. The result exposes a controllable accuracy--measurement frontier without retraining.
\end{minipage}
\hfill
\begin{minipage}[t]{0.54\linewidth}
\vspace{0pt}
\centering
\includegraphics[width=0.86\linewidth]{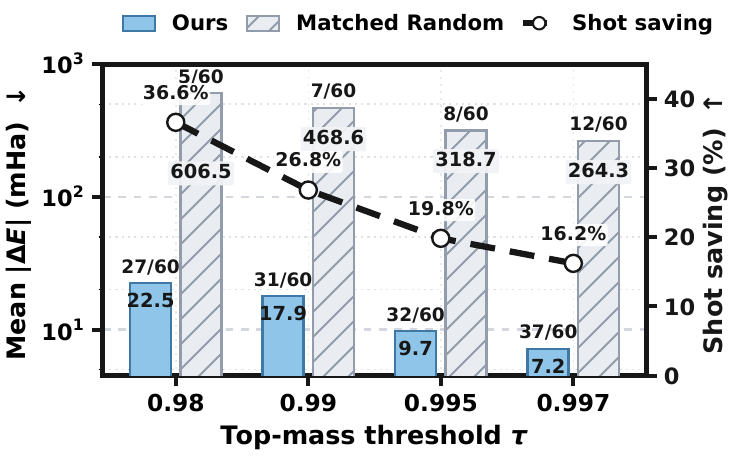}
\caption{\textbf{Depth-OOD threshold Pareto frontier.} Bars show mean $|\Delta E|$ to Full QWC; labels above them count no-worse trajectories out of 60 under the one-sided 1.6 mHa criterion. The line shows the proposed method's shot saving; Random uses an almost identical budget. Full values are in Appendix~\ref{sec:app_threshold_depth}, Table~\ref{tab:v42_depth_ood_frontier}.}
\label{fig:v42_depth_ood_pareto_bar}
\end{minipage}
\vspace{-0.5em}
\end{figure}

\clearpage
\subsection{Complete Numerical Threshold Frontier}
\begin{table}[H]
\centering
\vspace{-0.5em}
\caption{\textbf{Complete numerical counterpart of Figure~\ref{fig:v42_depth_ood_pareto_bar}.} A single model trained on depths 1--4 is reused across thresholds and evaluated on unseen depths 5--7. Only the inference top-mass threshold changes; energy gaps are in Ha and costs are nominal. This duplicates the depth-OOD aggregate in Table~\ref{tab:app_depth_ood_tau_summary}, not an independent replication. No-worse seeds have final energy at most 1.6 mHa above Full QWC; lower final energy is accepted. Further threshold, depth-wise, and molecule-wise results appear in Tables~\ref{tab:app_depth_ood_tau_summary}, \ref{tab:v42_depth_ood_by_depth}, and~\ref{tab:v42_tau0997_by_molecule}.}
\vspace{0.25em}
\setlength{\tabcolsep}{1.2mm}
\small
\resizebox{1.0\linewidth}{!}{
\begin{tabular}{c|ccccc|ccc}
\toprule
Threshold & \multicolumn{5}{c|}{\ours{}} & \multicolumn{3}{c}{Random same budget} \\
\cmidrule(lr){2-6}\cmidrule(lr){7-9}
$\tau$ & Mean \(|\dE|\downarrow\) & Mean \(\dE\) & No-worse seeds & Shot gain \(\uparrow\) & Groups/param. & Mean \(|\dE|\downarrow\) & No-worse seeds & Shot gain \(\uparrow\) \\
\midrule
0.98 & 0.02246 & \rbf{\(+0.0066\)} & 27/60 & \gbf{36.6\%} & 24.43 & 0.60645 & 5/60 & \gbf{37.0\%} \\
0.99 & 0.01791 & \rbf{\(+0.0025\)} & 31/60 & \gbf{26.8\%} & 30.47 & 0.46857 & 7/60 & \gbf{27.4\%} \\
0.995 & 0.00974 & \gbf{\(-0.0010\)} & 32/60 & \gbf{19.8\%} & 35.67 & 0.31867 & 8/60 & \gbf{20.4\%} \\
0.997 & 0.00716 & \gbf{\(-0.0012\)} & 37/60 & \gbf{16.2\%} & 38.90 & 0.26431 & 12/60 & \gbf{16.7\%} \\
\bottomrule
\end{tabular}
}
\label{tab:v42_depth_ood_frontier}
\vspace{-0.5em}
\end{table}

\clearpage
\section{Top-Mass Threshold Sweep Summary}
The current same-family ansatz-OOD table uses $\tau=0.997$. To make the budget dial explicit, Tables~\ref{tab:app_fixed_tau_summary} and~\ref{tab:app_depth_ood_tau_summary} summarize the existing threshold sweeps. In all cases, $\tau$ is the cumulative top-mass threshold used to convert the predicted parameter-by-QWC relevance matrix into selected QWC groups:
\begin{equation}
\sum_{k\in S_j(\tau)} \widehat w_{j,k} \ge \tau.
\end{equation}
For a fixed predicted row, larger $\tau$ retains a nested set of groups. This does not guarantee monotone final energy or closed-loop average cost, and the historical points below are not all evaluations of one checkpoint.

\begin{table}[H]
\centering
\vspace{-0.5em}
\caption{\textbf{Historical held-out-seed operating points, not one checkpoint's sweep.} Depths 1--3 use exact-expectation VQE and nominal shot accounting. Earlier 0.95/0.98/0.99 points belong to separately trained models; those models used the auxiliary count head. Only 0.995/0.997 reuse the $\tau_{\mathrm{train}}=0.99$ checkpoint with top-mass post-processing. Energy gaps are in Ha. Checkpoint and selection-mode provenance are in Appendix~\ref{sec:app_checkpoint_provenance}; detailed results follow.}
\label{tab:app_fixed_tau_summary}
\vspace{0.25em}
\setlength{\tabcolsep}{1.4mm}
\resizebox{1.0\linewidth}{!}{
\begin{tabular}{c|cccc|ccc}
\toprule
Threshold & \multicolumn{4}{c|}{\ours{}} & \multicolumn{3}{c}{Random same budget} \\
\cmidrule(lr){2-5}\cmidrule(lr){6-8}
$\tau$ & Mean $|\dE|\downarrow$ & Mean $\dE$ & Shot gain $\uparrow$ & Groups/param. & Mean $|\dE|\downarrow$ & Shot gain $\uparrow$ & Groups/param. \\
\midrule
\gcell{0.95} & 0.00917 & \rbf{+0.0073} & \gbf{62.3\%} & 12.19 & 1.33512 & \gbf{63.4\%} & 11.24 \\
\bcell{0.98} & 0.00984 & \rbf{+0.0040} & \gbf{52.8\%} & 17.59 & 1.07202 & \gbf{53.8\%} & 16.63 \\
\gcell{0.99} & 0.00439 & \gbf{+0.0011} & \gbf{45.5\%} & 21.54 & 0.93589 & \gbf{46.5\%} & 20.32 \\
\bcell{0.995} & 0.00567 & \gbf{+0.0014} & \gbf{49.6\%} & 17.62 & 1.17308 & \gbf{51.3\%} & 16.41 \\
\gcell{0.997} & 0.00352 & \gbf{+0.0000} & \gbf{45.5\%} & 19.84 & 1.10026 & \gbf{47.4\%} & 18.59 \\
\bottomrule
\end{tabular}}
\vspace{-0.5em}
\end{table}

\begin{table}[H]
\centering
\vspace{-0.5em}
\caption{\textbf{Depth-OOD threshold sweep.} A single depth-OOD model is trained on depths 1--4 and evaluated on unseen depths 5--7. Only the inference top-mass threshold changes; energy gaps are in Ha and shot costs are nominal. This differs from the historical table above (Appendix~\ref{sec:app_checkpoint_provenance}). No-worse seeds count trajectories whose final energy is not more than $1.6$ mHa above Full-QWC; lower energy is accepted.}
\label{tab:app_depth_ood_tau_summary}
\vspace{0.25em}
\setlength{\tabcolsep}{1.4mm}
\resizebox{1.0\linewidth}{!}{
\begin{tabular}{c|ccccc|ccc}
\toprule
Threshold & \multicolumn{5}{c|}{\ours{}} & \multicolumn{3}{c}{Random same budget} \\
\cmidrule(lr){2-6}\cmidrule(lr){7-9}
$\tau$ & Mean $|\dE|\downarrow$ & Mean $\dE$ & No-worse seeds & Shot gain $\uparrow$ & Groups/param. & Mean $|\dE|\downarrow$ & No-worse seeds & Shot gain $\uparrow$ \\
\midrule
\gcell{0.98} & 0.02246 & \rbf{+0.0066} & 27/60 & \gbf{36.6\%} & 24.43 & 0.60645 & 5/60 & \gbf{37.0\%} \\
\bcell{0.99} & 0.01791 & \rbf{+0.0025} & 31/60 & \gbf{26.8\%} & 30.47 & 0.46857 & 7/60 & \gbf{27.4\%} \\
\gcell{0.995} & 0.00974 & \gbf{-0.0010} & 32/60 & \gbf{19.8\%} & 35.67 & 0.31867 & 8/60 & \gbf{20.4\%} \\
\bcell{0.997} & 0.00716 & \gbf{-0.0012} & 37/60 & \gbf{16.2\%} & 38.90 & 0.26431 & 12/60 & \gbf{16.7\%} \\
\bottomrule
\end{tabular}}
\vspace{-0.5em}
\end{table}

\clearpage
\subsection{Stricter Point: $\tau=0.997$}

The $\tau=0.997$ point uses the same saved relevance model and increases only the top-mass post-processing threshold. It is intended as a conservative operating point for cases where $\tau=0.995$ is close but not always within the no-worse criterion.

\begin{table}[H]
\centering
\vspace{-0.5em}
\caption{\textbf{Matched-budget online VQE baseline comparison at $\tau=0.997$.} The same saved relevance model is reused; only the top-mass post-processing threshold is increased to $\tau=0.997$. For each depth and metric, \ourmethod{} is paired with a Random same-budget baseline that keeps the same number of selected QWC groups per parameter but chooses groups uniformly at random. Boldface marks the better method for each molecule-specific comparison.}
\vspace{0.25em}
\setlength{\tabcolsep}{0.8mm}
\resizebox{1.0\linewidth}{!}{
\begin{tabular}{lll|ccccc}
\toprule
Setting & Metric & Method & H2/STO-3G & HeH+/STO-3G & H3+/STO-3G & H4 chain/STO-3G & LiH/STO-3G \\
\midrule
\gcell{Depth 1} & \(\dE\) vs Full \(\downarrow\) & \ours{} & \textbf{\gbf{\(+0.0005\)}} & \textbf{\gbf{\(+0.0001\)}} & \textbf{\gbf{\(-0.0002\)}} & \textbf{\gbf{\(-0.0013\)}} & \textbf{\rbf{\(+0.0020\)}} \\
\gcell{Depth 1} & \(\dE\) vs Full \(\downarrow\) & Random & \rbf{\(+0.2192\)} & \rbf{\(+1.3164\)} & \rbf{\(+0.7630\)} & \rbf{\(+1.2211\)} & \rbf{\(+3.5105\)} \\
\bcell{Depth 1} & Shot gain \(\uparrow\) & \ours{} & \gbf{37.2\%} & \gbf{55.1\%} & \gbf{59.5\%} & \gbf{71.7\%} & \gbf{84.9\%} \\
\bcell{Depth 1} & Shot gain \(\uparrow\) & Random & \textbf{\gbf{41.3\%}} & \textbf{\gbf{56.6\%}} & \textbf{\gbf{62.7\%}} & \textbf{\gbf{75.5\%}} & \textbf{\gbf{88.9\%}} \\
\gcell{Depth 1} & No-worse vs Full & \ours{} & \cmark & \cmark & \cmark & \cmark & \xmark \\
\gcell{Depth 1} & No-worse vs Full & Random & \xmark & \xmark & \xmark & \xmark & \xmark \\
\midrule
\gcell{Depth 2} & \(\dE\) vs Full \(\downarrow\) & \ours{} & \textbf{\gbf{\(-0.0001\)}} & \textbf{\gbf{\(-0.0024\)}} & \textbf{\gbf{\(-0.0030\)}} & \textbf{\rbf{\(+0.0017\)}} & \textbf{\gbf{\(-0.0021\)}} \\
\gcell{Depth 2} & \(\dE\) vs Full \(\downarrow\) & Random & \rbf{\(+0.1092\)} & \rbf{\(+0.3799\)} & \rbf{\(+0.5018\)} & \rbf{\(+0.7784\)} & \rbf{\(+3.7420\)} \\
\bcell{Depth 2} & Shot gain \(\uparrow\) & \ours{} & \gbf{21.8\%} & \gbf{27.3\%} & \gbf{39.8\%} & \gbf{46.6\%} & \gbf{79.1\%} \\
\bcell{Depth 2} & Shot gain \(\uparrow\) & Random & \textbf{\gbf{23.0\%}} & \textbf{\gbf{29.3\%}} & \textbf{\gbf{42.3\%}} & \textbf{\gbf{47.8\%}} & \textbf{\gbf{80.0\%}} \\
\gcell{Depth 2} & No-worse vs Full & \ours{} & \cmark & \cmark & \cmark & \xmark & \cmark \\
\gcell{Depth 2} & No-worse vs Full & Random & \xmark & \xmark & \xmark & \xmark & \xmark \\
\midrule
\gcell{Depth 3} & \(\dE\) vs Full \(\downarrow\) & \ours{} & \textbf{\gbf{\(+0.0007\)}} & \textbf{\rbf{\(+0.0074\)}} & \textbf{\gbf{\(-0.0023\)}} & \textbf{\gbf{\(+0.0004\)}} & \textbf{\gbf{\(-0.0010\)}} \\
\gcell{Depth 3} & \(\dE\) vs Full \(\downarrow\) & Random & \rbf{\(+0.0092\)} & \rbf{\(+0.1772\)} & \rbf{\(+0.1448\)} & \rbf{\(+0.1285\)} & \rbf{\(+3.3875\)} \\
\bcell{Depth 3} & Shot gain \(\uparrow\) & \ours{} & \gbf{14.8\%} & \gbf{20.3\%} & \gbf{24.7\%} & \gbf{30.0\%} & \gbf{70.4\%} \\
\bcell{Depth 3} & Shot gain \(\uparrow\) & Random & \textbf{\gbf{14.9\%}} & \textbf{\gbf{20.6\%}} & \textbf{\gbf{25.2\%}} & \textbf{\gbf{30.7\%}} & \textbf{\gbf{72.6\%}} \\
\gcell{Depth 3} & No-worse vs Full & \ours{} & \cmark & \xmark & \cmark & \cmark & \cmark \\
\gcell{Depth 3} & No-worse vs Full & Random & \xmark & \xmark & \xmark & \xmark & \xmark \\
\bottomrule
\end{tabular}}
\label{tab:v39_tau0997_main}
\vspace{-0.5em}
\end{table}

\begin{table}[H]
\centering
\vspace{-0.5em}
\caption{\textbf{Depth-wise aggregate at $\tau=0.997$.} Mean $|\dE|$ averages the five molecule columns at each ansatz depth. No-worse marks whether the aggregate mean signed $\dE$ is at most $1.6$ mHa above Full.}
\vspace{0.25em}
\setlength{\tabcolsep}{1.2mm}
\small
\begin{tabular}{l|ccc}
\toprule
Setting & Mean \(|\dE|\downarrow\) & No-worse & Mean shot gain \(\uparrow\) \\
\midrule
\grow Full-QWC reference & 0.00000 & \cmark & \gray{0.0\%} \\
\brow \ours{} depth 1 & 0.00112 & \cmark & \gbf{61.7\%} \\
\brow \ours{} depth 2 & 0.00448 & \cmark & \gbf{42.9\%} \\
\brow \ours{} depth 3 & 0.00497 & \cmark & \gbf{32.0\%} \\
\midrule
Random depth 1 & 1.40602 & \xmark & \gbf{65.0\%} \\
Random depth 2 & 1.10226 & \xmark & \gbf{44.5\%} \\
Random depth 3 & 0.79248 & \xmark & \gbf{32.8\%} \\
\midrule
\brow \ours{} all & 0.00352 & \cmark & \gbf{45.5\%} \\
Random all & 1.10026 & \xmark & \gbf{47.4\%} \\
\bottomrule
\end{tabular}
\label{tab:v39_tau0997_agg}
\vspace{-0.5em}
\end{table}

\begin{table}[H]
\centering
\vspace{-0.5em}
\caption{\textbf{Detailed budget usage at $\tau=0.997$.} Groups/param. reports the mean selected QWC groups per parameter, shot gain is relative to full $2PGS$, and No-worse vs Full uses the one-sided $1.6$ mHa criterion ($\dE \le 1.6$ mHa; lower energy is accepted). Boldface marks the better \ourmethod{}-vs-Random entry for each molecule-depth comparison.}
\vspace{0.25em}
\setlength{\tabcolsep}{0.9mm}
\resizebox{1.0\linewidth}{!}{
\begin{tabular}{ll|c|ccc|ccc|ccc|c}
\toprule
Molecule & Method & Full QWC & \multicolumn{3}{c|}{Groups / param. \(\downarrow\)} & \multicolumn{3}{c|}{Shot gain \(\uparrow\)} & \multicolumn{3}{c|}{No-worse vs Full} & Best \(\dE\downarrow\) \\
\cmidrule(lr){4-6}\cmidrule(lr){7-9}\cmidrule(lr){10-12}
& & groups/param. & D1 & D2 & D3 & D1 & D2 & D3 & D1 & D2 & D3 & \\
\midrule
H2/STO-3G & \ours{} & \gcell{5} & \bcell{3.14} & \bcell{3.91} & \bcell{4.26} & \gbf{37.2\%} & \gbf{21.8\%} & \gbf{14.8\%} & \cmark & \cmark & \cmark & \textbf{\gbf{\(-0.0001\)}} \\
H2/STO-3G & Random & \gcell{5} & \bcell{\textbf{2.94}} & \bcell{\textbf{3.85}} & \bcell{\textbf{4.26}} & \textbf{\gbf{41.3\%}} & \textbf{\gbf{23.0\%}} & \textbf{\gbf{14.9\%}} & \xmark & \xmark & \xmark & \rbf{\(+0.0092\)} \\
\midrule
HeH+/STO-3G & \ours{} & \gcell{11} & \bcell{4.94} & \bcell{8.00} & \bcell{8.77} & \gbf{55.1\%} & \gbf{27.3\%} & \gbf{20.3\%} & \cmark & \cmark & \xmark & \textbf{\gbf{\(+0.0001\)}} \\
HeH+/STO-3G & Random & \gcell{11} & \bcell{\textbf{4.77}} & \bcell{\textbf{7.78}} & \bcell{\textbf{8.74}} & \textbf{\gbf{56.6\%}} & \textbf{\gbf{29.3\%}} & \textbf{\gbf{20.6\%}} & \xmark & \xmark & \xmark & \rbf{\(+0.1772\)} \\
\midrule
H3+/STO-3G & \ours{} & \gcell{19} & \bcell{7.69} & \bcell{11.44} & \bcell{14.31} & \gbf{59.5\%} & \gbf{39.8\%} & \gbf{24.7\%} & \cmark & \cmark & \cmark & \textbf{\gbf{\(-0.0002\)}} \\
H3+/STO-3G & Random & \gcell{19} & \bcell{\textbf{7.08}} & \bcell{\textbf{10.97}} & \bcell{\textbf{14.20}} & \textbf{\gbf{62.7\%}} & \textbf{\gbf{42.3\%}} & \textbf{\gbf{25.2\%}} & \xmark & \xmark & \xmark & \rbf{\(+0.1448\)} \\
\midrule
H4 chain/STO-3G & \ours{} & \gcell{75} & \bcell{21.26} & \bcell{40.03} & \bcell{52.52} & \gbf{71.7\%} & \gbf{46.6\%} & \gbf{30.0\%} & \cmark & \xmark & \cmark & \textbf{\gbf{\(+0.0004\)}} \\
H4 chain/STO-3G & Random & \gcell{75} & \bcell{\textbf{18.40}} & \bcell{\textbf{39.14}} & \bcell{\textbf{51.98}} & \textbf{\gbf{75.5\%}} & \textbf{\gbf{47.8\%}} & \textbf{\gbf{30.7\%}} & \xmark & \xmark & \xmark & \rbf{\(+0.1285\)} \\
\midrule
LiH/STO-3G & \ours{} & \gcell{179} & \bcell{26.99} & \bcell{37.46} & \bcell{52.92} & \gbf{84.9\%} & \gbf{79.1\%} & \gbf{70.4\%} & \xmark & \cmark & \cmark & \textbf{\gbf{\(-0.0010\)}} \\
LiH/STO-3G & Random & \gcell{179} & \bcell{\textbf{19.89}} & \bcell{\textbf{35.75}} & \bcell{\textbf{49.06}} & \textbf{\gbf{88.9\%}} & \textbf{\gbf{80.0\%}} & \textbf{\gbf{72.6\%}} & \xmark & \xmark & \xmark & \rbf{\(+3.3875\)} \\
\bottomrule
\end{tabular}}
\label{tab:v39_tau0997_budget}
\vspace{-0.5em}
\end{table}

\clearpage
\iffalse

\section{Additional Depth-OOD Pareto Sweep Tables}

\else

\section{Depth-OOD Pareto Sweep}

\fi

This sweep reuses the depth-OOD relevance model and changes only the top-mass threshold. This isolates the post-hoc budget dial: lower thresholds save more QWC measurements, while higher thresholds recover more of the Full-QWC gradient signal.

\iffalse

\begin{table}[H]
\centering
\vspace{-0.5em}
\caption{\textbf{Depth-OOD Pareto frontier from top-mass threshold sweep.} A single trained Transformer relevance model is reused across thresholds; only the cumulative top-mass threshold changes. Evaluation is closed-loop online VQE on unseen depths 5--7. No-worse seeds count held-out trajectories whose final energy is no more than 1.6 mHa above the Full-QWC reference; lower final energy is accepted.}
\vspace{0.25em}
\setlength{\tabcolsep}{1.2mm}
\small
\resizebox{1.0\linewidth}{!}{
\begin{tabular}{c|ccccc|ccc}
\toprule
Threshold & \multicolumn{5}{c|}{\ours{}} & \multicolumn{3}{c}{Random same budget} \\
\cmidrule(lr){2-6}\cmidrule(lr){7-9}
$\tau$ & Mean \(|\dE|\downarrow\) & Mean \(\dE\) & No-worse seeds & Shot gain \(\uparrow\) & Groups/param. & Mean \(|\dE|\downarrow\) & No-worse seeds & Shot gain \(\uparrow\) \\
\midrule
0.98 & 0.02246 & \rbf{\(+0.0066\)} & 27/60 & \gbf{36.6\%} & 24.43 & 0.60645 & 5/60 & \gbf{37.0\%} \\
0.99 & 0.01791 & \rbf{\(+0.0025\)} & 31/60 & \gbf{26.8\%} & 30.47 & 0.46857 & 7/60 & \gbf{27.4\%} \\
0.995 & 0.00974 & \gbf{\(-0.0010\)} & 32/60 & \gbf{19.8\%} & 35.67 & 0.31867 & 8/60 & \gbf{20.4\%} \\
0.997 & 0.00716 & \gbf{\(-0.0012\)} & 37/60 & \gbf{16.2\%} & 38.90 & 0.26431 & 12/60 & \gbf{16.7\%} \\
\bottomrule
\end{tabular}
}
\label{tab:v42_depth_ood_frontier}
\vspace{-0.5em}
\end{table}

\else

\begin{table}[H]
\centering
\vspace{-0.5em}
\caption{\textbf{Depth-wise view of the OOD Pareto sweep.} A single Transformer relevance model is trained on ansatz depths 1--4 and evaluated without retraining on unseen depths 5--7. Each row pools the five PySCF/STO-3G molecular Hamiltonians (H2, HeH+, H3+, H4 chain, and LiH) and four held-out seeds at one unseen evaluation depth. Mean \(\dE\) is reported as mean \(\pm\) standard deviation over the 20 paired trajectories. CA marks whether the mean signed \(\dE\) is no more than 1.6 mHa above the Full-QWC reference; lower final energy is accepted. Group gain is the relative reduction in selected QWC groups per parameter versus Full-QWC.}
\vspace{0.25em}
\setlength{\tabcolsep}{0.9mm}
\small
\resizebox{1.0\linewidth}{!}{
\begin{tabular}{cc|cccc|cccc}
\toprule
Threshold & Depth & \multicolumn{4}{c|}{\ours{}} & \multicolumn{4}{c}{Random same budget} \\
\cmidrule(lr){3-6}\cmidrule(lr){7-10}
$\tau$ & $d$ & Mean \(\dE \pm\) s.d. & CA & Shot gain \(\uparrow\) & Group gain \(\uparrow\) & Mean \(\dE \pm\) s.d. & CA & Shot gain \(\uparrow\) & Group gain \(\uparrow\) \\
\midrule
\gcell{0.98} & \gcell{5} & \rbf{\(+0.0219 \pm 0.0618\)} & \xmark & \gbf{39.4\%} & \gbf{62.3\%} & \rbf{\(+0.7969 \pm 1.2005\)} & \xmark & \gbf{39.9\%} & \gbf{64.4\%} \\
\gcell{0.98} & \gcell{6} & \gbf{\(-0.0054 \pm 0.0243\)} & \cmark & \gbf{36.5\%} & \gbf{57.4\%} & \rbf{\(+0.5774 \pm 0.9583\)} & \xmark & \gbf{36.9\%} & \gbf{59.3\%} \\
\gcell{0.98} & \gcell{7} & \rbf{\(+0.0035 \pm 0.0124\)} & \xmark & \gbf{33.9\%} & \gbf{53.5\%} & \rbf{\(+0.4024 \pm 0.7036\)} & \xmark & \gbf{34.1\%} & \gbf{55.0\%} \\
\midrule
\gcell{0.99} & \gcell{5} & \rbf{\(+0.0048 \pm 0.0555\)} & \xmark & \gbf{29.7\%} & \gbf{52.4\%} & \rbf{\(+0.7167 \pm 1.1610\)} & \xmark & \gbf{30.4\%} & \gbf{54.8\%} \\
\gcell{0.99} & \gcell{6} & \rbf{\(+0.0026 \pm 0.0256\)} & \xmark & \gbf{26.7\%} & \gbf{46.9\%} & \rbf{\(+0.4179 \pm 0.6588\)} & \xmark & \gbf{27.2\%} & \gbf{49.0\%} \\
\gcell{0.99} & \gcell{7} & \gbf{\(+0.0002 \pm 0.0130\)} & \cmark & \gbf{24.1\%} & \gbf{42.5\%} & \rbf{\(+0.2481 \pm 0.4441\)} & \xmark & \gbf{24.6\%} & \gbf{44.1\%} \\
\midrule
\gcell{0.995} & \gcell{5} & \gbf{\(-0.0016 \pm 0.0276\)} & \cmark & \gbf{22.6\%} & \gbf{43.8\%} & \rbf{\(+0.5389 \pm 0.9643\)} & \xmark & \gbf{23.5\%} & \gbf{46.2\%} \\
\gcell{0.995} & \gcell{6} & \gbf{\(-0.0014 \pm 0.0090\)} & \cmark & \gbf{19.6\%} & \gbf{37.8\%} & \rbf{\(+0.2580 \pm 0.4218\)} & \xmark & \gbf{20.1\%} & \gbf{39.2\%} \\
\gcell{0.995} & \gcell{7} & \gbf{\(-0.0001 \pm 0.0107\)} & \cmark & \gbf{17.3\%} & \gbf{33.2\%} & \rbf{\(+0.1470 \pm 0.2903\)} & \xmark & \gbf{17.7\%} & \gbf{34.5\%} \\
\midrule
\gcell{0.997} & \gcell{5} & \gbf{\(-0.0025 \pm 0.0155\)} & \cmark & \gbf{18.9\%} & \gbf{38.3\%} & \rbf{\(+0.4968 \pm 0.8809\)} & \xmark & \gbf{19.5\%} & \gbf{39.9\%} \\
\gcell{0.997} & \gcell{6} & \gbf{\(-0.0006 \pm 0.0105\)} & \cmark & \gbf{16.0\%} & \gbf{32.1\%} & \rbf{\(+0.1665 \pm 0.3072\)} & \xmark & \gbf{16.4\%} & \gbf{33.4\%} \\
\gcell{0.997} & \gcell{7} & \gbf{\(-0.0005 \pm 0.0098\)} & \cmark & \gbf{13.8\%} & \gbf{27.6\%} & \rbf{\(+0.0916 \pm 0.2070\)} & \xmark & \gbf{14.1\%} & \gbf{28.7\%} \\
\bottomrule
\end{tabular}
}
\label{tab:v42_depth_ood_by_depth}
\vspace{-0.5em}
\end{table}

\fi

\ifoursappendixcore

\begin{table}[H]
\centering
\vspace{-0.5em}
\caption{\textbf{Molecule-wise OOD summary at $\tau=0.997$.} This conservative point gives the best global mean $|\dE|$ in the sweep while still retaining nonzero shot savings, especially for richer Hamiltonians with many QWC groups.}
\vspace{0.25em}
\setlength{\tabcolsep}{1.0mm}
\small
\begin{tabular}{l|ccc|ccc}
\toprule
Molecule & \multicolumn{3}{c|}{\ours{}} & \multicolumn{3}{c}{Random same budget} \\
\cmidrule(lr){2-4}\cmidrule(lr){5-7}
& Mean \(|\dE|\downarrow\) & Mean \(\dE\) & Shot gain \(\uparrow\) & Mean \(|\dE|\downarrow\) & Mean \(\dE\) & Shot gain \(\uparrow\) \\
\midrule
H2/STO-3G & 0.00140 & \textbf{\gbf{\(+0.0003\)}} & \gbf{4.8\%} & 0.01690 & \rbf{\(+0.0066\)} & \gbf{4.7\%} \\
HeH+/STO-3G & 0.00735 & \textbf{\gbf{\(-0.0024\)}} & \gbf{7.6\%} & 0.06000 & \rbf{\(+0.0424\)} & \gbf{7.5\%} \\
H3+/STO-3G & 0.00668 & \textbf{\gbf{\(-0.0036\)}} & \gbf{8.9\%} & 0.03097 & \rbf{\(+0.0160\)} & \gbf{8.9\%} \\
H4 chain/STO-3G & 0.01298 & \textbf{\gbf{\(-0.0011\)}} & \gbf{14.9\%} & 0.09763 & \rbf{\(+0.0771\)} & \gbf{15.0\%} \\
LiH/STO-3G & 0.00742 & \textbf{\gbf{\(+0.0009\)}} & \gbf{45.0\%} & 1.11605 & \rbf{\(+1.1160\)} & \gbf{47.1\%} \\
\bottomrule
\end{tabular}
\label{tab:v42_tau0997_by_molecule}
\vspace{-0.5em}
\end{table}

\fi

\clearpage
\section{Previous Threshold Sweep Tables}
The main paper reports the stricter post-processing point $\tau=0.995$.  For completeness, we keep the earlier threshold-sweep tables below.  These tables use the same held-out-seed online VQE protocol and matched-budget random baseline, but reflect earlier trained operating points $\tau\in\{0.95,0.98,0.99\}$, including count-head selection rather than a single-checkpoint inference-threshold sweep. Appendix~\ref{sec:app_checkpoint_provenance} gives provenance. All energy gaps are in Ha and shot gains use nominal accounting.
\section{Previous Threshold: $\tau=0.98$}

\begin{table}[H]
    \centering
    \vspace{-0.5em}
    \caption{\textbf{Matched-budget online VQE baseline comparison (Main Results: threshold $\tau=0.98$).} The learned selector is trained on RY/RZ VQE trajectories from depths 1--3 using row-wise top-mass target threshold $\tau=0.98$ and evaluated in closed-loop online VQE on held-out random seeds for the same five PySCF molecules and depths. For each depth and metric, \ourmethod{} is paired with a Random same-budget baseline that keeps the same predicted number of QWC groups per parameter but chooses groups uniformly at random. Boldface marks the better method for each molecule-specific comparison.}
    \vspace{0.25em}
    \setlength{\tabcolsep}{0.8mm}
    \resizebox{1.0\linewidth}{!}{
    \begin{tabular}{lll|ccccc}
    \toprule
    Setting & Metric & Method & H2/STO-3G & HeH+/STO-3G & H3+/STO-3G & H4 chain/STO-3G & LiH/STO-3G \\
    \midrule
    \gcell{Depth 1} & \(\dE\) vs Full \(\downarrow\) & \ours{} & \textbf{\gbf{\(-0.0000\)}} & \textbf{\(+0.0028\)} & \textbf{\(+0.0040\)} & \textbf{\(-0.0021\)} & \textbf{\gbf{\(+0.0012\)}} \\
    \gcell{Depth 1} & \(\dE\) vs Full \(\downarrow\) & Random & \rbf{\(+0.2500\)} & \rbf{\(+1.3729\)} & \rbf{\(+0.7713\)} & \rbf{\(+1.1147\)} & \rbf{\(+3.2882\)} \\
    \bcell{Depth 1} & Shot gain \(\uparrow\) & \ours{} & \gbf{41.2\%} & \gbf{59.2\%} & \gbf{66.2\%} & \gbf{73.7\%} & \gbf{86.4\%} \\
    \bcell{Depth 1} & Shot gain \(\uparrow\) & Random & \textbf{\gbf{43.7\%}} & \textbf{\gbf{62.9\%}} & \textbf{\gbf{69.0\%}} & \textbf{\gbf{77.0\%}} & \textbf{\gbf{89.8\%}} \\
\gcell{Depth 1} & No-worse vs Full & \ours{} & \cmark & \xmark & \xmark & \cmark & \cmark \\
\gcell{Depth 1} & No-worse vs Full & Random & \xmark & \xmark & \xmark & \xmark & \xmark \\
    \midrule
    \gcell{Depth 2} & \(\dE\) vs Full \(\downarrow\) & \ours{} & \textbf{\(+0.0244\)} & \textbf{\(+0.0103\)} & \textbf{\(-0.0021\)} & \textbf{\(+0.0214\)} & \textbf{\(-0.0032\)} \\
    \gcell{Depth 2} & \(\dE\) vs Full \(\downarrow\) & Random & \rbf{\(+0.1560\)} & \rbf{\(+0.4481\)} & \rbf{\(+0.4674\)} & \rbf{\(+0.8634\)} & \rbf{\(+3.4465\)} \\
    \bcell{Depth 2} & Shot gain \(\uparrow\) & \ours{} & \textbf{\gbf{31.9\%}} & \textbf{\gbf{40.5\%}} & \gbf{49.6\%} & \gbf{56.3\%} & \gbf{79.9\%} \\
    \bcell{Depth 2} & Shot gain \(\uparrow\) & Random & \gbf{30.8\%} & \gbf{39.3\%} & \textbf{\gbf{49.7\%}} & \textbf{\gbf{56.7\%}} & \textbf{\gbf{81.2\%}} \\
\gcell{Depth 2} & No-worse vs Full & \ours{} & \xmark & \xmark & \cmark & \xmark & \cmark \\
\gcell{Depth 2} & No-worse vs Full & Random & \xmark & \xmark & \xmark & \xmark & \xmark \\
    \midrule
    \gcell{Depth 3} & \(\dE\) vs Full \(\downarrow\) & \ours{} & \textbf{\(-0.0022\)} & \textbf{\(+0.0121\)} & \textbf{\(-0.0026\)} & \textbf{\(-0.0020\)} & \textbf{\(-0.0014\)} \\
    \gcell{Depth 3} & \(\dE\) vs Full \(\downarrow\) & Random & \rbf{\(+0.0586\)} & \rbf{\(+0.1960\)} & \rbf{\(+0.2986\)} & \rbf{\(+0.2677\)} & \rbf{\(+3.0320\)} \\
    \bcell{Depth 3} & Shot gain \(\uparrow\) & \ours{} & \textbf{\gbf{24.9\%}} & \textbf{\gbf{31.3\%}} & \textbf{\gbf{37.4\%}} & \gbf{41.8\%} & \gbf{71.6\%} \\
    \bcell{Depth 3} & Shot gain \(\uparrow\) & Random & \gbf{24.0\%} & \gbf{30.6\%} & \gbf{36.7\%} & \textbf{\gbf{42.1\%}} & \textbf{\gbf{73.0\%}} \\
\gcell{Depth 3} & No-worse vs Full & \ours{} & \cmark & \xmark & \cmark & \cmark & \cmark \\
\gcell{Depth 3} & No-worse vs Full & Random & \xmark & \xmark & \xmark & \xmark & \xmark \\
    \bottomrule
    \end{tabular}}
    \label{tab:v38_tau098_main}
    \vspace{-0.5em}
\end{table}

Table~\ref{tab:v38_tau098_main} shows that at threshold $\tau=0.98$, \ourmethod{} reaches mean absolute final-energy gap $0.00984$ while saving $52.8\%$ of full-gradient measurement shots. The matched-budget random baseline has mean absolute gap $1.07202$ with $53.8\%$ shot saving, so the selected QWC groups matter beyond the budget alone.

\begin{table}[H]
    \centering
    \vspace{-0.5em}
    \caption{\textbf{Depth-wise aggregate (Main Results: threshold $\tau=0.98$).} Same protocol as Table~\ref{tab:v38_tau098_main} with top-mass target threshold $\tau=0.98$. Mean $|\dE|$ aggregates the five molecule columns at each ansatz depth; No-worse marks whether the aggregate mean signed $\dE$ is at most $1.6$ mHa above Full.}
    \vspace{0.25em}
    \setlength{\tabcolsep}{1.2mm}
    \small
    \begin{tabular}{l|ccc}
    \toprule
    Setting & Mean \(|\dE|\downarrow\) & No-worse & Mean shot gain \(\uparrow\) \\
    \midrule
    \grow Full-QWC reference & 0.00000 & \cmark & \gray{0.0\%} \\
\brow \ours{} depth 1 & 0.00217 & \cmark & \gbf{65.3\%} \\
\brow \ours{} depth 2 & 0.01616 & \xmark & \gbf{51.6\%} \\
\brow \ours{} depth 3 & 0.01119 & \cmark & \gbf{41.4\%} \\
    \midrule
Random depth 1 & 1.35940 & \xmark & \gbf{68.5\%} \\
Random depth 2 & 1.07628 & \xmark & \gbf{51.5\%} \\
Random depth 3 & 0.78037 & \xmark & \gbf{41.3\%} \\
    \midrule
\brow \ours{} all & 0.00984 & \xmark & \gbf{52.8\%} \\
Random all & 1.07202 & \xmark & \gbf{53.8\%} \\
    \bottomrule
    \end{tabular}
    \label{tab:v38_tau098_agg}
    \vspace{-0.5em}
\end{table}

\begin{table}[H]
    \centering
    \vspace{-0.5em}
    \caption{\textbf{Detailed budget usage (Main Results: threshold $\tau=0.98$).} Same table format with top-mass target threshold $\tau=0.98$. Groups/param. reports the mean selected QWC groups per parameter, shot gain is relative to full $2PGS$, and No-worse vs Full uses the one-sided $1.6$ mHa criterion ($\dE \le 1.6$ mHa; lower energy is accepted). Boldface marks the better Ours-vs-Random entry for each molecule-depth comparison.}
    \vspace{0.25em}
    \setlength{\tabcolsep}{0.9mm}
    \resizebox{1.0\linewidth}{!}{
    \begin{tabular}{ll|c|ccc|ccc|ccc|c}
    \toprule
    Molecule & Method & Full QWC & \multicolumn{3}{c|}{Groups / param. \(\downarrow\)} & \multicolumn{3}{c|}{Shot gain \(\uparrow\)} & \multicolumn{3}{c|}{No-worse vs Full} & Best \(\dE\downarrow\) \\
    \cmidrule(lr){4-6}\cmidrule(lr){7-9}\cmidrule(lr){10-12}
    & & groups/param. & D1 & D2 & D3 & D1 & D2 & D3 & D1 & D2 & D3 & \\
    \midrule
    H2/STO-3G & \ours{} & \gcell{5} & \bcell{2.94} & \bcell{\textbf{3.41}} & \bcell{\textbf{3.75}} & \gbf{41.2\%} & \textbf{\gbf{31.9\%}} & \textbf{\gbf{24.9\%}} & \cmark & \xmark & \cmark & \textbf{\gbf{\(-0.0000\)}} \\
    H2/STO-3G & Random & \gcell{5} & \bcell{\textbf{2.82}} & \bcell{3.46} & \bcell{3.80} & \textbf{\gbf{43.7\%}} & \gbf{30.8\%} & \gbf{24.0\%} & \xmark & \xmark & \xmark & \rbf{\(+0.0586\)} \\
    \midrule
    HeH+/STO-3G & \ours{} & \gcell{11} & \bcell{4.49} & \bcell{\textbf{6.55}} & \bcell{\textbf{7.56}} & \gbf{59.2\%} & \textbf{\gbf{40.5\%}} & \textbf{\gbf{31.3\%}} & \xmark & \xmark & \xmark & \textbf{\(+0.0028\)} \\
    HeH+/STO-3G & Random & \gcell{11} & \bcell{\textbf{4.08}} & \bcell{6.68} & \bcell{7.63} & \textbf{\gbf{62.9\%}} & \gbf{39.3\%} & \gbf{30.6\%} & \xmark & \xmark & \xmark & \rbf{\(+0.1960\)} \\
    \midrule
    H3+/STO-3G & \ours{} & \gcell{19} & \bcell{6.42} & \bcell{9.58} & \bcell{\textbf{11.89}} & \gbf{66.2\%} & \gbf{49.6\%} & \textbf{\gbf{37.4\%}} & \xmark & \cmark & \cmark & \textbf{\(-0.0021\)} \\
    H3+/STO-3G & Random & \gcell{19} & \bcell{\textbf{5.89}} & \bcell{\textbf{9.55}} & \bcell{12.03} & \textbf{\gbf{69.0\%}} & \textbf{\gbf{49.7\%}} & \gbf{36.7\%} & \xmark & \xmark & \xmark & \rbf{\(+0.2986\)} \\
    \midrule
    H4 chain/STO-3G & \ours{} & \gcell{75} & \bcell{19.72} & \bcell{32.77} & \bcell{43.63} & \gbf{73.7\%} & \gbf{56.3\%} & \gbf{41.8\%} & \cmark & \xmark & \cmark & \textbf{\(-0.0021\)} \\
    H4 chain/STO-3G & Random & \gcell{75} & \bcell{\textbf{17.22}} & \bcell{\textbf{32.46}} & \bcell{\textbf{43.46}} & \textbf{\gbf{77.0\%}} & \textbf{\gbf{56.7\%}} & \textbf{\gbf{42.1\%}} & \xmark & \xmark & \xmark & \rbf{\(+0.2677\)} \\
    \midrule
    LiH/STO-3G & \ours{} & \gcell{179} & \bcell{24.38} & \bcell{36.02} & \bcell{50.75} & \gbf{86.4\%} & \gbf{79.9\%} & \gbf{71.6\%} & \cmark & \cmark & \cmark & \textbf{\gbf{\(+0.0012\)}} \\
    LiH/STO-3G & Random & \gcell{179} & \bcell{\textbf{18.33}} & \bcell{\textbf{33.73}} & \bcell{\textbf{48.31}} & \textbf{\gbf{89.8\%}} & \textbf{\gbf{81.2\%}} & \textbf{\gbf{73.0\%}} & \xmark & \xmark & \xmark & \rbf{\(+3.0320\)} \\
    \bottomrule
    \end{tabular}}
    \label{tab:v38_tau098_budget}
    \vspace{-0.5em}
\end{table}

\clearpage
\section{Previous Threshold: $\tau=0.95$}
\begin{table}[H]
    \centering
    \vspace{-0.5em}
    \caption{\textbf{Matched-budget online VQE baseline comparison (previous threshold $\tau=0.95$).} The learned selector is trained on RY/RZ VQE trajectories from depths 1--3 using row-wise top-mass target threshold $\tau=0.95$ and evaluated in closed-loop online VQE on held-out random seeds for the same five PySCF molecules and depths. For each depth and metric, \ourmethod{} is paired with a Random same-budget baseline that keeps the same predicted number of QWC groups per parameter but chooses groups uniformly at random. Boldface marks the better method for each molecule-specific comparison.}
    \vspace{0.25em}
    \setlength{\tabcolsep}{0.8mm}
    \resizebox{1.0\linewidth}{!}{
    \begin{tabular}{lll|ccccc}
    \toprule
    Setting & Metric & Method & H2/STO-3G & HeH+/STO-3G & H3+/STO-3G & H4 chain/STO-3G & LiH/STO-3G \\
    \midrule
    \gcell{Depth 1} & \(\dE\) vs Full \(\downarrow\) & \ours{} & \textbf{\(+0.0023\)} & \textbf{\gbf{\(-0.0006\)}} & \textbf{\(+0.0072\)} & \textbf{\(+0.0017\)} & \textbf{\(+0.0028\)} \\
    \gcell{Depth 1} & \(\dE\) vs Full \(\downarrow\) & Random & \rbf{\(+0.2411\)} & \rbf{\(+1.4286\)} & \rbf{\(+1.0475\)} & \rbf{\(+1.6624\)} & \rbf{\(+3.4272\)} \\
    \bcell{Depth 1} & Shot gain \(\uparrow\) & \ours{} & \gbf{43.8\%} & \gbf{65.2\%} & \gbf{74.6\%} & \gbf{81.0\%} & \gbf{91.5\%} \\
    \bcell{Depth 1} & Shot gain \(\uparrow\) & Random & \textbf{\gbf{44.3\%}} & \textbf{\gbf{68.5\%}} & \textbf{\gbf{77.2\%}} & \textbf{\gbf{85.2\%}} & \textbf{\gbf{93.2\%}} \\
\gcell{Depth 1} & No-worse vs Full & \ours{} & \xmark & \cmark & \xmark & \xmark & \xmark \\
\gcell{Depth 1} & No-worse vs Full & Random & \xmark & \xmark & \xmark & \xmark & \xmark \\
    \midrule
    \gcell{Depth 2} & \(\dE\) vs Full \(\downarrow\) & \ours{} & \textbf{\(+0.0364\)} & \textbf{\(+0.0158\)} & \textbf{\(+0.0243\)} & \textbf{\(+0.0077\)} & \textbf{\(-0.0017\)} \\
    \gcell{Depth 2} & \(\dE\) vs Full \(\downarrow\) & Random & \rbf{\(+0.2499\)} & \rbf{\(+0.8874\)} & \rbf{\(+0.7010\)} & \rbf{\(+1.2117\)} & \rbf{\(+4.1073\)} \\
    \bcell{Depth 2} & Shot gain \(\uparrow\) & \ours{} & \textbf{\gbf{40.2\%}} & \textbf{\gbf{54.1\%}} & \textbf{\gbf{62.0\%}} & \gbf{66.4\%} & \gbf{88.3\%} \\
    \bcell{Depth 2} & Shot gain \(\uparrow\) & Random & \gbf{39.6\%} & \gbf{51.2\%} & \gbf{60.8\%} & \textbf{\gbf{68.8\%}} & \textbf{\gbf{89.1\%}} \\
\gcell{Depth 2} & No-worse vs Full & \ours{} & \xmark & \xmark & \xmark & \xmark & \cmark \\
\gcell{Depth 2} & No-worse vs Full & Random & \xmark & \xmark & \xmark & \xmark & \xmark \\
    \midrule
    \gcell{Depth 3} & \(\dE\) vs Full \(\downarrow\) & \ours{} & \textbf{\(-0.0029\)} & \textbf{\(+0.0234\)} & \textbf{\(-0.0047\)} & \textbf{\(+0.0022\)} & \textbf{\(-0.0039\)} \\
    \gcell{Depth 3} & \(\dE\) vs Full \(\downarrow\) & Random & \rbf{\(+0.1469\)} & \rbf{\(+0.3339\)} & \rbf{\(+0.5480\)} & \rbf{\(+0.5354\)} & \rbf{\(+3.4985\)} \\
    \bcell{Depth 3} & Shot gain \(\uparrow\) & \ours{} & \gbf{33.6\%} & \gbf{43.8\%} & \gbf{50.5\%} & \gbf{55.8\%} & \gbf{84.1\%} \\
    \bcell{Depth 3} & Shot gain \(\uparrow\) & Random & \textbf{\gbf{34.2\%}} & \textbf{\gbf{44.1\%}} & \textbf{\gbf{51.9\%}} & \textbf{\gbf{58.1\%}} & \textbf{\gbf{85.5\%}} \\
\gcell{Depth 3} & No-worse vs Full & \ours{} & \cmark & \xmark & \cmark & \xmark & \cmark \\
\gcell{Depth 3} & No-worse vs Full & Random & \xmark & \xmark & \xmark & \xmark & \xmark \\
    \bottomrule
    \end{tabular}}
    \label{tab:app_tau095_main}
    \vspace{-0.5em}
\end{table}

Table~\ref{tab:app_tau095_main} shows that at threshold $\tau=0.95$, \ourmethod{} reaches mean absolute final-energy gap $0.00917$ while saving $62.3\%$ of full-gradient measurement shots. The matched-budget random baseline has mean absolute gap $1.33512$ with $63.4\%$ shot saving, so the selected QWC groups matter beyond the budget alone.

\begin{table}[H]
    \centering
    \vspace{-0.5em}
    \caption{\textbf{Depth-wise aggregate (previous threshold $\tau=0.95$).} Same protocol as Table~\ref{tab:app_tau095_main} with top-mass target threshold $\tau=0.95$. Mean $|\dE|$ aggregates the five molecule columns at each ansatz depth; No-worse marks whether the aggregate mean signed $\dE$ is at most $1.6$ mHa above Full.}
    \vspace{0.25em}
    \setlength{\tabcolsep}{1.2mm}
    \small
    \begin{tabular}{l|ccc}
    \toprule
    Setting & Mean \(|\dE|\downarrow\) & No-worse & Mean shot gain \(\uparrow\) \\
    \midrule
    \grow Full-QWC reference & 0.00000 & \cmark & \gray{0.0\%} \\
\brow \ours{} depth 1 & 0.00290 & \xmark & \gbf{71.2\%} \\
\brow \ours{} depth 2 & 0.01719 & \xmark & \gbf{62.2\%} \\
\brow \ours{} depth 3 & 0.00742 & \xmark & \gbf{53.6\%} \\
    \midrule
Random depth 1 & 1.56136 & \xmark & \gbf{73.7\%} \\
Random depth 2 & 1.43145 & \xmark & \gbf{61.9\%} \\
Random depth 3 & 1.01255 & \xmark & \gbf{54.8\%} \\
    \midrule
\brow \ours{} all & 0.00917 & \xmark & \gbf{62.3\%} \\
Random all & 1.33512 & \xmark & \gbf{63.4\%} \\
    \bottomrule
    \end{tabular}
    \label{tab:app_tau095_agg}
    \vspace{-0.5em}
\end{table}

\begin{table}[H]
    \centering
    \vspace{-0.5em}
    \caption{\textbf{Detailed budget usage (previous threshold $\tau=0.95$).} Same table format with top-mass target threshold $\tau=0.95$. Groups/param. reports the mean selected QWC groups per parameter, shot gain is relative to full $2PGS$, and No-worse vs Full uses the one-sided $1.6$ mHa criterion ($\dE \le 1.6$ mHa; lower energy is accepted). Boldface marks the better Ours-vs-Random entry for each molecule-depth comparison.}
    \vspace{0.25em}
    \setlength{\tabcolsep}{0.9mm}
    \resizebox{1.0\linewidth}{!}{
    \begin{tabular}{ll|c|ccc|ccc|ccc|c}
    \toprule
    Molecule & Method & Full QWC & \multicolumn{3}{c|}{Groups / param. \(\downarrow\)} & \multicolumn{3}{c|}{Shot gain \(\uparrow\)} & \multicolumn{3}{c|}{No-worse vs Full} & Best \(\dE\downarrow\) \\
    \cmidrule(lr){4-6}\cmidrule(lr){7-9}\cmidrule(lr){10-12}
    & & groups/param. & D1 & D2 & D3 & D1 & D2 & D3 & D1 & D2 & D3 & \\
    \midrule
    H2/STO-3G & \ours{} & \gcell{5} & \bcell{2.81} & \bcell{\textbf{2.99}} & \bcell{3.32} & \gbf{43.8\%} & \textbf{\gbf{40.2\%}} & \gbf{33.6\%} & \xmark & \xmark & \cmark & \textbf{\(+0.0023\)} \\
    H2/STO-3G & Random & \gcell{5} & \bcell{\textbf{2.79}} & \bcell{3.02} & \bcell{\textbf{3.29}} & \textbf{\gbf{44.3\%}} & \gbf{39.6\%} & \textbf{\gbf{34.2\%}} & \xmark & \xmark & \xmark & \rbf{\(+0.1469\)} \\
    \midrule
    HeH+/STO-3G & \ours{} & \gcell{11} & \bcell{3.83} & \bcell{\textbf{5.04}} & \bcell{6.18} & \gbf{65.2\%} & \textbf{\gbf{54.1\%}} & \gbf{43.8\%} & \cmark & \xmark & \xmark & \textbf{\gbf{\(-0.0006\)}} \\
    HeH+/STO-3G & Random & \gcell{11} & \bcell{\textbf{3.47}} & \bcell{5.37} & \bcell{\textbf{6.15}} & \textbf{\gbf{68.5\%}} & \gbf{51.2\%} & \textbf{\gbf{44.1\%}} & \xmark & \xmark & \xmark & \rbf{\(+0.3339\)} \\
    \midrule
    H3+/STO-3G & \ours{} & \gcell{19} & \bcell{4.82} & \bcell{\textbf{7.23}} & \bcell{9.41} & \gbf{74.6\%} & \textbf{\gbf{62.0\%}} & \gbf{50.5\%} & \xmark & \xmark & \cmark & \textbf{\(-0.0047\)} \\
    H3+/STO-3G & Random & \gcell{19} & \bcell{\textbf{4.34}} & \bcell{7.44} & \bcell{\textbf{9.13}} & \textbf{\gbf{77.2\%}} & \gbf{60.8\%} & \textbf{\gbf{51.9\%}} & \xmark & \xmark & \xmark & \rbf{\(+0.5480\)} \\
    \midrule
    H4 chain/STO-3G & \ours{} & \gcell{75} & \bcell{14.28} & \bcell{25.16} & \bcell{33.14} & \gbf{81.0\%} & \gbf{66.4\%} & \gbf{55.8\%} & \xmark & \xmark & \xmark & \textbf{\(+0.0017\)} \\
    H4 chain/STO-3G & Random & \gcell{75} & \bcell{\textbf{11.08}} & \bcell{\textbf{23.37}} & \bcell{\textbf{31.45}} & \textbf{\gbf{85.2\%}} & \textbf{\gbf{68.8\%}} & \textbf{\gbf{58.1\%}} & \xmark & \xmark & \xmark & \rbf{\(+0.5354\)} \\
    \midrule
    LiH/STO-3G & \ours{} & \gcell{179} & \bcell{15.21} & \bcell{20.93} & \bcell{28.47} & \gbf{91.5\%} & \gbf{88.3\%} & \gbf{84.1\%} & \xmark & \cmark & \cmark & \textbf{\(-0.0017\)} \\
    LiH/STO-3G & Random & \gcell{179} & \bcell{\textbf{12.21}} & \bcell{\textbf{19.58}} & \bcell{\textbf{25.95}} & \textbf{\gbf{93.2\%}} & \textbf{\gbf{89.1\%}} & \textbf{\gbf{85.5\%}} & \xmark & \xmark & \xmark & \rbf{\(+3.4272\)} \\
    \bottomrule
    \end{tabular}}
    \label{tab:app_tau095_budget}
    \vspace{-0.5em}
\end{table}

\clearpage

\clearpage
\section{Previous Threshold: $\tau=0.99$}

\begin{table}[H]
    \centering
    \vspace{-0.5em}
    \caption{\textbf{Matched-budget online VQE baseline comparison (conservative threshold $\tau=0.99$).} The learned selector is trained on RY/RZ VQE trajectories from depths 1--3 using row-wise top-mass target threshold $\tau=0.99$ and evaluated in closed-loop online VQE on held-out random seeds for the same five PySCF molecules and depths. For each depth and metric, \ourmethod{} is paired with a Random same-budget baseline that keeps the same predicted number of QWC groups per parameter but chooses groups uniformly at random. Boldface marks the better method for each molecule-specific comparison.}
    \vspace{0.25em}
    \setlength{\tabcolsep}{0.8mm}
    \resizebox{1.0\linewidth}{!}{
    \begin{tabular}{lll|ccccc}
    \toprule
    Setting & Metric & Method & H2/STO-3G & HeH+/STO-3G & H3+/STO-3G & H4 chain/STO-3G & LiH/STO-3G \\
    \midrule
    \gcell{Depth 1} & \(\dE\) vs Full \(\downarrow\) & \ours{} & \textbf{\gbf{\(+0.0013\)}} & \textbf{\(+0.0013\)} & \textbf{\gbf{\(+0.0008\)}} & \textbf{\gbf{\(-0.0008\)}} & \textbf{\gbf{\(+0.0013\)}} \\
    \gcell{Depth 1} & \(\dE\) vs Full \(\downarrow\) & Random & \rbf{\(+0.2060\)} & \rbf{\(+1.0934\)} & \rbf{\(+0.5968\)} & \rbf{\(+1.0145\)} & \rbf{\(+3.0269\)} \\
    \bcell{Depth 1} & Shot gain \(\uparrow\) & \ours{} & \textbf{\gbf{36.6\%}} & \gbf{51.3\%} & \gbf{58.2\%} & \gbf{67.4\%} & \gbf{81.7\%} \\
    \bcell{Depth 1} & Shot gain \(\uparrow\) & Random & \gbf{35.5\%} & \textbf{\gbf{54.3\%}} & \textbf{\gbf{60.1\%}} & \textbf{\gbf{72.7\%}} & \textbf{\gbf{86.9\%}} \\
\gcell{Depth 1} & No-worse vs Full & \ours{} & \cmark & \cmark & \cmark & \cmark & \cmark \\
\gcell{Depth 1} & No-worse vs Full & Random & \xmark & \xmark & \xmark & \xmark & \xmark \\
    \midrule
    \gcell{Depth 2} & \(\dE\) vs Full \(\downarrow\) & \ours{} & \textbf{\(-0.0008\)} & \textbf{\(+0.0037\)} & \textbf{\(-0.0015\)} & \textbf{\(+0.0012\)} & \textbf{\(-0.0009\)} \\
    \gcell{Depth 2} & \(\dE\) vs Full \(\downarrow\) & Random & \rbf{\(+0.1683\)} & \rbf{\(+0.3132\)} & \rbf{\(+0.3132\)} & \rbf{\(+0.5964\)} & \rbf{\(+3.4087\)} \\
    \bcell{Depth 2} & Shot gain \(\uparrow\) & \ours{} & \textbf{\gbf{25.8\%}} & \gbf{32.1\%} & \gbf{41.2\%} & \textbf{\gbf{47.7\%}} & \gbf{74.7\%} \\
    \bcell{Depth 2} & Shot gain \(\uparrow\) & Random & \gbf{25.2\%} & \textbf{\gbf{32.3\%}} & \textbf{\gbf{41.6\%}} & \gbf{47.2\%} & \textbf{\gbf{75.7\%}} \\
\gcell{Depth 2} & No-worse vs Full & \ours{} & \cmark & \xmark & \cmark & \cmark & \cmark \\
\gcell{Depth 2} & No-worse vs Full & Random & \xmark & \xmark & \xmark & \xmark & \xmark \\
    \midrule
    \gcell{Depth 3} & \(\dE\) vs Full \(\downarrow\) & \ours{} & \textbf{\(+0.0028\)} & \textbf{\(+0.0087\)} & \textbf{\(-0.0006\)} & \textbf{\(-0.0009\)} & \textbf{\(+0.0008\)} \\
    \gcell{Depth 3} & \(\dE\) vs Full \(\downarrow\) & Random & \rbf{\(+0.1022\)} & \rbf{\(+0.2128\)} & \rbf{\(+0.1860\)} & \rbf{\(+0.1159\)} & \rbf{\(+2.6667\)} \\
    \bcell{Depth 3} & Shot gain \(\uparrow\) & \ours{} & \textbf{\gbf{18.0\%}} & \textbf{\gbf{23.6\%}} & \textbf{\gbf{28.5\%}} & \textbf{\gbf{33.1\%}} & \gbf{63.1\%} \\
    \bcell{Depth 3} & Shot gain \(\uparrow\) & Random & \gbf{17.4\%} & \gbf{23.4\%} & \gbf{28.2\%} & \gbf{33.0\%} & \textbf{\gbf{64.7\%}} \\
\gcell{Depth 3} & No-worse vs Full & \ours{} & \xmark & \xmark & \cmark & \cmark & \cmark \\
\gcell{Depth 3} & No-worse vs Full & Random & \xmark & \xmark & \xmark & \xmark & \xmark \\
    \bottomrule
    \end{tabular}}
    \label{tab:app_tau099_main}
    \vspace{-0.5em}
\end{table}

Table~\ref{tab:app_tau099_main} shows that at threshold $\tau=0.99$, \ourmethod{} reaches mean absolute final-energy gap $0.00439$ while saving $45.5\%$ of full-gradient measurement shots. The matched-budget random baseline has mean absolute gap $0.93589$ with $46.5\%$ shot saving, so the selected QWC groups matter beyond the budget alone.

\begin{table}[H]
    \centering
    \vspace{-0.5em}
    \caption{\textbf{Depth-wise aggregate (conservative threshold $\tau=0.99$).} Same protocol as Table~\ref{tab:app_tau099_main} with top-mass target threshold $\tau=0.99$. Mean $|\dE|$ aggregates the five molecule columns at each ansatz depth; No-worse marks whether the aggregate mean signed $\dE$ is at most $1.6$ mHa above Full.}
    \vspace{0.25em}
    \setlength{\tabcolsep}{1.2mm}
    \small
    \begin{tabular}{l|ccc}
    \toprule
    Setting & Mean \(|\dE|\downarrow\) & No-worse & Mean shot gain \(\uparrow\) \\
    \midrule
    \grow Full-QWC reference & 0.00000 & \cmark & \gray{0.0\%} \\
\brow \ours{} depth 1 & 0.00139 & \cmark & \gbf{59.0\%} \\
\brow \ours{} depth 2 & 0.00560 & \cmark & \gbf{44.3\%} \\
\brow \ours{} depth 3 & 0.00620 & \xmark & \gbf{33.2\%} \\
    \midrule
Random depth 1 & 1.18750 & \xmark & \gbf{61.9\%} \\
Random depth 2 & 0.95994 & \xmark & \gbf{44.4\%} \\
Random depth 3 & 0.66023 & \xmark & \gbf{33.3\%} \\
    \midrule
\brow \ours{} all & 0.00439 & \cmark & \gbf{45.5\%} \\
Random all & 0.93589 & \xmark & \gbf{46.5\%} \\
    \bottomrule
    \end{tabular}
    \label{tab:app_tau099_agg}
    \vspace{-0.5em}
\end{table}

\begin{table}[H]
    \centering
    \vspace{-0.5em}
    \caption{\textbf{Detailed budget usage (conservative threshold $\tau=0.99$).} Same table format with top-mass target threshold $\tau=0.99$. Groups/param. reports the mean selected QWC groups per parameter, shot gain is relative to full $2PGS$, and No-worse vs Full uses the one-sided $1.6$ mHa criterion ($\dE \le 1.6$ mHa; lower energy is accepted). Boldface marks the better Ours-vs-Random entry for each molecule-depth comparison.}
    \vspace{0.25em}
    \setlength{\tabcolsep}{0.9mm}
    \resizebox{1.0\linewidth}{!}{
    \begin{tabular}{ll|c|ccc|ccc|ccc|c}
    \toprule
    Molecule & Method & Full QWC & \multicolumn{3}{c|}{Groups / param. \(\downarrow\)} & \multicolumn{3}{c|}{Shot gain \(\uparrow\)} & \multicolumn{3}{c|}{No-worse vs Full} & Best \(\dE\downarrow\) \\
    \cmidrule(lr){4-6}\cmidrule(lr){7-9}\cmidrule(lr){10-12}
    & & groups/param. & D1 & D2 & D3 & D1 & D2 & D3 & D1 & D2 & D3 & \\
    \midrule
    H2/STO-3G & \ours{} & \gcell{5} & \bcell{\textbf{3.17}} & \bcell{\textbf{3.71}} & \bcell{\textbf{4.10}} & \textbf{\gbf{36.6\%}} & \textbf{\gbf{25.8\%}} & \textbf{\gbf{18.0\%}} & \cmark & \cmark & \xmark & \textbf{\gbf{\(+0.0013\)}} \\
    H2/STO-3G & Random & \gcell{5} & \bcell{3.22} & \bcell{3.74} & \bcell{4.13} & \gbf{35.5\%} & \gbf{25.2\%} & \gbf{17.4\%} & \xmark & \xmark & \xmark & \rbf{\(+0.1022\)} \\
    \midrule
    HeH+/STO-3G & \ours{} & \gcell{11} & \bcell{5.35} & \bcell{7.47} & \bcell{\textbf{8.41}} & \gbf{51.3\%} & \gbf{32.1\%} & \textbf{\gbf{23.6\%}} & \cmark & \xmark & \xmark & \textbf{\(+0.0013\)} \\
    HeH+/STO-3G & Random & \gcell{11} & \bcell{\textbf{5.03}} & \bcell{\textbf{7.45}} & \bcell{8.42} & \textbf{\gbf{54.3\%}} & \textbf{\gbf{32.3\%}} & \gbf{23.4\%} & \xmark & \xmark & \xmark & \rbf{\(+0.2128\)} \\
    \midrule
    H3+/STO-3G & \ours{} & \gcell{19} & \bcell{7.94} & \bcell{11.16} & \bcell{\textbf{13.59}} & \gbf{58.2\%} & \gbf{41.2\%} & \textbf{\gbf{28.5\%}} & \cmark & \cmark & \cmark & \textbf{\gbf{\(+0.0008\)}} \\
    H3+/STO-3G & Random & \gcell{19} & \bcell{\textbf{7.58}} & \bcell{\textbf{11.09}} & \bcell{13.65} & \textbf{\gbf{60.1\%}} & \textbf{\gbf{41.6\%}} & \gbf{28.2\%} & \xmark & \xmark & \xmark & \rbf{\(+0.1860\)} \\
    \midrule
    H4 chain/STO-3G & \ours{} & \gcell{75} & \bcell{24.45} & \bcell{\textbf{39.22}} & \bcell{\textbf{50.17}} & \gbf{67.4\%} & \textbf{\gbf{47.7\%}} & \textbf{\gbf{33.1\%}} & \cmark & \cmark & \cmark & \textbf{\gbf{\(-0.0008\)}} \\
    H4 chain/STO-3G & Random & \gcell{75} & \bcell{\textbf{20.45}} & \bcell{39.59} & \bcell{50.24} & \textbf{\gbf{72.7\%}} & \gbf{47.2\%} & \gbf{33.0\%} & \xmark & \xmark & \xmark & \rbf{\(+0.1159\)} \\
    \midrule
    LiH/STO-3G & \ours{} & \gcell{179} & \bcell{32.84} & \bcell{45.34} & \bcell{66.09} & \gbf{81.7\%} & \gbf{74.7\%} & \gbf{63.1\%} & \cmark & \cmark & \cmark & \textbf{\gbf{\(+0.0013\)}} \\
    LiH/STO-3G & Random & \gcell{179} & \bcell{\textbf{23.47}} & \bcell{\textbf{43.46}} & \bcell{\textbf{63.21}} & \textbf{\gbf{86.9\%}} & \textbf{\gbf{75.7\%}} & \textbf{\gbf{64.7\%}} & \xmark & \xmark & \xmark & \rbf{\(+2.6667\)} \\
    \bottomrule
    \end{tabular}}
    \label{tab:app_tau099_budget}
    \vspace{-0.5em}
\end{table}

\clearpage

\clearpage
\section{Shot-Allocation Details}
\label{sec:app_shot_allocation}
\subsection{Molecule-Wise In-Distribution Allocation}
\begin{table*}[t]
\centering
\vspace{-0.6em}
\caption{\textbf{Molecule-wise shot-allocation benchmark at $b=1.0$.} This expands Table~\ref{tab:v71_shot_allocation_pareto} using the layout of Table~\ref{tab:v39_tau0995_main}: five STO-3G molecules, depths 1--2, four held-out seeds, 28 online iterations, and 1024 nominal shots per shifted selected parameter--group setting. $\dE$ and shot gain are paired against Full-QWC uniform; adaptive allocators choose their realized shot counts. \cmark{}/\xmark{} applies the one-sided mean criterion $\dE\leq1.6$ mHa. Boldface marks the closest non-Full mean energy and bold-underlining marks the largest shot gain in each molecule--depth cell.}
\vspace{0.2em}
\setlength{\tabcolsep}{0.42mm}
\renewcommand{\arraystretch}{0.76}
\tiny
\resizebox{1.0\linewidth}{!}{
\begin{tabular}{lll|ccccc}
\toprule
Setting & Metric & Method & H2 (4q) & HeH+ (4q) & H3+ (6q) & H4 chain (8q) & LiH (12q) \\
\midrule
\gcell{Depth 1} & $\dE$ vs Full $\downarrow$ & Full-QWC uniform & \gray{\(+0.0000\)} & \gray{\(+0.0000\)} & \gray{\(+0.0000\)} & \gray{\(+0.0000\)} & \gray{\(+0.0000\)} \\
\gcell{Depth 1} & $\dE$ vs Full $\downarrow$ & iCANS~(\citeauthor{kubler2020adaptive}) & \rbf{\(+0.0038\)} & \rbf{\(+0.0793\)} & \rbf{\(+0.0321\)} & \rbf{\(+0.0086\)} & \rbf{\(+0.0177\)} \\
\gcell{Depth 1} & $\dE$ vs Full $\downarrow$ & gCANS~(\citeauthor{gu2021adaptive}) & \rbf{\(+0.0205\)} & \rbf{\(+0.0881\)} & \rbf{\(+0.0703\)} & \gbf{\(-0.0062\)} & \rbf{\(+0.0315\)} \\
\gcell{Depth 1} & $\dE$ vs Full $\downarrow$ & Rosalin~(\citeauthor{arrasmith2020operator}) & \rbf{\(+0.0070\)} & \textbf{\gbf{\(-0.0092\)}} & \textbf{\gbf{\(+0.0005\)}} & \rbf{\(+0.0060\)} & \rbf{\(+0.0076\)} \\
\gcell{Depth 1} & $\dE$ vs Full $\downarrow$ & Variance-aware~(\citeauthor{zhu2024optimizing}) & \gbf{\(-0.0028\)} & \gbf{\(-0.0116\)} & \rbf{\(+0.0020\)} & \textbf{\rbf{\(+0.0033\)}} & \rbf{\(+0.0136\)} \\
\gcell{Depth 1} & $\dE$ vs Full $\downarrow$ & \ours{} uniform & \textbf{\gbf{\(-0.0024\)}} & \gbf{\(-0.0221\)} & \gbf{\(-0.0157\)} & \gbf{\(-0.0082\)} & \textbf{\gbf{\(+0.0006\)}} \\
\gcell{Depth 1} & $\dE$ vs Full $\downarrow$ & \ours{}+variance-aware~(\citeauthor{zhu2024optimizing}) & \gbf{\(-0.0034\)} & \gbf{\(-0.0232\)} & \gbf{\(-0.0230\)} & \gbf{\(-0.0137\)} & \rbf{\(+0.0024\)} \\
\gcell{Depth 1} & $\dE$ vs Full $\downarrow$ & \ours{}+gCANS~(\citeauthor{gu2021adaptive}) & \rbf{\(+0.0108\)} & \rbf{\(+0.0335\)} & \rbf{\(+0.0145\)} & \rbf{\(+0.0099\)} & \gbf{\(-0.0020\)} \\
\cmidrule(lr){2-8}
\bcell{Depth 1} & Shot gain $\uparrow$ & Full-QWC uniform & \gray{0.0\%} & \gray{0.0\%} & \gray{0.0\%} & \gray{0.0\%} & \gray{0.0\%} \\
\bcell{Depth 1} & Shot gain $\uparrow$ & iCANS~(\citeauthor{kubler2020adaptive}) & $-73.0\%$ & $-64.3\%$ & $-73.8\%$ & $-89.1\%$ & $-108.9\%$ \\
\bcell{Depth 1} & Shot gain $\uparrow$ & gCANS~(\citeauthor{gu2021adaptive}) & 92.7\% & 93.8\% & 93.3\% & 93.0\% & 90.2\% \\
\bcell{Depth 1} & Shot gain $\uparrow$ & Rosalin~(\citeauthor{arrasmith2020operator}) & $-85.0\%$ & $-62.8\%$ & $-90.3\%$ & $-114.7\%$ & $-135.7\%$ \\
\bcell{Depth 1} & Shot gain $\uparrow$ & Variance-aware~(\citeauthor{zhu2024optimizing}) & 0.0\% & 0.0\% & 0.0\% & 0.0\% & 0.0\% \\
\bcell{Depth 1} & Shot gain $\uparrow$ & \ours{} uniform & 41.2\% & 59.6\% & 64.2\% & 76.2\% & 87.9\% \\
\bcell{Depth 1} & Shot gain $\uparrow$ & \ours{}+variance-aware~(\citeauthor{zhu2024optimizing}) & 41.5\% & 59.7\% & 63.7\% & 75.8\% & 87.7\% \\
\bcell{Depth 1} & Shot gain $\uparrow$ & \ours{}+gCANS~(\citeauthor{gu2021adaptive}) & \textbf{\underline{96.3\%}} & \textbf{\underline{97.6\%}} & \textbf{\underline{97.8\%}} & \textbf{\underline{98.5\%}} & \textbf{\underline{99.3\%}} \\
\cmidrule(lr){2-8}
\gcell{Depth 1} & No-worse vs Full & Full-QWC uniform & \cmark & \cmark & \cmark & \cmark & \cmark \\
\gcell{Depth 1} & No-worse vs Full & iCANS~(\citeauthor{kubler2020adaptive}) & \xmark & \xmark & \xmark & \xmark & \xmark \\
\gcell{Depth 1} & No-worse vs Full & gCANS~(\citeauthor{gu2021adaptive}) & \xmark & \xmark & \xmark & \cmark & \xmark \\
\gcell{Depth 1} & No-worse vs Full & Rosalin~(\citeauthor{arrasmith2020operator}) & \xmark & \cmark & \cmark & \xmark & \xmark \\
\gcell{Depth 1} & No-worse vs Full & Variance-aware~(\citeauthor{zhu2024optimizing}) & \cmark & \cmark & \xmark & \xmark & \xmark \\
\gcell{Depth 1} & No-worse vs Full & \ours{} uniform & \cmark & \cmark & \cmark & \cmark & \cmark \\
\gcell{Depth 1} & No-worse vs Full & \ours{}+variance-aware~(\citeauthor{zhu2024optimizing}) & \cmark & \cmark & \cmark & \cmark & \xmark \\
\gcell{Depth 1} & No-worse vs Full & \ours{}+gCANS~(\citeauthor{gu2021adaptive}) & \xmark & \xmark & \xmark & \xmark & \cmark \\
\midrule
\gcell{Depth 2} & $\dE$ vs Full $\downarrow$ & Full-QWC uniform & \gray{\(+0.0000\)} & \gray{\(+0.0000\)} & \gray{\(+0.0000\)} & \gray{\(+0.0000\)} & \gray{\(+0.0000\)} \\
\gcell{Depth 2} & $\dE$ vs Full $\downarrow$ & iCANS~(\citeauthor{kubler2020adaptive}) & \rbf{\(+0.1044\)} & \rbf{\(+0.1569\)} & \rbf{\(+0.0239\)} & \rbf{\(+0.1216\)} & \rbf{\(+0.2486\)} \\
\gcell{Depth 2} & $\dE$ vs Full $\downarrow$ & gCANS~(\citeauthor{gu2021adaptive}) & \textbf{\rbf{\(+0.0279\)}} & \gbf{\(-0.0474\)} & \rbf{\(+0.0853\)} & \rbf{\(+0.1612\)} & \rbf{\(+0.3004\)} \\
\gcell{Depth 2} & $\dE$ vs Full $\downarrow$ & Rosalin~(\citeauthor{arrasmith2020operator}) & \rbf{\(+0.0570\)} & \rbf{\(+0.1309\)} & \gbf{\(-0.0221\)} & \rbf{\(+0.0157\)} & \gbf{\(-0.1710\)} \\
\gcell{Depth 2} & $\dE$ vs Full $\downarrow$ & Variance-aware~(\citeauthor{zhu2024optimizing}) & \rbf{\(+0.0645\)} & \rbf{\(+0.0997\)} & \textbf{\rbf{\(+0.0050\)}} & \rbf{\(+0.0044\)} & \gbf{\(-0.1812\)} \\
\gcell{Depth 2} & $\dE$ vs Full $\downarrow$ & \ours{} uniform & \rbf{\(+0.0954\)} & \textbf{\rbf{\(+0.0222\)}} & \gbf{\(-0.0962\)} & \rbf{\(+0.0720\)} & \textbf{\rbf{\(+0.0427\)}} \\
\gcell{Depth 2} & $\dE$ vs Full $\downarrow$ & \ours{}+variance-aware~(\citeauthor{zhu2024optimizing}) & \rbf{\(+0.0336\)} & \rbf{\(+0.0237\)} & \gbf{\(-0.0175\)} & \textbf{\gbf{\(-0.0000\)}} & \gbf{\(-0.1850\)} \\
\gcell{Depth 2} & $\dE$ vs Full $\downarrow$ & \ours{}+gCANS~(\citeauthor{gu2021adaptive}) & \gbf{\(-0.0319\)} & \rbf{\(+0.0938\)} & \rbf{\(+0.0138\)} & \rbf{\(+0.1896\)} & \rbf{\(+0.0891\)} \\
\cmidrule(lr){2-8}
\bcell{Depth 2} & Shot gain $\uparrow$ & Full-QWC uniform & \gray{0.0\%} & \gray{0.0\%} & \gray{0.0\%} & \gray{0.0\%} & \gray{0.0\%} \\
\bcell{Depth 2} & Shot gain $\uparrow$ & iCANS~(\citeauthor{kubler2020adaptive}) & $-67.5\%$ & $-13.3\%$ & $-60.1\%$ & $-71.6\%$ & $-144.6\%$ \\
\bcell{Depth 2} & Shot gain $\uparrow$ & gCANS~(\citeauthor{gu2021adaptive}) & 92.8\% & 93.7\% & 93.1\% & 90.4\% & 84.6\% \\
\bcell{Depth 2} & Shot gain $\uparrow$ & Rosalin~(\citeauthor{arrasmith2020operator}) & $-68.1\%$ & $-42.6\%$ & $-86.8\%$ & $-110.4\%$ & $-181.2\%$ \\
\bcell{Depth 2} & Shot gain $\uparrow$ & Variance-aware~(\citeauthor{zhu2024optimizing}) & 0.0\% & 0.0\% & 0.0\% & 0.0\% & 0.0\% \\
\bcell{Depth 2} & Shot gain $\uparrow$ & \ours{} uniform & 27.8\% & 33.2\% & 46.3\% & 52.0\% & 82.4\% \\
\bcell{Depth 2} & Shot gain $\uparrow$ & \ours{}+variance-aware~(\citeauthor{zhu2024optimizing}) & 27.5\% & 31.9\% & 45.6\% & 52.1\% & 82.8\% \\
\bcell{Depth 2} & Shot gain $\uparrow$ & \ours{}+gCANS~(\citeauthor{gu2021adaptive}) & \textbf{\underline{95.3\%}} & \textbf{\underline{95.8\%}} & \textbf{\underline{96.6\%}} & \textbf{\underline{96.4\%}} & \textbf{\underline{98.3\%}} \\
\cmidrule(lr){2-8}
\gcell{Depth 2} & No-worse vs Full & Full-QWC uniform & \cmark & \cmark & \cmark & \cmark & \cmark \\
\gcell{Depth 2} & No-worse vs Full & iCANS~(\citeauthor{kubler2020adaptive}) & \xmark & \xmark & \xmark & \xmark & \xmark \\
\gcell{Depth 2} & No-worse vs Full & gCANS~(\citeauthor{gu2021adaptive}) & \xmark & \cmark & \xmark & \xmark & \xmark \\
\gcell{Depth 2} & No-worse vs Full & Rosalin~(\citeauthor{arrasmith2020operator}) & \xmark & \xmark & \cmark & \xmark & \cmark \\
\gcell{Depth 2} & No-worse vs Full & Variance-aware~(\citeauthor{zhu2024optimizing}) & \xmark & \xmark & \xmark & \xmark & \cmark \\
\gcell{Depth 2} & No-worse vs Full & \ours{} uniform & \xmark & \xmark & \cmark & \xmark & \xmark \\
\gcell{Depth 2} & No-worse vs Full & \ours{}+variance-aware~(\citeauthor{zhu2024optimizing}) & \xmark & \xmark & \cmark & \cmark & \cmark \\
\gcell{Depth 2} & No-worse vs Full & \ours{}+gCANS~(\citeauthor{gu2021adaptive}) & \cmark & \xmark & \xmark & \xmark & \xmark \\
\bottomrule
\end{tabular}}
\label{tab:v71_shot_allocation_detail}
\vspace{-0.5em}
\end{table*}

\clearpage
\section{Depth-OOD Shot-Allocation Benchmark}

\begin{table}[H]
\centering
\vspace{-0.6em}
\caption{\textbf{Shot-allocation benchmark in the Table~\ref{tab:v50_depth_ood_baselines_main} depth-OOD protocol at $b=1.0$.} The saved selector is trained on depths 1--4 and evaluated on unseen depths 5 and 6 over five PySCF/STO-3G molecules and four held-out seeds; depth 7 is reported in Appendix~\ref{sec:v75_depth_ood_allocation_depth7}. All methods use 28 common Adam updates and the same moment-matched Gaussian finite-shot estimator with 1024 nominal shots per selected parameter--group pair and $\tau=0.997$. iCANS~\citep{kubler2020adaptive}, gCANS~\citep{gu2021adaptive}, Rosalin~\citep{arrasmith2020operator}, and variance-aware allocation~\citep{zhu2024optimizing} are controlled allocation policies rather than reproductions of their complete optimizer stacks. Green/red $\dE$ and \cmark{}/\xmark{} use the one-sided 1.6 mHa criterion. Bold marks the closest non-Full mean $\dE$; bold-underlining marks the largest saving in each cell.}
\vspace{0.15em}
\setlength{\tabcolsep}{0.35mm}
\renewcommand{\arraystretch}{0.48}
\tiny
\resizebox{1.0\linewidth}{!}{
\begin{tabular}{lll|ccccc}
\toprule
Setting & Metric & Method & H2 (4q) & HeH+ (4q) & H3+ (6q) & H4 chain (8q) & LiH (12q) \\
\midrule
\gcell{Depth 5} & $\dE$ vs Full $\downarrow$ & Full-QWC uniform & \gray{\(+0.0000\)} & \gray{\(+0.0000\)} & \gray{\(+0.0000\)} & \gray{\(+0.0000\)} & \gray{\(+0.0000\)} \\
\gcell{Depth 5} & $\dE$ vs Full $\downarrow$ & iCANS & \rbf{\(+0.0380\)} & \rbf{\(+0.0300\)} & \rbf{\(+0.0334\)} & \gbf{\(-0.0279\)} & \gbf{\(-0.0371\)} \\
\gcell{Depth 5} & $\dE$ vs Full $\downarrow$ & gCANS & \rbf{\(+0.0690\)} & \rbf{\(+0.1606\)} & \rbf{\(+0.0775\)} & \rbf{\(+0.1232\)} & \rbf{\(+0.0538\)} \\
\gcell{Depth 5} & $\dE$ vs Full $\downarrow$ & Rosalin & \gbf{\(-0.0192\)} & \textbf{\gbf{\(-0.0242\)}} & \textbf{\gbf{\(-0.0014\)}} & \textbf{\gbf{\(-0.0212\)}} & \gbf{\(-0.2266\)} \\
\gcell{Depth 5} & $\dE$ vs Full $\downarrow$ & Variance-aware & \gbf{\(-0.0232\)} & \gbf{\(-0.0382\)} & \gbf{\(-0.0447\)} & \gbf{\(-0.1764\)} & \gbf{\(-0.2449\)} \\
\gcell{Depth 5} & $\dE$ vs Full $\downarrow$ & \ours{} uniform & \textbf{\gbf{\(-0.0026\)}} & \rbf{\(+0.0484\)} & \gbf{\(-0.0274\)} & \gbf{\(-0.0407\)} & \textbf{\gbf{\(-0.0139\)}} \\
\gcell{Depth 5} & $\dE$ vs Full $\downarrow$ & \ours{}+variance-aware & \gbf{\(-0.0094\)} & \gbf{\(-0.0339\)} & \gbf{\(-0.0543\)} & \gbf{\(-0.1588\)} & \gbf{\(-0.2644\)} \\
\gcell{Depth 5} & $\dE$ vs Full $\downarrow$ & \ours{}+gCANS & \rbf{\(+0.0158\)} & \rbf{\(+0.0569\)} & \gbf{\(-0.0597\)} & \gbf{\(-0.1106\)} & \gbf{\(-0.2085\)} \\
\addlinespace[0.10em]
\bcell{Depth 5} & Shot gain $\uparrow$ & Full-QWC uniform & \gray{0.0\%} & \gray{0.0\%} & \gray{0.0\%} & \gray{0.0\%} & \gray{0.0\%} \\
\bcell{Depth 5} & Shot gain $\uparrow$ & iCANS & $-37.2\%$ & $-8.0\%$ & $-79.9\%$ & $-116.5\%$ & $-190.5\%$ \\
\bcell{Depth 5} & Shot gain $\uparrow$ & gCANS & 89.9\% & 93.2\% & 73.7\% & 57.1\% & 39.5\% \\
\bcell{Depth 5} & Shot gain $\uparrow$ & Rosalin & $-43.2\%$ & $-21.3\%$ & $-104.4\%$ & $-162.1\%$ & $-229.2\%$ \\
\bcell{Depth 5} & Shot gain $\uparrow$ & Variance-aware & 0.0\% & 0.0\% & 0.0\% & 0.0\% & 0.0\% \\
\bcell{Depth 5} & Shot gain $\uparrow$ & \ours{} uniform & 5.6\% & 8.5\% & 10.6\% & 17.1\% & 55.7\% \\
\bcell{Depth 5} & Shot gain $\uparrow$ & \ours{}+variance-aware & 5.6\% & 8.5\% & 10.6\% & 17.0\% & 53.7\% \\
\bcell{Depth 5} & Shot gain $\uparrow$ & \ours{}+gCANS & \textbf{\underline{90.2\%}} & \textbf{\underline{93.6\%}} & \textbf{\underline{78.7\%}} & \textbf{\underline{61.6\%}} & \textbf{\underline{68.8\%}} \\
\addlinespace[0.10em]
\gcell{Depth 5} & No-worse vs Full & Full-QWC uniform & \cmark & \cmark & \cmark & \cmark & \cmark \\
\gcell{Depth 5} & No-worse vs Full & iCANS & \xmark & \xmark & \xmark & \cmark & \cmark \\
\gcell{Depth 5} & No-worse vs Full & gCANS & \xmark & \xmark & \xmark & \xmark & \xmark \\
\gcell{Depth 5} & No-worse vs Full & Rosalin & \cmark & \cmark & \cmark & \cmark & \cmark \\
\gcell{Depth 5} & No-worse vs Full & Variance-aware & \cmark & \cmark & \cmark & \cmark & \cmark \\
\gcell{Depth 5} & No-worse vs Full & \ours{} uniform & \cmark & \xmark & \cmark & \cmark & \cmark \\
\gcell{Depth 5} & No-worse vs Full & \ours{}+variance-aware & \cmark & \cmark & \cmark & \cmark & \cmark \\
\gcell{Depth 5} & No-worse vs Full & \ours{}+gCANS & \xmark & \xmark & \cmark & \cmark & \cmark \\
\midrule
\gcell{Depth 6} & $\dE$ vs Full $\downarrow$ & Full-QWC uniform & \gray{\(+0.0000\)} & \gray{\(+0.0000\)} & \gray{\(+0.0000\)} & \gray{\(+0.0000\)} & \gray{\(+0.0000\)} \\
\gcell{Depth 6} & $\dE$ vs Full $\downarrow$ & iCANS & \rbf{\(+0.0659\)} & \rbf{\(+0.0090\)} & \rbf{\(+0.0690\)} & \rbf{\(+0.1038\)} & \textbf{\rbf{\(+0.0150\)}} \\
\gcell{Depth 6} & $\dE$ vs Full $\downarrow$ & gCANS & \rbf{\(+0.0555\)} & \rbf{\(+0.0643\)} & \rbf{\(+0.1207\)} & \rbf{\(+0.0554\)} & \rbf{\(+0.2048\)} \\
\gcell{Depth 6} & $\dE$ vs Full $\downarrow$ & Rosalin & \rbf{\(+0.0101\)} & \textbf{\gbf{\(-0.0008\)}} & \rbf{\(+0.0200\)} & \gbf{\(-0.0183\)} & \gbf{\(-0.3586\)} \\
\gcell{Depth 6} & $\dE$ vs Full $\downarrow$ & Variance-aware & \textbf{\rbf{\(+0.0021\)}} & \gbf{\(-0.0258\)} & \textbf{\gbf{\(-0.0009\)}} & \textbf{\rbf{\(+0.0039\)}} & \gbf{\(-0.2956\)} \\
\gcell{Depth 6} & $\dE$ vs Full $\downarrow$ & \ours{} uniform & \rbf{\(+0.0320\)} & \gbf{\(-0.0262\)} & \rbf{\(+0.0307\)} & \rbf{\(+0.0214\)} & \rbf{\(+0.0280\)} \\
\gcell{Depth 6} & $\dE$ vs Full $\downarrow$ & \ours{}+variance-aware & \rbf{\(+0.0084\)} & \gbf{\(-0.0126\)} & \gbf{\(-0.0173\)} & \gbf{\(-0.0265\)} & \gbf{\(-0.3267\)} \\
\gcell{Depth 6} & $\dE$ vs Full $\downarrow$ & \ours{}+gCANS & \rbf{\(+0.0385\)} & \gbf{\(-0.0262\)} & \gbf{\(-0.0305\)} & \gbf{\(-0.0211\)} & \gbf{\(-0.2811\)} \\
\addlinespace[0.10em]
\bcell{Depth 6} & Shot gain $\uparrow$ & Full-QWC uniform & \gray{0.0\%} & \gray{0.0\%} & \gray{0.0\%} & \gray{0.0\%} & \gray{0.0\%} \\
\bcell{Depth 6} & Shot gain $\uparrow$ & iCANS & $-58.3\%$ & $-7.4\%$ & $-97.1\%$ & $-133.9\%$ & $-196.7\%$ \\
\bcell{Depth 6} & Shot gain $\uparrow$ & gCANS & 89.7\% & 92.8\% & 57.3\% & 38.0\% & 8.3\% \\
\bcell{Depth 6} & Shot gain $\uparrow$ & Rosalin & $-59.3\%$ & $-29.8\%$ & $-110.5\%$ & $-181.7\%$ & $-241.1\%$ \\
\bcell{Depth 6} & Shot gain $\uparrow$ & Variance-aware & 0.0\% & 0.0\% & 0.0\% & 0.0\% & 0.0\% \\
\bcell{Depth 6} & Shot gain $\uparrow$ & \ours{} uniform & 4.8\% & 7.6\% & 8.7\% & 14.5\% & \textbf{\underline{46.1\%}} \\
\bcell{Depth 6} & Shot gain $\uparrow$ & \ours{}+variance-aware & 4.8\% & 7.6\% & 8.7\% & 14.5\% & 45.1\% \\
\bcell{Depth 6} & Shot gain $\uparrow$ & \ours{}+gCANS & \textbf{\underline{89.9\%}} & \textbf{\underline{93.1\%}} & \textbf{\underline{61.6\%}} & \textbf{\underline{41.4\%}} & 36.4\% \\
\addlinespace[0.10em]
\gcell{Depth 6} & No-worse vs Full & Full-QWC uniform & \cmark & \cmark & \cmark & \cmark & \cmark \\
\gcell{Depth 6} & No-worse vs Full & iCANS & \xmark & \xmark & \xmark & \xmark & \xmark \\
\gcell{Depth 6} & No-worse vs Full & gCANS & \xmark & \xmark & \xmark & \xmark & \xmark \\
\gcell{Depth 6} & No-worse vs Full & Rosalin & \xmark & \cmark & \xmark & \cmark & \cmark \\
\gcell{Depth 6} & No-worse vs Full & Variance-aware & \xmark & \cmark & \cmark & \xmark & \cmark \\
\gcell{Depth 6} & No-worse vs Full & \ours{} uniform & \xmark & \cmark & \xmark & \xmark & \xmark \\
\gcell{Depth 6} & No-worse vs Full & \ours{}+variance-aware & \xmark & \cmark & \cmark & \cmark & \cmark \\
\gcell{Depth 6} & No-worse vs Full & \ours{}+gCANS & \xmark & \cmark & \cmark & \cmark & \cmark \\
\bottomrule
\end{tabular}}
\label{tab:v75_depth_ood_allocation_main}
\vspace{-0.8em}
\end{table}

Across all 60 held-out seed-level cases, \ours{} uniform is the most faithful non-Full policy, with mean $|\dE|=0.0568$ Ha while saving 16.6\% of shots. Composing selection with gCANS increases saving to 63.3\% and improves on gCANS alone in mean $|\dE|$ (0.0973 vs. 0.1014 Ha) and seed-level no-worse rate (60.0\% vs. 10.0\%). The variance-aware composition reaches the highest Ours-family no-worse rate (68.3\%) but with a larger absolute gap. Thus Ours-uniform is the clean fidelity--saving point, whereas Ours+gCANS is the aggressive Pareto point.

\clearpage
\section{Depth-7 Shot-Allocation Detail}
\label{sec:v75_depth_ood_allocation_depth7}

\begin{table}[H]
\centering
\vspace{-0.5em}
\caption{\textbf{Depth-7 continuation of the depth-OOD shot-allocation benchmark.} This table uses the same trained depth-1--4 selector, five PySCF/STO-3G molecules, four held-out seeds, finite-shot estimator, $b=1.0$, and $\tau=0.997$ as Table~\ref{tab:v75_depth_ood_allocation_main}.}
\vspace{0.15em}
\setlength{\tabcolsep}{0.45mm}
\renewcommand{\arraystretch}{0.58}
\tiny
\resizebox{1.0\linewidth}{!}{
\begin{tabular}{lll|ccccc}
\toprule
Setting & Metric & Method & H2 (4q) & HeH+ (4q) & H3+ (6q) & H4 chain (8q) & LiH (12q) \\
\midrule
\gcell{Depth 7} & $\dE$ vs Full $\downarrow$ & Full-QWC uniform & \gray{\(+0.0000\)} & \gray{\(+0.0000\)} & \gray{\(+0.0000\)} & \gray{\(+0.0000\)} & \gray{\(+0.0000\)} \\
\gcell{Depth 7} & $\dE$ vs Full $\downarrow$ & iCANS & \rbf{\(+0.1195\)} & \rbf{\(+0.0256\)} & \rbf{\(+0.0540\)} & \rbf{\(+0.0295\)} & \rbf{\(+0.0791\)} \\
\gcell{Depth 7} & $\dE$ vs Full $\downarrow$ & gCANS & \rbf{\(+0.0151\)} & \rbf{\(+0.1313\)} & \rbf{\(+0.0640\)} & \rbf{\(+0.0598\)} & \rbf{\(+0.1809\)} \\
\gcell{Depth 7} & $\dE$ vs Full $\downarrow$ & Rosalin & \rbf{\(+0.0517\)} & \rbf{\(+0.0421\)} & \gbf{\(-0.0208\)} & \gbf{\(-0.0153\)} & \gbf{\(-0.2483\)} \\
\gcell{Depth 7} & $\dE$ vs Full $\downarrow$ & Variance-aware & \gbf{\(-0.0276\)} & \textbf{\gbf{\(-0.0240\)}} & \gbf{\(-0.0199\)} & \gbf{\(-0.0429\)} & \gbf{\(-0.2975\)} \\
\gcell{Depth 7} & $\dE$ vs Full $\downarrow$ & \ours{} uniform & \gbf{\(-0.0109\)} & \gbf{\(-0.0544\)} & \rbf{\(+0.0231\)} & \textbf{\rbf{\(+0.0099\)}} & \textbf{\rbf{\(+0.0250\)}} \\
\gcell{Depth 7} & $\dE$ vs Full $\downarrow$ & \ours{}+variance-aware & \gbf{\(-0.0125\)} & \gbf{\(-0.0392\)} & \textbf{\gbf{\(-0.0045\)}} & \gbf{\(-0.0712\)} & \gbf{\(-0.2989\)} \\
\gcell{Depth 7} & $\dE$ vs Full $\downarrow$ & \ours{}+gCANS & \textbf{\gbf{\(-0.0025\)}} & \rbf{\(+0.0291\)} & \rbf{\(+0.0091\)} & \gbf{\(-0.0303\)} & \gbf{\(-0.3097\)} \\
\addlinespace[0.10em]
\bcell{Depth 7} & Shot gain $\uparrow$ & Full-QWC uniform & \gray{0.0\%} & \gray{0.0\%} & \gray{0.0\%} & \gray{0.0\%} & \gray{0.0\%} \\
\bcell{Depth 7} & Shot gain $\uparrow$ & iCANS & $-48.9\%$ & $-12.8\%$ & $-89.5\%$ & $-147.0\%$ & $-200.9\%$ \\
\bcell{Depth 7} & Shot gain $\uparrow$ & gCANS & \textbf{\underline{87.2\%}} & \textbf{\underline{92.0\%}} & \textbf{\underline{65.0\%}} & 5.3\% & $-31.4\%$ \\
\bcell{Depth 7} & Shot gain $\uparrow$ & Rosalin & $-52.0\%$ & $-31.4\%$ & $-111.2\%$ & $-191.9\%$ & $-249.3\%$ \\
\bcell{Depth 7} & Shot gain $\uparrow$ & Variance-aware & 0.0\% & 0.0\% & 0.0\% & 0.0\% & 0.0\% \\
\bcell{Depth 7} & Shot gain $\uparrow$ & \ours{} uniform & 3.8\% & 6.7\% & 7.5\% & 13.1\% & \textbf{\underline{39.2\%}} \\
\bcell{Depth 7} & Shot gain $\uparrow$ & \ours{}+variance-aware & 3.6\% & 6.6\% & 7.4\% & \textbf{\underline{13.1\%}} & 38.0\% \\
\bcell{Depth 7} & Shot gain $\uparrow$ & \ours{}+gCANS & 86.5\% & 91.6\% & 64.0\% & $-0.0\%$ & $-8.2\%$ \\
\addlinespace[0.10em]
\gcell{Depth 7} & No-worse vs Full & Full-QWC uniform & \cmark & \cmark & \cmark & \cmark & \cmark \\
\gcell{Depth 7} & No-worse vs Full & iCANS & \xmark & \xmark & \xmark & \xmark & \xmark \\
\gcell{Depth 7} & No-worse vs Full & gCANS & \xmark & \xmark & \xmark & \xmark & \xmark \\
\gcell{Depth 7} & No-worse vs Full & Rosalin & \xmark & \xmark & \cmark & \cmark & \cmark \\
\gcell{Depth 7} & No-worse vs Full & Variance-aware & \cmark & \cmark & \cmark & \cmark & \cmark \\
\gcell{Depth 7} & No-worse vs Full & \ours{} uniform & \cmark & \cmark & \xmark & \xmark & \xmark \\
\gcell{Depth 7} & No-worse vs Full & \ours{}+variance-aware & \cmark & \cmark & \cmark & \cmark & \cmark \\
\gcell{Depth 7} & No-worse vs Full & \ours{}+gCANS & \cmark & \xmark & \xmark & \cmark & \cmark \\
\bottomrule
\end{tabular}}
\label{tab:v75_depth_ood_allocation_depth7}
\vspace{-0.7em}
\end{table}

\clearpage
\subsection{Hamiltonian-Panel Allocation}
\begin{table*}[t]
\centering
\vspace{-0.6em}
\caption{\textbf{Shot-allocation benchmark on the Table~\ref{tab:v55_hamiltonian_panel} Hamiltonian panel at $b=1.0$.} We reuse the saved depth-1--4 selector and evaluate seven held-out 8-qubit Hamiltonians at depths 1 and 4 over seeds 97/101/107/109, with 28 Adam updates, $\tau=0.997$, and 1024 nominal shots per shifted selected parameter--group setting. All methods are rerun against Full-QWC uniform with the same moment-matched Gaussian finite-shot estimator and common Adam optimizer; iCANS~\citep{kubler2020adaptive}, gCANS~\citep{gu2021adaptive}, Rosalin~\citep{arrasmith2020operator}, and variance-aware allocation~\citep{zhu2024optimizing} are therefore controlled allocation policies rather than reproductions of their complete optimizer stacks. Green/red $\dE$ and \cmark{}/\xmark{} use the one-sided 1.6 mHa criterion. Bold marks the closest non-Full mean $\dE$; bold-underlining marks the largest shot gain in each cell.}
\vspace{0.15em}
\setlength{\tabcolsep}{0.38mm}
\renewcommand{\arraystretch}{0.53}
\tiny
\resizebox{\textwidth}{!}{
\begin{tabular}{lll|cccccccc}
\toprule
Testing Hamiltonian & Depth & Metric & Full-U & iCANS & gCANS & Rosalin & Var.-aware & \ours{}-U & \ours{}+Var. & \ours{}+gCANS \\
\midrule
\gcell{TFIM $h=0.5$} & Depth 1 & $\dE$ vs Full $\downarrow$ & \gray{\(+0.0000\)} & \gbf{\(-0.0120\)} & \textbf{\gbf{\(-0.0000\)}} & \gbf{\(-0.0050\)} & \rbf{\(+0.0041\)} & \gbf{\(-0.0013\)} & \gbf{\(-0.0012\)} & \gbf{\(+0.0009\)} \\
 & Depth 1 & Shot gain $\uparrow$ & \gray{0.0\%} & $-126.6\%$ & \textbf{\underline{72.1\%}} & $-106.4\%$ & 0.0\% & 2.1\% & 2.0\% & 69.9\% \\
 & Depth 1 & No-worse vs Full & \cmark & \cmark & \cmark & \cmark & \xmark & \cmark & \cmark & \cmark \\
\addlinespace[0.08em]
 & Depth 4 & $\dE$ vs Full $\downarrow$ & \gray{\(+0.0000\)} & \rbf{\(+0.0245\)} & \rbf{\(+0.0427\)} & \rbf{\(+0.0028\)} & \gbf{\(-0.0071\)} & \textbf{\gbf{\(-0.0002\)}} & \gbf{\(-0.0022\)} & \rbf{\(+0.0445\)} \\
 & Depth 4 & Shot gain $\uparrow$ & \gray{0.0\%} & $-88.5\%$ & 77.0\% & $-95.0\%$ & 0.0\% & 2.2\% & 1.9\% & \textbf{\underline{77.4\%}} \\
 & Depth 4 & No-worse vs Full & \cmark & \xmark & \xmark & \xmark & \cmark & \cmark & \cmark & \xmark \\
\midrule
\gcell{TFIM $h=0.7$} & Depth 1 & $\dE$ vs Full $\downarrow$ & \gray{\(+0.0000\)} & \gbf{\(-0.0122\)} & \rbf{\(+0.0081\)} & \gbf{\(-0.0031\)} & \rbf{\(+0.0093\)} & \gbf{\(-0.0020\)} & \textbf{\gbf{\(+0.0014\)}} & \rbf{\(+0.0192\)} \\
 & Depth 1 & Shot gain $\uparrow$ & \gray{0.0\%} & $-116.5\%$ & \textbf{\underline{81.4\%}} & $-91.5\%$ & 0.0\% & 3.2\% & 3.1\% & 80.9\% \\
 & Depth 1 & No-worse vs Full & \cmark & \cmark & \xmark & \cmark & \xmark & \cmark & \cmark & \xmark \\
\addlinespace[0.08em]
 & Depth 4 & $\dE$ vs Full $\downarrow$ & \gray{\(+0.0000\)} & \rbf{\(+0.0182\)} & \rbf{\(+0.0718\)} & \textbf{\rbf{\(+0.0087\)}} & \gbf{\(-0.0129\)} & \gbf{\(-0.0161\)} & \gbf{\(-0.0179\)} & \rbf{\(+0.0606\)} \\
 & Depth 4 & Shot gain $\uparrow$ & \gray{0.0\%} & $-88.5\%$ & 73.2\% & $-81.2\%$ & 0.0\% & 2.1\% & 1.9\% & \textbf{\underline{75.6\%}} \\
 & Depth 4 & No-worse vs Full & \cmark & \xmark & \xmark & \xmark & \cmark & \cmark & \cmark & \xmark \\
\midrule
\gcell{TFIM $h=0.9$} & Depth 1 & $\dE$ vs Full $\downarrow$ & \gray{\(+0.0000\)} & \rbf{\(+0.0041\)} & \rbf{\(+0.0300\)} & \textbf{\gbf{\(-0.0037\)}} & \rbf{\(+0.0163\)} & \rbf{\(+0.0811\)} & \rbf{\(+0.2016\)} & \rbf{\(+0.1387\)} \\
 & Depth 1 & Shot gain $\uparrow$ & \gray{0.0\%} & $-106.5\%$ & 82.7\% & $-80.4\%$ & 0.0\% & 25.0\% & 25.0\% & \textbf{\underline{88.7\%}} \\
 & Depth 1 & No-worse vs Full & \cmark & \xmark & \xmark & \cmark & \xmark & \xmark & \xmark & \xmark \\
\addlinespace[0.08em]
 & Depth 4 & $\dE$ vs Full $\downarrow$ & \gray{\(+0.0000\)} & \rbf{\(+0.0321\)} & \rbf{\(+0.0813\)} & \rbf{\(+0.0561\)} & \textbf{\gbf{\(-0.0146\)}} & \rbf{\(+0.0496\)} & \rbf{\(+0.0307\)} & \rbf{\(+0.0691\)} \\
 & Depth 4 & Shot gain $\uparrow$ & \gray{0.0\%} & $-88.3\%$ & 72.3\% & $-76.2\%$ & 0.0\% & 6.2\% & 6.2\% & \textbf{\underline{79.5\%}} \\
 & Depth 4 & No-worse vs Full & \cmark & \xmark & \xmark & \xmark & \cmark & \xmark & \xmark & \xmark \\
\midrule
\gcell{Heis. XXZ+X field} & Depth 1 & $\dE$ vs Full $\downarrow$ & \gray{\(+0.0000\)} & \rbf{\(+0.4624\)} & \rbf{\(+0.7663\)} & \rbf{\(+0.3730\)} & \gbf{\(-0.1707\)} & \gbf{\(-0.2163\)} & \textbf{\rbf{\(+0.1527\)}} & \rbf{\(+0.6480\)} \\
 & Depth 1 & Shot gain $\uparrow$ & \gray{0.0\%} & $-55.0\%$ & 89.8\% & $-62.1\%$ & 0.0\% & 16.7\% & 16.7\% & \textbf{\underline{92.8\%}} \\
 & Depth 1 & No-worse vs Full & \cmark & \xmark & \xmark & \xmark & \cmark & \cmark & \xmark & \xmark \\
\addlinespace[0.08em]
 & Depth 4 & $\dE$ vs Full $\downarrow$ & \gray{\(+0.0000\)} & \rbf{\(+0.5043\)} & \rbf{\(+1.1720\)} & \textbf{\gbf{\(-0.0388\)}} & \rbf{\(+0.0759\)} & \gbf{\(-0.0737\)} & \gbf{\(-0.0484\)} & \rbf{\(+0.8442\)} \\
 & Depth 4 & Shot gain $\uparrow$ & \gray{0.0\%} & $-59.6\%$ & 82.2\% & $-61.5\%$ & 0.0\% & 4.2\% & 4.2\% & \textbf{\underline{82.2\%}} \\
 & Depth 4 & No-worse vs Full & \cmark & \xmark & \xmark & \cmark & \xmark & \cmark & \cmark & \xmark \\
\midrule
\gcell{Heis. J1--J2} & Depth 1 & $\dE$ vs Full $\downarrow$ & \gray{\(+0.0000\)} & \rbf{\(+0.2970\)} & \rbf{\(+0.6094\)} & \textbf{\gbf{\(-0.0259\)}} & \gbf{\(-0.3164\)} & \gbf{\(-0.1839\)} & \gbf{\(-0.0655\)} & \rbf{\(+0.3957\)} \\
 & Depth 1 & Shot gain $\uparrow$ & \gray{0.0\%} & $-49.9\%$ & 91.4\% & $-53.7\%$ & 0.0\% & 16.7\% & 16.7\% & \textbf{\underline{93.9\%}} \\
 & Depth 1 & No-worse vs Full & \cmark & \xmark & \xmark & \cmark & \cmark & \cmark & \cmark & \xmark \\
\addlinespace[0.08em]
 & Depth 4 & $\dE$ vs Full $\downarrow$ & \gray{\(+0.0000\)} & \rbf{\(+0.1698\)} & \rbf{\(+1.0128\)} & \rbf{\(+0.3410\)} & \rbf{\(+0.2490\)} & \textbf{\rbf{\(+0.1508\)}} & \rbf{\(+0.3493\)} & \rbf{\(+0.8673\)} \\
 & Depth 4 & Shot gain $\uparrow$ & \gray{0.0\%} & $-52.0\%$ & 80.7\% & $-64.7\%$ & 0.0\% & 4.2\% & 4.2\% & \textbf{\underline{80.9\%}} \\
 & Depth 4 & No-worse vs Full & \cmark & \xmark & \xmark & \xmark & \xmark & \xmark & \xmark & \xmark \\
\midrule
\gcell{Heis. XXY+Z field} & Depth 1 & $\dE$ vs Full $\downarrow$ & \gray{\(+0.0000\)} & \rbf{\(+0.3060\)} & \rbf{\(+0.2416\)} & \rbf{\(+0.2757\)} & \gbf{\(-0.2531\)} & \textbf{\gbf{\(-0.0757\)}} & \rbf{\(+0.0825\)} & \rbf{\(+0.5081\)} \\
 & Depth 1 & Shot gain $\uparrow$ & \gray{0.0\%} & $-69.4\%$ & 86.0\% & $-63.3\%$ & 0.0\% & 16.7\% & 16.7\% & \textbf{\underline{89.5\%}} \\
 & Depth 1 & No-worse vs Full & \cmark & \xmark & \xmark & \xmark & \cmark & \cmark & \xmark & \xmark \\
\addlinespace[0.08em]
 & Depth 4 & $\dE$ vs Full $\downarrow$ & \gray{\(+0.0000\)} & \rbf{\(+0.3343\)} & \rbf{\(+0.4644\)} & \textbf{\rbf{\(+0.1376\)}} & \rbf{\(+0.3029\)} & \gbf{\(-0.4193\)} & \gbf{\(-0.1713\)} & \rbf{\(+0.8496\)} \\
 & Depth 4 & Shot gain $\uparrow$ & \gray{0.0\%} & $-69.7\%$ & \textbf{\underline{77.0\%}} & $-59.8\%$ & 0.0\% & 4.2\% & 4.2\% & 74.5\% \\
 & Depth 4 & No-worse vs Full & \cmark & \xmark & \xmark & \xmark & \xmark & \cmark & \cmark & \xmark \\
\midrule
\gcell{JW-like parity spin} & Depth 1 & $\dE$ vs Full $\downarrow$ & \gray{\(+0.0000\)} & \rbf{\(+0.0465\)} & \rbf{\(+0.1333\)} & \textbf{\rbf{\(+0.0172\)}} & \gbf{\(-0.0230\)} & \rbf{\(+0.0185\)} & \gbf{\(-0.0198\)} & \rbf{\(+0.0440\)} \\
 & Depth 1 & Shot gain $\uparrow$ & \gray{0.0\%} & $-94.4\%$ & 92.7\% & $-95.6\%$ & 0.0\% & 36.2\% & 36.1\% & \textbf{\underline{95.7\%}} \\
 & Depth 1 & No-worse vs Full & \cmark & \xmark & \xmark & \xmark & \cmark & \xmark & \cmark & \xmark \\
\addlinespace[0.08em]
 & Depth 4 & $\dE$ vs Full $\downarrow$ & \gray{\(+0.0000\)} & \textbf{\gbf{\(-0.0141\)}} & \rbf{\(+0.1421\)} & \gbf{\(-0.0724\)} & \gbf{\(-0.0814\)} & \gbf{\(-0.1151\)} & \gbf{\(-0.0997\)} & \rbf{\(+0.0190\)} \\
 & Depth 4 & Shot gain $\uparrow$ & \gray{0.0\%} & $-104.6\%$ & 72.2\% & $-118.7\%$ & 0.0\% & 11.5\% & 11.6\% & \textbf{\underline{75.2\%}} \\
 & Depth 4 & No-worse vs Full & \cmark & \cmark & \xmark & \cmark & \cmark & \cmark & \cmark & \xmark \\
\bottomrule
\end{tabular}}
\label{tab:v74_hamiltonian_allocation}
\vspace{-0.7em}
\end{table*}

Across all 56 paired seed-level cases, \ours{}-uniform saves 10.8\% of shots with mean $|\dE|=0.1889$ Ha, while Full-QWC variance-aware allocation has a smaller 0.1614 Ha gap but no shot saving. Composing \ourmethod{} with gCANS reaches an aggressive 82.6\% saving and slightly improves on gCANS alone in both saving (80.8\%) and mean absolute gap (0.3748 vs. 0.4001 Ha), but both pass only 14.3\% of seed-level comparisons. In contrast, iCANS and Rosalin spend 83.5\% and 79.3\% more shots than Full on average at this operating point. The panel therefore supports composability, while also showing that allocation policies calibrated on easier molecular regimes do not automatically preserve energy under Hamiltonian-family shift.

\clearpage
\section{Depth-OOD Transfer}
\label{sec:app_depth_ood}
\section{Depth-OOD Transfer Baselines}

This depth-OOD benchmark evaluates the conservative $\tau=0.997$ selector on unseen RY/RZ depths 5--7; the table reports depths 5 and 6. A matched-budget control adds a cell-wise cost-matched TITAN comparison: TITAN$^{\dagger}$ chooses the freezing threshold whose shot gain is closest to \ours{}, while TITAN$^{\dagger}$+\ours{} uses that same active-parameter mask and additionally sparsifies QWC groups.

\begin{table}[H]
\centering
\vspace{-0.6em}
\caption{\textbf{Depth-OOD online VQE transfer baselines at $\tau=0.997$.} The relevance model is trained on depths 1--4; the table reports unseen depths 5 and 6 over five PySCF/STO-3G molecules and four held-out seeds. Random is group-budget matched to \ourmethod{}. TITAN$^{\dagger}$ uses a cell-wise freezing threshold selected to match \ourmethod{}'s shot gain; TITAN$^{\dagger}$+\ours{} reuses that threshold and adds group selection. Green/red $\dE$ indicates the one-sided $1.6$ mHa no-worse criterion. Bold marks the closest mean $\dE$ to Full; bold-underlined shot gain marks the largest saving.}
\vspace{0.15em}
\setlength{\tabcolsep}{0.45mm}
\renewcommand{\arraystretch}{0.72}
\tiny
\resizebox{1.0\linewidth}{!}{
\begin{tabular}{lll|ccccc}
\toprule
Setting & Metric & Method & H2 (4q) & HeH+ (4q) & H3+ (6q) & H4 chain (8q) & LiH (12q) \\
\midrule
\gcell{Depth 5} & \(\dE\) vs Full \(\downarrow\) & \ours{} & \textbf{\gbf{-0.0003}} & \rbf{+0.0053} & \gbf{-0.0120} & \textbf{\gbf{+0.0012}} & \textbf{\gbf{-0.0068}} \\
\gcell{Depth 5} & \(\dE\) vs Full \(\downarrow\) & Random & \rbf{+0.0179} & \rbf{+0.0684} & \rbf{+0.0503} & \rbf{+0.1580} & \rbf{+2.1893} \\
\gcell{Depth 5} & \(\dE\) vs Full \(\downarrow\) & TITAN~(\citeauthor{peng2026titan}) & \rbf{+0.0239} & \rbf{+0.0130} & \rbf{+0.0146} & \rbf{+0.1011} & \rbf{+0.2254} \\
\gcell{Depth 5} & \(\dE\) vs Full \(\downarrow\) & TITAN+\ours{} & \rbf{+0.0251} & \textbf{\rbf{+0.0048}} & \rbf{+0.0111} & \rbf{+0.1009} & \rbf{+0.2250} \\
\gcell{Depth 5} & \(\dE\) vs Full \(\downarrow\) & TITAN$^{\dagger}$ & \gbf{-0.0028} & \gbf{-0.0059} & \textbf{\gbf{-0.0044}} & \rbf{+0.0416} & \rbf{+0.0914} \\
\gcell{Depth 5} & \(\dE\) vs Full \(\downarrow\) & TITAN$^{\dagger}$+\ours{} & \gbf{-0.0035} & \gbf{-0.0056} & \gbf{-0.0112} & \rbf{+0.0392} & \rbf{+0.0928} \\
\addlinespace[0.16em]
\bcell{Depth 5} & Shot gain \(\uparrow\) & \ours{} & 5.7\% & 8.4\% & 10.6\% & 17.1\% & 52.9\% \\
\bcell{Depth 5} & Shot gain \(\uparrow\) & Random & 5.7\% & 8.5\% & 10.7\% & 17.2\% & 55.4\% \\
\bcell{Depth 5} & Shot gain \(\uparrow\) & TITAN~(\citeauthor{peng2026titan}) & 32.0\% & 32.7\% & 38.7\% & 42.3\% & 73.4\% \\
\bcell{Depth 5} & Shot gain \(\uparrow\) & TITAN+\ours{} & \textbf{\underline{34.3\%}} & \textbf{\underline{37.9\%}} & \textbf{\underline{44.9\%}} & \textbf{\underline{52.2\%}} & \textbf{\underline{93.2\%}} \\
\bcell{Depth 5} & Shot gain \(\uparrow\) & TITAN$^{\dagger}$ & 6.9\% & 12.7\% & 13.1\% & 14.4\% & 52.4\% \\
\bcell{Depth 5} & Shot gain \(\uparrow\) & TITAN$^{\dagger}$+\ours{} & 10.3\% & 19.5\% & 22.2\% & 28.4\% & 83.2\% \\
\addlinespace[0.16em]
\gcell{Depth 5} & No-worse vs Full & \ours{} & \cmark & \xmark & \cmark & \cmark & \cmark \\
\gcell{Depth 5} & No-worse vs Full & Random & \xmark & \xmark & \xmark & \xmark & \xmark \\
\gcell{Depth 5} & No-worse vs Full & TITAN~(\citeauthor{peng2026titan}) & \xmark & \xmark & \xmark & \xmark & \xmark \\
\gcell{Depth 5} & No-worse vs Full & TITAN+\ours{} & \xmark & \xmark & \xmark & \xmark & \xmark \\
\gcell{Depth 5} & No-worse vs Full & TITAN$^{\dagger}$ & \cmark & \cmark & \cmark & \xmark & \xmark \\
\gcell{Depth 5} & No-worse vs Full & TITAN$^{\dagger}$+\ours{} & \cmark & \cmark & \cmark & \xmark & \xmark \\
\midrule
\gcell{Depth 6} & \(\dE\) vs Full \(\downarrow\) & \ours{} & \textbf{\gbf{+0.0000}} & \gbf{-0.0036} & \textbf{\gbf{+0.0003}} & \gbf{-0.0081} & \textbf{\rbf{+0.0086}} \\
\gcell{Depth 6} & \(\dE\) vs Full \(\downarrow\) & Random & \rbf{+0.0046} & \textbf{\gbf{+0.0005}} & \rbf{+0.0048} & \rbf{+0.0785} & \rbf{+0.7440} \\
\gcell{Depth 6} & \(\dE\) vs Full \(\downarrow\) & TITAN~(\citeauthor{peng2026titan}) & \rbf{+0.0139} & \gbf{-0.0088} & \gbf{-0.0426} & \rbf{+0.0070} & \rbf{+0.3083} \\
\gcell{Depth 6} & \(\dE\) vs Full \(\downarrow\) & TITAN+\ours{} & \rbf{+0.0133} & \gbf{-0.0113} & \gbf{-0.0429} & \rbf{+0.0082} & \rbf{+0.3092} \\
\gcell{Depth 6} & \(\dE\) vs Full \(\downarrow\) & TITAN$^{\dagger}$ & \gbf{-0.0070} & \gbf{-0.0242} & \rbf{+0.0130} & \textbf{\gbf{-0.0034}} & \rbf{+0.0825} \\
\gcell{Depth 6} & \(\dE\) vs Full \(\downarrow\) & TITAN$^{\dagger}$+\ours{} & \gbf{-0.0069} & \gbf{-0.0280} & \rbf{+0.0162} & \gbf{-0.0071} & \rbf{+0.0798} \\
\addlinespace[0.16em]
\bcell{Depth 6} & Shot gain \(\uparrow\) & \ours{} & 4.8\% & 7.6\% & 8.7\% & 14.6\% & 44.3\% \\
\bcell{Depth 6} & Shot gain \(\uparrow\) & Random & 4.7\% & 7.5\% & 8.7\% & 14.6\% & 46.3\% \\
\bcell{Depth 6} & Shot gain \(\uparrow\) & TITAN~(\citeauthor{peng2026titan}) & 32.8\% & 30.5\% & 28.1\% & 39.4\% & 71.2\% \\
\bcell{Depth 6} & Shot gain \(\uparrow\) & TITAN+\ours{} & \textbf{\underline{34.6\%}} & \textbf{\underline{34.4\%}} & \textbf{\underline{33.8\%}} & \textbf{\underline{47.6\%}} & \textbf{\underline{88.5\%}} \\
\bcell{Depth 6} & Shot gain \(\uparrow\) & TITAN$^{\dagger}$ & 7.0\% & 9.6\% & 7.5\% & 12.4\% & 37.0\% \\
\bcell{Depth 6} & Shot gain \(\uparrow\) & TITAN$^{\dagger}$+\ours{} & 10.1\% & 15.5\% & 15.0\% & 24.8\% & 67.6\% \\
\addlinespace[0.16em]
\gcell{Depth 6} & No-worse vs Full & \ours{} & \cmark & \cmark & \cmark & \cmark & \xmark \\
\gcell{Depth 6} & No-worse vs Full & Random & \xmark & \cmark & \xmark & \xmark & \xmark \\
\gcell{Depth 6} & No-worse vs Full & TITAN~(\citeauthor{peng2026titan}) & \xmark & \cmark & \cmark & \xmark & \xmark \\
\gcell{Depth 6} & No-worse vs Full & TITAN+\ours{} & \xmark & \cmark & \cmark & \xmark & \xmark \\
\gcell{Depth 6} & No-worse vs Full & TITAN$^{\dagger}$ & \cmark & \cmark & \xmark & \cmark & \xmark \\
\gcell{Depth 6} & No-worse vs Full & TITAN$^{\dagger}$+\ours{} & \cmark & \cmark & \xmark & \cmark & \xmark \\
\bottomrule
\end{tabular}}
\label{tab:v50_depth_ood_baselines_main}
\vspace{-0.8em}
\end{table}

Table~\ref{tab:v50_depth_ood_baselines_main} separates parameter-specific QWC selection from parameter freezing at both the original and cost-matched operating points. The dagger rows remove shot budget as an explanation for differences between \ourmethod{} and TITAN-only training; the composed row shows the additional saving available when the two orthogonal controls are combined.

\clearpage
\ifoursappendixcore
\section{Additional Depth-OOD Transfer Tables}

The tables below report aggregate performance and budget usage for depth-OOD transfer at $\tau=0.995$.
\else
\section{Depth-OOD Transfer: Train 1--4, Test 5--7}

This experiment tests whether the learned relevance selector transfers across ansatz depth. We train the Transformer relevance model on depths 1--4 and evaluate closed-loop online VQE on unseen depths 5--7. This is a stricter setting than the held-out-seed tables above because the number of parameters and repeated entangling layers both grow at test time.

\begin{table}[H]
\centering
\vspace{-0.5em}
\caption{\textbf{Depth-OOD online VQE transfer at $\tau=0.995$.} The Transformer relevance model is trained on depths 1--4 and evaluated on unseen deeper ans\"atze at depths 5--7 using the same five molecular Hamiltonians, RY/RZ ansatz family, and held-out seeds 97/101/107/109. All molecular Hamiltonians are generated with the STO-3G minimal basis; table headers report the resulting qubit counts. Random uses the same per-parameter QWC group count predicted by \ourmethod{} but samples groups uniformly. Boldface marks the better method for each molecule-specific comparison.}
\vspace{0.25em}
\setlength{\tabcolsep}{0.8mm}
\resizebox{1.0\linewidth}{!}{
\begin{tabular}{lll|ccccc}
\toprule
Setting & Metric & Method & H2 (4q) & HeH+ (4q) & H3+ (6q) & H4 chain (8q) & LiH (12q) \\
\midrule
\gcell{Depth 5} & \(\dE\) vs Full \(\downarrow\) & \ours{} & \textbf{\gbf{\(+0.0008\)}} & \textbf{\rbf{\(+0.0042\)}} & \textbf{\gbf{\(-0.0218\)}} & \textbf{\rbf{\(+0.0073\)}} & \textbf{\gbf{\(+0.0015\)}} \\
\gcell{Depth 5} & \(\dE\) vs Full \(\downarrow\) & Random & \rbf{\(+0.0365\)} & \rbf{\(+0.0717\)} & \rbf{\(+0.0941\)} & \rbf{\(+0.1296\)} & \rbf{\(+2.3623\)} \\
\bcell{Depth 5} & Shot gain \(\uparrow\) & \ours{} & \gbf{7.0\%} & \textbf{\gbf{11.5\%}} & \textbf{\gbf{14.0\%}} & \gbf{21.3\%} & \gbf{59.4\%} \\
\bcell{Depth 5} & Shot gain \(\uparrow\) & Random & \textbf{\gbf{7.0\%}} & \gbf{11.5\%} & \gbf{13.9\%} & \textbf{\gbf{21.8\%}} & \textbf{\gbf{63.2\%}} \\
\gcell{Depth 5} & No-worse vs Full & \ours{} & \cmark & \xmark & \cmark & \xmark & \cmark \\
\gcell{Depth 5} & No-worse vs Full & Random & \xmark & \xmark & \xmark & \xmark & \xmark \\
\midrule
\gcell{Depth 6} & \(\dE\) vs Full \(\downarrow\) & \ours{} & \textbf{\rbf{\(+0.0017\)}} & \textbf{\gbf{\(-0.0054\)}} & \textbf{\rbf{\(+0.0025\)}} & \textbf{\gbf{\(-0.0055\)}} & \textbf{\gbf{\(-0.0004\)}} \\
\gcell{Depth 6} & \(\dE\) vs Full \(\downarrow\) & Random & \rbf{\(+0.0026\)} & \rbf{\(+0.0433\)} & \rbf{\(+0.0675\)} & \rbf{\(+0.1191\)} & \rbf{\(+1.0575\)} \\
\bcell{Depth 6} & Shot gain \(\uparrow\) & \ours{} & \textbf{\gbf{6.0\%}} & \textbf{\gbf{10.7\%}} & \gbf{11.6\%} & \textbf{\gbf{18.7\%}} & \gbf{51.1\%} \\
\bcell{Depth 6} & Shot gain \(\uparrow\) & Random & \gbf{5.9\%} & \gbf{10.5\%} & \textbf{\gbf{11.9\%}} & \gbf{18.5\%} & \textbf{\gbf{53.5\%}} \\
\gcell{Depth 6} & No-worse vs Full & \ours{} & \xmark & \cmark & \xmark & \cmark & \cmark \\
\gcell{Depth 6} & No-worse vs Full & Random & \xmark & \xmark & \xmark & \xmark & \xmark \\
\midrule
\gcell{Depth 7} & \(\dE\) vs Full \(\downarrow\) & \ours{} & \rbf{\(+0.0026\)} & \textbf{\gbf{\(-0.0117\)}} & \textbf{\gbf{\(+0.0015\)}} & \textbf{\rbf{\(+0.0071\)}} & \textbf{\gbf{\(-0.0001\)}} \\
\gcell{Depth 7} & \(\dE\) vs Full \(\downarrow\) & Random & \textbf{\gbf{\(+0.0015\)}} & \rbf{\(+0.0207\)} & \rbf{\(+0.0481\)} & \rbf{\(+0.0151\)} & \rbf{\(+0.6498\)} \\
\bcell{Depth 7} & Shot gain \(\uparrow\) & \ours{} & \textbf{\gbf{4.7\%}} & \textbf{\gbf{9.6\%}} & \textbf{\gbf{10.2\%}} & \gbf{17.0\%} & \gbf{44.7\%} \\
\bcell{Depth 7} & Shot gain \(\uparrow\) & Random & \gbf{4.7\%} & \gbf{9.6\%} & \gbf{10.2\%} & \textbf{\gbf{17.2\%}} & \textbf{\gbf{46.7\%}} \\
\gcell{Depth 7} & No-worse vs Full & \ours{} & \xmark & \cmark & \cmark & \xmark & \cmark \\
\gcell{Depth 7} & No-worse vs Full & Random & \cmark & \xmark & \xmark & \xmark & \xmark \\
\bottomrule
\end{tabular}}
\label{tab:v41_depth_ood_main}
\vspace{-0.5em}
\end{table}
\fi

\ifoursappendixcore

\begin{table}[H]
\centering
\vspace{-0.5em}
\caption{\textbf{Depth-OOD aggregate at $\tau=0.995$.} Mean $|\dE|$ averages the five molecule columns at each unseen depth. No-worse uses the one-sided criterion $\dE \le 1.6$ mHa against the Full-QWC reference, so lower final energy is accepted.}
\vspace{0.25em}
\setlength{\tabcolsep}{1.2mm}
\small
\begin{tabular}{l|ccc}
\toprule
Setting & Mean \(|\dE|\downarrow\) & No-worse & Mean shot gain \(\uparrow\) \\
\midrule
\grow Full-QWC reference & 0.00000 & \cmark & \gray{0.0\%} \\
\brow \ours{} depth 5 & 0.01540 & \cmark & \gbf{22.6\%} \\
\brow \ours{} depth 6 & 0.00697 & \cmark & \gbf{19.6\%} \\
\brow \ours{} depth 7 & 0.00684 & \cmark & \gbf{17.3\%} \\
\midrule
Random depth 5 & 0.54065 & \xmark & \gbf{23.5\%} \\
Random depth 6 & 0.26326 & \xmark & \gbf{20.1\%} \\
Random depth 7 & 0.15211 & \xmark & \gbf{17.7\%} \\
\midrule
\brow \ours{} all & 0.00974 & \cmark & \gbf{19.8\%} \\
Random all & 0.31867 & \xmark & \gbf{20.4\%} \\
\bottomrule
\end{tabular}
\label{tab:v41_depth_ood_agg}
\vspace{-0.5em}
\end{table}

\begin{table}[H]
\centering
\vspace{-0.5em}
\caption{\textbf{Depth-OOD budget usage at $\tau=0.995$.} Groups/param. reports the mean selected QWC groups per parameter on unseen depths 5--7. Shot gain is relative to full $2PGS$ parameter-shift QWC measurement.}
\vspace{0.25em}
\setlength{\tabcolsep}{0.9mm}
\resizebox{1.0\linewidth}{!}{
\begin{tabular}{ll|c|ccc|ccc|ccc|c}
\toprule
Molecule & Method & Full QWC & \multicolumn{3}{c|}{Groups / param. \(\downarrow\)} & \multicolumn{3}{c|}{Shot gain \(\uparrow\)} & \multicolumn{3}{c|}{No-worse vs Full} & Best \(\dE\downarrow\) \\
\cmidrule(lr){4-6}\cmidrule(lr){7-9}\cmidrule(lr){10-12}
& & groups/param. & D5 & D6 & D7 & D5 & D6 & D7 & D5 & D6 & D7 & \\
\midrule
H2/STO-3G & \ours{} & \gcell{5} & \bcell{4.65} & \bcell{\textbf{4.70}} & \bcell{\textbf{4.77}} & \gbf{7.0\%} & \textbf{\gbf{6.0\%}} & \textbf{\gbf{4.7\%}} & \cmark & \xmark & \xmark & \textbf{\gbf{\(+0.0008\)}} \\
H2/STO-3G & Random & \gcell{5} & \bcell{\textbf{4.65}} & \bcell{4.70} & \bcell{4.77} & \textbf{\gbf{7.0\%}} & \gbf{5.9\%} & \gbf{4.7\%} & \xmark & \xmark & \cmark & \gbf{\(+0.0015\)} \\
\midrule
HeH+/STO-3G & \ours{} & \gcell{11} & \bcell{\textbf{9.73}} & \bcell{\textbf{9.82}} & \bcell{\textbf{9.95}} & \textbf{\gbf{11.5\%}} & \textbf{\gbf{10.7\%}} & \textbf{\gbf{9.6\%}} & \xmark & \cmark & \cmark & \textbf{\rbf{\(+0.0042\)}} \\
HeH+/STO-3G & Random & \gcell{11} & \bcell{9.73} & \bcell{9.85} & \bcell{9.95} & \gbf{11.5\%} & \gbf{10.5\%} & \gbf{9.6\%} & \xmark & \xmark & \xmark & \rbf{\(+0.0207\)} \\
\midrule
H3+/STO-3G & \ours{} & \gcell{19} & \bcell{\textbf{16.34}} & \bcell{16.79} & \bcell{\textbf{17.05}} & \textbf{\gbf{14.0\%}} & \gbf{11.6\%} & \textbf{\gbf{10.2\%}} & \cmark & \xmark & \cmark & \textbf{\gbf{\(+0.0015\)}} \\
H3+/STO-3G & Random & \gcell{19} & \bcell{16.36} & \bcell{\textbf{16.74}} & \bcell{17.06} & \gbf{13.9\%} & \textbf{\gbf{11.9\%}} & \gbf{10.2\%} & \xmark & \xmark & \xmark & \rbf{\(+0.0481\)} \\
\midrule
H4 chain/STO-3G & \ours{} & \gcell{75} & \bcell{58.99} & \bcell{\textbf{60.96}} & \bcell{62.25} & \gbf{21.3\%} & \textbf{\gbf{18.7\%}} & \gbf{17.0\%} & \xmark & \cmark & \xmark & \textbf{\gbf{\(-0.0055\)}} \\
H4 chain/STO-3G & Random & \gcell{75} & \bcell{\textbf{58.69}} & \bcell{61.11} & \bcell{\textbf{62.13}} & \textbf{\gbf{21.8\%}} & \gbf{18.5\%} & \textbf{\gbf{17.2\%}} & \xmark & \xmark & \xmark & \rbf{\(+0.0151\)} \\
\midrule
LiH/STO-3G & \ours{} & \gcell{179} & \bcell{72.65} & \bcell{87.48} & \bcell{98.91} & \gbf{59.4\%} & \gbf{51.1\%} & \gbf{44.7\%} & \cmark & \cmark & \cmark & \textbf{\gbf{\(-0.0001\)}} \\
LiH/STO-3G & Random & \gcell{179} & \bcell{\textbf{65.92}} & \bcell{\textbf{83.27}} & \bcell{\textbf{95.42}} & \textbf{\gbf{63.2\%}} & \textbf{\gbf{53.5\%}} & \textbf{\gbf{46.7\%}} & \xmark & \xmark & \xmark & \rbf{\(+0.6498\)} \\
\bottomrule
\end{tabular}}
\label{tab:v41_depth_ood_budget}
\vspace{-0.5em}
\end{table}
\fi

\clearpage
\section{Depth-OOD Baselines by Molecule and Depth}

\begin{table}[H]
\centering
\vspace{-0.5em}
\caption{\textbf{Aggregate depth-OOD baselines at $\tau=0.997$.} This appendix table summarizes the same experiment as Table~\ref{tab:v50_depth_ood_baselines_main} across all 15 molecule-depth cells and 60 held-out trajectories. CA seeds count held-out trajectories with $|\dE|\le 1.6$ mHa against Full-QWC. Cells won counts molecule-depth cells where the method has the smallest mean $|\dE|$ among all listed methods.}
\vspace{0.25em}
\setlength{\tabcolsep}{1.0mm}
\small
\resizebox{1.0\linewidth}{!}{
\begin{tabular}{l|cccccc}
\toprule
Method & Mean $\dE$ & Mean $|\dE|\downarrow$ & CA seeds & Shot gain $\uparrow$ & Groups/param. & Cells won \\
\midrule
\brow \ours{} & \gbf{-0.00118} & \gbf{0.00716} & \gbf{18/60} & 16.2\% & 38.90 & \gbf{12/15} \\
Global coeff. same budget & \rbf{+0.00226} & 0.01139 & 11/60 & 16.2\% & 38.88 & 3/15 \\
Random same budget & \rbf{+0.25163} & 0.26431 & 2/60 & 16.7\% & 38.13 & 0/15 \\
TITAN~(\citeauthor{peng2026titan}) & \rbf{+0.05683} & 0.09985 & 3/60 & 41.1\% & 24.45 & 0/15 \\
TITAN+\ours{} & \rbf{+0.05647} & 0.09981 & 4/60 & \underline{\gbf{48.6\%}} & 16.52 & 0/15 \\
\bottomrule
\end{tabular}}
\label{tab:v50_depth_ood_baselines_aggregate}
\vspace{-0.5em}
\end{table}

\begin{table}[H]
\centering
\vspace{-0.5em}
\caption{\textbf{Molecule-depth detail for depth-OOD baselines at $\tau=0.997$.} Each entry reports mean $|\dE|$ in Hartree followed by shot saving. Rows are grouped by molecular Hamiltonian and unseen evaluation depth. The model is trained on depths 1--4 and evaluated on depths 5--7 with four held-out seeds per molecule-depth cell. Green bold marks the lowest mean $|\dE|$ in each row.}
\vspace{0.25em}
\setlength{\tabcolsep}{0.75mm}
\scriptsize
\resizebox{1.0\linewidth}{!}{
\begin{tabular}{lc|ccccc}
\toprule
Molecule & Depth & \ours{} & Global coeff. & TITAN~(\citeauthor{peng2026titan}) & TITAN+\ours{} & Random \\
\midrule
\gcell{H2 (4q)} & \gcell{5} & \gbf{0.00119 / 5.7\%} & 0.00786 / 5.7\% & 0.03143 / 32.0\% & 0.03200 / 34.3\% & 0.01792 / 5.7\% \\
\gcell{H2 (4q)} & \gcell{6} & \gbf{0.00101 / 4.8\%} & 0.00915 / 4.7\% & 0.01577 / 32.8\% & 0.01678 / 34.6\% & 0.02628 / 4.7\% \\
\gcell{H2 (4q)} & \gcell{7} & \gbf{0.00199 / 3.8\%} & 0.00352 / 3.8\% & 0.01781 / 29.5\% & 0.01761 / 31.3\% & 0.00650 / 3.8\% \\
\midrule
\bcell{HeH+ (4q)} & \bcell{5} & \gbf{0.00572 / 8.4\%} & 0.01661 / 8.4\% & 0.08943 / 32.7\% & 0.08406 / 37.9\% & 0.06841 / 8.5\% \\
\bcell{HeH+ (4q)} & \bcell{6} & \gbf{0.00484 / 7.6\%} & 0.01717 / 7.6\% & 0.11144 / 30.5\% & 0.11416 / 34.4\% & 0.03489 / 7.5\% \\
\bcell{HeH+ (4q)} & \bcell{7} & 0.01149 / 6.7\% & \gbf{0.00857 / 6.7\%} & 0.09757 / 31.3\% & 0.09613 / 34.6\% & 0.07671 / 6.6\% \\
\midrule
\gcell{H3+ (6q)} & \gcell{5} & \gbf{0.01204 / 10.6\%} & 0.01437 / 10.7\% & 0.05297 / 38.7\% & 0.05186 / 44.9\% & 0.05029 / 10.7\% \\
\gcell{H3+ (6q)} & \gcell{6} & \gbf{0.00422 / 8.7\%} & 0.01294 / 8.6\% & 0.04326 / 28.1\% & 0.04484 / 33.8\% & 0.02628 / 8.7\% \\
\gcell{H3+ (6q)} & \gcell{7} & \gbf{0.00378 / 7.4\%} & 0.00473 / 7.4\% & 0.04070 / 31.6\% & 0.04348 / 35.8\% & 0.01635 / 7.3\% \\
\midrule
\bcell{H4 chain (8q)} & \bcell{5} & \gbf{0.02352 / 17.1\%} & 0.04048 / 17.0\% & 0.12680 / 42.3\% & 0.12740 / 52.2\% & 0.15799 / 17.2\% \\
\bcell{H4 chain (8q)} & \bcell{6} & \gbf{0.00929 / 14.6\%} & 0.01119 / 14.6\% & 0.05705 / 39.4\% & 0.05890 / 47.6\% & 0.07854 / 14.6\% \\
\bcell{H4 chain (8q)} & \bcell{7} & \gbf{0.00614 / 13.0\%} & 0.00704 / 13.1\% & 0.04519 / 36.1\% & 0.04317 / 43.9\% & 0.05636 / 13.0\% \\
\midrule
\gcell{LiH (12q)} & \gcell{5} & 0.00877 / 52.9\% & \gbf{0.00208 / 52.9\%} & 0.22542 / 73.4\% & 0.22498 / 93.2\% & 2.18927 / 55.4\% \\
\gcell{LiH (12q)} & \gcell{6} & \gbf{0.01041 / 44.3\%} & 0.01414 / 44.4\% & 0.34075 / 71.2\% & 0.33967 / 88.5\% & 0.74399 / 46.3\% \\
\gcell{LiH (12q)} & \gcell{7} & 0.00307 / 37.9\% & \gbf{0.00095 / 38.0\%} & 0.20216 / 66.6\% & 0.20205 / 82.5\% & 0.41488 / 39.7\% \\
\bottomrule
\end{tabular}}
\label{tab:v50_depth_ood_baselines_detail}
\vspace{-0.5em}
\end{table}

\clearpage
\section{Controlled Structural OOD Transfer Summary}
\label{sec:app_structural_ood_summary}
Figure~\ref{fig:ood_transfer_bars} presents the matched-budget energy comparison compactly; Table~\ref{tab:ood_transfer_summary} gives the corresponding exact energy gaps, success counts, shot gains, and selected-group budgets.
\begin{table}[H]
\centering
\vspace{-0.5em}
\caption{\textbf{Controlled OOD transfer at $\tau=0.997$.} \textbf{(a)} Each target Hamiltonian is absent from training; no-worse counts final energies at most $1.6$ mHa above Full-QWC, accepting lower energies. \textbf{(b)} Each named ansatz is unseen during training; strict CA counts seeds with $|\dE|\leq1.6$ mHa, and \ourmethod{} wins counts molecule--depth cells with a smaller mean absolute gap than Random. In both panels, Random uses the same parameter-specific QWC-group count as \ourmethod{}, isolating relevance ordering at a matched measurement budget. Within each paired setting, bold marks the lower mean $|\dE|$ and larger success count; underline marks the runner-up. Budget columns are intentionally unranked. Complete TITAN controls are Tables~\ref{tab:v54_hamiltonian_generalization_full} and~\ref{tab:v43_named_ansatz_transfer_full} in Appendices~\ref{sec:app_molecule_hamiltonian_ood} and~\ref{sec:app_ansatz_ood}.}
\label{tab:ood_transfer_summary}
\label{tab:v54_hamiltonian_generalization}
\label{tab:v43_named_ansatz_transfer}
\vspace{0.25em}
\begingroup
\newcommand{\oodbest}[1]{\textbf{#1}}
\newcommand{\oodrunner}[1]{\underline{#1}}

\textbf{(a) Hamiltonian-family OOD}
\par\vspace{0.15em}
\setlength{\tabcolsep}{1.25mm}
\scriptsize
\resizebox{\linewidth}{!}{
\begin{tabular}{ll|c|ccc|cc}
\toprule
Test Hamiltonian & Method & Full groups & Mean $\dE$ & Mean $|\dE|\downarrow$ & No-worse vs Full & Shot gain $\uparrow$ & Groups/param. \\
\midrule
\brow TFIM $h=0.7$ & \ours{} & 2 & $-1.55{\times}10^{-15}$ & \oodbest{$3.66{\times}10^{-15}$} & \oodbest{16/16} & 6.13\% & 1.877 \\
TFIM $h=0.7$ & Random & 2 & $-0.00253$ & \oodrunner{$0.00710$} & \oodrunner{6/16} & 6.20\% & 1.876 \\
\midrule
\brow TFIM $h=1.2$ & \ours{} & 2 & $0.21181$ & \oodrunner{$0.22777$} & \oodrunner{0/16} & 13.02\% & 1.740 \\
TFIM $h=1.2$ & Random & 2 & $0.03008$ & \oodbest{$0.03988$} & \oodbest{1/16} & 13.02\% & 1.740 \\
\midrule
\brow Heisenberg XXZ+field & \ours{} & 3 & $-0.00011$ & \oodbest{$0.00011$} & \oodbest{16/16} & 8.69\% & 2.739 \\
Heisenberg XXZ+field & Random & 3 & $0.37128$ & \oodrunner{$0.43000$} & \oodrunner{0/16} & 8.69\% & 2.739 \\
\midrule
\brow Heisenberg J1--J2 & \ours{} & 3 & $0.00563$ & \oodbest{$0.00567$} & \oodbest{14/16} & 8.72\% & 2.738 \\
Heisenberg J1--J2 & Random & 3 & $0.22592$ & \oodrunner{$0.30762$} & \oodrunner{0/16} & 8.70\% & 2.739 \\
\midrule
\grow All targets & \ours{} & -- & $0.05433$ & \oodbest{$0.05839$} & \oodbest{46/64} & 9.14\% & 2.274 \\
All targets & Random & -- & $0.15624$ & \oodrunner{$0.19615$} & \oodrunner{7/64} & 9.15\% & 2.273 \\
\bottomrule
\end{tabular}}

\normalsize
\vspace{0.55em}
\textbf{(b) Named-ansatz OOD}
\par\vspace{0.15em}
\setlength{\tabcolsep}{1.15mm}
\scriptsize
\resizebox{0.94\linewidth}{!}{
\begin{tabular}{ll|ccccc}
\toprule
OOD ansatz & Method & Mean $\dE$ & Mean $|\dE|\downarrow$ & Strict CA seeds & Shot gain $\uparrow$ & \ours{} wins \\
\midrule
\gcell{EfficientSU2} & \ours{} & $+0.0030$ & \oodbest{0.0119} & \oodbest{13/60} & 9.5\% & \gray{--} \\
\gcell{EfficientSU2} & Random & $+0.0344$ & \oodrunner{0.0630} & \oodrunner{2/60} & 9.4\% & 14/15 \\
\midrule
\bcell{HEA-RYRZ-Circular} & \ours{} & $+0.0033$ & \oodbest{0.0088} & \oodbest{12/60} & 12.9\% & \gray{--} \\
\bcell{HEA-RYRZ-Circular} & Random & $+0.0627$ & \oodrunner{0.0690} & \oodrunner{2/60} & 13.2\% & 15/15 \\
\midrule
\gcell{RealAmplitudes} & \ours{} & $-0.0006$ & \oodbest{0.0030} & \oodbest{29/60} & 13.5\% & \gray{--} \\
\gcell{RealAmplitudes} & Random & $+0.1867$ & \oodrunner{0.2182} & \oodrunner{1/60} & 13.8\% & 15/15 \\
\midrule
\bcell{StronglyEntanglingLayers} & \ours{} & $+0.0395$ & \oodrunner{0.0453} & \oodrunner{0/60} & 11.1\% & \gray{--} \\
\bcell{StronglyEntanglingLayers} & Random & $+0.0271$ & \oodbest{0.0390} & \oodbest{2/60} & 11.2\% & 4/15 \\
\bottomrule
\end{tabular}}
\endgroup
\vspace{-0.5em}
\end{table}

\clearpage
\section{Molecule- and Hamiltonian-Family OOD Transfer}
\label{sec:app_molecule_hamiltonian_ood}
The compact comparison between \ourmethod{} and Random on Hamiltonian transfer appears in panel~(a) of Table~\ref{tab:ood_transfer_summary} in Appendix~\ref{sec:app_structural_ood_summary}; Table~\ref{tab:v54_hamiltonian_generalization_full} below preserves the corresponding TITAN controls.
\section{Clean Molecule-OOD Hamiltonian Transfer}

For each split, one molecule is held out and the relevance model is trained on the remaining four molecules at depths 1--4. The matched TITAN comparison uses the corresponding holdout-specific checkpoint, so neither \ourmethod{} nor either dagger composition has seen the testing molecule.

\begin{table}[H]
\centering
\vspace{-0.7em}
\caption{\textbf{Leave-one-molecule-out Hamiltonian transfer at $\tau=0.997$.} Representative depths 1 and 4 are shown; depths 2 and 3 appear in Appendix Table~\ref{tab:v53_leave_one_molecule_ood_depth23}. TITAN$^{\dagger}$ uses a cell-wise freezing threshold selected to match \ourmethod{}'s shot gain; TITAN$^{\dagger}$+\ours{} uses the same active-parameter mask and the matching holdout-specific selector. Green/red $\dE$ and \cmark{}/\xmark{} follow the one-sided $1.6$ mHa no-worse criterion. Bold marks the closest mean $\dE$ to Full; bold-underlined shot gain marks the largest saving.}
\vspace{0.15em}
\setlength{\tabcolsep}{0.38mm}
\renewcommand{\arraystretch}{0.55}
\tiny
\resizebox{1.0\linewidth}{!}{
\begin{tabular}{lll|cccccc}
\toprule
Molecule & Depth & Metric & \ours{} & Random & TITAN~(\citeauthor{peng2026titan}) & TITAN+\ours{} & TITAN$^{\dagger}$ & TITAN$^{\dagger}$+\ours{} \\
\midrule
\gcell{H2 (4q)} & Depth 1 & $\dE$ vs Full $\downarrow$ & \textbf{\rbf{+0.0016}} & \rbf{+0.3323} & \rbf{+0.0228} & \rbf{+0.0258} & \rbf{+0.0262} & \rbf{+0.0272} \\
 & Depth 1 & Shot gain $\uparrow$ & 55.2\% & 56.1\% & 61.8\% & \textbf{\underline{88.8\%}} & 56.0\% & 86.0\% \\
 & Depth 1 & Chem. acc. & \xmark & \xmark & \xmark & \xmark & \xmark & \xmark \\
\addlinespace[0.08em]
 & Depth 4 & $\dE$ vs Full $\downarrow$ & \textbf{\gbf{-0.0002}} & \rbf{+0.0249} & \gbf{-0.0003} & \gbf{-0.0029} & \gbf{+0.0010} & \gbf{+0.0007} \\
 & Depth 4 & Shot gain $\uparrow$ & 9.9\% & 9.7\% & 40.8\% & \textbf{\underline{45.3\%}} & 10.2\% & 16.0\% \\
 & Depth 4 & Chem. acc. & \cmark & \xmark & \cmark & \cmark & \cmark & \cmark \\
\midrule
\gcell{HeH+ (4q)} & Depth 1 & $\dE$ vs Full $\downarrow$ & \textbf{\rbf{+0.0021}} & \rbf{+0.5282} & \rbf{+0.1040} & \rbf{+0.1063} & \rbf{+0.0090} & \rbf{+0.0114} \\
 & Depth 1 & Shot gain $\uparrow$ & 32.2\% & 30.4\% & 56.2\% & \textbf{\underline{75.6\%}} & 30.9\% & 58.3\% \\
 & Depth 1 & Chem. acc. & \xmark & \xmark & \xmark & \xmark & \xmark & \xmark \\
\addlinespace[0.08em]
 & Depth 4 & $\dE$ vs Full $\downarrow$ & \textbf{\gbf{-0.0022}} & \rbf{+0.0452} & \gbf{-0.0642} & \gbf{-0.0650} & \rbf{+0.0111} & \rbf{+0.0112} \\
 & Depth 4 & Shot gain $\uparrow$ & 8.1\% & 8.1\% & 38.3\% & \textbf{\underline{42.7\%}} & 4.7\% & 12.5\% \\
 & Depth 4 & Chem. acc. & \cmark & \xmark & \cmark & \cmark & \xmark & \xmark \\
\midrule
\gcell{H3+ (6q)} & Depth 1 & $\dE$ vs Full $\downarrow$ & \textbf{\rbf{+0.0051}} & \rbf{+0.9872} & \rbf{+0.0871} & \rbf{+0.0867} & \rbf{+0.0871} & \rbf{+0.0867} \\
 & Depth 1 & Shot gain $\uparrow$ & 56.2\% & 60.6\% & 58.2\% & \textbf{\underline{94.3\%}} & 58.2\% & 94.3\% \\
 & Depth 1 & Chem. acc. & \xmark & \xmark & \xmark & \xmark & \xmark & \xmark \\
\addlinespace[0.08em]
 & Depth 4 & $\dE$ vs Full $\downarrow$ & \gbf{-0.0052} & \rbf{+0.0504} & \rbf{+0.0116} & \rbf{+0.0128} & \rbf{+0.0049} & \textbf{\rbf{+0.0017}} \\
 & Depth 4 & Shot gain $\uparrow$ & 10.0\% & 10.0\% & 45.9\% & \textbf{\underline{52.0\%}} & 6.2\% & 16.0\% \\
 & Depth 4 & Chem. acc. & \cmark & \xmark & \xmark & \xmark & \xmark & \xmark \\
\midrule
\gcell{H4 chain (8q)} & Depth 1 & $\dE$ vs Full $\downarrow$ & \textbf{\rbf{+0.0038}} & \rbf{+1.5543} & \rbf{+0.0220} & \rbf{+0.0236} & \rbf{+0.0220} & \rbf{+0.0236} \\
 & Depth 1 & Shot gain $\uparrow$ & 63.7\% & 68.3\% & 62.7\% & \textbf{\underline{97.4\%}} & 62.7\% & 97.4\% \\
 & Depth 1 & Chem. acc. & \xmark & \xmark & \xmark & \xmark & \xmark & \xmark \\
\addlinespace[0.08em]
 & Depth 4 & $\dE$ vs Full $\downarrow$ & \textbf{\rbf{+0.0652}} & \rbf{+0.3159} & \rbf{+0.1904} & \rbf{+0.1887} & \rbf{+0.0770} & \rbf{+0.0752} \\
 & Depth 4 & Shot gain $\uparrow$ & 28.5\% & 29.4\% & 44.7\% & \textbf{\underline{63.2\%}} & 28.6\% & 50.4\% \\
 & Depth 4 & Chem. acc. & \xmark & \xmark & \xmark & \xmark & \xmark & \xmark \\
\midrule
\gcell{LiH (12q)} & Depth 1 & $\dE$ vs Full $\downarrow$ & \textbf{\rbf{+0.0022}} & \rbf{+0.4404} & \gbf{-0.0881} & \gbf{-0.0880} & \rbf{+0.0387} & \rbf{+0.0385} \\
 & Depth 1 & Shot gain $\uparrow$ & 40.4\% & 37.9\% & 71.9\% & \textbf{\underline{88.5\%}} & 47.1\% & 71.9\% \\
 & Depth 1 & Chem. acc. & \xmark & \xmark & \cmark & \cmark & \xmark & \xmark \\
\addlinespace[0.08em]
 & Depth 4 & $\dE$ vs Full $\downarrow$ & \textbf{\gbf{-0.0023}} & \rbf{+0.2301} & \gbf{-0.0100} & \gbf{-0.0092} & \rbf{+0.0516} & \rbf{+0.0510} \\
 & Depth 4 & Shot gain $\uparrow$ & 17.9\% & 17.7\% & 76.2\% & \textbf{\underline{81.9\%}} & 20.2\% & 33.8\% \\
 & Depth 4 & Chem. acc. & \cmark & \xmark & \cmark & \cmark & \xmark & \xmark \\
\bottomrule
\end{tabular}}
\label{tab:v53_leave_one_molecule_ood}
\vspace{-0.8em}
\end{table}

Table~\ref{tab:v53_leave_one_molecule_ood} directly compares group selection and parameter freezing under Hamiltonian shift. The dagger columns control the TITAN-only shot budget cell by cell, while the final column measures the additional savings from composing the matched active-parameter mask with learned QWC-group selection.

\clearpage
\subsection{Leave-One-Molecule-Out Uncertainty and Intermediate Depths}
\begin{table}[p]
\centering
\vspace{-0.7em}
\caption{\textbf{Uncertainty view for the leave-one-molecule-out comparison at $\tau=0.997$.} This appendix table reports mean $\dE\pm$ s.d. for the representative depths 1 and 4 shown in Table~\ref{tab:v53_leave_one_molecule_ood}. Green/red coloring follows the mean $\dE$ rule used in Table~\ref{tab:v53_leave_one_molecule_ood}.}
\vspace{0.15em}
\setlength{\tabcolsep}{0.75mm}
\renewcommand{\arraystretch}{0.72}
\tiny
\resizebox{1.0\linewidth}{!}{
\begin{tabular}{ll|cccc}
\toprule
Molecule & Depth & \ours{} & Random & TITAN~(\citeauthor{peng2026titan}) & TITAN+\ours{} \\
\midrule
\gcell{H2 (4q)} & Depth 1 & \textbf{\gbf{\(+0.0016\pm0.0035\)}} & \rbf{\(+0.3323\pm0.0518\)} & \rbf{\(+0.0228\pm0.0259\)} & \rbf{\(+0.0258\pm0.0255\)} \\
 & Depth 4 & \textbf{\gbf{\(-0.0002\pm0.0099\)}} & \rbf{\(+0.0249\pm0.0284\)} & \gbf{\(-0.0003\pm0.0206\)} & \gbf{\(-0.0029\pm0.0213\)} \\
\midrule
\gcell{HeH+ (4q)} & Depth 1 & \textbf{\rbf{\(+0.0021\pm0.0034\)}} & \rbf{\(+0.5282\pm0.1378\)} & \rbf{\(+0.1040\pm0.0586\)} & \rbf{\(+0.1063\pm0.0578\)} \\
 & Depth 4 & \textbf{\gbf{\(-0.0022\pm0.0065\)}} & \rbf{\(+0.0452\pm0.0291\)} & \gbf{\(-0.0642\pm0.0614\)} & \gbf{\(-0.0650\pm0.0628\)} \\
\midrule
\gcell{H3+ (6q)} & Depth 1 & \textbf{\rbf{\(+0.0051\pm0.0041\)}} & \rbf{\(+0.9872\pm0.0979\)} & \rbf{\(+0.0871\pm0.0466\)} & \rbf{\(+0.0867\pm0.0453\)} \\
 & Depth 4 & \textbf{\gbf{\(-0.0052\pm0.0072\)}} & \rbf{\(+0.0504\pm0.0457\)} & \rbf{\(+0.0116\pm0.0819\)} & \rbf{\(+0.0128\pm0.0771\)} \\
\midrule
\gcell{H4 chain (8q)} & Depth 1 & \textbf{\rbf{\(+0.0038\pm0.0051\)}} & \rbf{\(+1.5543\pm0.1952\)} & \rbf{\(+0.0220\pm0.0193\)} & \rbf{\(+0.0236\pm0.0171\)} \\
 & Depth 4 & \textbf{\rbf{\(+0.0652\pm0.0666\)}} & \rbf{\(+0.3159\pm0.1015\)} & \rbf{\(+0.1904\pm0.0841\)} & \rbf{\(+0.1887\pm0.0819\)} \\
\midrule
\gcell{LiH (12q)} & Depth 1 & \textbf{\rbf{\(+0.0022\pm0.0031\)}} & \rbf{\(+0.4404\pm0.1245\)} & \gbf{\(-0.0881\pm0.0470\)} & \gbf{\(-0.0880\pm0.0466\)} \\
 & Depth 4 & \textbf{\gbf{\(-0.0023\pm0.0049\)}} & \rbf{\(+0.2301\pm0.1702\)} & \gbf{\(-0.0100\pm0.2342\)} & \gbf{\(-0.0092\pm0.2339\)} \\
\bottomrule
\end{tabular}}
\label{tab:v53_leave_one_molecule_ood_depth14_uncertainty}
\vspace{-0.8em}
\end{table}

\clearpage
\begin{table}[p]
\centering
\vspace{-0.7em}
\caption{\textbf{Appendix depth-2/3 view for leave-one-molecule-out Hamiltonian transfer at $\tau=0.997$.} This table uses the same protocol and layout as Table~\ref{tab:v53_leave_one_molecule_ood}: one molecule is held out for testing, the remaining four molecules train the selector, and depths are kept fixed between train and test. $\dE$ entries report mean $\pm$ s.d. across held-out seeds. \cmark{} denotes the one-sided no-worse chemical-accuracy criterion $\dE\le 1.6$ mHa relative to Full-QWC training; \xmark{} denotes a larger positive energy gap. In $\dE$ rows, green means $\dE\le 1.6$ mHa (including negative gaps) and red means a larger positive gap.}
\vspace{0.15em}
\setlength{\tabcolsep}{0.55mm}
\renewcommand{\arraystretch}{0.55}
\tiny
\resizebox{1.0\linewidth}{!}{
\begin{tabular}{lll|cccc}
\toprule
Molecule & Depth & Metric & \ours{} & Random & TITAN~(\citeauthor{peng2026titan}) & TITAN+\ours{} \\
\midrule
\gcell{H2 (4q)} & Depth 2 & $\dE$ vs Full $\downarrow$ & \textbf{\rbf{\(+0.0046\pm0.0063\)}} & \rbf{\(+0.1387\pm0.1420\)} & \gbf{\(-0.0200\pm0.1784\)} & \gbf{\(-0.0180\pm0.1775\)} \\
 & Depth 2 & Shot gain $\uparrow$ & 29.9\% & 30.1\% & 54.0\% & \textbf{\underline{72.3\%}} \\
 & Depth 2 & Chem. acc. & \xmark & \xmark & \cmark & \cmark \\
\addlinespace[0.08em]
 & Depth 3 & $\dE$ vs Full $\downarrow$ & \rbf{\(+0.0576\pm0.1057\)} & \rbf{\(+0.0814\pm0.0453\)} & \rbf{\(+0.0223\pm0.0453\)} & \textbf{\rbf{\(+0.0205\pm0.0409\)}} \\
 & Depth 3 & Shot gain $\uparrow$ & 19.3\% & 21.3\% & 39.7\% & \textbf{\underline{52.0\%}} \\
 & Depth 3 & Chem. acc. & \xmark & \xmark & \xmark & \xmark \\
\midrule
\gcell{HeH+ (4q)} & Depth 2 & $\dE$ vs Full $\downarrow$ & \textbf{\rbf{\(+0.0167\pm0.0183\)}} & \rbf{\(+0.2844\pm0.0547\)} & \gbf{\(-0.1194\pm0.1021\)} & \gbf{\(-0.1191\pm0.1008\)} \\
 & Depth 2 & Shot gain $\uparrow$ & 22.2\% & 23.7\% & 54.4\% & \textbf{\underline{70.5\%}} \\
 & Depth 2 & Chem. acc. & \xmark & \xmark & \cmark & \cmark \\
\addlinespace[0.08em]
 & Depth 3 & $\dE$ vs Full $\downarrow$ & \rbf{\(+0.0133\pm0.0145\)} & \rbf{\(+0.0677\pm0.1411\)} & \rbf{\(+0.0112\pm0.0424\)} & \textbf{\rbf{\(+0.0102\pm0.0444\)}} \\
 & Depth 3 & Shot gain $\uparrow$ & 13.2\% & 13.6\% & 43.3\% & \textbf{\underline{52.2\%}} \\
 & Depth 3 & Chem. acc. & \xmark & \xmark & \xmark & \xmark \\
\midrule
\gcell{H3+ (6q)} & Depth 2 & $\dE$ vs Full $\downarrow$ & \textbf{\gbf{\(+0.0013\pm0.0018\)}} & \rbf{\(+0.3811\pm0.0782\)} & \gbf{\(-0.0031\pm0.0610\)} & \gbf{\(-0.0036\pm0.0629\)} \\
 & Depth 2 & Shot gain $\uparrow$ & 27.6\% & 27.3\% & 50.0\% & \textbf{\underline{71.0\%}} \\
 & Depth 2 & Chem. acc. & \cmark & \xmark & \cmark & \cmark \\
\addlinespace[0.08em]
 & Depth 3 & $\dE$ vs Full $\downarrow$ & \textbf{\rbf{\(+0.0017\pm0.0076\)}} & \rbf{\(+0.1683\pm0.1216\)} & \rbf{\(+0.0338\pm0.0777\)} & \rbf{\(+0.0332\pm0.0800\)} \\
 & Depth 3 & Shot gain $\uparrow$ & 16.2\% & 16.1\% & 47.1\% & \textbf{\underline{57.7\%}} \\
 & Depth 3 & Chem. acc. & \xmark & \xmark & \xmark & \xmark \\
\midrule
\gcell{H4 chain (8q)} & Depth 2 & $\dE$ vs Full $\downarrow$ & \textbf{\rbf{\(+0.0065\pm0.0112\)}} & \rbf{\(+1.1356\pm0.1572\)} & \rbf{\(+0.1521\pm0.1283\)} & \rbf{\(+0.1556\pm0.1269\)} \\
 & Depth 2 & Shot gain $\uparrow$ & 52.3\% & 57.1\% & 59.1\% & \textbf{\underline{88.9\%}} \\
 & Depth 2 & Chem. acc. & \xmark & \xmark & \xmark & \xmark \\
\addlinespace[0.08em]
 & Depth 3 & $\dE$ vs Full $\downarrow$ & \textbf{\rbf{\(+0.0107\pm0.0102\)}} & \rbf{\(+0.3099\pm0.0388\)} & \gbf{\(-0.0650\pm0.1313\)} & \gbf{\(-0.0615\pm0.1294\)} \\
 & Depth 3 & Shot gain $\uparrow$ & 39.3\% & 40.6\% & 48.3\% & \textbf{\underline{72.9\%}} \\
 & Depth 3 & Chem. acc. & \xmark & \xmark & \cmark & \cmark \\
\midrule
\gcell{LiH (12q)} & Depth 2 & $\dE$ vs Full $\downarrow$ & \textbf{\rbf{\(+0.0026\pm0.0026\)}} & \rbf{\(+1.0068\pm0.1733\)} & \rbf{\(+0.0806\pm0.0923\)} & \rbf{\(+0.0803\pm0.0926\)} \\
 & Depth 2 & Shot gain $\uparrow$ & 32.1\% & 30.6\% & 77.1\% & \textbf{\underline{87.1\%}} \\
 & Depth 2 & Chem. acc. & \xmark & \xmark & \xmark & \xmark \\
\addlinespace[0.08em]
 & Depth 3 & $\dE$ vs Full $\downarrow$ & \textbf{\rbf{\(+0.0038\pm0.0133\)}} & \rbf{\(+0.5742\pm0.2506\)} & \rbf{\(+0.3972\pm0.2246\)} & \rbf{\(+0.3975\pm0.2246\)} \\
 & Depth 3 & Shot gain $\uparrow$ & 24.1\% & 23.6\% & 76.4\% & \textbf{\underline{84.2\%}} \\
 & Depth 3 & Chem. acc. & \xmark & \xmark & \xmark & \xmark \\
\bottomrule
\end{tabular}}
\label{tab:v53_leave_one_molecule_ood_depth23}
\vspace{-0.8em}
\end{table}

\clearpage
\section{Full Hamiltonian-Family Baseline Comparison}
\label{sec:v54_hamiltonian_generalization_full}

This Hamiltonian-transfer experiment asks whether the learned relevance selector transfers beyond molecular electronic-structure Hamiltonians. We train one shared selector on five PySCF/OpenFermion molecular Hamiltonians (H2, HeH$^+$, H3$^+$, H4 chain, and LiH; STO-3G basis) plus one 8-qubit Heisenberg-chain Hamiltonian, using ansatz depths 1--4. We then evaluate closed-loop online VQE at the same depths on four unseen spin Hamiltonians: two transverse-field Ising models (TFIM) and two Heisenberg variants. The ansatz is the same layered RY/RZ circuit with a linear entangler, and the top-mass threshold is fixed at $\tau=0.997$.

\begin{table}[H]
\centering
\vspace{-0.5em}
\caption{\textbf{Full Hamiltonian-family OOD baseline comparison at $\tau=0.997$.} The selector is trained on molecular Hamiltonians plus one Heisenberg Hamiltonian and evaluated on unseen TFIM/Heisenberg-family Hamiltonians at the same depths 1--4. This appendix table preserves the TITAN and TITAN+\ours{} controls omitted from the compact comparison in Table~\ref{tab:ood_transfer_summary}. No-worse vs Full reports seeds whose final energy is not more than $1.6$ mHa above full-QWC training, accepting lower energies. Full groups is the number of QWC groups in the target Hamiltonian. Bold green marks the lowest mean $|\dE|$ within each test Hamiltonian; underlined green marks the largest shot gain.}
\vspace{0.25em}
\setlength{\tabcolsep}{1.1mm}
\resizebox{\linewidth}{!}{
\begin{tabular}{ll|c|ccc|cc}
\toprule
Test Hamiltonian & Method & Full groups & Mean $\dE$ & Mean $|\dE|\downarrow$ & No-worse vs Full & Shot gain $\uparrow$ & Groups/param. \\
\midrule
\brow TFIM $h=0.7$ & \ours{} & 2 & $-1.55{\times}10^{-15}$ & \gbf{$3.66{\times}10^{-15}$} & \gbf{16/16} & 6.13\% & 1.877 \\
TFIM $h=0.7$ & Random & 2 & $-0.00253$ & $0.00710$ & 6/16 & 6.20\% & 1.876 \\
TFIM $h=0.7$ & TITAN~\citep{peng2026titan} & 2 & $-0.00587$ & $0.01964$ & 1/16 & 56.11\% & 0.878 \\
TFIM $h=0.7$ & TITAN+\ours{} & 2 & $-0.00587$ & $0.01964$ & 1/16 & \underline{\gbf{57.37\%}} & 0.853 \\
\midrule
\brow TFIM $h=1.2$ & \ours{} & 2 & $0.21181$ & $0.22777$ & 0/16 & 13.02\% & 1.740 \\
TFIM $h=1.2$ & Random & 2 & $0.03008$ & \gbf{$0.03988$} & 1/16 & 13.02\% & 1.740 \\
TFIM $h=1.2$ & TITAN~\citep{peng2026titan} & 2 & $0.02293$ & $0.05889$ & 1/16 & 49.22\% & 1.016 \\
TFIM $h=1.2$ & TITAN+\ours{} & 2 & $0.22750$ & $0.24289$ & 0/16 & \underline{\gbf{53.81\%}} & 0.924 \\
\midrule
\brow Heisenberg XXZ+field & \ours{} & 3 & $-0.00011$ & \gbf{$0.00011$} & \gbf{16/16} & 8.69\% & 2.739 \\
Heisenberg XXZ+field & Random & 3 & $0.37128$ & $0.43000$ & 0/16 & 8.69\% & 2.739 \\
Heisenberg XXZ+field & TITAN~\citep{peng2026titan} & 3 & $0.50014$ & $0.71973$ & 0/16 & 45.31\% & 1.641 \\
Heisenberg XXZ+field & TITAN+\ours{} & 3 & $0.49967$ & $0.71926$ & 0/16 & \underline{\gbf{46.79\%}} & 1.596 \\
\midrule
\brow Heisenberg J1--J2 & \ours{} & 3 & $0.00563$ & \gbf{$0.00567$} & \gbf{14/16} & 8.72\% & 2.738 \\
Heisenberg J1--J2 & Random & 3 & $0.22592$ & $0.30762$ & 0/16 & 8.70\% & 2.739 \\
Heisenberg J1--J2 & TITAN~\citep{peng2026titan} & 3 & $0.58258$ & $0.67115$ & 1/16 & 44.13\% & 1.676 \\
Heisenberg J1--J2 & TITAN+\ours{} & 3 & $0.58165$ & $0.67070$ & 1/16 & \underline{\gbf{45.64\%}} & 1.631 \\
\midrule
\grow All targets & \ours{} & -- & $0.05433$ & \gbf{$0.05839$} & \gbf{46/64} & 9.14\% & 2.274 \\
All targets & Random & -- & $0.15624$ & $0.19615$ & 7/64 & 9.15\% & 2.273 \\
All targets & TITAN~\citep{peng2026titan} & -- & $0.27494$ & $0.36735$ & 3/64 & 48.69\% & 1.303 \\
All targets & TITAN+\ours{} & -- & $0.32574$ & $0.41312$ & 2/64 & \underline{\gbf{50.90\%}} & 1.251 \\
\bottomrule
\end{tabular}}
\label{tab:v54_hamiltonian_generalization_full}
\vspace{-0.5em}
\end{table}

The result separates Hamiltonian-family transfer from ordinary molecular holdout. Our method remains the most accurate method on three of the four unseen spin Hamiltonians, including both Heisenberg variants and TFIM at $h=0.7$. In contrast, TITAN-style parameter freezing obtains much larger shot savings but usually changes the optimization trajectory enough to lose final-energy fidelity. TFIM at $h=1.2$ is the failure case: the learned molecular-plus-Heisenberg relevance ordering is not yet robust to this field regime. This suggests that Hamiltonian-family coverage, not just more molecular seeds, is the next dataset axis for a stronger general-purpose selector.

\clearpage
\section{Hamiltonian-Panel Stress Test}
\label{sec:v55_hamiltonian_panel}

This expanded panel systematically tests Hamiltonian-family OOD transfer. We train one shared selector on five molecular Hamiltonians plus two 8-qubit Heisenberg-chain Hamiltonians, using depths 1--4 and the same layered RY/RZ ansatz. We then evaluate the saved model on held-out transverse-field Ising models, unseen Heisenberg variants, and a larger-group JW-like parity spin Hamiltonian at depths 1--4 with $\tau=0.997$. The baseline comparison adds cell-wise cost-matched TITAN controls so that parameter freezing and QWC-group selection are compared at similar measurement budgets.

\begin{table}[H]
\centering
\vspace{-0.7em}
\caption{\textbf{Hamiltonian-panel transfer at $\tau=0.997$.} Each testing Hamiltonian is held out from training; representative depths 1 and 4 are shown. TITAN$^{\dagger}$ uses the cell-wise freezing threshold closest to \ourmethod{}'s shot gain, and TITAN$^{\dagger}$+\ours{} reuses that active-parameter mask before group selection. Green/red $\dE$ and \cmark{}/\xmark{} follow the one-sided $1.6$ mHa no-worse criterion. Bold marks the closest mean $\dE$ to Full; bold-underlined shot gain marks the largest saving.}
\vspace{0.15em}
\setlength{\tabcolsep}{0.38mm}
\renewcommand{\arraystretch}{0.50}
\tiny
\resizebox{1.0\linewidth}{!}{
\begin{tabular}{lll|cccccc}
\toprule
Testing Hamiltonian & Depth & Metric & \ours{} & Random & TITAN~(\citeauthor{peng2026titan}) & TITAN+\ours{} & TITAN$^{\dagger}$ & TITAN$^{\dagger}$+\ours{} \\
\midrule
\gcell{TFIM $h=0.5$} & Depth 1 & $\dE$ vs Full $\downarrow$ & \textbf{\gbf{+0.0000}} & \gbf{-0.0043} & \gbf{-0.0139} & \gbf{-0.0139} & \textbf{\gbf{+0.0000}} & \textbf{\gbf{+0.0000}} \\
 & Depth 1 & Shot gain $\uparrow$ & 2.0\% & 2.0\% & 56.8\% & \textbf{\underline{57.5\%}} & 0.0\% & 2.0\% \\
 & Depth 1 & Chem. acc. & \cmark & \cmark & \cmark & \cmark & \cmark & \cmark \\
\addlinespace[0.08em]
 & Depth 4 & $\dE$ vs Full $\downarrow$ & \gbf{-0.0002} & \gbf{-0.0008} & \rbf{+0.0062} & \rbf{+0.0063} & \textbf{\gbf{+0.0000}} & \gbf{-0.0002} \\
 & Depth 4 & Shot gain $\uparrow$ & 2.6\% & 2.6\% & 59.0\% & \textbf{\underline{59.3\%}} & 0.0\% & 2.6\% \\
 & Depth 4 & Chem. acc. & \cmark & \cmark & \xmark & \xmark & \cmark & \cmark \\
\midrule
\gcell{TFIM $h=0.7$} & Depth 1 & $\dE$ vs Full $\downarrow$ & \textbf{\gbf{+0.0000}} & \gbf{-0.0138} & \gbf{-0.0227} & \gbf{-0.0227} & \textbf{\gbf{+0.0000}} & \textbf{\gbf{+0.0000}} \\
 & Depth 1 & Shot gain $\uparrow$ & 2.9\% & 2.9\% & 56.4\% & \textbf{\underline{57.6\%}} & 0.0\% & 2.9\% \\
 & Depth 1 & Chem. acc. & \cmark & \cmark & \cmark & \cmark & \cmark & \cmark \\
\addlinespace[0.08em]
 & Depth 4 & $\dE$ vs Full $\downarrow$ & \textbf{\gbf{+0.0000}} & \gbf{-0.0018} & \rbf{+0.0107} & \rbf{+0.0107} & \textbf{\gbf{+0.0000}} & \textbf{\gbf{+0.0000}} \\
 & Depth 4 & Shot gain $\uparrow$ & 2.4\% & 2.4\% & 53.0\% & \textbf{\underline{53.5\%}} & 0.0\% & 2.4\% \\
 & Depth 4 & Chem. acc. & \cmark & \cmark & \xmark & \xmark & \cmark & \cmark \\
\midrule
\gcell{TFIM $h=0.9$} & Depth 1 & $\dE$ vs Full $\downarrow$ & \rbf{+0.0771} & \gbf{-0.0010} & \gbf{-0.0018} & \rbf{+0.0784} & \textbf{\gbf{-0.0002}} & \rbf{+0.0771} \\
 & Depth 1 & Shot gain $\uparrow$ & 25.0\% & 25.0\% & 54.4\% & \textbf{\underline{59.8\%}} & 27.1\% & 37.6\% \\
 & Depth 1 & Chem. acc. & \xmark & \cmark & \cmark & \xmark & \cmark & \xmark \\
\addlinespace[0.08em]
 & Depth 4 & $\dE$ vs Full $\downarrow$ & \rbf{+0.0302} & \rbf{+0.0064} & \rbf{+0.0028} & \rbf{+0.0351} & \textbf{\gbf{-0.0012}} & \rbf{+0.0306} \\
 & Depth 4 & Shot gain $\uparrow$ & 6.2\% & 6.2\% & 47.1\% & \textbf{\underline{49.0\%}} & 6.0\% & 11.1\% \\
 & Depth 4 & Chem. acc. & \xmark & \xmark & \xmark & \xmark & \cmark & \xmark \\
\midrule
\gcell{Heis. XXZ+X field} & Depth 1 & $\dE$ vs Full $\downarrow$ & \textbf{\gbf{+0.0000}} & \rbf{+0.8893} & \rbf{+1.2891} & \rbf{+1.2891} & \rbf{+0.2391} & \rbf{+0.2391} \\
 & Depth 1 & Shot gain $\uparrow$ & 16.7\% & 16.7\% & 52.2\% & \textbf{\underline{53.5\%}} & 19.4\% & 29.6\% \\
 & Depth 1 & Chem. acc. & \cmark & \xmark & \xmark & \xmark & \xmark & \xmark \\
\addlinespace[0.08em]
 & Depth 4 & $\dE$ vs Full $\downarrow$ & \textbf{\gbf{+0.0000}} & \rbf{+0.3996} & \rbf{+0.5969} & \rbf{+0.5969} & \gbf{-0.1146} & \gbf{-0.1146} \\
 & Depth 4 & Shot gain $\uparrow$ & 4.2\% & 4.2\% & 36.3\% & \textbf{\underline{38.1\%}} & 3.7\% & 7.7\% \\
 & Depth 4 & Chem. acc. & \cmark & \xmark & \xmark & \xmark & \cmark & \cmark \\
\midrule
\gcell{Heis. J1--J2} & Depth 1 & $\dE$ vs Full $\downarrow$ & \textbf{\gbf{+0.0000}} & \rbf{+0.6602} & \rbf{+0.9835} & \rbf{+0.9835} & \rbf{+0.0863} & \rbf{+0.0863} \\
 & Depth 1 & Shot gain $\uparrow$ & 16.7\% & 16.7\% & 50.6\% & \textbf{\underline{52.2\%}} & 12.7\% & 25.1\% \\
 & Depth 1 & Chem. acc. & \cmark & \xmark & \xmark & \xmark & \xmark & \xmark \\
\addlinespace[0.08em]
 & Depth 4 & $\dE$ vs Full $\downarrow$ & \textbf{\gbf{+0.0000}} & \gbf{-0.0010} & \rbf{+0.5678} & \rbf{+0.5678} & \rbf{+0.1550} & \rbf{+0.1550} \\
 & Depth 4 & Shot gain $\uparrow$ & 4.2\% & 4.2\% & 34.2\% & \textbf{\underline{36.0\%}} & 3.5\% & 7.4\% \\
 & Depth 4 & Chem. acc. & \cmark & \cmark & \xmark & \xmark & \xmark & \xmark \\
\midrule
\gcell{Heis. XXY+Z field} & Depth 1 & $\dE$ vs Full $\downarrow$ & \textbf{\gbf{+0.0000}} & \rbf{+0.9213} & \rbf{+0.8904} & \rbf{+0.8904} & \rbf{+0.1382} & \rbf{+0.1382} \\
 & Depth 1 & Shot gain $\uparrow$ & 16.7\% & 16.7\% & 49.6\% & \textbf{\underline{51.9\%}} & 18.9\% & 29.4\% \\
 & Depth 1 & Chem. acc. & \cmark & \xmark & \xmark & \xmark & \xmark & \xmark \\
\addlinespace[0.08em]
 & Depth 4 & $\dE$ vs Full $\downarrow$ & \textbf{\gbf{+0.0000}} & \rbf{+0.0742} & \gbf{-0.0137} & \gbf{-0.0137} & \rbf{+0.0219} & \rbf{+0.0219} \\
 & Depth 4 & Shot gain $\uparrow$ & 4.2\% & 4.2\% & 31.1\% & \textbf{\underline{33.3\%}} & 3.6\% & 7.5\% \\
 & Depth 4 & Chem. acc. & \cmark & \xmark & \cmark & \cmark & \xmark & \xmark \\
\midrule
\gcell{JW-like parity spin} & Depth 1 & $\dE$ vs Full $\downarrow$ & \rbf{+0.0312} & \rbf{+0.6318} & \rbf{+0.1336} & \rbf{+0.1387} & \textbf{\rbf{+0.0049}} & \rbf{+0.0350} \\
 & Depth 1 & Shot gain $\uparrow$ & 35.7\% & 35.6\% & 55.5\% & \textbf{\underline{76.9\%}} & 31.4\% & 59.5\% \\
 & Depth 1 & Chem. acc. & \xmark & \xmark & \xmark & \xmark & \xmark & \xmark \\
\addlinespace[0.08em]
 & Depth 4 & $\dE$ vs Full $\downarrow$ & \gbf{-0.0046} & \rbf{+0.0076} & \rbf{+0.0822} & \rbf{+0.0738} & \textbf{\gbf{+0.0012}} & \gbf{-0.0198} \\
 & Depth 4 & Shot gain $\uparrow$ & 11.6\% & 11.5\% & 45.7\% & \textbf{\underline{52.1\%}} & 9.8\% & 19.6\% \\
 & Depth 4 & Chem. acc. & \cmark & \xmark & \xmark & \xmark & \cmark & \cmark \\
\bottomrule
\end{tabular}}
\label{tab:v55_hamiltonian_panel}
\vspace{-0.8em}
\end{table}

Table~\ref{tab:v55_hamiltonian_panel} shows a useful but bounded generalization result. Our method is full-QWC-equivalent on the held-out Heisenberg variants and low-field TFIM, whereas shot-matched parameter freezing still incurs large positive drift on most Heisenberg cells. On low-field TFIM, our method saves only 2--3\%, so the nearest TITAN-only operating point is no freezing; the composed dagger column then reduces exactly to \ourmethod{} group selection. The JW-like target remains a harder family shift, although matched TITAN substantially narrows the gap relative to aggressive freezing. These controls show that the proposed method's stability is not explained by a more conservative parameter-freezing threshold.

\clearpage
\section{Joint Molecule+Depth OOD Stress Test}

This joint-OOD experiment is preserved as a strict stress test rather than the clean molecule-OOD experiment. The model is trained on H2, HeH+, H3+, and LiH at RY/RZ depths 1--4, then evaluated without retraining on H4 chain at unseen depths 5--7. Thus both the molecular Hamiltonian and the ansatz depth are held out. We do not claim this is the first cross-molecule VQE transfer experiment broadly: prior work such as Meta-VQE~\citep{cervera2021meta} studies learning over parameterized Hamiltonians. Here the role of this appendix result is narrower: it probes whether a learned parameter-by-QWC measurement policy survives simultaneous Hamiltonian and depth shift.

\begin{table}[H]
\centering
\vspace{-0.5em}
\caption{\textbf{Joint molecule+depth OOD transfer to H4 at $\tau=0.997$.} The relevance model is trained on H2, HeH+, H3+, and LiH with RY/RZ depths 1--4 and evaluated on held-out H4 chain at unseen depths 5--7 with seeds 97/101/107/109. Each depth cell reports mean $\dE$ versus full-QWC followed by mean shot saving. Mean $|\dE|$ averages the three H4 depths. No-worse counts held-out seeds with $\dE \le 1.6$ mHa, so lower final energy is accepted. Global coefficient uses the same per-parameter group count as \ours{} but ranks QWC groups by Hamiltonian coefficient magnitude.}
\vspace{0.25em}
\setlength{\tabcolsep}{1.0mm}
\small
\resizebox{1.0\linewidth}{!}{
\begin{tabular}{l|ccc|ccc}
\toprule
Method & Depth 5 & Depth 6 & Depth 7 & Mean \(|\dE|\downarrow\) & No-worse seeds & Shot gain \(\uparrow\) \\
\midrule
\brow \ours{} & \textbf{\gbf{\(+0.0342\) / 23.5\%}} & \(+0.0030\) / 20.3\% & \(+0.0060\) / 18.8\% & 0.0251 & \textbf{\gbf{6/12}} & 20.9\% \\
Random & \rbf{\(+0.1682\) / 23.6\%} & \rbf{\(+0.1171\) / 20.4\%} & \rbf{\(+0.0322\) / 18.9\%} & \rbf{0.1064} & 1/12 & 21.0\% \\
Global coeff. & \(+0.0380\) / 23.8\% & \textbf{\gbf{\(-0.0020\) / 20.2\%}} & \textbf{\gbf{\(+0.0049\) / 18.9\%}} & \textbf{\gbf{0.0201}} & 5/12 & 20.9\% \\
TITAN~(\citeauthor{peng2026titan}) & \rbf{\(+0.1011\) / 42.3\%} & \(+0.0070\) / 39.4\% & \(-0.0289\) / 36.1\% & \rbf{0.0763} & 5/12 & 39.3\% \\
TITAN+\ours{} & \rbf{\(+0.1040\) / 56.8\%} & \(+0.0071\) / 52.7\% & \(-0.0300\) / 48.3\% & \rbf{0.0791} & 5/12 & \underline{\gbf{52.6\%}} \\
\bottomrule
\end{tabular}}
\label{tab:v51_molecule_ood}
\vspace{-0.7em}
\end{table}

Table~\ref{tab:v51_molecule_ood} shows that joint molecule+depth OOD transfer is substantially harder than depth-OOD within the same molecule panel. Our method is much more stable than matched-budget random selection and wins the hardest H4 depth-5 cell, but the static Global coefficient baseline is competitive on depths 6--7 and obtains the lowest aggregate mean absolute gap. We therefore treat this joint shift as a stress test and diagnostic rather than the clean molecule-OOD result: the current learner captures useful transferable structure, but stronger Hamiltonian/QWC group features are needed for robust cross-Hamiltonian generalization.

\clearpage
\section{Ansatz-Family OOD Transfer}
\label{sec:app_ansatz_ood}
The compact comparison between \ourmethod{} and Random on named-ansatz transfer appears in panel~(b) of Table~\ref{tab:ood_transfer_summary} in Appendix~\ref{sec:app_structural_ood_summary}; Table~\ref{tab:v43_named_ansatz_transfer_full} below preserves the corresponding cost-matched TITAN controls.
\section{Full Named-Ansatz Zero-Shot Baselines}
The preceding depth-OOD experiments keep the ansatz family fixed and increase depth. We next test a harder transfer axis: the saved depth-OOD model is trained only on the HEA-RYRZ-linear ansatz at depths 1--4, then evaluated \emph{zero-shot} on named ansatz families at depths 5--7. The molecule panel, held-out seeds, top-mass threshold, and online VQE protocol are unchanged: H2, HeH+, H3+, H4 chain, and LiH in STO-3G; four held-out seeds per molecule-depth cell; $\tau=0.997$; and 1024 shots per parameter-shift measurement setting.

\begin{table}[H]
\centering
\vspace{-0.5em}
\caption{\textbf{Full zero-shot baseline comparison on named ansatz families.} The saved depth-OOD selector is trained on HEA-RYRZ-linear depths 1--4 and evaluated without retraining on depths 5--7, five STO-3G molecules, and four held-out seeds per molecule-depth cell at $\tau=0.997$. This appendix table preserves the TITAN$^{\dagger}$ and TITAN$^{\dagger}$+\ours{} controls omitted from the compact comparison in Table~\ref{tab:ood_transfer_summary}. Random uses the same per-parameter group budget as \ourmethod{}. TITAN$^{\dagger}$ directly chooses the active-parameter count in each molecule-depth cell so its shot gain matches \ourmethod{}; TITAN$^{\dagger}$+\ours{} applies group sparsification inside that active set. Strict CA counts seeds with $|\dE|\leq1.6$ mHa. The final column reports cells, out of 15, where \ourmethod{} has a smaller mean absolute gap than the method in that row.}
\vspace{0.25em}
\setlength{\tabcolsep}{0.85mm}
\resizebox{1.0\linewidth}{!}{
\begin{tabular}{ll|ccccc}
\toprule
OOD ansatz & Method & Mean \(\dE\) & Mean \(|\dE|\downarrow\) & Strict CA seeds & Shot gain \(\uparrow\) & \ours{} wins \\
\midrule
\gcell{EfficientSU2} & \ours{} & \(+0.0030\) & 0.0119 & 13/60 & 9.5\% & \gray{--} \\
\gcell{EfficientSU2} & Random & \rbf{\(+0.0344\)} & 0.0630 & 2/60 & 9.4\% & 14/15 \\
\gcell{EfficientSU2} & TITAN$^{\dagger}$ & \gbf{\(-0.0048\)} & \textbf{\gbf{0.0118}} & \textbf{\gbf{14/60}} & 9.5\% & 7/15 \\
\gcell{EfficientSU2} & TITAN$^{\dagger}$+\ours{} & \gbf{\(-0.0010\)} & 0.0175 & 6/60 & \textbf{\underline{17.4\%}} & 9/15 \\
\midrule
\bcell{HEA-RYRZ-Circular} & \ours{} & \rbf{\(+0.0033\)} & \textbf{\gbf{0.0088}} & 12/60 & 12.9\% & \gray{--} \\
\bcell{HEA-RYRZ-Circular} & Random & \rbf{\(+0.0627\)} & 0.0690 & 2/60 & 13.2\% & 15/15 \\
\bcell{HEA-RYRZ-Circular} & TITAN$^{\dagger}$ & \gbf{\(-0.0012\)} & 0.0197 & \textbf{\gbf{14/60}} & 13.0\% & 9/15 \\
\bcell{HEA-RYRZ-Circular} & TITAN$^{\dagger}$+\ours{} & \gbf{\(+0.0011\)} & 0.0209 & 8/60 & \textbf{\underline{23.0\%}} & 11/15 \\
\midrule
\gcell{RealAmplitudes} & \ours{} & \gbf{\(-0.0006\)} & \textbf{\gbf{0.0030}} & \textbf{\gbf{29/60}} & 13.5\% & \gray{--} \\
\gcell{RealAmplitudes} & Random & \rbf{\(+0.1867\)} & 0.2182 & 1/60 & 13.8\% & 15/15 \\
\gcell{RealAmplitudes} & TITAN$^{\dagger}$ & \rbf{\(+0.0151\)} & 0.0429 & 8/60 & 13.5\% & 15/15 \\
\gcell{RealAmplitudes} & TITAN$^{\dagger}$+\ours{} & \rbf{\(+0.0155\)} & 0.0435 & 5/60 & \textbf{\underline{23.4\%}} & 15/15 \\
\midrule
\bcell{StronglyEntanglingLayers} & \ours{} & \rbf{\(+0.0395\)} & 0.0453 & 0/60 & 11.1\% & \gray{--} \\
\bcell{StronglyEntanglingLayers} & Random & \rbf{\(+0.0271\)} & 0.0390 & 2/60 & 11.2\% & 4/15 \\
\bcell{StronglyEntanglingLayers} & TITAN$^{\dagger}$ & \gbf{\(-0.0032\)} & \textbf{\gbf{0.0080}} & \textbf{\gbf{34/60}} & 11.1\% & 0/15 \\
\bcell{StronglyEntanglingLayers} & TITAN$^{\dagger}$+\ours{} & \rbf{\(+0.0343\)} & 0.0407 & 0/60 & \textbf{\underline{20.0\%}} & 5/15 \\
\bottomrule
\end{tabular}}
\label{tab:v43_named_ansatz_transfer_full}
\vspace{-0.5em}
\end{table}

Table~\ref{tab:v43_named_ansatz_transfer_full} shows that the learned relevance ordering transfers well to EfficientSU2, HEA-RYRZ-Circular, and RealAmplitudes, where \ourmethod{} beats Random on 44 of 45 cells. Cost matching rules out a universal advantage from parameter freezing: TITAN$^{\dagger}$ is comparable on EfficientSU2, but \ourmethod{} has smaller gaps on 9/15 HEA-RYRZ-Circular cells and all 15 RealAmplitudes cells. The ordering reverses on StronglyEntanglingLayers, where TITAN$^{\dagger}$ reaches a 0.0080 Ha mean absolute gap while \ourmethod{} fails. Composing the controls adds 8.9--10.0 percentage points of shot saving beyond the matched TITAN arm, but generally increases energy error. Thus parameter freezing and group selection are complementary cost axes, while their reliability remains ansatz dependent.

\clearpage
\subsection{Predictor Architecture Under Named-Ansatz OOD}
\begin{table}[H]
\centering
\vspace{-0.5em}
\caption{\textbf{Predictor comparison under named-ansatz OOD at $\tau=0.997$.} Transformer denotes the depth-OOD predictor used by \ourmethod{} in Table~\ref{tab:v43_named_ansatz_transfer}. MLP, ResNet-18, and GNN use the same labels, token features, train/evaluation splits, and online VQE protocol; only the relevance predictor changes. The final column counts cells, out of 15, where Transformer has a smaller mean absolute gap than the method in that row. Boldface marks the smallest mean absolute gap and largest strict-CA count per ansatz; bold-underlined values mark the largest shot gain.}
\vspace{0.25em}
\setlength{\tabcolsep}{0.85mm}
\resizebox{1.0\linewidth}{!}{
\begin{tabular}{ll|ccccc}
\toprule
OOD ansatz & Predictor & Mean \(\dE\) & Mean \(|\dE|\downarrow\) & Strict CA seeds & Shot gain \(\uparrow\) & Transformer wins \\
\midrule
\gcell{EfficientSU2} & Transformer & \rbf{\(+0.0030\)} & 0.0119 & 13/60 & 9.5\% & \gray{--} \\
\gcell{EfficientSU2} & MLP & \gbf{\(+0.0005\)} & 0.0088 & 9/60 & 8.7\% & 6/15 \\
\gcell{EfficientSU2} & ResNet-18 & \gbf{\(-0.0003\)} & \textbf{\gbf{0.0066}} & \textbf{\gbf{23/60}} & \textbf{\underline{11.0\%}} & 3/15 \\
\gcell{EfficientSU2} & GNN & \gbf{\(+0.0009\)} & 0.0089 & 22/60 & 5.1\% & 2/15 \\
\midrule
\bcell{HEA-RYRZ-Circular} & Transformer & \rbf{\(+0.0033\)} & 0.0088 & 12/60 & \textbf{\underline{12.9\%}} & \gray{--} \\
\bcell{HEA-RYRZ-Circular} & MLP & \rbf{\(+0.0038\)} & 0.0102 & 15/60 & 12.0\% & 8/15 \\
\bcell{HEA-RYRZ-Circular} & ResNet-18 & \gbf{\(+0.0014\)} & \textbf{\gbf{0.0048}} & \textbf{\gbf{31/60}} & 12.7\% & 4/15 \\
\bcell{HEA-RYRZ-Circular} & GNN & \rbf{\(+0.0038\)} & 0.0094 & 16/60 & 10.7\% & 5/15 \\
\midrule
\gcell{RealAmplitudes} & Transformer & \gbf{\(-0.0006\)} & \textbf{\gbf{0.0030}} & \textbf{\gbf{29/60}} & 13.5\% & \gray{--} \\
\gcell{RealAmplitudes} & MLP & \gbf{\(-0.0033\)} & 0.0106 & 17/60 & \textbf{\underline{17.6\%}} & 14/15 \\
\gcell{RealAmplitudes} & ResNet-18 & \gbf{\(-0.0020\)} & 0.0080 & 23/60 & 15.5\% & 9/15 \\
\gcell{RealAmplitudes} & GNN & \rbf{\(+0.0031\)} & 0.0149 & 9/60 & 17.0\% & 14/15 \\
\midrule
\bcell{StronglyEntanglingLayers} & Transformer & \rbf{\(+0.0395\)} & 0.0453 & 0/60 & 11.1\% & \gray{--} \\
\bcell{StronglyEntanglingLayers} & MLP & \rbf{\(+0.0148\)} & 0.0243 & 5/60 & 11.0\% & 1/15 \\
\bcell{StronglyEntanglingLayers} & ResNet-18 & \gbf{\(-0.0002\)} & \textbf{\gbf{0.0073}} & \textbf{\gbf{27/60}} & \textbf{\underline{11.3\%}} & 0/15 \\
\bcell{StronglyEntanglingLayers} & GNN & \rbf{\(+0.0018\)} & 0.0116 & 10/60 & 9.8\% & 0/15 \\
\bottomrule
\end{tabular}}
\label{tab:v66_named_ansatz_model_zoo}
\vspace{-0.5em}
\end{table}

ResNet-18 is the strongest ansatz-OOD predictor on three of four targets: it reduces the mean absolute gap to 0.0066 Ha on EfficientSU2, 0.0048 Ha on HEA-RYRZ-Circular, and 0.0073 Ha on StronglyEntanglingLayers while retaining 11.0--12.7\% shot savings. The Transformer remains best on RealAmplitudes at 0.0030 Ha. This result refines the earlier architecture conclusion: token cross-attention is strongest for molecule OOD, whereas convolutional grid inductive bias is unexpectedly robust to unseen ansatz topology. No single predictor dominates every transfer axis.

\clearpage
\section{Multi-Ansatz Training Improves Transfer}
Single-ansatz training isolates a hard transfer question: a selector trained only on HEA-RYRZ-linear circuits can transfer to some named ansatz families, but fails on StronglyEntanglingLayers. Multi-ansatz training keeps the architecture, molecule panel, online VQE protocol, threshold, held-out seeds, and depth-OOD setting fixed, but changes the training distribution. The learner is now trained on 11{,}520 snapshots from four ansatz families---HEA-RYRZ-linear, HEA-RYRZ-circular, EfficientSU2, and RealAmplitudes---across depths 1--4, then evaluated on depths 5--7.

\begin{table}[H]
\centering
\vspace{-0.5em}
\caption{\textbf{Multi-ansatz training improves named-ansatz transfer.} The single-ansatz baseline trains the relevance selector only on HEA-RYRZ-linear circuits. The multi-ansatz variant trains the same model architecture on a mixed ansatz panel: HEA-RYRZ-linear, HEA-RYRZ-circular, EfficientSU2, and RealAmplitudes. Both settings evaluate online VQE on the same five STO-3G molecular Hamiltonians, depths 5--7, four held-out seeds per molecule-depth cell, $\tau=0.997$, and 1024 shots per parameter-shift measurement setting. ``Abs-gap gain'' is the reduction in mean $|\dE|$ from single-ansatz to multi-ansatz training; positive values mean multi-ansatz training improved the final-energy match to Full-QWC.}
\vspace{0.25em}
\setlength{\tabcolsep}{1.0mm}
\resizebox{1.0\linewidth}{!}{
\begin{tabular}{ll|cccccc}
\toprule
Target ansatz & Training regime & Mean \(\dE\) & Mean \(|\dE|\downarrow\) & Abs-gap gain \(\uparrow\) & Strict CA seeds & Shot gain \(\uparrow\) & Cells won vs Random \\
\midrule
\gcell{HEA-RYRZ-Circular} & Single ansatz & \(+0.0033\) & \rbf{0.0088} & \gray{--} & 12/60 & 12.9\% & \textbf{\gbf{15/15}} \\
\gcell{HEA-RYRZ-Circular} & \ours{} multi-ansatz & \textbf{\(+0.0015\)} & \textbf{\gbf{0.0032}} & \textbf{\gbf{+0.0057}} & \textbf{\gbf{34/60}} & \textbf{\gbf{14.0\%}} & \gbf{14/15} \\
\midrule
\bcell{EfficientSU2} & Single ansatz & \(+0.0030\) & \rbf{0.0119} & \gray{--} & 13/60 & 9.5\% & \gbf{14/15} \\
\bcell{EfficientSU2} & \ours{} multi-ansatz & \textbf{\(+0.0002\)} & \textbf{\gbf{0.0069}} & \textbf{\gbf{+0.0050}} & \textbf{\gbf{24/60}} & \textbf{\gbf{15.1\%}} & \textbf{\gbf{15/15}} \\
\midrule
\gcell{RealAmplitudes} & Single ansatz & \(-0.0006\) & \textbf{\gbf{0.0030}} & \gray{--} & \textbf{\gbf{29/60}} & 13.5\% & \textbf{\gbf{15/15}} \\
\gcell{RealAmplitudes} & \ours{} multi-ansatz & \textbf{\(-0.0008\)} & \rbf{0.0034} & \rbf{-0.0004} & 22/60 & \textbf{\gbf{20.8\%}} & \textbf{\gbf{15/15}} \\
\midrule
\bcell{StronglyEntanglingLayers} & Single ansatz & \(+0.0395\) & \textbf{\rbf{0.0453}} & \gray{--} & 0/60 & 11.1\% & \textbf{\rbf{4/15}} \\
\bcell{StronglyEntanglingLayers} & \ours{} multi-ansatz & \textbf{\(+0.0360\)} & \rbf{0.0476} & \rbf{-0.0023} & \textbf{2/60} & \textbf{\gbf{13.4\%}} & \rbf{3/15} \\
\bottomrule
\end{tabular}}
\label{tab:v44_multi_ansatz_training}
\vspace{-0.5em}
\end{table}

The multi-ansatz training distribution substantially improves transfer to HEA-RYRZ-Circular and EfficientSU2: mean absolute gap drops from 0.0088 to 0.0032 Ha for HEA-RYRZ-Circular and from 0.0119 to 0.0069 Ha for EfficientSU2, while strict chemical-accuracy seed counts increase from 12/60 to 34/60 and from 13/60 to 24/60, respectively. RealAmplitudes remains a strong case but shifts toward a more measurement-frugal operating point, increasing shot saving from 13.5\% to 20.8\% at a small cost in mean absolute gap. StronglyEntanglingLayers remains the hard OOD failure case. This suggests that our method learns transferable structure-conditioned relevance rules when the training panel covers comparable entangler families, but substantially different entangling layers require either explicit gate/topology embeddings or inclusion in the training distribution.

\clearpage
\section{Controlled Same-Family Ansatz OOD Transfer With Baselines}
This comparison extends the controlled same-family ansatz-transfer setting with TITAN-style parameter-freezing baselines under the identical online VQE protocol. For each ansatz family, the learned selector is trained on two topology variants and evaluated on held-out same-family topology variants. The test ansatz is unseen during training, but the transfer is controlled: same rotation family, same molecular panel, train depths 1--4, unseen evaluation depths 5--7, four held-out seeds per molecule-depth cell, $\tau=0.997$, and 1024 shots per parameter-shift measurement setting.

\begin{table}[H]
\centering
\vspace{-0.5em}
\caption{\textbf{Controlled same-family ansatz OOD transfer with matched baselines at $\tau=0.997$.} Each block trains \ourmethod{} on two topology variants from one ansatz family and evaluates on held-out same-family topologies that do not appear in the training labels. TITAN~\citep{peng2026titan} applies active-parameter freezing and full-QWC measurements for active parameters; TITAN+\ours{} composes the same active-parameter mask with \ourmethod{} group selection. CA excess reports $1000\max(\overline{\dE}-0.0016,0)$ in mHa, so lower energies are accepted and smaller is better. ``Cells won'' counts molecule-depth cells, out of 15, where the method has smaller mean $|\dE|$ than matched-budget Random.}
\vspace{0.25em}
\setlength{\tabcolsep}{0.75mm}
\scriptsize
\resizebox{1.0\linewidth}{!}{
\begin{tabular}{lll|ccccc}
\toprule
Family / train topologies & Held-out target & Method & Mean $\dE$ & Mean $|\dE|\downarrow$ & Shot gain $\uparrow$ & CA excess (mHa) $\downarrow$ & Cells won \\
\midrule
\gcell{HEA-RYRZ / linear + reverse-linear} & \gcell{Circular} & \ours{} & \textbf{\rbf{+0.0017}} & \textbf{\gbf{0.0055}} & 12.40\% & \gbf{0.09} & \textbf{\gbf{14/15}} \\
\gcell{HEA-RYRZ / linear + reverse-linear} & \gcell{Circular} & Random & \rbf{+0.0614} & \rbf{0.0716} & 12.70\% & 59.80 & \gray{--} \\
\gcell{HEA-RYRZ / linear + reverse-linear} & \gcell{Circular} & TITAN~(\citeauthor{peng2026titan}) & \rbf{+0.0074} & \gbf{0.0458} & 33.00\% & 5.77 & 6/15 \\
\gcell{HEA-RYRZ / linear + reverse-linear} & \gcell{Circular} & TITAN+\ours{} & \rbf{+0.0074} & \gbf{0.0451} & \underline{\gbf{40.10\%}} & 5.82 & 6/15 \\
\midrule
\bcell{HEA-RYRZ / linear + reverse-linear} & \bcell{Full} & \ours{} & \textbf{\gbf{-0.0002}} & \textbf{\gbf{0.0085}} & 8.50\% & \gbf{0.00} & \textbf{\gbf{14/15}} \\
\bcell{HEA-RYRZ / linear + reverse-linear} & \bcell{Full} & Random & \rbf{+0.0280} & \rbf{0.0460} & 8.50\% & 26.45 & \gray{--} \\
\bcell{HEA-RYRZ / linear + reverse-linear} & \bcell{Full} & TITAN~(\citeauthor{peng2026titan}) & \rbf{+0.0484} & \rbf{0.1007} & 40.80\% & 46.82 & 2/15 \\
\bcell{HEA-RYRZ / linear + reverse-linear} & \bcell{Full} & TITAN+\ours{} & \rbf{+0.0481} & \rbf{0.1010} & \underline{\gbf{44.60\%}} & 46.54 & 2/15 \\
\midrule
\gcell{EfficientSU2 / linear + reverse-linear} & \gcell{Circular} & \ours{} & \textbf{\rbf{+0.0019}} & \textbf{\gbf{0.0056}} & 12.40\% & \gbf{0.27} & \textbf{\gbf{15/15}} \\
\gcell{EfficientSU2 / linear + reverse-linear} & \gcell{Circular} & Random & \rbf{+0.0612} & \rbf{0.0716} & 12.60\% & 59.59 & \gray{--} \\
\gcell{EfficientSU2 / linear + reverse-linear} & \gcell{Circular} & TITAN~(\citeauthor{peng2026titan}) & \rbf{+0.0074} & \gbf{0.0458} & 33.00\% & 5.77 & 6/15 \\
\gcell{EfficientSU2 / linear + reverse-linear} & \gcell{Circular} & TITAN+\ours{} & \rbf{+0.0073} & \gbf{0.0450} & \underline{\gbf{40.10\%}} & 5.69 & 6/15 \\
\midrule
\bcell{EfficientSU2 / linear + reverse-linear} & \bcell{Full} & \ours{} & \textbf{\gbf{-0.0002}} & \textbf{\gbf{0.0089}} & 8.60\% & \gbf{0.00} & \textbf{\gbf{12/15}} \\
\bcell{EfficientSU2 / linear + reverse-linear} & \bcell{Full} & Random & \rbf{+0.0377} & \rbf{0.0648} & 8.50\% & 36.11 & \gray{--} \\
\bcell{EfficientSU2 / linear + reverse-linear} & \bcell{Full} & TITAN~(\citeauthor{peng2026titan}) & \rbf{+0.0484} & \rbf{0.1007} & 40.80\% & 46.82 & 3/15 \\
\bcell{EfficientSU2 / linear + reverse-linear} & \bcell{Full} & TITAN+\ours{} & \rbf{+0.0481} & \rbf{0.1009} & \underline{\gbf{44.60\%}} & 46.51 & 3/15 \\
\midrule
\gcell{RealAmplitudes / linear + reverse-linear} & \gcell{Circular} & \ours{} & \textbf{\rbf{+0.0052}} & \textbf{\gbf{0.0177}} & 19.60\% & 3.63 & \textbf{\gbf{13/15}} \\
\gcell{RealAmplitudes / linear + reverse-linear} & \gcell{Circular} & Random & \rbf{+0.0684} & \rbf{0.0858} & 19.90\% & 66.75 & \gray{--} \\
\gcell{RealAmplitudes / linear + reverse-linear} & \gcell{Circular} & TITAN~(\citeauthor{peng2026titan}) & \gbf{-0.0245} & \gbf{0.0512} & 18.00\% & \gbf{0.00} & 11/15 \\
\gcell{RealAmplitudes / linear + reverse-linear} & \gcell{Circular} & TITAN+\ours{} & \gbf{-0.0240} & \gbf{0.0527} & \underline{\gbf{33.40\%}} & \gbf{0.00} & 10/15 \\
\midrule
\bcell{RealAmplitudes / linear + reverse-linear} & \bcell{Full} & \ours{} & \textbf{\gbf{+0.0012}} & \textbf{\gbf{0.0235}} & 13.90\% & \gbf{0.00} & \textbf{\gbf{14/15}} \\
\bcell{RealAmplitudes / linear + reverse-linear} & \bcell{Full} & Random & \rbf{+0.0648} & \rbf{0.0880} & 14.00\% & 63.18 & \gray{--} \\
\bcell{RealAmplitudes / linear + reverse-linear} & \bcell{Full} & TITAN~(\citeauthor{peng2026titan}) & \gbf{-0.0014} & \rbf{0.0885} & 22.80\% & \gbf{0.00} & 9/15 \\
\bcell{RealAmplitudes / linear + reverse-linear} & \bcell{Full} & TITAN+\ours{} & \gbf{-0.0014} & \rbf{0.0899} & \underline{\gbf{33.90\%}} & \gbf{0.00} & 9/15 \\
\midrule
\gcell{StronglyEntangling / range-1 + range-2} & \gcell{Layer-dependent} & \ours{} & \textbf{\gbf{+0.0012}} & \textbf{\gbf{0.0077}} & 12.40\% & \gbf{0.00} & \textbf{\gbf{13/15}} \\
\gcell{StronglyEntangling / range-1 + range-2} & \gcell{Layer-dependent} & Random & \rbf{+0.0343} & \rbf{0.0447} & 12.60\% & 32.69 & \gray{--} \\
\gcell{StronglyEntangling / range-1 + range-2} & \gcell{Layer-dependent} & TITAN~(\citeauthor{peng2026titan}) & \rbf{+0.0104} & \rbf{0.0454} & 39.60\% & 8.80 & 4/15 \\
\gcell{StronglyEntangling / range-1 + range-2} & \gcell{Layer-dependent} & TITAN+\ours{} & \rbf{+0.0106} & \rbf{0.0456} & \underline{\gbf{45.50\%}} & 9.01 & 5/15 \\
\midrule
\bcell{StronglyEntangling / range-1 + range-2} & \bcell{Full} & \ours{} & \textbf{\rbf{+0.0023}} & \textbf{\gbf{0.0206}} & 8.90\% & \gbf{0.66} & \textbf{\gbf{13/15}} \\
\bcell{StronglyEntangling / range-1 + range-2} & \bcell{Full} & Random & \rbf{+0.0596} & \rbf{0.0968} & 9.10\% & 57.97 & \gray{--} \\
\bcell{StronglyEntangling / range-1 + range-2} & \bcell{Full} & TITAN~(\citeauthor{peng2026titan}) & \rbf{+0.0493} & \rbf{0.1071} & 47.30\% & 47.68 & 4/15 \\
\bcell{StronglyEntangling / range-1 + range-2} & \bcell{Full} & TITAN+\ours{} & \rbf{+0.0502} & \rbf{0.1066} & \underline{\gbf{50.70\%}} & 48.59 & 4/15 \\
\bottomrule
\end{tabular}}
\label{tab:v45_within_family_ansatz_ood}
\vspace{-0.5em}
\end{table}

Table~\ref{tab:v45_within_family_ansatz_ood} shows that the controlled same-family transfer result is not explained by measurement budget alone. Matched-budget Random keeps similar shot savings but is substantially less accurate. TITAN-style freezing can reduce cost by dropping parameters, and TITAN+\ours{} can compound this saving, but these rows often trade away final-energy fidelity. The cleanest read is that \ourmethod{} targets a different axis from parameter freezing: for parameters that are still measured, it learns which QWC groups are worth measuring.

\clearpage
\section{Cell-Wise Breakdown of Table~\ref{tab:v45_within_family_ansatz_ood}}
Table~\ref{tab:v45_cellwise_wins} expands the ``Cells won'' column from Table~\ref{tab:v45_within_family_ansatz_ood}. Each entry reports the molecule-depth cell improvement in mean absolute final-energy gap, measured in mHa:
\begin{equation}
1000\times\left(|\Delta E|_{\mathrm{Random}}-|\Delta E|_{\mathrm{Ours}}\right).
\end{equation}
Positive green values mean Ours is closer to Full-QWC for that exact molecule-depth cell; negative red values indicate a random same-budget win.

\begin{table}[H]
\centering
\vspace{-0.5em}
\caption{\textbf{Cell-wise breakdown of the same-family ansatz-OOD wins.} Entries are Random-minus-Ours mean $|\dE|$ in mHa for each molecule and depth. Green positive entries are cells won by Ours; red negative entries are cells won by the matched-budget random baseline.}
\label{tab:v45_cellwise_wins}
\vspace{0.25em}
\setlength{\tabcolsep}{0.8mm}
\resizebox{1.0\linewidth}{!}{
\begin{tabular}{l|ccc|ccc|ccc|ccc|ccc|c}
\toprule
Held-out target & \multicolumn{3}{c|}{H2} & \multicolumn{3}{c|}{HeH+} & \multicolumn{3}{c|}{H3+} & \multicolumn{3}{c|}{H4} & \multicolumn{3}{c|}{LiH} & Wins \\
\cmidrule(lr){2-4}\cmidrule(lr){5-7}\cmidrule(lr){8-10}\cmidrule(lr){11-13}\cmidrule(lr){14-16}
 & D5 & D6 & D7 & D5 & D6 & D7 & D5 & D6 & D7 & D5 & D6 & D7 & D5 & D6 & D7 &  \\
\midrule
\gcell{HEA-RYRZ / Circular} & \gbf{+11.4} & \gbf{+4.8} & \rbf{-0.4} & \gbf{+30.9} & \gbf{+5.0} & \gbf{+3.0} & \gbf{+18.0} & \gbf{+16.0} & \gbf{+8.7} & \gbf{+22.7} & \gbf{+47.7} & \gbf{+36.8} & \gbf{+320.2} & \gbf{+198.4} & \gbf{+268.8} & \textbf{\gbf{14/15}} \\
\midrule
\bcell{HEA-RYRZ / Full} & \gbf{+18.1} & \gbf{+22.6} & \rbf{-0.0} & \gbf{+37.0} & \gbf{+10.2} & \gbf{+10.8} & \gbf{+46.5} & \gbf{+32.6} & \gbf{+0.7} & \gbf{+22.9} & \gbf{+8.3} & \gbf{+51.8} & \gbf{+42.9} & \gbf{+109.2} & \gbf{+148.2} & \textbf{\gbf{14/15}} \\
\midrule
\gcell{EfficientSU2 / Circular} & \gbf{+3.5} & \gbf{+5.3} & \gbf{+0.3} & \gbf{+28.0} & \gbf{+5.7} & \gbf{+0.9} & \gbf{+14.5} & \gbf{+13.3} & \gbf{+16.6} & \gbf{+25.2} & \gbf{+53.8} & \gbf{+111.6} & \gbf{+280.9} & \gbf{+215.2} & \gbf{+214.5} & \textbf{\gbf{15/15}} \\
\midrule
\bcell{EfficientSU2 / Full} & \gbf{+18.1} & \gbf{+14.8} & \rbf{-1.8} & \gbf{+31.5} & \rbf{-4.5} & \gbf{+22.9} & \gbf{+38.6} & \gbf{+13.8} & \rbf{-0.2} & \gbf{+125.8} & \gbf{+20.9} & \gbf{+51.5} & \gbf{+109.9} & \gbf{+206.7} & \gbf{+190.7} & \textbf{\gbf{12/15}} \\
\midrule
\gcell{RealAmplitudes / Circular} & \gbf{+19.9} & \gbf{+7.8} & \rbf{-2.1} & \gbf{+94.3} & \gbf{+0.5} & \gbf{+42.1} & \gbf{+24.5} & \gbf{+18.2} & \rbf{-8.1} & \gbf{+80.6} & \gbf{+33.9} & \gbf{+63.8} & \gbf{+188.2} & \gbf{+190.3} & \gbf{+267.0} & \textbf{\gbf{13/15}} \\
\midrule
\bcell{RealAmplitudes / Full} & \gbf{+34.1} & \rbf{-1.6} & \gbf{+0.0} & \gbf{+33.1} & \gbf{+27.2} & \gbf{+99.1} & \gbf{+48.6} & \gbf{+121.0} & \gbf{+39.5} & \gbf{+121.7} & \gbf{+18.1} & \gbf{+5.6} & \gbf{+148.9} & \gbf{+75.9} & \gbf{+195.2} & \textbf{\gbf{14/15}} \\
\midrule
\gcell{StronglyEntangling / Layer-dependent} & \rbf{-3.6} & \gbf{+12.9} & \gbf{+2.7} & \gbf{+15.4} & \gbf{+18.1} & \gbf{+4.9} & \gbf{+13.7} & \rbf{-1.7} & \gbf{+27.0} & \gbf{+39.2} & \gbf{+2.6} & \gbf{+29.1} & \gbf{+183.9} & \gbf{+84.9} & \gbf{+125.5} & \textbf{\gbf{13/15}} \\
\midrule
\bcell{StronglyEntangling / Full} & \gbf{+10.7} & \gbf{+9.0} & \gbf{+21.3} & \gbf{+83.2} & \rbf{-14.9} & \gbf{+11.3} & \gbf{+4.9} & \gbf{+14.1} & \gbf{+20.1} & \gbf{+62.3} & \rbf{-23.3} & \gbf{+5.3} & \gbf{+473.9} & \gbf{+249.6} & \gbf{+215.9} & \textbf{\gbf{13/15}} \\
\bottomrule
\end{tabular}}
\vspace{-0.5em}
\end{table}

\clearpage
\section{Molecule-Depth Breakdown of Table~\ref{tab:v45_within_family_ansatz_ood}}
\begin{figure}[!t]
\centering
\includegraphics[width=0.98\linewidth]{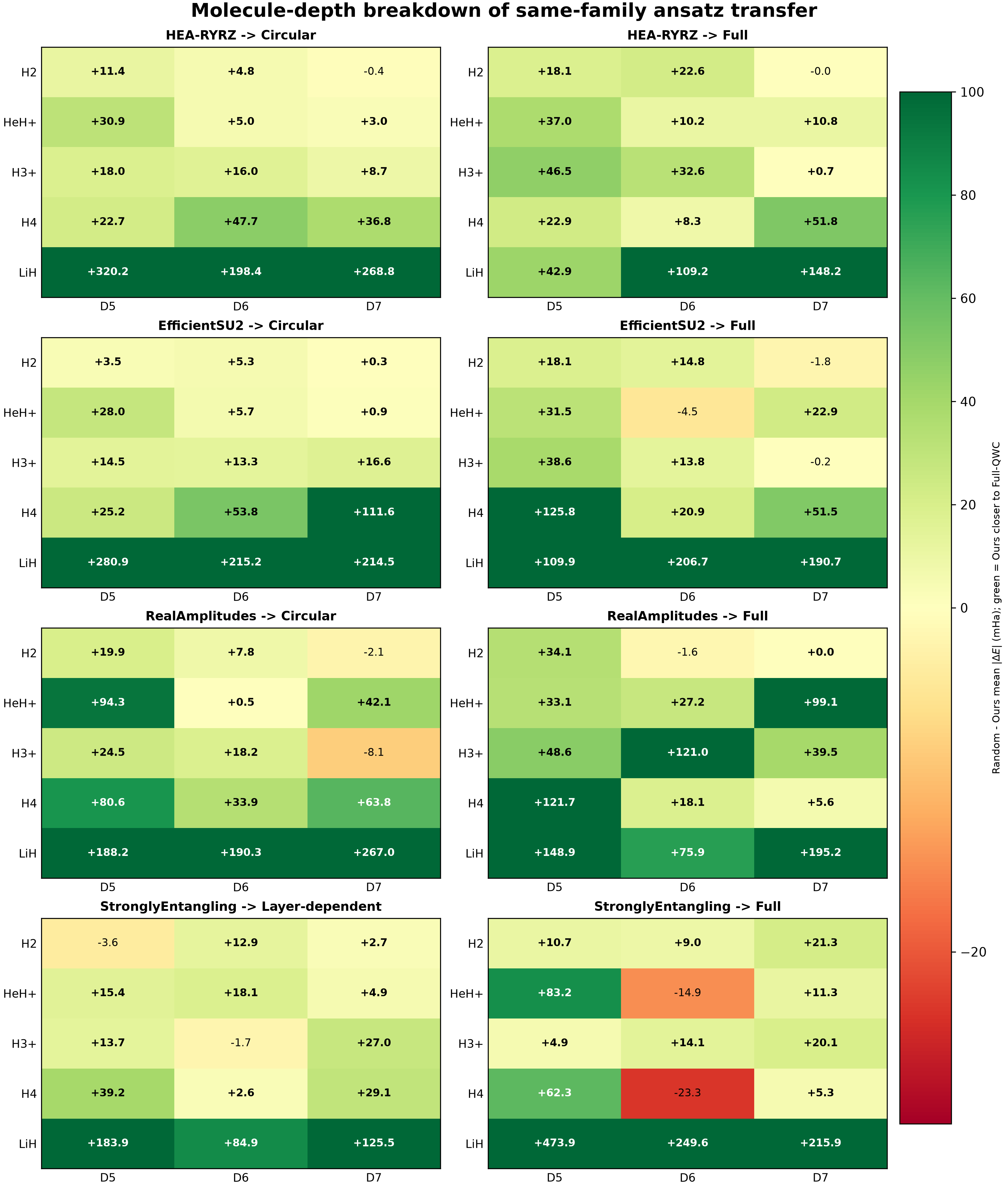}
\caption{\textbf{Molecule-depth breakdown for Table~\ref{tab:v45_within_family_ansatz_ood}.} Each panel corresponds to one controlled same-family ansatz-OOD held-out target. Rows are molecules, columns are unseen evaluation depths 5--7, and entries are Random-minus-\ours{} mean $|\dE|$ in mHa. Green positive entries are cells where our method is closer to Full-QWC than matched-budget Random. The color scale is clipped for readability, but numeric labels show the unclipped values rounded to one decimal place.}
\label{fig:v49_table8_molecule_detail_heatmap}
\vspace{-0.5em}
\end{figure}

Figure~\ref{fig:v49_table8_molecule_detail_heatmap} expands the ``Cells won'' column from Table~\ref{tab:v45_within_family_ansatz_ood} into individual molecule-depth cells. Each entry reports
\begin{equation}
1000\times\left(|\Delta E|_{\mathrm{Random}}-|\Delta E|_{\mathrm{Ours}}\right),
\end{equation}
measured in mHa. Positive green values mean our method is closer to Full-QWC for that exact molecule-depth cell; negative red values mean the matched-budget random selector is closer.

\clearpage
\section{Predictor Architecture Tables}
\label{sec:app_predictor_tables}
\begin{figure}[H]
\centering
\includegraphics[width=0.94\linewidth]{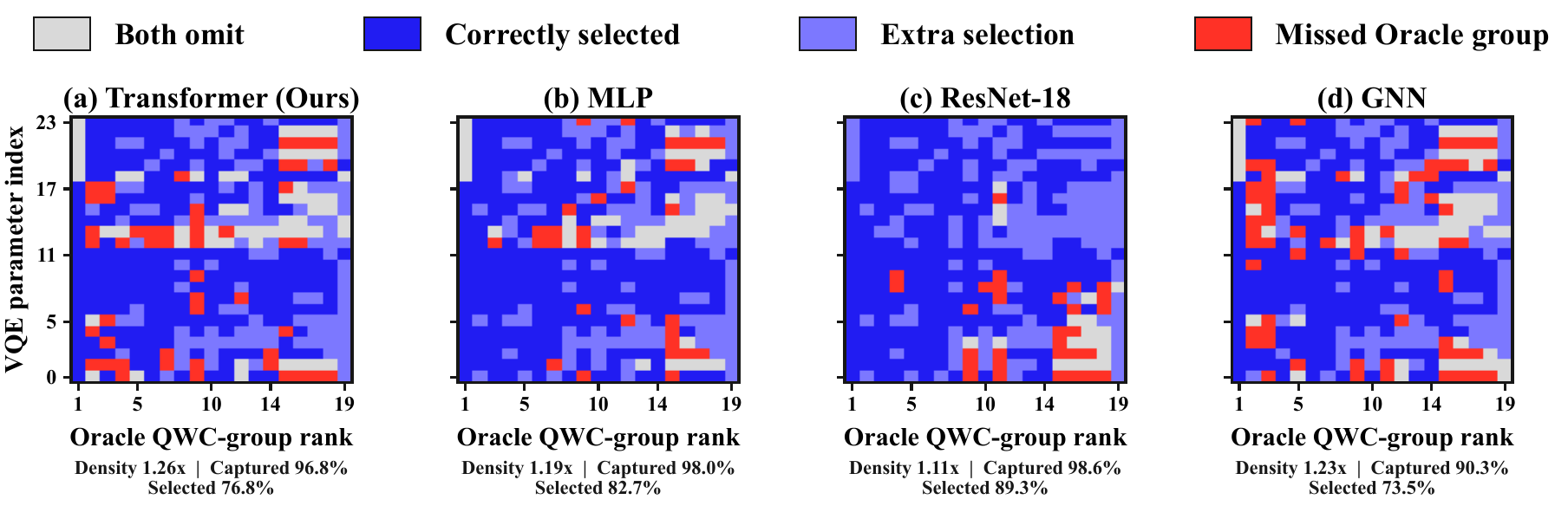}
\caption{\textbf{Oracle-relative parameter--QWC masks for four relevance predictors.} All predictors are evaluated on the same held-out $\mathrm{H}_3^+$ molecule-OOD snapshot (depth 2, seed 97, optimization iteration 14) at $\tau=0.997$. Rows index VQE parameters and columns index QWC groups sorted by oracle relevance. Dark-blue cells are selected by both the predictor and oracle; gray cells are omitted by both; light-blue cells are predictor-only extra measurements; and red cells are oracle-selected groups missed by the predictor. ``Captured'' is the mean exact relevance mass covered by a predicted row mask, ``Selected'' is its mean selected-group fraction, and ``Density'' is captured mass divided by selected fraction. The snapshot is a qualitative structural diagnostic; aggregate performance is reported in main-text Figure~\ref{fig:ml_predictor_comparison} and Tables~\ref{tab:v62_lomo_model_zoo} and~\ref{tab:v64_hamiltonian_model_zoo}.}
\label{fig:ml_qwc_mask_comparison}
\vspace{-0.5em}
\end{figure}

Figure~\ref{fig:ml_predictor_comparison} gives the aggregate architecture
comparison referenced in the main text. Figure~\ref{fig:ml_qwc_mask_comparison}
provides its cell-level counterpart. On this held-out snapshot, the Transformer
captures 96.8\% of the exact relevance mass while selecting 76.8\% of
parameter--group cells. ResNet-18 captures slightly more mass (98.6\%) by
selecting 89.3\% of cells, whereas the GNN is sparser but misses more
oracle-relevant groups. Similar scalar budgets can therefore arise from
materially different parameter-specific masks.

\appMLPredictorComparison
\clearpage

\clearpage
\subsection{Held-Out-Seed Predictor Comparison}
\begin{table*}[t]
\centering
\vspace{-0.5em}
\caption{\textbf{Predictor architecture comparison at $\tau=0.995$.} This table copies the Table~\ref{tab:v39_tau0995_main} held-out-seed online VQE setting and changes only the learned relevance predictor. Our method uses the Transformer predictor throughout the paper; MLP, ResNet-18, and GNN use the same labels, features, threshold, molecules, depths, and online protocol. Boldface marks the closest final energy; shot-gain values are green/red according to the no-worse criterion, and bold-underlined values mark the largest shot gain per cell.}
\vspace{0.25em}
\setlength{\tabcolsep}{0.45mm}
\renewcommand{\arraystretch}{0.88}
\tiny
\resizebox{1.0\linewidth}{!}{
\begin{tabular}{lll|ccccc}
\toprule
Setting & Metric & Method & H2 (4q) & HeH+ (4q) & H3+ (6q) & H4 chain (8q) & LiH (12q) \\
\midrule
\gcell{Depth 1} & \(\dE\) vs Full \(\downarrow\) & \ours{} & \gbf{\(+0.0011\)} & \gbf{\(+0.0015\)} & \gbf{\(+0.0005\)} & \textbf{\gbf{\(-0.0006\)}} & \rbf{\(+0.0023\)} \\
\gcell{Depth 1} & \(\dE\) vs Full \(\downarrow\) & MLP & \gbf{\(+0.0008\)} & \rbf{\(+0.0028\)} & \gbf{\(+0.0015\)} & \gbf{\(-0.0021\)} & \rbf{\(+0.0025\)} \\
\gcell{Depth 1} & \(\dE\) vs Full \(\downarrow\) & ResNet-18 & \textbf{\gbf{\(+0.0000\)}} & \textbf{\gbf{\(-0.0010\)}} & \rbf{\(+0.0029\)} & \rbf{\(+0.0024\)} & \gbf{\(+0.0016\)} \\
\gcell{Depth 1} & \(\dE\) vs Full \(\downarrow\) & GNN & \gbf{\(+0.0012\)} & \rbf{\(+0.0022\)} & \textbf{\gbf{\(+0.0004\)}} & \gbf{\(-0.0021\)} & \textbf{\gbf{\(+0.0013\)}} \\
\cmidrule(lr){2-8}
\bcell{Depth 1} & Shot gain \(\uparrow\) & \ours{} & \textcolor{green!55!black}{40.4\%} & \textcolor{green!55!black}{58.9\%} & \textcolor{green!55!black}{63.0\%} & \textcolor{green!55!black}{74.7\%} & \textcolor{red!70!black}{86.9\%} \\
\bcell{Depth 1} & Shot gain \(\uparrow\) & MLP & \textbf{\underline{\textcolor{green!55!black}{47.3\%}}} & \textbf{\underline{\textcolor{red!70!black}{62.5\%}}} & \textbf{\underline{\textcolor{green!55!black}{68.2\%}}} & \textbf{\underline{\textcolor{green!55!black}{75.9\%}}} & \textbf{\underline{\textcolor{red!70!black}{87.6\%}}} \\
\bcell{Depth 1} & Shot gain \(\uparrow\) & ResNet-18 & \textcolor{green!55!black}{0.0\%} & \textcolor{green!55!black}{7.5\%} & \textcolor{red!70!black}{38.1\%} & \textcolor{red!70!black}{46.1\%} & \textcolor{green!55!black}{77.9\%} \\
\bcell{Depth 1} & Shot gain \(\uparrow\) & GNN & \textcolor{green!55!black}{42.7\%} & \textcolor{red!70!black}{57.7\%} & \textcolor{green!55!black}{60.3\%} & \textcolor{green!55!black}{66.6\%} & \textcolor{green!55!black}{82.5\%} \\
\cmidrule(lr){2-8}
\gcell{Depth 1} & No-worse vs Full & \ours{} & \cmark & \cmark & \cmark & \cmark & \xmark \\
\gcell{Depth 1} & No-worse vs Full & MLP & \cmark & \xmark & \cmark & \cmark & \xmark \\
\gcell{Depth 1} & No-worse vs Full & ResNet-18 & \cmark & \cmark & \xmark & \xmark & \cmark \\
\gcell{Depth 1} & No-worse vs Full & GNN & \cmark & \xmark & \cmark & \cmark & \cmark \\
\midrule
\gcell{Depth 2} & \(\dE\) vs Full \(\downarrow\) & \ours{} & \gbf{\(+0.0003\)} & \textbf{\gbf{\(-0.0011\)}} & \gbf{\(-0.0015\)} & \rbf{\(+0.0158\)} & \gbf{\(-0.0049\)} \\
\gcell{Depth 2} & \(\dE\) vs Full \(\downarrow\) & MLP & \rbf{\(+0.0127\)} & \gbf{\(-0.0014\)} & \gbf{\(-0.0029\)} & \textbf{\gbf{\(+0.0006\)}} & \gbf{\(-0.0037\)} \\
\gcell{Depth 2} & \(\dE\) vs Full \(\downarrow\) & ResNet-18 & \textbf{\gbf{\(+0.0000\)}} & \gbf{\(+0.0011\)} & \textbf{\gbf{\(+0.0007\)}} & \rbf{\(+0.0221\)} & \rbf{\(+0.0065\)} \\
\gcell{Depth 2} & \(\dE\) vs Full \(\downarrow\) & GNN & \gbf{\(+0.0010\)} & \gbf{\(-0.0134\)} & \rbf{\(+0.0031\)} & \rbf{\(+0.0254\)} & \textbf{\gbf{\(-0.0026\)}} \\
\cmidrule(lr){2-8}
\bcell{Depth 2} & Shot gain \(\uparrow\) & \ours{} & \textcolor{green!55!black}{25.6\%} & \textcolor{green!55!black}{31.6\%} & \textcolor{green!55!black}{44.9\%} & \textbf{\underline{\textcolor{red!70!black}{51.8\%}}} & \textbf{\underline{\textcolor{green!55!black}{82.6\%}}} \\
\bcell{Depth 2} & Shot gain \(\uparrow\) & MLP & \textbf{\underline{\textcolor{red!70!black}{28.3\%}}} & \textbf{\underline{\textcolor{green!55!black}{33.4\%}}} & \textbf{\underline{\textcolor{green!55!black}{46.1\%}}} & \textcolor{green!55!black}{48.4\%} & \textcolor{green!55!black}{78.0\%} \\
\bcell{Depth 2} & Shot gain \(\uparrow\) & ResNet-18 & \textcolor{green!55!black}{0.0\%} & \textcolor{green!55!black}{2.2\%} & \textcolor{green!55!black}{24.2\%} & \textcolor{red!70!black}{35.2\%} & \textcolor{red!70!black}{73.3\%} \\
\bcell{Depth 2} & Shot gain \(\uparrow\) & GNN & \textcolor{green!55!black}{21.6\%} & \textcolor{green!55!black}{28.3\%} & \textcolor{red!70!black}{41.7\%} & \textcolor{red!70!black}{45.9\%} & \textcolor{green!55!black}{77.6\%} \\
\cmidrule(lr){2-8}
\gcell{Depth 2} & No-worse vs Full & \ours{} & \cmark & \cmark & \cmark & \xmark & \cmark \\
\gcell{Depth 2} & No-worse vs Full & MLP & \xmark & \cmark & \cmark & \cmark & \cmark \\
\gcell{Depth 2} & No-worse vs Full & ResNet-18 & \cmark & \cmark & \cmark & \xmark & \xmark \\
\gcell{Depth 2} & No-worse vs Full & GNN & \cmark & \cmark & \xmark & \xmark & \cmark \\
\midrule
\gcell{Depth 3} & \(\dE\) vs Full \(\downarrow\) & \ours{} & \gbf{\(+0.0005\)} & \rbf{\(+0.0092\)} & \gbf{\(+0.0005\)} & \gbf{\(-0.0014\)} & \gbf{\(-0.0017\)} \\
\gcell{Depth 3} & \(\dE\) vs Full \(\downarrow\) & MLP & \rbf{\(+0.0035\)} & \gbf{\(-0.0045\)} & \gbf{\(-0.0008\)} & \textbf{\gbf{\(-0.0000\)}} & \gbf{\(-0.0039\)} \\
\gcell{Depth 3} & \(\dE\) vs Full \(\downarrow\) & ResNet-18 & \textbf{\gbf{\(+0.0000\)}} & \rbf{\(+0.0044\)} & \textbf{\gbf{\(-0.0004\)}} & \gbf{\(-0.0064\)} & \rbf{\(+0.0053\)} \\
\gcell{Depth 3} & \(\dE\) vs Full \(\downarrow\) & GNN & \rbf{\(+0.0039\)} & \textbf{\gbf{\(-0.0017\)}} & \rbf{\(+0.0020\)} & \gbf{\(-0.0025\)} & \textbf{\gbf{\(+0.0001\)}} \\
\cmidrule(lr){2-8}
\bcell{Depth 3} & Shot gain \(\uparrow\) & \ours{} & \textcolor{green!55!black}{18.1\%} & \textbf{\underline{\textcolor{red!70!black}{24.9\%}}} & \textbf{\underline{\textcolor{green!55!black}{29.9\%}}} & \textbf{\underline{\textcolor{green!55!black}{35.2\%}}} & \textbf{\underline{\textcolor{green!55!black}{75.4\%}}} \\
\bcell{Depth 3} & Shot gain \(\uparrow\) & MLP & \textbf{\underline{\textcolor{red!70!black}{20.8\%}}} & \textcolor{green!55!black}{23.7\%} & \textcolor{green!55!black}{28.9\%} & \textcolor{green!55!black}{32.7\%} & \textcolor{green!55!black}{71.5\%} \\
\bcell{Depth 3} & Shot gain \(\uparrow\) & ResNet-18 & \textcolor{green!55!black}{0.0\%} & \textcolor{red!70!black}{2.0\%} & \textcolor{green!55!black}{19.9\%} & \textcolor{green!55!black}{28.4\%} & \textcolor{red!70!black}{73.0\%} \\
\bcell{Depth 3} & Shot gain \(\uparrow\) & GNN & \textcolor{red!70!black}{16.8\%} & \textcolor{green!55!black}{22.1\%} & \textcolor{red!70!black}{28.4\%} & \textcolor{green!55!black}{31.7\%} & \textcolor{green!55!black}{66.9\%} \\
\cmidrule(lr){2-8}
\gcell{Depth 3} & No-worse vs Full & \ours{} & \cmark & \xmark & \cmark & \cmark & \cmark \\
\gcell{Depth 3} & No-worse vs Full & MLP & \xmark & \cmark & \cmark & \cmark & \cmark \\
\gcell{Depth 3} & No-worse vs Full & ResNet-18 & \cmark & \xmark & \cmark & \cmark & \xmark \\
\gcell{Depth 3} & No-worse vs Full & GNN & \xmark & \cmark & \xmark & \cmark & \cmark \\
\bottomrule
\end{tabular}}
\label{tab:v60_model_zoo_table1}
\vspace{-0.5em}
\end{table*}

\clearpage
\subsection{Leave-One-Molecule-Out Predictor Comparison}
\begin{table}[H]
\centering
\vspace{-0.7em}
\caption{\textbf{Predictor comparison under leave-one-molecule-out transfer at $\tau=0.997$.} The predictor is the only changed component: all methods use identical features, labels, molecule splits, depths, seeds, and closed-loop online VQE. Representative depths 1 and 4 are shown. Green/red follows the one-sided $1.6$ mHa no-worse criterion; bold marks the closest mean energy to Full, and bold-underlined shot gain marks the largest saving per cell.}
\vspace{0.15em}
\setlength{\tabcolsep}{0.55mm}
\renewcommand{\arraystretch}{0.58}
\tiny
\resizebox{1.0\linewidth}{!}{
\begin{tabular}{lll|cccc}
\toprule
Molecule & Depth & Metric & \ours{} & MLP & ResNet-18 & GNN \\
\midrule
\gcell{H2 (4q)} & Depth 1 & $\dE$ vs Full $\downarrow$ & \rbf{+0.0016} & \rbf{+0.0030} & \textbf{\gbf{+0.0000}} & \rbf{+0.0018} \\
 & Depth 1 & Shot gain $\uparrow$ & \textcolor{red!70!black}{55.2\%} & \textbf{\underline{\textcolor{red!70!black}{56.7\%}}} & \textcolor{green!55!black}{0.0\%} & \textcolor{red!70!black}{46.9\%} \\
 & Depth 1 & Chem. acc. & \xmark & \xmark & \cmark & \xmark \\
\addlinespace[0.08em]
 & Depth 4 & $\dE$ vs Full $\downarrow$ & \gbf{-0.0002} & \gbf{-0.0020} & \textbf{\gbf{+0.0000}} & \gbf{-0.0000} \\
 & Depth 4 & Shot gain $\uparrow$ & \textcolor{green!55!black}{9.9\%} & \textbf{\underline{\textcolor{green!55!black}{13.7\%}}} & \textcolor{green!55!black}{0.0\%} & \textcolor{green!55!black}{12.2\%} \\
 & Depth 4 & Chem. acc. & \cmark & \cmark & \cmark & \cmark \\
\midrule
\gcell{HeH+ (4q)} & Depth 1 & $\dE$ vs Full $\downarrow$ & \textbf{\rbf{+0.0021}} & \gbf{-0.0038} & \rbf{+0.0139} & \gbf{-0.0032} \\
 & Depth 1 & Shot gain $\uparrow$ & \textcolor{red!70!black}{32.2\%} & \textbf{\underline{\textcolor{green!55!black}{36.2\%}}} & \textcolor{red!70!black}{27.6\%} & \textcolor{green!55!black}{33.6\%} \\
 & Depth 1 & Chem. acc. & \xmark & \cmark & \xmark & \cmark \\
\addlinespace[0.08em]
 & Depth 4 & $\dE$ vs Full $\downarrow$ & \gbf{-0.0022} & \gbf{-0.0025} & \gbf{-0.0071} & \textbf{\gbf{-0.0018}} \\
 & Depth 4 & Shot gain $\uparrow$ & \textcolor{green!55!black}{8.1\%} & \textcolor{green!55!black}{6.0\%} & \textbf{\underline{\textcolor{green!55!black}{9.8\%}}} & \textcolor{green!55!black}{6.0\%} \\
 & Depth 4 & Chem. acc. & \cmark & \cmark & \cmark & \cmark \\
\midrule
\gcell{H3+ (6q)} & Depth 1 & $\dE$ vs Full $\downarrow$ & \rbf{+0.0051} & \textbf{\gbf{-0.0004}} & \rbf{+0.0078} & \rbf{+0.0034} \\
 & Depth 1 & Shot gain $\uparrow$ & \textbf{\underline{\textcolor{red!70!black}{56.2\%}}} & \textcolor{green!55!black}{55.2\%} & \textcolor{red!70!black}{45.3\%} & \textcolor{red!70!black}{52.3\%} \\
 & Depth 1 & Chem. acc. & \xmark & \cmark & \xmark & \xmark \\
\addlinespace[0.08em]
 & Depth 4 & $\dE$ vs Full $\downarrow$ & \gbf{-0.0052} & \gbf{-0.0057} & \gbf{-0.0320} & \textbf{\rbf{+0.0029}} \\
 & Depth 4 & Shot gain $\uparrow$ & \textcolor{green!55!black}{10.0\%} & \textcolor{green!55!black}{8.9\%} & \textbf{\underline{\textcolor{green!55!black}{16.1\%}}} & \textcolor{red!70!black}{12.3\%} \\
 & Depth 4 & Chem. acc. & \cmark & \cmark & \cmark & \xmark \\
\midrule
\gcell{H4 chain (8q)} & Depth 1 & $\dE$ vs Full $\downarrow$ & \rbf{+0.0038} & \rbf{+0.0052} & \textbf{\rbf{+0.0030}} & \rbf{+0.0072} \\
 & Depth 1 & Shot gain $\uparrow$ & \textcolor{red!70!black}{63.7\%} & \textbf{\underline{\textcolor{red!70!black}{67.1\%}}} & \textcolor{red!70!black}{59.8\%} & \textcolor{red!70!black}{66.3\%} \\
 & Depth 1 & Chem. acc. & \xmark & \xmark & \xmark & \xmark \\
\addlinespace[0.08em]
 & Depth 4 & $\dE$ vs Full $\downarrow$ & \rbf{+0.0652} & \rbf{+0.0195} & \rbf{+0.0335} & \textbf{\gbf{-0.0043}} \\
 & Depth 4 & Shot gain $\uparrow$ & \textcolor{red!70!black}{28.5\%} & \textcolor{red!70!black}{35.7\%} & \textbf{\underline{\textcolor{red!70!black}{46.5\%}}} & \textcolor{green!55!black}{28.9\%} \\
 & Depth 4 & Chem. acc. & \xmark & \xmark & \xmark & \cmark \\
\midrule
\gcell{LiH (12q)} & Depth 1 & $\dE$ vs Full $\downarrow$ & \textbf{\rbf{+0.0022}} & \rbf{+0.8939} & \gbf{-0.0314} & \rbf{+0.0101} \\
 & Depth 1 & Shot gain $\uparrow$ & \textcolor{red!70!black}{40.4\%} & \textcolor{red!70!black}{47.1\%} & \textcolor{green!55!black}{41.4\%} & \textbf{\underline{\textcolor{red!70!black}{51.9\%}}} \\
 & Depth 1 & Chem. acc. & \xmark & \xmark & \cmark & \xmark \\
\addlinespace[0.08em]
 & Depth 4 & $\dE$ vs Full $\downarrow$ & \textbf{\gbf{-0.0023}} & \rbf{+0.0037} & \rbf{+0.1054} & \rbf{+0.0073} \\
 & Depth 4 & Shot gain $\uparrow$ & \textcolor{green!55!black}{17.9\%} & \textcolor{red!70!black}{19.4\%} & \textcolor{red!70!black}{13.5\%} & \textbf{\underline{\textcolor{red!70!black}{19.5\%}}} \\
 & Depth 4 & Chem. acc. & \cmark & \xmark & \xmark & \xmark \\
\bottomrule
\end{tabular}}
\label{tab:v62_lomo_model_zoo}
\vspace{-0.8em}
\end{table}

\clearpage
\subsection{Hamiltonian-Family Predictor Comparison}
\begin{table}[H]
\centering
\vspace{-0.7em}
\caption{\textbf{Predictor comparison under Hamiltonian-family transfer at $\tau=0.997$.} One shared predictor per architecture is trained on the same five molecular and two Heisenberg Hamiltonians, then evaluated on seven held-out Hamiltonians with identical features, depths, seeds, and online VQE. Representative depths 1 and 4 are shown. A zero-gap ResNet-18 cell with zero shot gain denotes Full-QWC fallback rather than successful sparsification.}
\vspace{0.15em}
\setlength{\tabcolsep}{0.55mm}
\renewcommand{\arraystretch}{0.49}
\tiny
\resizebox{1.0\linewidth}{!}{
\begin{tabular}{lll|cccc}
\toprule
Testing Hamiltonian & Depth & Metric & \ours{} & MLP & ResNet-18 & GNN \\
\midrule
\gcell{TFIM $h=0.5$} & Depth 1 & $\dE$ vs Full $\downarrow$ & \textbf{\gbf{-0.0000}} & \textbf{\gbf{-0.0000}} & \textbf{\gbf{+0.0000}} & \textbf{\gbf{+0.0000}} \\
 & Depth 1 & Shot gain $\uparrow$ & \textcolor{green!55!black}{2.0\%} & \textbf{\underline{\textcolor{green!55!black}{5.7\%}}} & \textcolor{green!55!black}{0.0\%} & \textcolor{green!55!black}{2.1\%} \\
 & Depth 1 & Chem. acc. & \cmark & \cmark & \cmark & \cmark \\
\addlinespace[0.08em]
 & Depth 4 & $\dE$ vs Full $\downarrow$ & \gbf{-0.0002} & \textbf{\gbf{+0.0000}} & \textbf{\gbf{+0.0000}} & \gbf{+0.0000} \\
 & Depth 4 & Shot gain $\uparrow$ & \textcolor{green!55!black}{2.6\%} & \textbf{\underline{\textcolor{green!55!black}{4.5\%}}} & \textcolor{green!55!black}{0.0\%} & \textcolor{green!55!black}{1.3\%} \\
 & Depth 4 & Chem. acc. & \cmark & \cmark & \cmark & \cmark \\
\midrule
\gcell{TFIM $h=0.7$} & Depth 1 & $\dE$ vs Full $\downarrow$ & \textbf{\gbf{-0.0000}} & \textbf{\gbf{-0.0000}} & \textbf{\gbf{+0.0000}} & \textbf{\gbf{+0.0000}} \\
 & Depth 1 & Shot gain $\uparrow$ & \textcolor{green!55!black}{2.9\%} & \textbf{\underline{\textcolor{green!55!black}{4.9\%}}} & \textcolor{green!55!black}{0.0\%} & \textcolor{green!55!black}{1.7\%} \\
 & Depth 1 & Chem. acc. & \cmark & \cmark & \cmark & \cmark \\
\addlinespace[0.08em]
 & Depth 4 & $\dE$ vs Full $\downarrow$ & \textbf{\gbf{+0.0000}} & \textbf{\gbf{-0.0000}} & \textbf{\gbf{+0.0000}} & \gbf{-0.0014} \\
 & Depth 4 & Shot gain $\uparrow$ & \textcolor{green!55!black}{2.4\%} & \textbf{\underline{\textcolor{green!55!black}{4.3\%}}} & \textcolor{green!55!black}{0.0\%} & \textcolor{green!55!black}{1.2\%} \\
 & Depth 4 & Chem. acc. & \cmark & \cmark & \cmark & \cmark \\
\midrule
\gcell{TFIM $h=0.9$} & Depth 1 & $\dE$ vs Full $\downarrow$ & \rbf{+0.0771} & \rbf{+0.0771} & \textbf{\gbf{+0.0000}} & \rbf{+0.0496} \\
 & Depth 1 & Shot gain $\uparrow$ & \textbf{\underline{\textcolor{red!70!black}{25.0\%}}} & \textbf{\underline{\textcolor{red!70!black}{25.0\%}}} & \textcolor{green!55!black}{0.0\%} & \textcolor{red!70!black}{14.1\%} \\
 & Depth 1 & Chem. acc. & \xmark & \xmark & \cmark & \xmark \\
\addlinespace[0.08em]
 & Depth 4 & $\dE$ vs Full $\downarrow$ & \rbf{+0.0302} & \rbf{+0.0302} & \textbf{\gbf{+0.0000}} & \rbf{+0.0255} \\
 & Depth 4 & Shot gain $\uparrow$ & \textbf{\underline{\textcolor{red!70!black}{6.2\%}}} & \textbf{\underline{\textcolor{red!70!black}{6.2\%}}} & \textcolor{green!55!black}{0.0\%} & \textcolor{red!70!black}{3.8\%} \\
 & Depth 4 & Chem. acc. & \xmark & \xmark & \cmark & \xmark \\
\midrule
\gcell{Heis. XXZ+X field} & Depth 1 & $\dE$ vs Full $\downarrow$ & \textbf{\gbf{+0.0000}} & \textbf{\gbf{+0.0000}} & \textbf{\gbf{+0.0000}} & \textbf{\gbf{+0.0000}} \\
 & Depth 1 & Shot gain $\uparrow$ & \textbf{\underline{\textcolor{green!55!black}{16.7\%}}} & \textbf{\underline{\textcolor{green!55!black}{16.7\%}}} & \textcolor{green!55!black}{0.0\%} & \textcolor{green!55!black}{16.4\%} \\
 & Depth 1 & Chem. acc. & \cmark & \cmark & \cmark & \cmark \\
\addlinespace[0.08em]
 & Depth 4 & $\dE$ vs Full $\downarrow$ & \textbf{\gbf{+0.0000}} & \textbf{\gbf{+0.0000}} & \textbf{\gbf{+0.0000}} & \textbf{\gbf{+0.0000}} \\
 & Depth 4 & Shot gain $\uparrow$ & \textbf{\underline{\textcolor{green!55!black}{4.2\%}}} & \textbf{\underline{\textcolor{green!55!black}{4.2\%}}} & \textcolor{green!55!black}{0.0\%} & \textbf{\underline{\textcolor{green!55!black}{4.2\%}}} \\
 & Depth 4 & Chem. acc. & \cmark & \cmark & \cmark & \cmark \\
\midrule
\gcell{Heis. J1--J2} & Depth 1 & $\dE$ vs Full $\downarrow$ & \textbf{\gbf{-0.0000}} & \gbf{+0.0000} & \textbf{\gbf{+0.0000}} & \gbf{+0.0001} \\
 & Depth 1 & Shot gain $\uparrow$ & \textcolor{green!55!black}{16.7\%} & \textbf{\underline{\textcolor{green!55!black}{16.8\%}}} & \textcolor{green!55!black}{0.0\%} & \textcolor{green!55!black}{15.9\%} \\
 & Depth 1 & Chem. acc. & \cmark & \cmark & \cmark & \cmark \\
\addlinespace[0.08em]
 & Depth 4 & $\dE$ vs Full $\downarrow$ & \textbf{\gbf{+0.0000}} & \textbf{\gbf{+0.0000}} & \textbf{\gbf{+0.0000}} & \gbf{+0.0001} \\
 & Depth 4 & Shot gain $\uparrow$ & \textcolor{green!55!black}{4.2\%} & \textcolor{green!55!black}{4.2\%} & \textcolor{green!55!black}{0.0\%} & \textbf{\underline{\textcolor{green!55!black}{4.2\%}}} \\
 & Depth 4 & Chem. acc. & \cmark & \cmark & \cmark & \cmark \\
\midrule
\gcell{Heis. XXY+Z field} & Depth 1 & $\dE$ vs Full $\downarrow$ & \textbf{\gbf{+0.0000}} & \gbf{+0.0001} & \textbf{\gbf{+0.0000}} & \textbf{\gbf{+0.0000}} \\
 & Depth 1 & Shot gain $\uparrow$ & \textcolor{green!55!black}{16.7\%} & \textbf{\underline{\textcolor{green!55!black}{16.7\%}}} & \textcolor{green!55!black}{0.0\%} & \textcolor{green!55!black}{15.8\%} \\
 & Depth 1 & Chem. acc. & \cmark & \cmark & \cmark & \cmark \\
\addlinespace[0.08em]
 & Depth 4 & $\dE$ vs Full $\downarrow$ & \textbf{\gbf{+0.0000}} & \textbf{\gbf{+0.0000}} & \textbf{\gbf{+0.0000}} & \textbf{\gbf{+0.0000}} \\
 & Depth 4 & Shot gain $\uparrow$ & \textbf{\underline{\textcolor{green!55!black}{4.2\%}}} & \textbf{\underline{\textcolor{green!55!black}{4.2\%}}} & \textcolor{green!55!black}{0.0\%} & \textbf{\underline{\textcolor{green!55!black}{4.2\%}}} \\
 & Depth 4 & Chem. acc. & \cmark & \cmark & \cmark & \cmark \\
\midrule
\gcell{JW-like parity spin} & Depth 1 & $\dE$ vs Full $\downarrow$ & \rbf{+0.0312} & \textbf{\rbf{+0.0290}} & \rbf{+0.0452} & \rbf{+0.0337} \\
 & Depth 1 & Shot gain $\uparrow$ & \textcolor{red!70!black}{35.7\%} & \textcolor{red!70!black}{23.0\%} & \textbf{\underline{\textcolor{red!70!black}{65.8\%}}} & \textcolor{red!70!black}{31.9\%} \\
 & Depth 1 & Chem. acc. & \xmark & \xmark & \xmark & \xmark \\
\addlinespace[0.08em]
 & Depth 4 & $\dE$ vs Full $\downarrow$ & \gbf{-0.0046} & \textbf{\gbf{-0.0014}} & \rbf{+0.1674} & \rbf{+0.0049} \\
 & Depth 4 & Shot gain $\uparrow$ & \textcolor{green!55!black}{11.6\%} & \textcolor{green!55!black}{8.8\%} & \textbf{\underline{\textcolor{red!70!black}{28.2\%}}} & \textcolor{red!70!black}{15.1\%} \\
 & Depth 4 & Chem. acc. & \cmark & \cmark & \xmark & \xmark \\
\bottomrule
\end{tabular}}
\label{tab:v64_hamiltonian_model_zoo}
\vspace{-0.8em}
\end{table}

\clearpage
\section{Frozen-LLM and Frontier-Model Details}
\label{sec:app_llm_prompt_protocol}

\paragraph{Controlled intervention.}
The local-LLM experiment does not replace our full controller and does not fine-tune an LLM.  Instead, it isolates one decision: ranking QWC groups for each parameter.  At a refresh iteration $t$, our saved controller first predicts a binary mask and therefore a per-parameter cardinality
\begin{equation}
k_{j,t}=\sum_{g=1}^{G}M^{\mathrm{Ours}}_{j,g}(t).
\end{equation}
The LLM receives $k_{j,t}$ and the same structured parameter, group, and parameter--group features, but it never receives the oracle contribution matrix.  It returns a ranking $r^{\mathrm{LLM}}_{j,t}$, from which we form
\begin{equation}
M^{\mathrm{LLM}}_{j,g}(t)
=\mathbb{I}\!\left[g\in\operatorname{Top}_{k_{j,t}}
\bigl(r^{\mathrm{LLM}}_{j,t}\bigr)\right].
\end{equation}
Thus the comparison holds the count controller fixed at each query and tests whether a frozen language model can replace the learned group-ranking function.

\paragraph{Exact system prompt for the repaired rank-only protocol.}
All four models receive the following system message without demonstrations:
\begin{quote}
\small\ttfamily\raggedright
You are a measurement-selection controller for variational quantum eigensolvers.\\
For each requested circuit parameter, choose exactly k distinct QWC group IDs and order them from\\
most to least relevant to that parameter's energy gradient. Use coefficient, circuit-location,\\
theta, iteration, and parameter--group light-cone features jointly. A globally large coefficient\\
does not imply equal relevance to every parameter. Every returned ID must be an integer present in\\
the candidate groups. Return only the JSON object required by the supplied schema.
\end{quote}

\paragraph{Compact request and response.}
The request contains only feature columns that are nonzero in the current instance.  The ellipses below suppress repeated rows, not additional natural-language instructions.
\begin{verbatim}
{
  "task": "rank_parameter_specific_qwc_groups",
  "context": {
    "ansatz": "ryrz", "depth": 2,
    "parameters": 48, "qwc_groups": 19,
    "iteration": 7, "total_iterations": 28
  },
  "columns": {
    "parameter_x": [...],
    "group_rows": ["group_id", ...],
    "pair_nonzero_rows": ["group_id", ...]
  },
  "groups": [[0, ...], [1, ...], ...],
  "parameters_to_rank": [
    {"id": 3, "k": 5, "x": [...],
     "pair_nonzero": [[0, ...], ...]}
  ]
}
\end{verbatim}
The response is constrained by Ollama's JSON schema and contains one row for every requested parameter:
\begin{verbatim}
{"selections": [
  {"parameter_id": 3,
   "ranked_group_ids": [7, 2, 5, 1, 8]}
]}
\end{verbatim}

\paragraph{Validation and repair in the rank-only protocol.}
Later clean/explanation output contracts are different; Appendix~\ref{sec:app_trace_provenance} documents the clean and mechanism-audit protocols and their prompt-specific trace records.
Boolean values, repeated IDs, noninteger values, and group IDs outside $[0,G-1]$ are discarded.  If fewer than $k_{j,t}$ valid distinct IDs remain, the missing positions are filled deterministically using descending global group-coefficient magnitude.  We log the requested and repaired group counts, invalid batches, context splits, prompt/output tokens, and inference latency.  This repair prevents malformed JSON from changing the requested cardinality while preserving a measurable repair-rate diagnostic.

\paragraph{Sparse-refresh execution.}
The LLM is queried only at $t\in\{0,7,14,21\}$ during each 28-step online trajectory.  The resulting mask is reused for the following seven optimizer steps.  Because the count targets predicted by \ourmethod{} can evolve between refreshes while the LLM mask is held fixed, an LLM's realized cumulative shot gain need not equal the continuously updated reference gain from \ourmethod{}.  We therefore treat the reference-gain rows in Table~\ref{tab:v76_ollama_table2} as count-controller context rather than as measured LLM cost.

\begin{table}[H]
\centering
\caption{Frozen local-LLM inference settings used in the sparse-refresh comparison.}
\label{tab:v76b_prompt_settings}
\setlength{\tabcolsep}{3.5pt}
\renewcommand{\arraystretch}{1.06}
\small
\begin{tabular}{p{0.30\linewidth}p{0.62\linewidth}}
\toprule
Component & Setting \\
\midrule
Models & Qwen3-8B, Llama-3.1-8B, DeepSeek-R1-8B, GPT-OSS-20B \\
Adaptation & Zero-shot; no demonstrations; no fine-tuning \\
Decoding & Temperature $0$, seed $123$, \texttt{think=false} \\
Context/output & \texttt{num\_ctx}$=32768$, \texttt{num\_predict}$=1536$ \\
Parameter batching & Up to 16 parameters; adaptive halving on context overflow \\
Refresh schedule & Iterations $\{0,7,14,21\}$ of 28 \\
VQE setting & 1024 shots per shift, Adam, layered RY/RZ ansatz \\
Evaluation panel & Five STO-3G molecules, depths 1--3, seeds 97/101/107/109 \\
Count controller & Saved \ours{} model at $\tau=0.995$ \\
\bottomrule
\end{tabular}
\end{table}

\label{sec:app_llm_details}
\appVSeventySixPromptProtocol

\clearpage
\subsection{Clean Frozen-LLM Audit}
\begin{table*}[t]
\centering
\vspace{-0.5em}
\caption{\textbf{Clean frozen local-LLM QWC-selection audit.} This table duplicates the molecule--depth protocol and layout of Table~\ref{tab:v76_ollama_table2}, but replaces its repaired rank-only outputs with the stricter exact select/exclude contract. Each frozen LLM must emit unique QWC IDs at the requested per-parameter cardinality; invalid responses receive schema feedback for at most two retries, with no coefficient-based repair. Persistent failures use a logged Full-QWC safety fallback and pay the corresponding shots. Random uses the same per-parameter cardinality predicted by our method and $K{=}7$ held-mask cadence, but samples group identities uniformly. All four models complete 60 online trajectories over seeds 97/101/107/109; their aggregate realized shot savings lie between 52.0\% and 52.4\%, versus 52.7\% for Random. Energy entries are mean $\pm$ standard deviation in Ha relative to Full-QWC, using exact-expectation gradients and nominal shot accounting (Appendix~\ref{sec:app_estimator_scope}). Random cells are intentionally unshaded. For every other method, green/red satisfies/violates the one-sided 1.6 mHa no-worse criterion ($\dE\leq1.6$ mHa; lower energy is accepted); deeper color indicates a lower energy/larger violation. The black check/cross separately indicates whether the absolute mean $\dE$ is smaller/not smaller than Random. Bold/underline mark the smallest/second-smallest absolute mean $\dE$ per cell. Figure~\ref{fig:v79_clean_llm_protocol} summarizes the audited protocol.}
\vspace{0.25em}
\setlength{\tabcolsep}{0.4mm}
\renewcommand{\arraystretch}{0.90}
\scriptsize
\begingroup
\definecolor{vistaheatgreen}{HTML}{A8D8A8}
\definecolor{vistaheatred}{HTML}{F3A6B7}
\newcommand{\llmlogo}[1]{\raisebox{-0.28ex}{\includegraphics[height=1.15em]{llm_logos/#1.png}}}
\newcommand{\llmmethod}[2]{\llmlogo{#1}\,#2}
\newcommand{\vpass}[2]{\cellcolor{vistaheatgreen!#1!white}\textcolor{black}{#2}}
\newcommand{\vfail}[2]{\cellcolor{vistaheatred!#1!white}\textcolor{black}{#2}}
\newcommand{\epos}[1]{+#1}
\newcommand{\eneg}[1]{\textminus#1}
\newcommand{\estat}[2]{#1\,\(\pm\)\,#2}
\newcommand{\pstat}[2]{#1\,\(\pm\)\,#2\%}
\newcommand{\best}[1]{\textbf{#1}}
\newcommand{\runnerup}[1]{\underline{#1}}
\newcommand{\randwin}{\textcolor{black}{\ding{51}}}
\newcommand{\randloss}{\textcolor{black}{\ding{55}}}
\newcommand{\grayrow}{\specialrule{0.20pt}{0.35mm}{0.35mm}}
\arrayrulecolor{black!18}
\begin{tabular}{llccccc}
\arrayrulecolor{black}\toprule
Setting & Selection method & H2 (4q) & HeH+ (4q) & H3+ (6q) & H4 chain (8q) & LiH (12q) \\
\midrule\arrayrulecolor{black!18}
\bcell{Depth 1} & \ours{} gain, $K{=}1$ $\uparrow$ & \pstat{40.4}{6.3} & \pstat{58.9}{3.5} & \pstat{63.0}{2.8} & \pstat{74.7}{1.1} & \pstat{86.9}{0.5} \\
\grayrow
\bcell{Depth 1} & \ours{} gain, $K{=}7$ $\uparrow$ & \pstat{41.2}{5.3} & \pstat{60.7}{3.1} & \pstat{65.1}{2.6} & \pstat{77.3}{0.7} & \pstat{88.2}{0.4} \\
\arrayrulecolor{black}\cmidrule(lr){1-7}\arrayrulecolor{black!18}
\gcell{Depth 1} & Random ($K{=}7$) & \estat{\epos{0.1847}}{0.0868} & \estat{\epos{1.2191}}{0.2759} & \estat{\epos{0.7247}}{0.2856} & \estat{\epos{0.9736}}{0.3101} & \estat{\epos{2.9030}}{0.6058} \\
\grayrow
\gcell{Depth 1} & \llmmethod{qwen}{Qwen3-8B} & \vfail{33}{\estat{\epos{0.2012}}{0.1085}\,\randloss} & \vfail{50}{\estat{\epos{0.9101}}{0.4857}\,\randwin} & \vfail{23}{\estat{\epos{0.0200}}{0.0203}\,\randwin} & \vfail{50}{\estat{\epos{0.9166}}{0.5180}\,\randwin} & \vfail{61}{\estat{\epos{1.6190}}{0.9510}\,\randwin} \\
\grayrow
\gcell{Depth 1} & \llmmethod{meta}{Llama-3.1-8B} & \vfail{37}{\estat{\epos{0.3306}}{0.0615}\,\randloss} & \vfail{48}{\estat{\epos{0.8140}}{0.2673}\,\randwin} & \vfail{43}{\estat{\epos{0.5751}}{0.2637}\,\randwin} & \vfail{41}{\estat{\epos{0.4804}}{0.3100}\,\randwin} & \vfail{44}{\estat{\epos{0.6219}}{0.5640}\,\randwin} \\
\grayrow
\gcell{Depth 1} & \llmmethod{deepseek}{DeepSeek-R1-8B} & \vfail{34}{\estat{\epos{0.2245}}{0.0696}\,\randloss} & \vfail{52}{\estat{\epos{1.0313}}{0.3889}\,\randwin} & \vfail{42}{\estat{\epos{0.5259}}{0.0514}\,\randwin} & \vfail{45}{\estat{\epos{0.6296}}{0.1083}\,\randwin} & \vfail{56}{\estat{\epos{1.2793}}{0.7197}\,\randwin} \\
\grayrow
\gcell{Depth 1} & \llmmethod{openai}{GPT-OSS-20B} & \vfail{39}{\estat{\epos{0.3930}}{0.1936}\,\randloss} & \vfail{48}{\estat{\epos{0.8238}}{0.2303}\,\randwin} & \vfail{47}{\estat{\epos{0.7300}}{0.4383}\,\randloss} & \vfail{43}{\estat{\epos{0.5671}}{0.3164}\,\randwin} & \vfail{60}{\estat{\epos{1.5503}}{0.7644}\,\randwin} \\
\grayrow
\gcell{Depth 1} & \ours{} ($K{=}7$) & \vpass{18}{\runnerup{\estat{\epos{0.0013}}{0.0003}}\,\randwin} & \vfail{19}{\runnerup{\estat{\epos{0.0035}}{0.0015}}\,\randwin} & \vpass{18}{\best{\estat{\epos{0.0004}}{0.0006}}\,\randwin} & \vpass{18}{\best{\estat{\epos{0.0001}}{0.0060}}\,\randwin} & \vfail{19}{\runnerup{\estat{\epos{0.0030}}{0.0016}}\,\randwin} \\
\grayrow
\gcell{Depth 1} & \ours{} ($K{=}1$) & \vpass{18}{\best{\estat{\epos{0.0011}}{0.0008}}\,\randwin} & \vpass{18}{\best{\estat{\epos{0.0015}}{0.0012}}\,\randwin} & \vpass{18}{\runnerup{\estat{\epos{0.0005}}{0.0006}}\,\randwin} & \vpass{23}{\runnerup{\estat{\eneg{0.0006}}{0.0024}}\,\randwin} & \vfail{19}{\best{\estat{\epos{0.0023}}{0.0016}}\,\randwin} \\
\arrayrulecolor{black}\midrule\arrayrulecolor{black!18}
\bcell{Depth 2} & \ours{} gain, $K{=}1$ $\uparrow$ & \pstat{25.6}{2.9} & \pstat{31.6}{2.6} & \pstat{44.9}{1.3} & \pstat{51.8}{0.5} & \pstat{82.6}{0.4} \\
\grayrow
\bcell{Depth 2} & \ours{} gain, $K{=}7$ $\uparrow$ & \pstat{27.3}{2.5} & \pstat{32.5}{2.4} & \pstat{47.9}{1.4} & \pstat{55.1}{0.5} & \pstat{83.7}{0.4} \\
\arrayrulecolor{black}\cmidrule(lr){1-7}\arrayrulecolor{black!18}
\gcell{Depth 2} & Random ($K{=}7$) & \estat{\epos{0.0930}}{0.0297} & \estat{\epos{0.5668}}{0.4463} & \estat{\epos{0.3054}}{0.1633} & \estat{\epos{0.9327}}{0.3273} & \estat{\epos{2.6969}}{0.2816} \\
\grayrow
\gcell{Depth 2} & \llmmethod{qwen}{Qwen3-8B} & \vfail{28}{\estat{\epos{0.0861}}{0.0637}\,\randwin} & \vfail{43}{\estat{\epos{0.5614}}{0.3196}\,\randwin} & \vfail{23}{\estat{\epos{0.0206}}{0.0054}\,\randwin} & \vfail{39}{\estat{\epos{0.3825}}{0.1254}\,\randwin} & \vfail{74}{\estat{\epos{2.7674}}{0.5685}\,\randloss} \\
\grayrow
\gcell{Depth 2} & \llmmethod{meta}{Llama-3.1-8B} & \vfail{22}{\estat{\epos{0.0184}}{0.2098}\,\randwin} & \vfail{50}{\estat{\epos{0.8950}}{0.1642}\,\randloss} & \vfail{43}{\estat{\epos{0.5715}}{0.0934}\,\randloss} & \vfail{36}{\estat{\epos{0.2758}}{0.1151}\,\randwin} & \vfail{46}{\estat{\epos{0.7222}}{0.3456}\,\randwin} \\
\grayrow
\gcell{Depth 2} & \llmmethod{deepseek}{DeepSeek-R1-8B} & \vfail{30}{\estat{\epos{0.1366}}{0.1334}\,\randloss} & \vfail{37}{\estat{\epos{0.3298}}{0.3566}\,\randwin} & \vfail{43}{\estat{\epos{0.5573}}{0.1298}\,\randloss} & \vfail{48}{\estat{\epos{0.7876}}{0.2497}\,\randwin} & \vfail{70}{\estat{\epos{2.4135}}{0.3556}\,\randwin} \\
\grayrow
\gcell{Depth 2} & \llmmethod{openai}{GPT-OSS-20B} & \vfail{33}{\estat{\epos{0.2148}}{0.1220}\,\randloss} & \vfail{46}{\estat{\epos{0.6807}}{0.7081}\,\randloss} & \vfail{41}{\estat{\epos{0.4691}}{0.2024}\,\randloss} & \vfail{43}{\estat{\epos{0.5616}}{0.2675}\,\randwin} & \vfail{56}{\estat{\epos{1.2894}}{0.3290}\,\randwin} \\
\grayrow
\gcell{Depth 2} & \ours{} ($K{=}7$) & \vpass{23}{\runnerup{\estat{\eneg{0.0007}}{0.0015}}\,\randwin} & \vfail{21}{\runnerup{\estat{\epos{0.0076}}{0.0176}}\,\randwin} & \vpass{18}{\best{\estat{\epos{0.0006}}{0.0031}}\,\randwin} & \vfail{22}{\runnerup{\estat{\epos{0.0162}}{0.0438}}\,\randwin} & \vpass{31}{\best{\estat{\eneg{0.0040}}{0.0118}}\,\randwin} \\
\grayrow
\gcell{Depth 2} & \ours{} ($K{=}1$) & \vpass{18}{\best{\estat{\epos{0.0003}}{0.0007}}\,\randwin} & \vpass{25}{\best{\estat{\eneg{0.0011}}{0.0026}}\,\randwin} & \vpass{26}{\runnerup{\estat{\eneg{0.0015}}{0.0041}}\,\randwin} & \vfail{22}{\best{\estat{\epos{0.0158}}{0.0443}}\,\randwin} & \vpass{32}{\runnerup{\estat{\eneg{0.0049}}{0.0097}}\,\randwin} \\
\arrayrulecolor{black}\midrule\arrayrulecolor{black!18}
\bcell{Depth 3} & \ours{} gain, $K{=}1$ $\uparrow$ & \pstat{18.1}{2.7} & \pstat{24.9}{1.3} & \pstat{29.9}{1.0} & \pstat{35.2}{0.9} & \pstat{75.4}{1.8} \\
\grayrow
\bcell{Depth 3} & \ours{} gain, $K{=}7$ $\uparrow$ & \pstat{19.0}{2.4} & \pstat{26.1}{1.2} & \pstat{32.1}{1.2} & \pstat{38.6}{0.7} & \pstat{76.6}{1.3} \\
\arrayrulecolor{black}\cmidrule(lr){1-7}\arrayrulecolor{black!18}
\gcell{Depth 3} & Random ($K{=}7$) & \estat{\epos{0.0730}}{0.0859} & \estat{\epos{0.2617}}{0.2130} & \estat{\epos{0.2750}}{0.0765} & \estat{\epos{0.2773}}{0.2665} & \estat{\epos{2.3928}}{0.4840} \\
\grayrow
\gcell{Depth 3} & \llmmethod{qwen}{Qwen3-8B} & \vpass{60}{\estat{\eneg{0.0435}}{0.1539}\,\randwin} & \vfail{30}{\estat{\epos{0.1295}}{0.1997}\,\randwin} & \vfail{27}{\estat{\epos{0.0721}}{0.0496}\,\randwin} & \vpass{75}{\estat{\eneg{0.0803}}{0.1909}\,\randwin} & \vfail{66}{\estat{\epos{2.0654}}{0.2623}\,\randwin} \\
\grayrow
\gcell{Depth 3} & \llmmethod{meta}{Llama-3.1-8B} & \vfail{28}{\estat{\epos{0.0883}}{0.2356}\,\randloss} & \vfail{36}{\estat{\epos{0.2766}}{0.1085}\,\randloss} & \vfail{33}{\estat{\epos{0.2083}}{0.1798}\,\randwin} & \vfail{33}{\estat{\epos{0.2084}}{0.2366}\,\randwin} & \vfail{35}{\estat{\epos{0.2603}}{0.2330}\,\randwin} \\
\grayrow
\gcell{Depth 3} & \llmmethod{deepseek}{DeepSeek-R1-8B} & \vfail{23}{\estat{\epos{0.0220}}{0.1312}\,\randwin} & \vfail{33}{\estat{\epos{0.2061}}{0.0660}\,\randwin} & \vfail{32}{\estat{\epos{0.1788}}{0.1528}\,\randwin} & \vfail{37}{\estat{\epos{0.3246}}{0.1489}\,\randloss} & \vfail{60}{\estat{\epos{1.6082}}{0.4131}\,\randwin} \\
\grayrow
\gcell{Depth 3} & \llmmethod{openai}{GPT-OSS-20B} & \vfail{31}{\estat{\epos{0.1633}}{0.1815}\,\randloss} & \vfail{38}{\estat{\epos{0.3438}}{0.1193}\,\randloss} & \vfail{32}{\estat{\epos{0.1786}}{0.0759}\,\randwin} & \vfail{42}{\estat{\epos{0.5089}}{0.3596}\,\randloss} & \vfail{43}{\estat{\epos{0.5647}}{0.4252}\,\randwin} \\
\grayrow
\gcell{Depth 3} & \ours{} ($K{=}7$) & \vpass{18}{\runnerup{\estat{\epos{0.0010}}{0.0027}}\,\randwin} & \vfail{21}{\runnerup{\estat{\epos{0.0108}}{0.0118}}\,\randwin} & \vfail{19}{\runnerup{\estat{\epos{0.0029}}{0.0095}}\,\randwin} & \vfail{19}{\runnerup{\estat{\epos{0.0029}}{0.0113}}\,\randwin} & \vpass{32}{\runnerup{\estat{\eneg{0.0046}}{0.0040}}\,\randwin} \\
\grayrow
\gcell{Depth 3} & \ours{} ($K{=}1$) & \vpass{18}{\best{\estat{\epos{0.0005}}{0.0015}}\,\randwin} & \vfail{21}{\best{\estat{\epos{0.0092}}{0.0131}}\,\randwin} & \vpass{18}{\best{\estat{\epos{0.0005}}{0.0077}}\,\randwin} & \vpass{26}{\best{\estat{\eneg{0.0014}}{0.0097}}\,\randwin} & \vpass{26}{\best{\estat{\eneg{0.0017}}{0.0036}}\,\randwin} \\
\arrayrulecolor{black}\bottomrule
\end{tabular}
\endgroup
\label{tab:v79_clean_llm_audit}
\vspace{-0.5em}
\end{table*}

\clearpage
\subsection{Frontier Models at Deeper Circuits}
\begingroup
\def\OursFrontierAppendix{1}

\endgroup
\clearpage
\begin{table*}[t]
\centering
\caption{\textbf{Online controller token traffic and API-equivalent cost.}
The accounting covers the same 60 held-out VQE cells used in
Tables~\ref{tab:v87_frontier_depth1}--\ref{tab:v87_frontier_depth3}:
28 optimizer steps per cell and four selector refreshes (240 total).
Token categories come from run logs. OpenAI dollar entries are historical
API-equivalent estimates retained from the original version; Claude entries
are CLI-reported API totals. Dollar values describe historical API-equivalent
accounting; token traffic is directly recorded in the experiment logs
(Appendix~\ref{sec:app_estimator_scope}).
Our method uses local checkpoint inference and no LLM tokens or API credits.
Lower is better. Bold marks minima; underline marks the lowest nonzero
entry in the total-token and cost columns.}
\vspace{0.25em}
\setlength{\tabcolsep}{1.45mm}
\renewcommand{\arraystretch}{1.10}
\fontsize{8.1}{9.7}\selectfont
\begingroup
\definecolor{resourcegreen}{HTML}{DDEED8}
\definecolor{resourcered}{HTML}{F9E1E5}
\newcommand{\resourcelogo}[1]{\raisebox{-0.28ex}{\includegraphics[height=1.10em]{llm_logos/#1.png}}}
\newcommand{\resourcemethod}[2]{\resourcelogo{#1}\,#2}
\newcommand{\llmcost}[1]{\cellcolor{resourcered}#1}
\newcommand{\ourmethodcost}[1]{\cellcolor{resourcegreen}#1}
\newcommand{\best}[1]{\textbf{#1}}
\newcommand{\runnerup}[1]{\underline{#1}}
\arrayrulecolor{black!18}
\begin{tabular*}{\textwidth}{@{\extracolsep{\fill}}lcccccc@{}}
\arrayrulecolor{black}\toprule
Method & \multicolumn{4}{c}{Token traffic (million)}
& \multicolumn{2}{c}{API-equivalent cost (USD)} \\
\cmidrule(lr){2-5}\cmidrule(lr){6-7}
& Non-cached input & Cached input & Output & Total
& 60-cell total & Per cell \\
\midrule\arrayrulecolor{black!18}
\resourcemethod{openai}{GPT-5.4}
& 12.71 & 3.64 & 0.78 & 17.13 & \llmcost{44.43} & \llmcost{0.74} \\
\resourcemethod{openai}{GPT-5.5}
& 12.58 & 4.42 & 0.45 & 17.45 & \llmcost{78.67} & \llmcost{1.31} \\
\resourcemethod{openai}{GPT-5.6 Sol}
& 16.50 & 1.46 & 0.57 & 18.52 & \llmcost{100.35} & \llmcost{1.67} \\
\resourcemethod{openai}{GPT-5.6 Terra}
& 13.46 & 3.62 & 0.61 & 17.70 & \llmcost{43.76} & \llmcost{0.73} \\
\resourcemethod{openai}{GPT-5.6 Luna}
& 14.74 & 1.74 & 0.52 & \runnerup{17.00}
& \llmcost{\runnerup{18.01}} & \llmcost{\runnerup{0.30}} \\
\arrayrulecolor{black!18}\cmidrule(lr){1-7}
\resourcemethod{claude}{Claude Haiku 4.5}
& 24.58 & 1.08 & 6.75 & 32.40 & \llmcost{71.29} & \llmcost{1.19} \\
\resourcemethod{claude}{Claude Sonnet 5}
& 25.10 & 4.81 & 1.43 & 31.35 & \llmcost{183.34} & \llmcost{3.06} \\
\resourcemethod{claude}{Claude Opus 5}
& 24.90 & 2.75 & 0.55 & 28.20 & \llmcost{159.01} & \llmcost{2.65} \\
\resourcemethod{claude}{Claude Fable 5}
& 24.24 & 1.33 & 1.15 & 26.72 & \llmcost{327.15} & \llmcost{5.45} \\
\arrayrulecolor{black!18}\cmidrule(lr){1-7}
\ours{}
& \best{0.00} & \best{0.00} & \best{0.00} & \best{0.00}
& \ourmethodcost{\best{0.00}} & \ourmethodcost{\best{0.00}} \\
\arrayrulecolor{black}\bottomrule
\end{tabular*}
\endgroup
\label{tab:v87_frontier_token_cost}
\vspace{0.25em}

\begin{minipage}{0.96\textwidth}
\footnotesize
\textit{Accounting note.} OpenAI input totals include cached tokens, so the
non-cached column subtracts the logged cached subset before pricing; Claude
cache-read tokens are logged separately.
The OpenAI reconstruction uses
\href{https://openai.com/api/pricing/}{OpenAI API pricing}, while Claude uses
each CLI response's provider-reported \texttt{total\_cost\_usd}
(see \href{https://docs.anthropic.com/en/docs/about-claude/pricing}{Anthropic pricing}).
These are normalized API-equivalent costs, not necessarily the subscription invoice.
The one-time training cost of \ourmethod{} is excluded from this
online-inference accounting; its 984,454-parameter checkpoint is 3.78 MiB and
averages 7.29 ms per refresh on the local CPU.
\end{minipage}

\vspace{0.75em}
\definecolor{frontierblue}{HTML}{4F7EA8}
\definecolor{frontierbluebg}{HTML}{EDF5FB}
\definecolor{agentgold}{HTML}{C8952D}
\definecolor{agentgoldbg}{HTML}{FFF7DE}
\begingroup
\setlength{\fboxsep}{5.5pt}
\noindent\fcolorbox{frontierblue}{frontierbluebg}{%
\parbox{0.94\textwidth}{%
\raggedright\fontsize{8.0}{9.6}\selectfont
\textcolor{frontierblue}{\textbf{Current takeaway}}
\begin{itemize}
\setlength{\itemsep}{0.20em}\setlength{\parskip}{0pt}\setlength{\topsep}{0.25em}
\item Stronger frontier LLMs better integrate long structured contexts and
exploit retrieved task supervision.
\item No model consistently wins across molecules, depths, and
fidelity--measurement trade-offs: current in-context policies are capable
but not yet stable.
\item The progression from clean prompts to physics context and data exposure
shows substantial potential to identify task-relevant physical structure.
\end{itemize}
}}
\par\vspace{0.35em}
\noindent\fcolorbox{agentgold}{agentgoldbg}{%
\parbox{0.94\textwidth}{%
\raggedright\fontsize{8.0}{9.6}\selectfont
\textcolor{agentgold}{\textbf{Future implication: Physics-aware agent}}
\begin{itemize}
\setlength{\itemsep}{0.20em}\setlength{\parskip}{0pt}\setlength{\topsep}{0.25em}
\item Scale alone is insufficient; a physics-aware agent should combine
retrieval, executable physics tools, memory, and verification.
\item A promising hybrid uses \ourmethod{} as an efficient prior and invokes LLM
reasoning only for OOD or low-confidence states.
\item Crossed clean/data-exposed evaluations remain necessary before
attributing the gain to frontier-model scale alone.
\end{itemize}
}}
\endgroup
\end{table*}

\clearpage
\subsection{Observable Decision Traces}
The main text retains the representative Llama trace. The complementary model
traces below use the same fixed-state audit and expose only observable outputs,
schema checks, selected-group evidence, and counterfactual feature tests.
\appTraceQwen
\clearpage
\appTraceGPTOss

\clearpage
\subsection{Prompt and Context Cost}
\begin{table*}[t]
\centering
\caption{\textbf{Prompt, output, and context ablations with token cost.}
Values are macro-averaged across Qwen3-8B, Llama-3.1-8B,
DeepSeek-R1-8B, and GPT-OSS-20B on the 12 fixed audit snapshots per model.
Token and call costs include retries, context splits, and invalid outputs,
and are divided by all 48 attempted model--snapshot cases. Performance
metrics are averaged over protocol-successful cases, whose identities can
change across conditions; this is not an all-case matched accuracy comparison. Each ablation is
compared with the first-row 32K reference; that reference is repeated in
the context block for a direct 8K/16K/32K comparison. Within each numeric
column, \textbf{bold} marks the best reported value and \underline{underlining}
marks the second-best distinct value, following the indicated arrow; tied
values share the same formatting, and the repeated reference is counted
once. These column-wise marks do not imply one overall-best protocol. The
full-ranking row ($^{\dagger}$) succeeds in only 22/48 cases and is
survivor-biased.}
\label{tab:v80_prompt_context_cost}
\vspace{0.2em}
\setlength{\tabcolsep}{1.25mm}
\renewcommand{\arraystretch}{1.08}
\small
\begin{tabular}{@{}llrrrrrr@{}}
\toprule
\shortstack[l]{Ablation\\axis} &
\shortstack[l]{Setting} &
\shortstack[r]{Captured\\mass $\uparrow$} &
\shortstack[r]{Grad.\\cosine $\uparrow$} &
\shortstack[r]{Success\\$\uparrow$} &
\shortstack[r]{Prompt\\kTok $\downarrow$} &
\shortstack[r]{Output\\kTok $\downarrow$} &
\shortstack[r]{Calls\\$\downarrow$} \\
\midrule
Reference & Domain + compressed, 32K & 0.554 & 0.628 & \underline{97.9\%} & 160.2 & 12.0 & 10.1 \\
\midrule
Prompt & Minimal + compressed, 32K & 0.549 & \underline{0.655} & \textbf{100.0\%} & 140.6 & \underline{8.7} & \underline{9.0} \\
 & Domain + 2-shot, 32K & 0.510 & 0.509 & \underline{97.9\%} & 189.1 & 13.4 & 10.8 \\
\midrule
Output & Domain + direct IDs, 32K & \underline{0.594} & 0.633 & 95.8\% & 169.8 & 13.2 & 11.5 \\
 & Domain + full ranking, 32K$^{\dagger}$ & \textbf{0.677} & \textbf{0.789} & 45.8\% & 114.6 & 16.5 & 12.5 \\
\midrule
Context & Domain + compressed, 8K & 0.562 & 0.592 & 60.4\% & \textbf{56.7} & \textbf{5.9} & \textbf{8.0} \\
 & Domain + compressed, 16K & 0.564 & 0.583 & 81.2\% & \underline{113.5} & 10.2 & 9.5 \\
 & Domain + compressed, 32K & 0.554 & 0.628 & \underline{97.9\%} & 160.2 & 12.0 & 10.1 \\
\bottomrule
\end{tabular}
\vspace{-0.45em}
\end{table*}

\clearpage
\subsection{Token- and Feature-Level Attribution}
\appQwenAttribution
\clearpage
\appLlamaAttribution
\clearpage
\appDeepSeekAttribution
\clearpage
\appGPTOssAttribution
\clearpage
\appVistaAttribution

\clearpage
\subsection{Best-Prompt Online VQE and Explanation Faithfulness}
\begin{table*}[t]
\centering
\vspace{-0.5em}
\caption{\textbf{Does the selected prompt improve clean frozen-LLM selection online?} GPT-OSS (clean) is the clean exact select/exclude baseline; GPT-OSS (best) uses the selected domain prompt, direct-ID output, 32K context, and all physical features. Both remain frozen, receive the same count $k_j$ requested by \ourmethod{}, and refresh group identities every $K{=}7$ steps. Gain and energy entries are mean $\pm$ standard deviation over seeds 97/101/107/109, with $\dE$ in Ha relative to Full-QWC; gradients use exact expectations and shot costs are nominal (Appendix~\ref{sec:app_estimator_scope}). An energy cell for the best prompt is green when its absolute mean gap is smaller than that of the clean baseline and red otherwise. The complete Random and \ourmethod{} controls are reported in Appendix Table~\ref{tab:v80b_best_prompt_online_full}.}
\label{tab:v80b_best_prompt_online}
\vspace{0.25em}
\setlength{\tabcolsep}{0.5mm}
\renewcommand{\arraystretch}{0.92}
\scriptsize
\begingroup
\definecolor{vistaheatgreen}{HTML}{A8D8A8}
\definecolor{vistaheatred}{HTML}{F3A6B7}
\newcommand{\llmlogo}[1]{\raisebox{-0.28ex}{\includegraphics[height=1.15em]{llm_logos/#1.png}}}
\newcommand{\llmmethod}[2]{\llmlogo{#1}\,#2}
\newcommand{\better}[1]{\cellcolor{vistaheatgreen!45!white}\textcolor{black}{#1}}
\newcommand{\worse}[1]{\cellcolor{vistaheatred!45!white}\textcolor{black}{#1}}
\newcommand{\epos}[1]{+#1}
\newcommand{\eneg}[1]{\textminus#1}
\newcommand{\estat}[2]{#1\,\(\pm\)\,#2}
\newcommand{\pstat}[2]{#1\,\(\pm\)\,#2\%}
\newcommand{\grayrow}{\specialrule{0.20pt}{0.35mm}{0.35mm}}
\arrayrulecolor{black!18}
\begin{tabular}{llccccc}
\arrayrulecolor{black}\toprule
Setting & Protocol & H2 (4q) & HeH+ (4q) & H3+ (6q) & H4 chain (8q) & LiH (12q) \\
\midrule\arrayrulecolor{black!18}
\bcell{Depth 1} & GPT clean gain, $K{=}7$ $\uparrow$ & \pstat{39.4}{1.8} & \pstat{62.6}{1.3} & \pstat{67.0}{1.5} & \pstat{79.6}{1.8} & \pstat{89.9}{0.8} \\
\grayrow
\bcell{Depth 1} & GPT best gain, $K{=}7$ $\uparrow$ & \pstat{42.5}{4.6} & \pstat{61.7}{3.4} & \pstat{66.7}{1.5} & \pstat{79.3}{1.8} & \pstat{90.5}{0.4} \\
\arrayrulecolor{black}\cmidrule(lr){1-7}\arrayrulecolor{black!18}
\gcell{Depth 1} & \llmmethod{openai}{GPT-OSS (clean)}, $\dE$ & \estat{\epos{0.3930}}{0.1936} & \estat{\epos{0.8238}}{0.2303} & \estat{\epos{0.7300}}{0.4383} & \estat{\epos{0.5671}}{0.3164} & \estat{\epos{1.5503}}{0.7644} \\
\grayrow
\gcell{Depth 1} & \llmmethod{openai}{GPT-OSS (best)}, $\dE$ & \better{\estat{\epos{0.3392}}{0.1856}} & \worse{\estat{\epos{0.9079}}{0.4907}} & \better{\estat{\epos{0.3882}}{0.2525}} & \worse{\estat{\epos{0.6337}}{0.2763}} & \worse{\estat{\epos{1.9964}}{0.9332}} \\
\arrayrulecolor{black}\midrule\arrayrulecolor{black!18}
\bcell{Depth 2} & GPT clean gain, $K{=}7$ $\uparrow$ & \pstat{29.5}{3.5} & \pstat{34.3}{1.7} & \pstat{49.8}{1.9} & \pstat{55.5}{1.6} & \pstat{84.1}{0.5} \\
\grayrow
\bcell{Depth 2} & GPT best gain, $K{=}7$ $\uparrow$ & \pstat{29.7}{2.2} & \pstat{33.0}{2.3} & \pstat{49.8}{0.8} & \pstat{55.4}{0.4} & \pstat{84.3}{0.5} \\
\arrayrulecolor{black}\cmidrule(lr){1-7}\arrayrulecolor{black!18}
\gcell{Depth 2} & \llmmethod{openai}{GPT-OSS (clean)}, $\dE$ & \estat{\epos{0.2148}}{0.1220} & \estat{\epos{0.6807}}{0.7081} & \estat{\epos{0.4691}}{0.2024} & \estat{\epos{0.5616}}{0.2675} & \estat{\epos{1.2894}}{0.3290} \\
\grayrow
\gcell{Depth 2} & \llmmethod{openai}{GPT-OSS (best)}, $\dE$ & \better{\estat{\eneg{0.0510}}{0.0402}} & \better{\estat{\epos{0.1302}}{0.0822}} & \worse{\estat{\epos{0.4776}}{0.3130}} & \better{\estat{\epos{0.4682}}{0.1297}} & \worse{\estat{\epos{1.5294}}{0.3237}} \\
\arrayrulecolor{black}\midrule\arrayrulecolor{black!18}
\bcell{Depth 3} & GPT clean gain, $K{=}7$ $\uparrow$ & \pstat{19.8}{1.9} & \pstat{25.1}{1.8} & \pstat{32.1}{1.4} & \pstat{39.0}{0.5} & \pstat{76.8}{1.5} \\
\grayrow
\bcell{Depth 3} & GPT best gain, $K{=}7$ $\uparrow$ & \pstat{19.5}{1.6} & \pstat{26.3}{1.2} & \pstat{32.1}{1.0} & \pstat{38.7}{1.1} & \pstat{76.6}{1.3} \\
\arrayrulecolor{black}\cmidrule(lr){1-7}\arrayrulecolor{black!18}
\gcell{Depth 3} & \llmmethod{openai}{GPT-OSS (clean)}, $\dE$ & \estat{\epos{0.1633}}{0.1815} & \estat{\epos{0.3438}}{0.1193} & \estat{\epos{0.1786}}{0.0759} & \estat{\epos{0.5089}}{0.3596} & \estat{\epos{0.5647}}{0.4252} \\
\grayrow
\gcell{Depth 3} & \llmmethod{openai}{GPT-OSS (best)}, $\dE$ & \better{\estat{\epos{0.0478}}{0.0637}} & \better{\estat{\epos{0.0162}}{0.1316}} & \better{\estat{\epos{0.1222}}{0.1216}} & \better{\estat{\epos{0.0449}}{0.2110}} & \better{\estat{\epos{0.5331}}{0.3351}} \\
\arrayrulecolor{black}\bottomrule
\end{tabular}
\endgroup
\vspace{-0.8em}
\end{table*}

\clearpage
\appReasonConsistency

\clearpage
\subsection{Complete Mechanism-Audit Summary}
\label{sec:app_v80_full}
Table~\ref{tab:v80_llm_ablation} preserves the aggregate intervention
summary behind the main-text attribution and cost views. 
\begin{table*}[t]
\centering
\caption{\textbf{What drives frozen-LLM QWC selection?}
The audit evaluates four frozen local LLMs on the same 12 Full-QWC snapshots
(H3+, H4 chain, and LiH; depths 2 and 3; seed 97; iterations 0 and 14),
with no fine-tuning and identical per-parameter
cardinalities. The reference uses a domain prompt, compressed output,
32K context, and all non-oracle coefficient, light-cone, parameter, and
trajectory features. $\Delta$ captured mass is the paired,
model-macro-averaged change from this reference in percentage points;
positive is better. Protocol success is the fraction of the 48
model--snapshot cases yielding a valid exact-cardinality mask. The
apparent ranking-output gain is conditional on only 22 successful cases
and should not be interpreted as a reliable improvement.}
\label{tab:v80_llm_ablation}
\vspace{0.25em}
\setlength{\tabcolsep}{1.25mm}
\renewcommand{\arraystretch}{1.06}
\scriptsize
\begingroup
\arrayrulecolor{black!18}
\begin{tabular}{llrrp{0.39\textwidth}}
\arrayrulecolor{black}\toprule
Factor & Intervention & $\Delta$ mass (pp) $\uparrow$ &
Protocol success $\uparrow$ & Main observation \\
\midrule\arrayrulecolor{black!18}
\bcell{Prompt} & Minimal instruction & $-1.0$ & $100.0\%$ &
Shorter wording improves validity but not relevance quality. \\
\bcell{Prompt} & Two demonstrations & $-4.8$ & $97.9\%$ &
Few-shot prompting does not provide a universal gain. \\
\arrayrulecolor{black}\cmidrule(lr){1-5}\arrayrulecolor{black!18}
\gcell{Output} & Direct selected IDs & $\mathbf{+3.5}$ & $95.8\%$ &
The strongest reliable output contract on average. \\
\gcell{Output} & Full ranking & $+7.9^{\dagger}$ & $45.8\%$ &
Long structured outputs are brittle and strongly survivor-biased. \\
\arrayrulecolor{black}\cmidrule(lr){1-5}\arrayrulecolor{black!18}
\bcell{Context} & 8K tokens & $-1.2$ & $60.4\%$ &
Frequent context splitting and incomplete structured outputs. \\
\bcell{Context} & 16K tokens & $-2.4$ & $81.2\%$ &
Higher reliability than 8K, without a quality gain on valid cases. \\
\bcell{Context} & 32K reference & $0.0$ & $\mathbf{97.9\%}$ &
Long context primarily prevents protocol failure. \\
\arrayrulecolor{black}\cmidrule(lr){1-5}\arrayrulecolor{black!18}
\gcell{Physics} & Remove coefficients & $\mathbf{-6.4}$ & $97.9\%$ &
Coefficient information is useful overall, especially to Qwen and Llama. \\
\gcell{Physics} & Remove light-cone features & $\mathbf{-4.9}$ & $93.8\%$ &
Light-cone information is useful overall but strongly model-dependent. \\
\gcell{Physics} & Remove $\theta$/iteration features & $+1.8$ & $97.9\%$ &
No stable aggregate benefit from current-state fields. \\
\gcell{Physics} & Shuffle coefficients & $-1.5$ & $91.7\%$ &
The largest shuffle effect is in Qwen rather than universal. \\
\gcell{Physics} & Shuffle light-cone features & $+2.3$ & $93.8\%$ &
Opposite model-level effects cancel in the aggregate. \\
\arrayrulecolor{black}\bottomrule
\end{tabular}
\endgroup
\vspace{-0.5em}
\end{table*}

\subsection{Complete Best-Prompt Online-VQE Controls}
Table~\ref{tab:v80b_best_prompt_online_full} retains the Random baseline
and the refresh-matched and per-step \ourmethod{} controls omitted from the
compact clean-versus-best comparison in the main text.
\begin{table*}[t]
\centering
\vspace{-0.5em}
\caption{\textbf{Complete best-prompt closed-loop online-VQE comparison.} This appendix table preserves the full matched-control version of Table~\ref{tab:v80b_best_prompt_online}. GPT-OSS (clean) is the clean exact select/exclude baseline, while GPT-OSS (best) uses the audit-selected domain prompt, direct-ID output, 32K context, and all physical features. Both models remain frozen, receive only the count $k_j$ requested by the proposed method, and refresh group identities every $K{=}7$ steps. Random uses the same cardinality and refresh cadence; \ourmethod{} is reported at both $K{=}7$ and per-step $K{=}1$. Gain rows report realized mean $\pm$ standard deviation over seeds 97/101/107/109; all safety fallbacks pay Full-QWC shots. Energy entries are mean $\pm$ standard deviation relative to Full-QWC. Green/red satisfies/violates the one-sided 1.6 mHa no-worse criterion; the black check/cross compares absolute mean gap with Random. Bold/underline mark the smallest/second-smallest absolute mean gap among the five methods in each molecule--depth cell.}
\label{tab:v80b_best_prompt_online_full}
\vspace{0.25em}
\setlength{\tabcolsep}{0.1mm}
\renewcommand{\arraystretch}{0.90}
\scriptsize
\begingroup
\definecolor{vistaheatgreen}{HTML}{A8D8A8}
\definecolor{vistaheatred}{HTML}{F3A6B7}
\newcommand{\llmlogo}[1]{\raisebox{-0.28ex}{\includegraphics[height=1.15em]{llm_logos/#1.png}}}
\newcommand{\llmmethod}[2]{\llmlogo{#1}\,#2}
\newcommand{\vpass}[2]{\cellcolor{vistaheatgreen!#1!white}\textcolor{black}{#2}}
\newcommand{\vfail}[2]{\cellcolor{vistaheatred!#1!white}\textcolor{black}{#2}}
\newcommand{\epos}[1]{+#1}
\newcommand{\eneg}[1]{\textminus#1}
\newcommand{\estat}[2]{#1\,\(\pm\)\,#2}
\newcommand{\pstat}[2]{#1\,\(\pm\)\,#2\%}
\newcommand{\best}[1]{\textbf{#1}}
\newcommand{\runnerup}[1]{\underline{#1}}
\newcommand{\randwin}{\textcolor{black}{\ding{51}}}
\newcommand{\randloss}{\textcolor{black}{\ding{55}}}
\newcommand{\grayrow}{\specialrule{0.20pt}{0.35mm}{0.35mm}}
\arrayrulecolor{black!18}
\begin{tabular}{llccccc}
\arrayrulecolor{black}\toprule
Setting & Selection method & H2 (4q) & HeH+ (4q) & H3+ (6q) & H4 chain (8q) & LiH (12q) \\
\midrule\arrayrulecolor{black!18}
\bcell{Depth 1} & GPT clean gain, $K{=}7$ $\uparrow$ & \pstat{39.4}{1.8} & \pstat{62.6}{1.3} & \pstat{67.0}{1.5} & \pstat{79.6}{1.8} & \pstat{89.9}{0.8} \\
\grayrow
\bcell{Depth 1} & GPT best gain, $K{=}7$ $\uparrow$ & \pstat{42.5}{4.6} & \pstat{61.7}{3.4} & \pstat{66.7}{1.5} & \pstat{79.3}{1.8} & \pstat{90.5}{0.4} \\
\grayrow
\bcell{Depth 1} & \ours{} gain, $K{=}7$ $\uparrow$ & \pstat{41.2}{5.3} & \pstat{60.7}{3.1} & \pstat{65.1}{2.6} & \pstat{77.3}{0.7} & \pstat{88.2}{0.4} \\
\grayrow
\bcell{Depth 1} & \ours{} gain, $K{=}1$ $\uparrow$ & \pstat{40.4}{6.3} & \pstat{58.9}{3.5} & \pstat{63.0}{2.8} & \pstat{74.7}{1.1} & \pstat{86.9}{0.5} \\
\arrayrulecolor{black}\cmidrule(lr){1-7}\arrayrulecolor{black!18}
\gcell{Depth 1} & Random ($K{=}7$) & \estat{\epos{0.1847}}{0.0868} & \estat{\epos{1.2191}}{0.2759} & \estat{\epos{0.7247}}{0.2856} & \estat{\epos{0.9736}}{0.3101} & \estat{\epos{2.9030}}{0.6058} \\
\grayrow
\gcell{Depth 1} & \llmmethod{openai}{GPT-OSS (clean)} & \vfail{39}{\estat{\epos{0.3930}}{0.1936}\,\randloss} & \vfail{48}{\estat{\epos{0.8238}}{0.2303}\,\randwin} & \vfail{47}{\estat{\epos{0.7300}}{0.4383}\,\randloss} & \vfail{43}{\estat{\epos{0.5671}}{0.3164}\,\randwin} & \vfail{60}{\estat{\epos{1.5503}}{0.7644}\,\randwin} \\
\grayrow
\gcell{Depth 1} & \llmmethod{openai}{GPT-OSS (best)} & \vfail{37}{\estat{\epos{0.3392}}{0.1856}\,\randloss} & \vfail{50}{\estat{\epos{0.9079}}{0.4907}\,\randwin} & \vfail{39}{\estat{\epos{0.3882}}{0.2525}\,\randwin} & \vfail{45}{\estat{\epos{0.6337}}{0.2763}\,\randwin} & \vfail{65}{\estat{\epos{1.9964}}{0.9332}\,\randwin} \\
\grayrow
\gcell{Depth 1} & \ours{} ($K{=}7$) & \vpass{18}{\runnerup{\estat{\epos{0.0013}}{0.0003}}\,\randwin} & \vfail{19}{\runnerup{\estat{\epos{0.0035}}{0.0015}}\,\randwin} & \vpass{18}{\best{\estat{\epos{0.0004}}{0.0006}}\,\randwin} & \vpass{18}{\best{\estat{\epos{0.0001}}{0.0060}}\,\randwin} & \vfail{19}{\runnerup{\estat{\epos{0.0030}}{0.0016}}\,\randwin} \\
\grayrow
\gcell{Depth 1} & \ours{} ($K{=}1$) & \vpass{18}{\best{\estat{\epos{0.0011}}{0.0008}}\,\randwin} & \vpass{18}{\best{\estat{\epos{0.0015}}{0.0012}}\,\randwin} & \vpass{18}{\runnerup{\estat{\epos{0.0005}}{0.0006}}\,\randwin} & \vpass{24}{\runnerup{\estat{\eneg{0.0006}}{0.0024}}\,\randwin} & \vfail{19}{\best{\estat{\epos{0.0023}}{0.0016}}\,\randwin} \\
\arrayrulecolor{black}\midrule\arrayrulecolor{black!18}
\bcell{Depth 2} & GPT clean gain, $K{=}7$ $\uparrow$ & \pstat{29.5}{3.5} & \pstat{34.3}{1.7} & \pstat{49.8}{1.9} & \pstat{55.5}{1.6} & \pstat{84.1}{0.5} \\
\grayrow
\bcell{Depth 2} & GPT best gain, $K{=}7$ $\uparrow$ & \pstat{29.7}{2.2} & \pstat{33.0}{2.3} & \pstat{49.8}{0.8} & \pstat{55.4}{0.4} & \pstat{84.3}{0.5} \\
\grayrow
\bcell{Depth 2} & \ours{} gain, $K{=}7$ $\uparrow$ & \pstat{27.3}{2.5} & \pstat{32.5}{2.4} & \pstat{47.9}{1.4} & \pstat{55.1}{0.5} & \pstat{83.7}{0.4} \\
\grayrow
\bcell{Depth 2} & \ours{} gain, $K{=}1$ $\uparrow$ & \pstat{25.6}{2.9} & \pstat{31.6}{2.6} & \pstat{44.9}{1.3} & \pstat{51.8}{0.5} & \pstat{82.6}{0.4} \\
\arrayrulecolor{black}\cmidrule(lr){1-7}\arrayrulecolor{black!18}
\gcell{Depth 2} & Random ($K{=}7$) & \estat{\epos{0.0930}}{0.0297} & \estat{\epos{0.5668}}{0.4463} & \estat{\epos{0.3054}}{0.1633} & \estat{\epos{0.9327}}{0.3273} & \estat{\epos{2.6969}}{0.2816} \\
\grayrow
\gcell{Depth 2} & \llmmethod{openai}{GPT-OSS (clean)} & \vfail{33}{\estat{\epos{0.2148}}{0.1220}\,\randloss} & \vfail{46}{\estat{\epos{0.6807}}{0.7081}\,\randloss} & \vfail{41}{\estat{\epos{0.4691}}{0.2024}\,\randloss} & \vfail{43}{\estat{\epos{0.5616}}{0.2675}\,\randwin} & \vfail{56}{\estat{\epos{1.2894}}{0.3290}\,\randwin} \\
\grayrow
\gcell{Depth 2} & \llmmethod{openai}{GPT-OSS (best)} & \vpass{75}{\estat{\eneg{0.0510}}{0.0402}\,\randwin} & \vfail{30}{\estat{\epos{0.1302}}{0.0822}\,\randwin} & \vfail{41}{\estat{\epos{0.4776}}{0.3130}\,\randloss} & \vfail{41}{\estat{\epos{0.4682}}{0.1297}\,\randwin} & \vfail{59}{\estat{\epos{1.5294}}{0.3237}\,\randwin} \\
\grayrow
\gcell{Depth 2} & \ours{} ($K{=}7$) & \vpass{25}{\runnerup{\estat{\eneg{0.0007}}{0.0015}}\,\randwin} & \vfail{21}{\runnerup{\estat{\epos{0.0076}}{0.0176}}\,\randwin} & \vpass{18}{\best{\estat{\epos{0.0006}}{0.0031}}\,\randwin} & \vfail{22}{\runnerup{\estat{\epos{0.0162}}{0.0438}}\,\randwin} & \vpass{34}{\best{\estat{\eneg{0.0040}}{0.0118}}\,\randwin} \\
\grayrow
\gcell{Depth 2} & \ours{} ($K{=}1$) & \vpass{18}{\best{\estat{\epos{0.0003}}{0.0007}}\,\randwin} & \vpass{26}{\best{\estat{\eneg{0.0011}}{0.0026}}\,\randwin} & \vpass{28}{\runnerup{\estat{\eneg{0.0015}}{0.0041}}\,\randwin} & \vfail{22}{\best{\estat{\epos{0.0158}}{0.0443}}\,\randwin} & \vpass{36}{\runnerup{\estat{\eneg{0.0049}}{0.0097}}\,\randwin} \\
\arrayrulecolor{black}\midrule\arrayrulecolor{black!18}
\bcell{Depth 3} & GPT clean gain, $K{=}7$ $\uparrow$ & \pstat{19.8}{1.9} & \pstat{25.1}{1.8} & \pstat{32.1}{1.4} & \pstat{39.0}{0.5} & \pstat{76.8}{1.5} \\
\grayrow
\bcell{Depth 3} & GPT best gain, $K{=}7$ $\uparrow$ & \pstat{19.5}{1.6} & \pstat{26.3}{1.2} & \pstat{32.1}{1.0} & \pstat{38.7}{1.1} & \pstat{76.6}{1.3} \\
\grayrow
\bcell{Depth 3} & \ours{} gain, $K{=}7$ $\uparrow$ & \pstat{19.0}{2.4} & \pstat{26.1}{1.2} & \pstat{32.1}{1.2} & \pstat{38.6}{0.7} & \pstat{76.6}{1.3} \\
\grayrow
\bcell{Depth 3} & \ours{} gain, $K{=}1$ $\uparrow$ & \pstat{18.1}{2.7} & \pstat{24.9}{1.3} & \pstat{29.9}{1.0} & \pstat{35.2}{0.9} & \pstat{75.4}{1.8} \\
\arrayrulecolor{black}\cmidrule(lr){1-7}\arrayrulecolor{black!18}
\gcell{Depth 3} & Random ($K{=}7$) & \estat{\epos{0.0730}}{0.0859} & \estat{\epos{0.2617}}{0.2130} & \estat{\epos{0.2750}}{0.0765} & \estat{\epos{0.2773}}{0.2665} & \estat{\epos{2.3928}}{0.4840} \\
\grayrow
\gcell{Depth 3} & \llmmethod{openai}{GPT-OSS (clean)} & \vfail{31}{\estat{\epos{0.1633}}{0.1815}\,\randloss} & \vfail{38}{\estat{\epos{0.3438}}{0.1193}\,\randloss} & \vfail{32}{\estat{\epos{0.1786}}{0.0759}\,\randwin} & \vfail{42}{\estat{\epos{0.5089}}{0.3596}\,\randloss} & \vfail{43}{\estat{\epos{0.5647}}{0.4252}\,\randwin} \\
\grayrow
\gcell{Depth 3} & \llmmethod{openai}{GPT-OSS (best)} & \vfail{25}{\estat{\epos{0.0478}}{0.0637}\,\randwin} & \vfail{22}{\estat{\epos{0.0162}}{0.1316}\,\randwin} & \vfail{30}{\estat{\epos{0.1222}}{0.1216}\,\randwin} & \vfail{25}{\estat{\epos{0.0449}}{0.2110}\,\randwin} & \vfail{42}{\estat{\epos{0.5331}}{0.3351}\,\randwin} \\
\grayrow
\gcell{Depth 3} & \ours{} ($K{=}7$) & \vpass{18}{\runnerup{\estat{\epos{0.0010}}{0.0027}}\,\randwin} & \vfail{21}{\runnerup{\estat{\epos{0.0108}}{0.0118}}\,\randwin} & \vfail{19}{\runnerup{\estat{\epos{0.0029}}{0.0095}}\,\randwin} & \vfail{19}{\runnerup{\estat{\epos{0.0029}}{0.0113}}\,\randwin} & \vpass{35}{\runnerup{\estat{\eneg{0.0046}}{0.0040}}\,\randwin} \\
\grayrow
\gcell{Depth 3} & \ours{} ($K{=}1$) & \vpass{18}{\best{\estat{\epos{0.0005}}{0.0015}}\,\randwin} & \vfail{21}{\best{\estat{\epos{0.0092}}{0.0131}}\,\randwin} & \vpass{18}{\best{\estat{\epos{0.0005}}{0.0077}}\,\randwin} & \vpass{27}{\best{\estat{\eneg{0.0014}}{0.0097}}\,\randwin} & \vpass{28}{\best{\estat{\eneg{0.0017}}{0.0036}}\,\randwin} \\
\arrayrulecolor{black}\bottomrule
\end{tabular}
\endgroup
\vspace{-0.5em}
\end{table*}

\begin{figure}[t]
\centering
\includegraphics[width=0.5\linewidth]{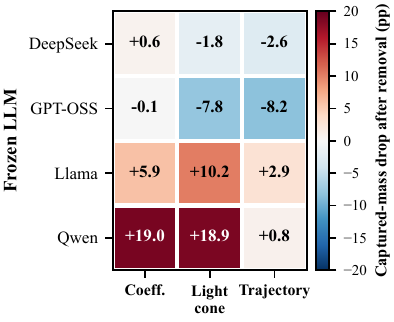}
\caption{\textbf{Frozen LLMs use physical feature families differently.}
Each cell is the paired captured-mass drop, in percentage points, after
removing one feature family from the 32K full-feature reference. Positive
red values indicate useful dependence; negative blue values indicate that
removal improves selection. The aggregate benefit is model dependent, and
per-snapshot effects may reverse sign.}
\label{fig:v80_feature_attribution}
\vspace{-0.45em}
\end{figure}

\clearpage
\section{Calibration-Derived Noise Diagnostics}
\label{sec:app_noise}
\begin{figure}[!t]
\centering
\includegraphics[width=\linewidth]{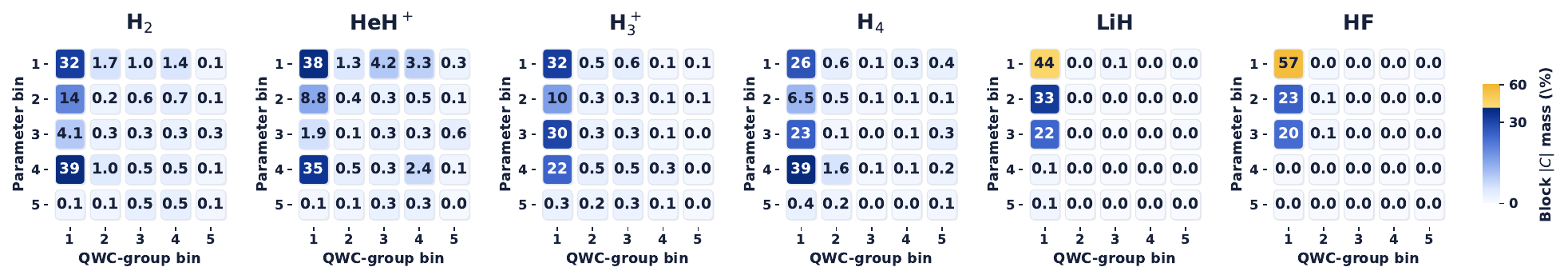}
\caption{\textbf{The same contribution matrices under IBM calibration-derived noise.} Molecules, circuits, initialization, parameter shifts, block aggregation, and color normalization are identical to Figure~\ref{fig:vista_qwc_sparsity}. We extract median gate and readout errors, gate durations, and coherence metadata from the fixed Qiskit \texttt{FakeGuadalupeV2} system snapshot~\citep{javadi2024quantum,ibm2026fakeprovider}. Following standard backend-derived noise modeling~\citep{ibm2026noisemodels}, our scalable Pauli-channel surrogate combines gate depolarization, $T_2$-limited decay, and readout attenuation. This is a calibrated noise simulation, not execution on a physical IBM QPU.}
\label{fig:vista_qwc_sparsity_ibm_noise}
\vspace{-0.5em}
\end{figure}

Across 120 surrogate snapshots, clean and noisy contribution grids have 50.6\% and 51.5\% mean sparsity, respectively. Their masks have 0.970 mean Jaccard overlap, their relevance rankings have 0.996 mean Spearman correlation, and a clean top-mass mask retains 99.7\% of noisy contribution mass. The smaller gate-level Aer check is deliberately more conservative: finite-shot transpiled circuits reduce mask overlap and rank correlation, but 15--26\% of parameter--group cells remain unselected at $\tau=0.995$. Neither experiment is a physical-QPU run.
\begin{table}[H]
\centering
\vspace{-0.5em}
\caption{\textbf{Noisy online VQE comparison at $\tau=0.995$.} This repeats the Table~\ref{tab:v39_tau0995_main} protocol under a finite-shot Pauli-channel surrogate derived from the fixed IBM FakeGuadalupeV2 calibration snapshot~\citep{javadi2024quantum,ibm2026fakeprovider,ibm2026noisemodels}. Final energies are evaluated noiselessly and paired against noisy Full-QWC training; $^{\dagger}$ rows retain the noiseless cell-wise TITAN thresholds. The same noise study additionally tests contribution-mask stability (Figure~\ref{fig:vista_qwc_sparsity_ibm_noise} and Table~\ref{tab:v72_noise_sparsity}), a gate-level Aer implementation (Table~\ref{tab:v72_aer_spotcheck}), and Hamiltonian-family transfer (Table~\ref{tab:v73_noisy_hamiltonian_panel}).}
\vspace{0.25em}
\setlength{\tabcolsep}{0.45mm}
\renewcommand{\arraystretch}{0.88}
\tiny
\resizebox{1.0\linewidth}{!}{
\begin{tabular}{lll|ccccc}
\toprule
Setting & Metric & Method & H2 (4q) & HeH+ (4q) & H3+ (6q) & H4 chain (8q) & LiH (12q) \\
\midrule
\gcell{Depth 1} & \(\dE\) vs Full \(\downarrow\) & \ours{} & \gbf{\(-0.0076\)} & \gbf{\(-0.0219\)} & \gbf{\(-0.0252\)} & \textbf{\gbf{\(+0.0001\)}} & \gbf{\(-0.0304\)} \\
\gcell{Depth 1} & \(\dE\) vs Full \(\downarrow\) & Random & \rbf{\(+0.2170\)} & \rbf{\(+1.2823\)} & \rbf{\(+0.9091\)} & \rbf{\(+1.3470\)} & \rbf{\(+3.3622\)} \\
\gcell{Depth 1} & \(\dE\) vs Full \(\downarrow\) & Global coeff. & \gbf{\(-0.0050\)} & \textbf{\rbf{\(+0.0155\)}} & \textbf{\gbf{\(-0.0022\)}} & \gbf{\(-0.0020\)} & \textbf{\rbf{\(+0.0041\)}} \\
\gcell{Depth 1} & \(\dE\) vs Full \(\downarrow\) & TITAN~(\citeauthor{peng2026titan}) & \rbf{\(+0.0150\)} & \rbf{\(+0.0723\)} & \rbf{\(+0.0413\)} & \rbf{\(+0.0116\)} & \gbf{\(-0.1104\)} \\
\gcell{Depth 1} & \(\dE\) vs Full \(\downarrow\) & TITAN+\ours{} & \rbf{\(+0.0138\)} & \rbf{\(+0.0746\)} & \rbf{\(+0.0422\)} & \rbf{\(+0.0138\)} & \gbf{\(-0.1065\)} \\
\gcell{Depth 1} & \(\dE\) vs Full \(\downarrow\) & TITAN$^{\dagger}$ & \rbf{\(+0.0053\)} & \rbf{\(+0.0985\)} & \rbf{\(+0.0293\)} & \gbf{\(-0.0851\)} & \gbf{\(-0.0705\)} \\
\gcell{Depth 1} & \(\dE\) vs Full \(\downarrow\) & TITAN$^{\dagger}$+\ours{} & \textbf{\gbf{\(+0.0011\)}} & \rbf{\(+0.0988\)} & \rbf{\(+0.0291\)} & \gbf{\(-0.0864\)} & \gbf{\(-0.0693\)} \\
\cmidrule(lr){2-8}
\bcell{Depth 1} & Shot gain \(\uparrow\) & \ours{} & 43.2\% & 59.9\% & 64.0\% & 75.9\% & 88.1\% \\
\bcell{Depth 1} & Shot gain \(\uparrow\) & Random & 42.6\% & 60.5\% & 67.9\% & 79.1\% & 90.3\% \\
\bcell{Depth 1} & Shot gain \(\uparrow\) & Global coeff. & 41.5\% & 61.1\% & 64.8\% & 76.1\% & 88.3\% \\
\bcell{Depth 1} & Shot gain \(\uparrow\) & TITAN~(\citeauthor{peng2026titan}) & 65.0\% & 56.2\% & 58.6\% & 63.1\% & 73.7\% \\
\bcell{Depth 1} & Shot gain \(\uparrow\) & TITAN+\ours{} & \textbf{\underline{81.7\%}} & 90.5\% & 93.6\% & 96.1\% & 99.7\% \\
\bcell{Depth 1} & Shot gain \(\uparrow\) & TITAN$^{\dagger}$ & 43.6\% & 59.4\% & 64.3\% & 71.6\% & 82.3\% \\
\bcell{Depth 1} & Shot gain \(\uparrow\) & TITAN$^{\dagger}$+\ours{} & 65.9\% & \textbf{\underline{91.7\%}} & \textbf{\underline{95.5\%}} & \textbf{\underline{98.0\%}} & \textbf{\underline{99.9\%}} \\
\cmidrule(lr){2-8}
\gcell{Depth 1} & No-worse vs Full & \ours{} & \cmark & \cmark & \cmark & \cmark & \cmark \\
\gcell{Depth 1} & No-worse vs Full & Random & \xmark & \xmark & \xmark & \xmark & \xmark \\
\gcell{Depth 1} & No-worse vs Full & Global coeff. & \cmark & \xmark & \cmark & \cmark & \xmark \\
\gcell{Depth 1} & No-worse vs Full & TITAN~(\citeauthor{peng2026titan}) & \xmark & \xmark & \xmark & \xmark & \cmark \\
\gcell{Depth 1} & No-worse vs Full & TITAN+\ours{} & \xmark & \xmark & \xmark & \xmark & \cmark \\
\gcell{Depth 1} & No-worse vs Full & TITAN$^{\dagger}$ & \xmark & \xmark & \xmark & \cmark & \cmark \\
\gcell{Depth 1} & No-worse vs Full & TITAN$^{\dagger}$+\ours{} & \cmark & \xmark & \xmark & \cmark & \cmark \\
\midrule
\gcell{Depth 2} & \(\dE\) vs Full \(\downarrow\) & \ours{} & \gbf{\(-0.0632\)} & \gbf{\(-0.1081\)} & \gbf{\(-0.0765\)} & \gbf{\(-0.0583\)} & \gbf{\(-0.1241\)} \\
\gcell{Depth 2} & \(\dE\) vs Full \(\downarrow\) & Random & \rbf{\(+0.2196\)} & \rbf{\(+0.5977\)} & \rbf{\(+0.5653\)} & \rbf{\(+0.8670\)} & \rbf{\(+3.6871\)} \\
\gcell{Depth 2} & \(\dE\) vs Full \(\downarrow\) & Global coeff. & \gbf{\(-0.0965\)} & \gbf{\(-0.1624\)} & \textbf{\gbf{\(-0.0270\)}} & \gbf{\(-0.0491\)} & \textbf{\rbf{\(+0.0889\)}} \\
\gcell{Depth 2} & \(\dE\) vs Full \(\downarrow\) & TITAN~(\citeauthor{peng2026titan}) & \gbf{\(-0.0650\)} & \gbf{\(-0.2074\)} & \gbf{\(-0.0688\)} & \rbf{\(+0.0172\)} & \gbf{\(-0.1387\)} \\
\gcell{Depth 2} & \(\dE\) vs Full \(\downarrow\) & TITAN+\ours{} & \gbf{\(-0.0603\)} & \gbf{\(-0.2072\)} & \gbf{\(-0.0736\)} & \textbf{\rbf{\(+0.0162\)}} & \gbf{\(-0.1369\)} \\
\gcell{Depth 2} & \(\dE\) vs Full \(\downarrow\) & TITAN$^{\dagger}$ & \gbf{\(-0.0819\)} & \textbf{\gbf{\(-0.0621\)}} & \gbf{\(-0.0283\)} & \gbf{\(-0.0596\)} & \rbf{\(+0.0998\)} \\
\gcell{Depth 2} & \(\dE\) vs Full \(\downarrow\) & TITAN$^{\dagger}$+\ours{} & \textbf{\rbf{\(+0.0120\)}} & \gbf{\(-0.0710\)} & \gbf{\(-0.0400\)} & \gbf{\(-0.0239\)} & \rbf{\(+0.1058\)} \\
\cmidrule(lr){2-8}
\bcell{Depth 2} & Shot gain \(\uparrow\) & \ours{} & 28.8\% & 32.8\% & 46.2\% & 52.6\% & 82.6\% \\
\bcell{Depth 2} & Shot gain \(\uparrow\) & Random & 29.9\% & 34.7\% & 47.5\% & 53.4\% & 83.9\% \\
\bcell{Depth 2} & Shot gain \(\uparrow\) & Global coeff. & 27.9\% & 33.2\% & 46.7\% & 52.0\% & 83.0\% \\
\bcell{Depth 2} & Shot gain \(\uparrow\) & TITAN~(\citeauthor{peng2026titan}) & 53.2\% & 54.3\% & 51.2\% & 61.7\% & 78.7\% \\
\bcell{Depth 2} & Shot gain \(\uparrow\) & TITAN+\ours{} & \textbf{\underline{70.5\%}} & \textbf{\underline{75.5\%}} & \textbf{\underline{80.9\%}} & \textbf{\underline{87.6\%}} & 98.8\% \\
\bcell{Depth 2} & Shot gain \(\uparrow\) & TITAN$^{\dagger}$ & 25.3\% & 40.2\% & 42.0\% & 48.2\% & 84.3\% \\
\bcell{Depth 2} & Shot gain \(\uparrow\) & TITAN$^{\dagger}$+\ours{} & 43.1\% & 64.2\% & 74.8\% & 80.9\% & \textbf{\underline{99.4\%}} \\
\cmidrule(lr){2-8}
\gcell{Depth 2} & No-worse vs Full & \ours{} & \cmark & \cmark & \cmark & \cmark & \cmark \\
\gcell{Depth 2} & No-worse vs Full & Random & \xmark & \xmark & \xmark & \xmark & \xmark \\
\gcell{Depth 2} & No-worse vs Full & Global coeff. & \cmark & \cmark & \cmark & \cmark & \xmark \\
\gcell{Depth 2} & No-worse vs Full & TITAN~(\citeauthor{peng2026titan}) & \cmark & \cmark & \cmark & \xmark & \cmark \\
\gcell{Depth 2} & No-worse vs Full & TITAN+\ours{} & \cmark & \cmark & \cmark & \xmark & \cmark \\
\gcell{Depth 2} & No-worse vs Full & TITAN$^{\dagger}$ & \cmark & \cmark & \cmark & \cmark & \xmark \\
\gcell{Depth 2} & No-worse vs Full & TITAN$^{\dagger}$+\ours{} & \xmark & \cmark & \cmark & \cmark & \xmark \\
\bottomrule
\end{tabular}}
\label{tab:v72_realistic_noise_table1}
\vspace{-0.5em}
\end{table}

\begin{table}[H]
\centering
\caption{\textbf{Relevance sparsity under calibration-derived noise.} Metrics average four seeds at iterations 0, 14, and 27. Cross-mass is the noisy contribution mass retained by the clean top-mass mask.}
\setlength{\tabcolsep}{2.0mm}
\small
\begin{tabular}{lrrrrrr}
\toprule
System & Depth & Clean sparsity & Noisy sparsity & Mask Jaccard & Rank $\rho$ & Cross-mass \\
\midrule
H2 (4q) & 1 & 47.3\% & 47.5\% & 0.996 & 0.986 & 99.9\%  \\
H2 (4q) & 2 & 30.3\% & 30.7\% & 0.994 & 0.998 & 99.9\%  \\
HeH+ (4q) & 1 & 49.9\% & 50.7\% & 0.981 & 0.996 & 99.7\%  \\
HeH+ (4q) & 2 & 25.5\% & 26.0\% & 0.987 & 0.998 & 99.7\%  \\
H3+ (6q) & 1 & 54.2\% & 54.6\% & 0.986 & 0.996 & 99.7\%  \\
H3+ (6q) & 2 & 35.4\% & 36.1\% & 0.981 & 0.999 & 99.7\%  \\
H4 chain (8q) & 1 & 68.4\% & 69.2\% & 0.967 & 0.996 & 99.6\%  \\
H4 chain (8q) & 2 & 43.1\% & 44.5\% & 0.969 & 0.999 & 99.6\%  \\
LiH (12q) & 1 & 82.3\% & 83.3\% & 0.938 & 0.995 & 99.7\%  \\
LiH (12q) & 2 & 69.5\% & 72.4\% & 0.899 & 0.998 & 99.7\%  \\
\bottomrule
\end{tabular}
\label{tab:v72_noise_sparsity}
\end{table}

\begin{table}[H]
\centering
\caption{\textbf{Gate-level Aer spot-check under the FakeGuadalupeV2 snapshot}~\citep{ibm2026fakeprovider,ibm2026noisemodels}. Each depth-1 contribution grid uses seed 97 and 4096 shots for every shifted QWC-group circuit. The selected-cell counts use $\tau=0.995$. This is a calibrated Aer simulation, not a physical-QPU run.}
\setlength{\tabcolsep}{2.4mm}
\small
\resizebox{0.94\linewidth}{!}{
\begin{tabular}{lrrrrrrr}
\toprule
System & Params. & QWC groups & Circuits & Clean selected & Noisy selected & Noisy sparsity & Jaccard / Rank $\rho$ \\
\midrule
H2 (4q) & 8 & 5 & 80 & 21 & 34 & 15.0\% & 0.571 / 0.635 \\
HeH+ (4q) & 8 & 11 & 176 & 39 & 65 & 26.1\% & 0.529 / 0.441 \\
H3+ (6q) & 12 & 19 & 456 & 88 & 172 & 24.6\% & 0.477 / 0.335 \\
\bottomrule
\end{tabular}}
\label{tab:v72_aer_spotcheck}
\end{table}

\clearpage
\begin{table}[H]
\centering
\vspace{-0.7em}
\caption{\textbf{Calibration-derived noisy Hamiltonian-panel transfer at $\tau=0.997$.} This table repeats Table~\ref{tab:v55_hamiltonian_panel} under a fixed IBM \texttt{FakeGuadalupeV2}-derived Pauli-channel surrogate~\citep{javadi2024quantum,ibm2026fakeprovider,ibm2026noisemodels}. $\dE$ is paired against noisy Full-QWC at the same seed; final energies are evaluated noiselessly. TITAN$^{\dagger}$ reuses the clean, cell-wise cost-matched thresholds without noise-specific tuning. Green/red $\dE$ and \cmark{}/\xmark{} follow the one-sided $1.6$ mHa no-worse criterion; bold marks the closest mean $\dE$ to noisy Full, and bold-underlined shot gain marks the largest saving. This is calibrated simulation, not physical-QPU execution.}
\vspace{0.15em}
\setlength{\tabcolsep}{0.38mm}
\renewcommand{\arraystretch}{0.50}
\tiny
\resizebox{1.0\linewidth}{!}{
\begin{tabular}{lll|cccccc}
\toprule
Testing Hamiltonian & Depth & Metric & \ours{} & Random & TITAN~(\citeauthor{peng2026titan}) & TITAN+\ours{} & TITAN$^{\dagger}$ & TITAN$^{\dagger}$+\ours{} \\
\midrule
\gcell{TFIM $h=0.5$} & Depth 1 & $\dE$ vs noisy Full $\downarrow$ & \gbf{-0.0030} & \gbf{-0.0112} & \gbf{-0.0146} & \gbf{-0.0113} & \textbf{\gbf{-0.0002}} & \gbf{-0.0004} \\
 & Depth 1 & Shot gain $\uparrow$ & 2.5\% & 2.3\% & 55.3\% & \textbf{\underline{55.6\%}} & 0.0\% & 2.1\% \\
 & Depth 1 & Chem. acc. & \cmark & \cmark & \cmark & \cmark & \cmark & \cmark \\
\addlinespace[0.08em]
 & Depth 4 & $\dE$ vs noisy Full $\downarrow$ & \rbf{+0.0203} & \rbf{+0.0104} & \rbf{+0.0033} & \rbf{+0.0040} & \textbf{\gbf{-0.0032}} & \rbf{+0.0110} \\
 & Depth 4 & Shot gain $\uparrow$ & 2.1\% & 1.9\% & 55.2\% & \textbf{\underline{57.1\%}} & 0.0\% & 1.9\% \\
 & Depth 4 & Chem. acc. & \xmark & \xmark & \xmark & \xmark & \cmark & \xmark \\
\midrule
\gcell{TFIM $h=0.7$} & Depth 1 & $\dE$ vs noisy Full $\downarrow$ & \textbf{\gbf{+0.0010}} & \gbf{-0.0061} & \gbf{-0.0157} & \gbf{-0.0133} & \rbf{+0.0075} & \rbf{+0.0036} \\
 & Depth 1 & Shot gain $\uparrow$ & 3.2\% & 3.0\% & 53.3\% & \textbf{\underline{54.3\%}} & 0.0\% & 2.8\% \\
 & Depth 1 & Chem. acc. & \cmark & \cmark & \cmark & \cmark & \xmark & \xmark \\
\addlinespace[0.08em]
 & Depth 4 & $\dE$ vs noisy Full $\downarrow$ & \rbf{+0.0063} & \rbf{+0.0123} & \textbf{\gbf{+0.0013}} & \rbf{+0.0028} & \gbf{-0.0077} & \gbf{-0.0021} \\
 & Depth 4 & Shot gain $\uparrow$ & 1.9\% & 2.0\% & 47.6\% & \textbf{\underline{49.8\%}} & 0.0\% & 1.8\% \\
 & Depth 4 & Chem. acc. & \xmark & \xmark & \cmark & \xmark & \cmark & \cmark \\
\midrule
\gcell{TFIM $h=0.9$} & Depth 1 & $\dE$ vs noisy Full $\downarrow$ & \rbf{+0.1398} & \textbf{\rbf{+0.0032}} & \gbf{-0.0214} & \rbf{+0.0998} & \rbf{+0.0048} & \rbf{+0.1120} \\
 & Depth 1 & Shot gain $\uparrow$ & 25.0\% & 25.0\% & 53.4\% & \textbf{\underline{58.8\%}} & 24.8\% & 38.5\% \\
 & Depth 1 & Chem. acc. & \xmark & \xmark & \cmark & \xmark & \xmark & \xmark \\
\addlinespace[0.08em]
 & Depth 4 & $\dE$ vs noisy Full $\downarrow$ & \rbf{+0.0909} & \rbf{+0.0181} & \textbf{\gbf{-0.0138}} & \rbf{+0.0887} & \gbf{-0.0154} & \rbf{+0.0357} \\
 & Depth 4 & Shot gain $\uparrow$ & 6.2\% & 6.2\% & 42.9\% & \textbf{\underline{46.2\%}} & 4.5\% & 9.8\% \\
 & Depth 4 & Chem. acc. & \xmark & \xmark & \cmark & \xmark & \cmark & \xmark \\
\midrule
\gcell{Heis. XXZ+X field} & Depth 1 & $\dE$ vs noisy Full $\downarrow$ & \gbf{-0.1190} & \rbf{+0.3294} & \rbf{+0.6037} & \rbf{+0.6111} & \rbf{+0.0398} & \textbf{\gbf{-0.0175}} \\
 & Depth 1 & Shot gain $\uparrow$ & 16.7\% & 16.7\% & 52.2\% & \textbf{\underline{53.7\%}} & 18.9\% & 27.6\% \\
 & Depth 1 & Chem. acc. & \cmark & \xmark & \xmark & \xmark & \xmark & \cmark \\
\addlinespace[0.08em]
 & Depth 4 & $\dE$ vs noisy Full $\downarrow$ & \rbf{+0.3771} & \textbf{\rbf{+0.0446}} & \rbf{+0.2029} & \rbf{+0.1974} & \gbf{-0.1167} & \rbf{+0.7431} \\
 & Depth 4 & Shot gain $\uparrow$ & 4.2\% & 4.2\% & 36.4\% & \textbf{\underline{37.8\%}} & 4.3\% & 7.5\% \\
 & Depth 4 & Chem. acc. & \xmark & \xmark & \xmark & \xmark & \cmark & \xmark \\
\midrule
\gcell{Heis. J1--J2} & Depth 1 & $\dE$ vs noisy Full $\downarrow$ & \gbf{-0.0907} & \rbf{+0.4036} & \rbf{+0.5352} & \rbf{+0.5145} & \textbf{\gbf{-0.0779}} & \gbf{-0.2654} \\
 & Depth 1 & Shot gain $\uparrow$ & 16.7\% & 16.7\% & 50.5\% & \textbf{\underline{52.2\%}} & 11.3\% & 22.8\% \\
 & Depth 1 & Chem. acc. & \cmark & \xmark & \xmark & \xmark & \cmark & \cmark \\
\addlinespace[0.08em]
 & Depth 4 & $\dE$ vs noisy Full $\downarrow$ & \rbf{+0.6422} & \textbf{\gbf{-0.0145}} & \rbf{+0.4860} & \rbf{+0.2510} & \rbf{+0.4632} & \rbf{+0.6444} \\
 & Depth 4 & Shot gain $\uparrow$ & 4.2\% & 4.2\% & 34.2\% & \textbf{\underline{36.1\%}} & 4.1\% & 7.3\% \\
 & Depth 4 & Chem. acc. & \xmark & \cmark & \xmark & \xmark & \xmark & \xmark \\
\midrule
\gcell{Heis. XXY+Z field} & Depth 1 & $\dE$ vs noisy Full $\downarrow$ & \textbf{\rbf{+0.0889}} & \rbf{+0.3424} & \rbf{+0.3749} & \rbf{+0.2889} & \gbf{-0.1381} & \gbf{-0.2234} \\
 & Depth 1 & Shot gain $\uparrow$ & 16.7\% & 16.7\% & 49.3\% & \textbf{\underline{52.1\%}} & 14.1\% & 25.9\% \\
 & Depth 1 & Chem. acc. & \xmark & \xmark & \xmark & \xmark & \cmark & \cmark \\
\addlinespace[0.08em]
 & Depth 4 & $\dE$ vs noisy Full $\downarrow$ & \gbf{-0.0345} & \gbf{-0.1825} & \gbf{-0.2512} & \textbf{\gbf{-0.0173}} & \gbf{-0.0943} & \rbf{+0.1709} \\
 & Depth 4 & Shot gain $\uparrow$ & 4.2\% & 4.2\% & 31.8\% & \textbf{\underline{32.7\%}} & 3.3\% & 7.0\% \\
 & Depth 4 & Chem. acc. & \cmark & \cmark & \cmark & \cmark & \cmark & \xmark \\
\midrule
\gcell{JW-like parity spin} & Depth 1 & $\dE$ vs noisy Full $\downarrow$ & \textbf{\rbf{+0.0102}} & \rbf{+0.7538} & \rbf{+0.1011} & \rbf{+0.0976} & \gbf{-0.0114} & \rbf{+0.0106} \\
 & Depth 1 & Shot gain $\uparrow$ & 36.4\% & 36.1\% & 56.2\% & \textbf{\underline{77.1\%}} & 29.6\% & 59.4\% \\
 & Depth 1 & Chem. acc. & \xmark & \xmark & \xmark & \xmark & \cmark & \xmark \\
\addlinespace[0.08em]
 & Depth 4 & $\dE$ vs noisy Full $\downarrow$ & \textbf{\gbf{-0.0024}} & \rbf{+0.0546} & \rbf{+0.0515} & \rbf{+0.0092} & \gbf{-0.0346} & \gbf{-0.0106} \\
 & Depth 4 & Shot gain $\uparrow$ & 11.6\% & 11.6\% & 47.1\% & \textbf{\underline{51.0\%}} & 10.7\% & 19.3\% \\
 & Depth 4 & Chem. acc. & \cmark & \xmark & \xmark & \xmark & \cmark & \cmark \\
\bottomrule
\end{tabular}
}
\label{tab:v73_noisy_hamiltonian_panel}
\end{table}

\clearpage
\section{Long-Horizon Convergence and FCI-Anchored Validation}
\label{sec:app_v82_converged_fci}
Table~\ref{tab:v82_converged_fci} reports the complete long-horizon
benchmark referenced in Section~\ref{sec:converged_fci}. The exact-gradient
Adam trajectories define attainable, seed-specific references for the fixed
depth-3 ansatz; FCI is used only after optimization to quantify the remaining
variational gap.
\begin{table}[H]
\centering
\caption{\textbf{Long-horizon shot-noise-simulated VQE with an FCI-anchored diagnostic.} Gradients use Gaussian sampling from exact grouped-observable variances, not physical-QPU execution.
All methods use the same depth-3 layered RY/RZ ansatz, three paired held-out
seeds, 400 Adam steps, and 4096 shots per measured shifted QWC setting.
Our method uses $\tau=0.997$; Random receives its per-parameter group count,
and TITAN~\citep{peng2026titan} uses privileged exact-gradient ranking with the nearest Full-QWC
parameter cost. The particle-number-penalized objective is $H+(N-N_e)^2$,
while reported energies are those of $H$. FCI and exact statevector energies
are diagnostics only and are never supplied to the finite-shot optimizer or
stopping rule. \emph{Exact floor} is the best exact-gradient Adam gap to FCI.
\emph{Parity} counts seeds reaching within 1.6~mHa of their exact-Adam
reference at any checkpoint. No method reaches FCI chemical accuracy because
the depth-3 variational floor remains 9.459--51.371~mHa above FCI. Green/red
final-error and parity entries satisfy/violate the attainable-reference
criterion. Within each molecule, bold marks the best final FCI gap, absolute
final error, or shot saving; underline marks the corresponding runner-up.
Shot savings remain black.}
\label{tab:v82_converged_fci}
\vspace{0.25em}
\setlength{\tabcolsep}{1.45mm}
\renewcommand{\arraystretch}{0.94}
\scriptsize
\begingroup
\definecolor{vistatextgreen}{HTML}{278A57}
\definecolor{vistatextred}{HTML}{D84D68}
\newcommand{\tgood}[1]{\textcolor{vistatextgreen}{#1}}
\newcommand{\tbad}[1]{\textcolor{vistatextred}{#1}}
\newcommand{\best}[1]{\textbf{#1}}
\newcommand{\runnerup}[1]{\underline{#1}}
\newcommand{\mstat}[2]{\ensuremath{#1\mathbin{\pm}#2}}
\newcommand{\pstat}[2]{\ensuremath{#1\mathbin{\pm}#2\%}}
\newcommand{\grayrow}{\specialrule{0.20pt}{0.30mm}{0.30mm}}
\arrayrulecolor{black!18}
\begin{tabular*}{\textwidth}{@{\extracolsep{\fill}}llrrrrr@{}}
\arrayrulecolor{black}\toprule
Molecule & Method
& Exact floor (mHa)
& Final FCI gap (mHa)
& $\Delta E$ to exact ref. (mHa)
& Shot saving
& Parity \\
\midrule\arrayrulecolor{black!18}
H2
& Full-QWC
& \mstat{20.283}{0.303} & \runnerup{\mstat{20.672}{0.106}} & \tgood{\runnerup{\mstat{+0.389}{0.350}}} & \pstat{0.0}{0.0} & \tgood{3/3} \\
\grayrow
& Random, same budget
& \mstat{20.283}{0.303} & \mstat{21.027}{0.222} & \tgood{\mstat{+0.744}{0.305}} & \best{\pstat{26.3}{1.7}} & \tgood{3/3} \\
\grayrow
& TITAN, matched cost
& \mstat{20.283}{0.303} & \best{\mstat{20.238}{0.718}} & \tgood{\best{\mstat{-0.045}{0.415}}} & \runnerup{\pstat{26.0}{1.9}} & \tgood{3/3} \\
\grayrow
& \ours{}
& \mstat{20.283}{0.303} & \mstat{20.673}{0.072} & \tgood{\mstat{+0.390}{0.370}} & \pstat{24.7}{1.0} & \tgood{3/3} \\
\arrayrulecolor{black}\midrule\arrayrulecolor{black!18}
HeH+
& Full-QWC
& \mstat{9.459}{0.016} & \mstat{9.613}{0.159} & \tgood{\mstat{+0.154}{0.158}} & \pstat{0.0}{0.0} & \tgood{3/3} \\
\grayrow
& Random, same budget
& \mstat{9.459}{0.016} & \mstat{9.663}{0.057} & \tgood{\mstat{+0.204}{0.072}} & \pstat{29.4}{0.4} & \tgood{3/3} \\
\grayrow
& TITAN, matched cost
& \mstat{9.459}{0.016} & \best{\mstat{9.446}{0.206}} & \tgood{\best{\mstat{-0.014}{0.197}}} & \best{\pstat{30.5}{0.2}} & \tgood{3/3} \\
\grayrow
& \ours{}
& \mstat{9.459}{0.016} & \runnerup{\mstat{9.499}{0.111}} & \tgood{\runnerup{\mstat{+0.040}{0.103}}} & \runnerup{\pstat{29.8}{0.9}} & \tgood{3/3} \\
\arrayrulecolor{black}\midrule\arrayrulecolor{black!18}
H3+
& Full-QWC
& \mstat{11.939}{0.000} & \mstat{12.415}{0.187} & \tgood{\mstat{+0.477}{0.187}} & \pstat{0.0}{0.0} & \tgood{3/3} \\
\grayrow
& Random, same budget
& \mstat{11.939}{0.000} & \mstat{13.127}{0.699} & \tgood{\mstat{+1.188}{0.699}} & \pstat{56.8}{0.8} & \tgood{3/3} \\
\grayrow
& TITAN, matched cost
& \mstat{11.939}{0.000} & \best{\mstat{12.220}{0.019}} & \tgood{\best{\mstat{+0.281}{0.019}}} & \best{\pstat{58.5}{1.1}} & \tgood{3/3} \\
\grayrow
& \ours{}
& \mstat{11.939}{0.000} & \runnerup{\mstat{12.274}{0.134}} & \tgood{\runnerup{\mstat{+0.335}{0.134}}} & \runnerup{\pstat{57.1}{1.3}} & \tgood{3/3} \\
\arrayrulecolor{black}\midrule\arrayrulecolor{black!18}
H4 chain
& Full-QWC
& \mstat{51.371}{0.000} & \runnerup{\mstat{51.981}{0.034}} & \tgood{\runnerup{\mstat{+0.610}{0.034}}} & \pstat{0.0}{0.0} & \tgood{3/3} \\
\grayrow
& Random, same budget
& \mstat{51.371}{0.000} & \mstat{53.116}{0.897} & \tbad{\mstat{+1.745}{0.897}} & \pstat{76.7}{0.3} & \tbad{2/3} \\
\grayrow
& TITAN, matched cost
& \mstat{51.371}{0.000} & \mstat{52.075}{0.161} & \tgood{\mstat{+0.704}{0.161}} & \best{\pstat{77.6}{0.1}} & \tgood{3/3} \\
\grayrow
& \ours{}
& \mstat{51.371}{0.000} & \best{\mstat{51.849}{0.100}} & \tgood{\best{\mstat{+0.479}{0.100}}} & \runnerup{\pstat{76.9}{0.7}} & \tgood{3/3} \\
\arrayrulecolor{black}\midrule\arrayrulecolor{black!18}
LiH
& Full-QWC
& \mstat{18.415}{0.000} & \runnerup{\mstat{19.921}{0.437}} & \tgood{\runnerup{\mstat{+1.507}{0.437}}} & \pstat{0.0}{0.0} & \tgood{3/3} \\
\grayrow
& Random, same budget
& \mstat{18.415}{0.000} & \mstat{728.571}{144.340} & \tbad{\mstat{+710.156}{144.340}} & \pstat{92.0}{0.1} & \tbad{0/3} \\
\grayrow
& TITAN, matched cost
& \mstat{18.415}{0.000} & \best{\mstat{19.673}{0.646}} & \tgood{\best{\mstat{+1.259}{0.646}}} & \best{\pstat{92.3}{0.0}} & \tgood{3/3} \\
\grayrow
& \ours{}
& \mstat{18.415}{0.000} & \mstat{20.098}{0.455} & \tbad{\mstat{+1.684}{0.455}} & \runnerup{\pstat{92.1}{0.2}} & \tgood{3/3} \\
\arrayrulecolor{black}\bottomrule
\end{tabular*}
\endgroup
\vspace{-0.5em}
\end{table}


\clearpage
\section{Implementation and Estimator Details}
\label{sec:app_reproducibility}

This appendix distinguishes the implemented learning objective, deployment
inputs, and estimator used by each experiment. Classical predictor weights
are fixed throughout quantum-parameter optimization. Updating circuit angles
$\boldsymbol\theta$ with Adam is not fine-tuning the classical predictor or LLM.

\subsection{Architecture and Feature Construction}
The main predictor has two independent Transformer encoders, one for
parameter tokens and one for QWC-group tokens. Each encoder has two
pre-normalized layers, width 128, four attention heads, GELU activation,
feed-forward width 512, and dropout 0.14. There is no cross-attention block:
interaction occurs in the pairwise scorer described in Section 2. The scorer concatenates parameter and group embeddings, their
elementwise product and absolute difference, and two pair features. Its
LayerNorm--MLP has hidden widths 256 and 128, GELU nonlinearities, dropout,
and a scalar output. A separate parameter-only LayerNorm--MLP with hidden
width 128 predicts a sigmoid selected-group fraction. Variable-length
examples are padded within a batch; padding masks exclude invalid groups and
parameters from attention, normalization of row probabilities, and losses.
The main saved state contains 984,454 trainable scalar parameters.

\paragraph{Actual inputs, not the entire historical feature interface.}
The saved interface has 35 parameter, 30 group, and two pair channels.
Parameter descriptors encode normalized parameter, layer, and qubit indices,
sinusoidal index features, rotation channel, and last-layer status. Group
descriptors encode group index, coefficient $\ell_1$ magnitude and its
log/rank transforms, group/term counts, and Pauli-support fractions.
Structural parameter/group summaries aggregate light-cone reachability.
Pair channels are the fraction of group terms overlapping the parameter's
light cone and the coefficient-weighted fraction of those terms.
Current-state channels are exactly
$[\sin\theta_j,\cos\theta_j,\tanh\theta_j,t/T]$; they are not a history of
energies, gradients, previous masks, or optimizer moments. Historical cache
and probe slots remain in the common interface but are not populated with
oracle contributions in the direct relevance learner: token construction
passes an empty cache and uses fixed action descriptors. Thus a feature name
containing ``history'' does not imply access to past exact gradients. For each token type, training-set channel means and
standard deviations standardize inputs; scales below $10^{-6}$ are replaced
by one. Held-out examples use these saved statistics without refitting.

\subsection{Exact Training Objective}
Let $\mathcal R$ be valid rows with $\sum_k|C_{j,k}|>10^{-10}$,
$a_{j,k}=s_{j,k}/T_s$, and $q_j=\operatorname{softmax}(a_j)$ over valid groups.
Here a row indexes a snapshot and parameter, so reductions below cover all
valid rows in a batch. Let $m_{j,k}$ be the binary oracle top-mass mask at the
\emph{training} threshold $\tau_{\rm train}$, let
$b_j=G_j^{-1}\sum_k m_{j,k}$, and let $u_j$ be the auxiliary budget-head logit.
The five terms in Eq.~\ref{eq:training-loss} are implemented as
\begin{align}
\mathcal L_{\rm CE}
 &= -\frac{1}{|\mathcal R|}\sum_{j\in\mathcal R}\sum_{k=1}^{G_j}w_{j,k}\log q_{j,k},\\
\mathcal L_{\rm mask}
 &= \frac{\sum_{j\in\mathcal R}\sum_{k=1}^{G_j}
       \operatorname{BCEWithLogits}(a_{j,k},m_{j,k})}
      {\sum_{j\in\mathcal R}G_j},\\
\mathcal L_{\rm rank}
 &= \frac{\sum_{j\in\mathcal R}\sum_{k,l}
       (w_{j,k}-w_{j,l})_+
       \operatorname{softplus}(-(a_{j,k}-a_{j,l})/T_r)}
      {\max\{1,\sum_{j\in\mathcal R}\sum_{k,l}(w_{j,k}-w_{j,l})_+\}},\\
\mathcal L_{\rm budget}
 &= \frac{1}{|\mathcal R|}\sum_{j\in\mathcal R}
       \operatorname{SmoothL1}(\sigma(u_j),b_j),\\
\mathcal L_{\rm entropy}
 &= -\frac{1}{|\mathcal R|}\sum_{j\in\mathcal R}\sum_k q_{j,k}\log q_{j,k}.
\end{align}
The entropy term has a positive coefficient and therefore penalizes diffuse
predictions. BCE supervises the logits independently; it is not BCE applied
to row-softmax probabilities. The budget head is an auxiliary training loss
for the main top-mass results, not their deployment rule. Historical
count-head results instead round its predicted fraction times $G_j$ and keep
that many top-ranked groups. These two selection modes must not be conflated.

\begin{table}[ht]
\centering\small
\caption{Verified settings of the main molecular checkpoint and the separate
depth-OOD checkpoint. Training and inference thresholds are different controls.}
\label{tab:verified_training_settings}
\begin{tabular}{p{0.31\linewidth}p{0.61\linewidth}}
\toprule
Component & Saved configuration / implementation \\
\midrule
Training panel & Five molecular Hamiltonians; layered RY/RZ; depths 1--3 \\
Training seeds & 11, 23, 37, 41, 53, 67, 79, 83 \\
Held-out seeds & 97, 101, 107, 109 \\
Label generation & 18 exact-gradient Adam snapshots per cell; 2,160 snapshots \\
Angle initialization & Independent $\mathcal N(0,0.35^2)$ angles \\
Quantum optimizer & Adam; learning rate 0.035; no gradient-norm clipping in the main 28-step runner \\
Classical training & 260 epochs; batch size 8; AdamW; learning rate $3.5\times10^{-4}$; weight decay $2.5\times10^{-4}$ \\
Classical gradient clip & Norm 1.0; initialization seed 123 \\
Loss coefficients & $(\lambda_{\rm m},\lambda_{\rm r},\lambda_{\rm b},\lambda_{\rm e})=(0.15,0.35,0.60,0.010)$ \\
Temperatures & Training $T_s=0.75$; ranking $T_r=0.75$; inference 0.55 \\
Thresholds & Main training mask 0.99; main inference top-mass 0.995 or 0.997 \\
Checkpoint selection & Lowest epoch-average training loss; no held-out evaluation labels used for selection in this runner \\
Depth-OOD checkpoint & Separately trained on depths 1--4; 2,880 snapshots; training mask 0.995; evaluated at depths 5--7 \\
\bottomrule
\end{tabular}
\end{table}

\subsection{Alternative Predictor Implementations}
\label{sec:app_baseline_implementations}
All four predictors receive the same standardized parameter, group, and
pair features and the same training relevance labels in the main
architecture comparison. The MLP, GNN, and ResNet-18 use separately trained checkpoints.
Their saved settings agree with
Table~\ref{tab:verified_training_settings}: width setting 128, dropout 0.14,
260 epochs, batch size eight, AdamW learning rate $3.5\times10^{-4}$,
weight decay $2.5\times10^{-4}$, and initialization seed 123.
The initialization and architecture differ, while the objective,
normalization procedure, data split, and online protocol are fixed.
Other transfer panels use their separately trained checkpoints.

\paragraph{Pairwise MLP.}
Independent parameter and group encoders each apply
Linear--LayerNorm--GELU--Dropout--Linear--GELU with hidden width 128.
The pair scorer concatenates the two embeddings, their product, their
absolute difference, and the two pair features, then uses the same
$514\to256\to128\to1$ head as the Transformer. The budget head uses
LayerNorm and a $128\to128\to1$ MLP. This model has 224,902 parameters
and no token-to-token attention.

\paragraph{Bipartite GNN.}
Parameter and group inputs are projected to width 128. A
$2\to128\to128$ GELU MLP followed by a sigmoid produces a vector gate
for every parameter--group edge. Two message-passing rounds alternately
update parameter nodes from the mean gated valid-group embeddings, then
group nodes from the updated valid-parameter embeddings.
Each residual update applies LayerNorm and a $256\to128\to128$
GELU/dropout MLP to the node and its aggregated message.
The pair and budget heads match the MLP above. Padding is excluded from
message sums and denominators. This model has 407,942 parameters.

\paragraph{ResNet-18-style dense predictor.}
Linear projections map parameter, group, and pair inputs to 64 channels,
which are added to form a $P\times G\times64$ grid. A stride-one
$3\times3$ convolution, one-group GroupNorm, and ReLU replace the
image-classification stem. Four stages have two BasicBlocks each, channel
widths $64,128,256,512$, and first-block strides $1,2,2,2$.
Blocks contain two $3\times3$ convolutions, GroupNorm, ReLU,
Dropout2d, and a projected skip when dimensions change.
The final feature map is bilinearly interpolated to the input grid size
with \texttt{align\_corners=False}; GroupNorm--ReLU--$1\times1$ convolution
produces one score per pair. A separate parameter-only
$64\to64\to1$ budget head is used. There is no ImageNet pretraining or
fixed-resolution image resize. The model has 11,214,338 parameters.

\paragraph{Variable shapes and masking.}
Training pads each batch to its maximum parameter and group counts after
feature standardization. Invalid output logits are masked and invalid rows
are omitted from the loss. The Transformer masks padded attention keys;
the GNN masks messages. ResNet convolutions and GroupNorm operate on the
padded grid before output masking, so this implementation should not be
assumed invariant to padding. Prediction uses one example at a time.
These are architecture substitutions in the implemented common protocol,
not independently optimized best configurations for each architecture.

\paragraph{Classical scaling.}
Let $d$ be width and $L_e$ the number of layers per encoder. With a fixed
feed-forward multiplier, dense Transformer encoding requires
\[
O\!\left(L_e[(P^2+G^2)d+(P+G)d^2]\right)
\]
operations; the all-pairs MLP adds $O(PGd^2)$.
Materialized attention and pair activations require
$O(L_e(P^2+G^2)+PGd+L_e(P+G)d)$ memory for fixed head count,
apart from weights and optimizer state. Sorting all rows for top-mass
selection costs $O(PG\log G)$. These bounds describe inference after
features and groups are constructed; they exclude offline simulation,
grouping, and training. Efficient attention implementations can reduce
materialized memory without removing the dense interaction cost.

\subsection{Metric Definitions and Aggregation}
\label{sec:app_metrics}
For a molecule/ansatz/depth cell $c$ and paired seed $s$, define
\[
\Delta E_{c,s}=E^{\rm method}_{c,s}(T)-E^{\rm Full}_{c,s}(T),\qquad
\overline{\Delta E}_c=\frac1{n_c}\sum_s\Delta E_{c,s}.
\]
Energy entries are in Hartree (Ha), with $1$ mHa $=10^{-3}$ Ha.
Energy $\pm$ values in the molecular, local-LLM, and frontier tables
use population standard deviation over the recorded seeds (denominator
$n_c$). Shot-gain uncertainties in Table~\ref{tab:v39_tau0995_main}
use sample standard deviation (denominator $n_c-1$); its reused $K=1$
rows retain that convention. The sparse-refresh and best-prompt shot-gain
entries use population standard deviation. These values describe variation
over the four paired seeds, not standard errors or confidence intervals.

\paragraph{Absolute gaps and table rankings.}
The main 5.67-mHa headline averages seed-level absolute gaps,
$\frac1{|\mathcal C|}\sum_c\frac1{n_c}\sum_s|\Delta E_{c,s}|$.
It is not the absolute value of the signed mean and need not equal
the average of $|\overline{\Delta E}_c|$.
The same seed-level convention is used for the frontier method summaries
quoted in the main text. Bold/underline rankings and Random check/cross
symbols in detailed tables compare unrounded cell means as specified
by each caption. Coloring also uses unrounded means, so values printed
identically can have different ranks or threshold outcomes. These are
descriptive comparisons, not significance tests.
\emph{Best $\Delta E$} in historical budget tables selects a depth-wise
reported endpoint; it is not an average over depths.

\paragraph{Passing criteria and reference energies.}
\emph{No worse} is one-sided: $\Delta E_{c,s}\leq0.0016$ Ha, so lower
energy is accepted even outside the symmetric tolerance.
Cell-level coloring applies this threshold to $\overline{\Delta E}_c$;
seed counts apply it to individual trajectories. FCI error uses the
separate exact ground-state reference. Legacy raw fields named \texttt{ca\_vs\_full} use the symmetric
test $|\Delta E|\leq0.0016$ Ha; they must not be substituted for the
one-sided no-worse counts. The long-horizon benchmark instead records whether a trajectory
enters the 1.6-mHa neighborhood of its paired attainable exact-Adam
reference at any checkpoint. Its final reference error, best observed
reference attainment, and gap to FCI are separate quantities.
The best exact-Adam energy found defines the empirical attainable reference
for the chosen ansatz, initialization, and optimization schedule.

\paragraph{Measurement cost.}
For constant shots $S$ per shifted selected group,
$B=2S\sum_t\sum_j|S_{j,t}|$ and
$\mathrm{gain}=100(1-B_{\rm method}/B_{\rm Full})\%$.
With adaptive allocation $n_{j,k,t}$, use
$B=2\sum_{t,j,k}n_{j,k,t}$.
Reported aggregate gains average the per-trajectory ratios.
A negative gain means more shots than the paired reference.
Groups/parameter averages selected pairs over iterations and parameters;
it is neither a count of distinct bases across the whole run nor the
fraction of optimizer steps skipped. Refresh reuse holds the selection
mask, not previously measured expectation values.
Nominal counts for exact-expectation runs and sampled-gradient protocols
are distinguished in Appendix~\ref{sec:app_estimator_scope}.

\paragraph{Feature-audit quantities.}
Captured mass for row $j$ is
$m_j=\sum_kM_{j,k}|C_{j,k}|/\sum_k|C_{j,k}|$ on nonzero rows;
mean mass averages valid rows. The gradient cosine compares
$\sum_kM_{j,k}C_{j,k}$ with the full gradient over all parameters.
These quantities need not improve together because signed contributions
can cancel. Jaccard compares the selected group sets against the
full-feature reference action. A deletion effect in percentage points
is $100(m^{\rm reference}-m^{\rm deletion})$ at matched snapshot and
protocol; it is not a percentage relative improvement.
Audit quality and token costs are conditional on the validity rules of
the corresponding table, so context-window comparisons also report the
number of valid cases.

\subsection{Deterministic Truncation Bias}
Condition on the current angles and predicted subset $S_j$. If measurements
of retained groups are unbiased, the sparse estimator obeys
\begin{equation}
\mathbb E[\widehat g_j\mid\boldsymbol\theta,S_j]-g_j
 =-\sum_{k\notin S_j}C_{j,k},\qquad
\left|\mathbb E[\widehat g_j]-g_j\right|
 \leq\sum_{k\notin S_j}|C_{j,k}|.
\end{equation}
If the subset captures at least $\tau$ of the \emph{exact} absolute mass,
the last quantity is at most $(1-\tau)\sum_k|C_{j,k}|$. This bound is
conditional on exact retained mass; deployment applies the threshold to
predicted mass. The implemented estimator uses deterministic truncation,
whose bias is the omitted signed contribution above. The experiments
evaluate the resulting closed-loop energy--cost trade-off. Offline
exact-statevector label construction scales exponentially in qubit count,
setting the computational challenge for larger training systems.

\clearpage
\section{Protocol Provenance and Numerical Diagnostics}
\label{sec:app_evidence_audit}

\subsection{Estimator Fidelity and Resource Accounting}
\label{sec:app_estimator_scope}
The main 28-step molecular, transfer, and local/frontier-LLM ranking runners
form gradients as $\sum_k M_{j,k}C_{j,k}$ using exact simulated expectations.
Their shot counts are \emph{nominal}: each retained parameter--group pair is
charged two shifted settings times the specified shots per setting. The
shot-count argument does not inject sampling noise in these runners.
Randomness there comes from angle initialization and, for Random, sampled
group identities, not from quantum shot sampling. Their energy differences
therefore measure closed-loop truncation effects, not finite-shot robustness.
Exact contributions may be calculated inside the simulator, but are not
provided to our learned selector or to an LLM prompt. TITAN-style freezing
controls additionally use privileged exact-gradient importance in simulation;
their ranking information is stronger than a deployable zero-cost controller.

The separate shot-allocation panel samples gradient noise using exact grouped
observable moments under the specified allocation, and the long-horizon benchmark uses a Gaussian
finite-shot approximation with variance determined by those moments. These
are simulated noisy estimators, not hardware measurements or exact sampling
of every bitstring. The long-horizon benchmark uses 4096 shots per shifted selected group, depth 3,
three paired seeds, 400 steps, and objective $H+(N-N_e)^2$, reporting energy
of $H$. It is a distinct protocol, not a continuation of each 28-step run.
It trains a separate depth-3 controller on 80-label-snapshot trajectories
per cell, with training seeds 11/23/37 and validation seed 41; evaluation
uses 97/101/107. Its model stays frozen during the 400-step quantum
optimization. Its exact-Adam attainable references remain 9.459--51.371 mHa above FCI;
Reference parity and absolute FCI error therefore quantify distinct aspects
of the optimization outcome.

The calibration-derived Pauli-channel surrogate, finite-shot Aer grid spot
check, and shot-noise VQE are also distinct experiments. The Aer check uses
depth 1, seed 97, and 4096 shots per shifted group. Its 15--26\% omitted
cells describe \emph{grid sparsity}, not realized learned-controller savings
in a complete noisy online VQE. All reported noise studies use simulation.
Online shot gain counts gradient measurement settings; it excludes offline
label generation, classical training/inference, retrieval, compilation,
energy diagnostics, and LLM token cost. Parameter counts, file sizes, tokens,
and wall times are separate resource axes, not directly interchangeable costs.
Historical dollar estimates use the original run-time accounting assumptions. Token traffic is the directly logged resource measure.
The current scale comprises small molecular systems, four-seed
short-horizon panels, and compact spin-model QWC partitions; larger systems
require broader data generation and greater simulation resources.

\paragraph{Allocation controls and comparison scope.}
iCANS, gCANS, and Rosalin labels denote adapted allocation/operator-sampling
policies inside a common Adam harness, with the caps, warm-up, and running
moment estimates specified by the corresponding experiment. They do not
use each paper's native optimizer or stopping rule: Adam, the horizon,
and the experimental budget are shared controls. The panels evaluate
allocation and operator-sampling policies, including their composition with
learned QWC selection. The empirical baseline scope is this fixed-partition
selection/allocation setting; related work separately discusses learned
POVM and shadow-based measurement design.

\paragraph{Recorded allocation settings.}
\label{sec:app_allocation_protocol}
The main allocation panel uses five molecules, depths 1--2,
seeds 97/101/107/109, 28 Adam steps at learning rate 0.035, and no
gradient-norm clipping. Nominal shots are $1024b$ per shifted group,
with $b\in\{0.25,0.50,1.00\}$.
The learned mask uses the main saved checkpoint, inference temperature
0.55, and top-mass threshold 0.995.
The allocation state starts with zero gradient and coefficient-based
variance prior $\max(\|h_{\mathcal G_k}\|_1,10^{-4})^2$.
After observations, gradient and group-variance forecasts use
exponential averaging with decay 0.9. Variance observations are simulated
from the exact grouped moments using a chi-square sample with $n-1$
degrees of freedom; the gradients use a Gaussian moment approximation.

The adapted CANS policies use $L=10$, $\eta=0.035$, and
$a=2\eta L/(2-\eta L)$. Their unrounded per-pair requests are
\[
s_j^{\rm iCANS}
=a\frac{\sum_k\widehat v_{j,k}}{\widehat g_j^2+10^{-8}},
\qquad
s_j^{\rm gCANS}
=a\frac{\sigma_j\sum_i\sigma_i}
{\sum_i\widehat g_i^2+10^{-8}},\quad
\sigma_j^2=\sum_{k\in S_j}\widehat v_{j,k}.
\]
Requests are multiplied by $b$, rounded upward, and clipped to
$[\lceil64b\rceil,\lceil4096b\rceil]$. The first uninitialized step
uses the lower bound. These are shots per selected pair, so the row
cost also scales with its number of retained groups.

Variance-aware allocation keeps each row's nominal total and distributes
it in proportion to predicted standard deviations, with a minimum of
two shots per selected group and stable largest-remainder integer rounding.
The combined learned-mask/gCANS control additionally uses this within-mask
reallocation. Rosalin uses the adapted iCANS row request times the full
group count, multinomial group sampling proportional to coefficient
$\ell_1$ magnitude, and inverse-probability correction.
All methods then take the same Adam update.

Small estimated gradients can drive the CANS request toward its cap;
therefore shared nominal scale does not enforce equal realized shot cost,
and negative gains are possible. This explains how overspending arises
in this implementation without establishing why a native algorithm
would perform poorly. The stored run records one configuration for
these constants, not a dedicated hyperparameter search for each native
optimizer. The comparison supports composition under the stated harness.

\subsection{Threshold and Checkpoint Provenance}
\label{sec:app_checkpoint_provenance}
Table~\ref{tab:app_fixed_tau_summary} is a historical aggregate, not a
single-checkpoint threshold sweep. Its 0.98 and 0.99 rows use separately
trained checkpoints with different oracle training-mask thresholds and
the historical count-head selection mode. Its 0.95 row is an earlier trained
operating point. Only the 0.995 and 0.997 rows use the same saved $\tau_{\mathrm{train}}=0.99$
checkpoint with top-mass post-processing. The main molecular baseline
panel also loads that checkpoint. The Transformer
checkpoint used for LLM cardinalities and native attribution has identical
model-state and normalization values; only its serialized artifact/configuration
hash differs. In contrast,
Table~\ref{tab:app_depth_ood_tau_summary} reuses one separately trained depth-OOD
checkpoint over the displayed inference thresholds. Thus the
nonmonotone group counts between historical 0.99 and 0.995 rows are not a
violation of fixed-prediction top-mass nesting. Even for a fixed checkpoint,
different masks change subsequent angles, so closed-loop average group counts
and final energies need not be monotone in the threshold. A checkpoint
transferred within a named study should not be described as one universal
checkpoint across separately trained depth-, molecule-, and Hamiltonian-family
panels. ``Best $\Delta E$'' columns are selected descriptive endpoints, not
estimates of average reliability; depth-wise and seed-wise results remain
the relevant evidence for that question.

\subsection{Train-Only Retrieval for Frontier Models}
\label{sec:app_frontier_retrieval}
The retrieval bank is restricted to the same molecule, ansatz, depth, and
parameter ID, using training seeds only. For a parameter row, distances are
squared Euclidean distances between
$[\sin\theta_j,\cos\theta_j,\tanh\theta_j,t/T]$, standardized by the
training bank's per-channel standard deviation (values below $10^{-4}$
are replaced by one). Invalid zero-mass rows are excluded and ties use a
stable sort. The prompt receives two explicit nearest-neighbor records and
a summary over the eight-neighbor pool, including average training relevance
and selection frequencies, with a limited list of salient group signals.
It is not an unbounded dump of the full training dataset. Retrieval uses
training relevance labels, never the current held-out contribution matrix.
Both use exact offline relevance supervision, but the frontier bank uses
20 snapshots per training cell (2,400 total), versus 18 (2,160 total) for
the main learned checkpoint. Data access is therefore not matched in
snapshot count, and in-context examples are not equivalent to gradient-based training. Unlike the local fixed-cardinality tests, frontier models choose both
identities and counts: the request explicitly removes each parameter's
\texttt{k} field and accepts either a select or exclude list with at least one
measured group. This differs in both data access and count policy from the
local tests, so the comparison evaluates the combined model and data interface.
The requested 0.995 mass is an objective informed by retrieved training
labels; the validator checks output structure, range, and cardinality.
The four-refresh schedule is shared with our $K=7$ control, but budgets are
not matched; comparisons are accuracy--cost operating points. This evaluates
data-exposed frozen selection within an external harness, not a fine-tuned
LLM or autonomous code-executing agent.

\clearpage
\subsection{Decision Traces, Prompt Identity, and Reason Tags}
\label{sec:app_trace_provenance}
The action and reason tags in trace Box (4) come from the audit panel's
\texttt{explain\_tags} prompt. The full-feature reference and deletions in
Box (7) come from \texttt{baseline}, \texttt{no\_coeff},
\texttt{no\_lightcone}, and \texttt{no\_theta\_time}. All four decision
traces use H$_3^+$, depth 2, seed 97, iteration 0, parameter 14, and
cardinality six. The decision-trace state is initialized with
$\theta_j\sim\mathcal N(0,0.35^2)$, giving
$\theta_{14}=+0.1067889170$ rad. The 24 circuit parameters and 19 QWC groups
describe problem dimensions, not tokenizer counts. Local contribution
shares, mean captured mass, and gradient cosine are evaluated on this same
state. The integrated-gradient studies teacher-force the reference action
using their separately serialized attribution inputs
(Appendix~\ref{sec:app_ig_quality}). The prompt-specific groups are:
\begin{center}\small
\begin{tabular}{lll}
\toprule
Model & Explanation-prompt action & Reference action / IG target \\
\midrule
Qwen3-8B & $[0,1,2,3,4,5]$ & $[0,1,2,3,4,5]$ \\
Llama-3.1-8B & $[7,8,9,10,11,12]$ & $[4,5,6,7,8,9]$ \\
DeepSeek-R1-8B & $[0,2,3,4,6,7]$ & $[2,5,9,6,0,7]$ \\
GPT-OSS-20B & $[18,0,17,16,15,14]$ & $[0,1,2,6,7,3]$ \\
\bottomrule
\end{tabular}
\end{center}
The raw Qwen explanation tags are \texttt{lightcone}, \texttt{theta}, and
\texttt{circuit\_location}; Llama emits \texttt{coefficient}, \texttt{theta},
\texttt{iteration}; DeepSeek emits \texttt{coefficient}, \texttt{lightcone};
GPT-OSS emits \texttt{coefficient}. Box (3) reproduces these logged tags.
Each action's exact contribution shares are computed \emph{after} inference,
outside the model, solely for evaluation. An intervention difference must
be subtracted from the reference-prompt score, not from a different
explanation-prompt score. Llama's explanation action captures 1.797\%; its
reference action captures 1.442\%, and deleting coefficients raises that
local reference mass to 94.843\% (+93.401 percentage points). Over valid
parameters, mean mass instead falls from 0.483 to 0.395, distinguishing
the local intervention effect from aggregate selection quality. 

\paragraph{Dependence, usefulness, and faithfulness are distinct.}
A changed mask establishes input sensitivity in the reference protocol.
A positive captured-mass drop means the deleted feature was useful there;
a negative drop means it was harmful there, not that it was unused.
Unchanged outputs can reflect redundancy or feature interactions.
Reason tags and deletion effects are linked by snapshot while retaining
their distinct prompt identities: \texttt{explain\_tags} produces the
reported tags, and reference-prompt interventions measure feature utility.
The verdicts in the drawings are interpreted through this distinction.
These traces record observable interaction and externally measured
sensitivity. Local LLMs run frozen text inference with code-execution tools
disabled; \texttt{think=false} specifies the direct-output configuration
used in this comparison.

\paragraph{Protocol-specific output contracts.}
The rank-only/repaired local protocol in
Appendix~\ref{sec:app_llm_prompt_protocol} emits a ranking. The clean protocol instead uses
\texttt{selections.p14 = \{mode,group\_ids\}}, emitting either the selected
IDs or their shorter excluded complement at exact cardinality. The audit's optional tags use
the controlled vocabulary \texttt{coefficient}, \texttt{lightcone},
\texttt{theta}, \texttt{circuit\_location}, \texttt{iteration}, and
\texttt{group\_support}, with one to three unique tags per row. The clean validator
checks integer type (excluding booleans), range, uniqueness, mode, and length;
schema feedback permits retries, but no coefficient completion repairs
IDs. Persistent failure invokes logged Full-QWC safety fallback and charges
all groups. The audit additionally varies output contracts and optionally returns
\texttt{reason\_codes}; the best-prompt evaluation tests its selected prompt online. The example
\texttt{\{parameter\_id,mode,group\_ids,reason\_codes\}} in a trace is a
normalized logged record, not the complete literal clean-protocol wire schema.
The best prompt was selected on the fixed audit panel, which includes
seed 97, also present in the online evaluation. It is the best tested
configuration on that tuning panel; the shared seed links prompt selection
to the corresponding online evaluation cells. The \texttt{pair\_nonzero\_rows} key names the column dictionary, whereas
\texttt{pair\_nonzero} holds each parameter's rows; the different names are
intentional. These are frozen LLM inference protocols within our external
Python harness, not learned tool-use agents.

\paragraph{Clean system instruction and explanation extension.}
The clean instruction, also used by the domain-prompt audit baseline, is:
\begin{quote}\small\ttfamily\raggedright
You are a measurement-selection controller for variational quantum eigensolvers.
For every requested circuit parameter, follow its fixed output mode and emit
exactly the requested number of distinct QWC group IDs. In select mode, emit
the groups most relevant to that parameter's energy gradient, ordered from
most to least relevant. In exclude mode, emit the groups least relevant to
that gradient, ordered from least to most relevant; every unlisted candidate
will be measured. Use Hamiltonian coefficient, circuit location, theta,
iteration, and parameter--group light-cone features jointly. A globally large
coefficient does not imply equal relevance to every parameter. Return only the
JSON object required by the schema. Never omit a requested parameter and
never invent an ID. The exact cardinality is mandatory.
\end{quote}
For the explanation condition the appended instruction is:
\begin{quote}\small\ttfamily\raggedright
For each parameter, also emit one to three reason\_codes chosen only from
['coefficient', 'lightcone', 'theta', 'circuit\_location', 'iteration',
'group\_support']. These are concise evidence tags, not chain-of-thought.
\end{quote}
The domain reference uses a 32,768-token context window and the compressed
select/exclude contract. A row with requested count $k_j\leq G-k_j$
uses select mode and emits $k_j$ IDs; otherwise it emits the $G-k_j$
excluded IDs. For the shared trace with $G=19$ and $k_{14}=6$, the
explanation response has the following wire structure (shown with the
accepted Qwen action):
{\small
\begin{verbatim}
{"selections": {
  "p14": {
    "mode": "select",
    "group_ids": [0, 1, 2, 3, 4, 5],
    "reason_codes": ["lightcone", "theta", "circuit_location"]
  }
}}
\end{verbatim}
}
Without explanation tags the same reference contract omits
\texttt{reason\_codes}. The request also specifies each row's
\texttt{selected\_count}, \texttt{output\_mode}, and \texttt{emit\_count}.
The returned object is validated against those values, range and uniqueness
constraints, and the complete set of requested parameter keys.

\paragraph{What the context ablation varies.}
The context ablation changes the configured context-window capacity to 8,192, 16,384,
or 32,768 tokens. This is a capacity setting, not a promise that every
request fills that many tokens. Its separate prompt variants use a minimal
instruction or add two training demonstrations; output variants emit a
direct selected list or a complete ranking. Deletion and shuffling
conditions hold the Full-QWC snapshot and requested cardinalities fixed.
Explanation tags are an additional condition. Reported differences compare
configurations under this harness; context capacity, protocol success, and
valid-case mask quality describe distinct parts of controller performance.

\paragraph{Observable computation boundary.}
The models serialize a selection as text/JSON; Python constructs features,
retrieves training records where enabled, validates outputs, applies masks,
and performs the VQE update. Code-execution and external tool calls are not
available to the evaluated local models, and frontier tool use is disabled
in the saved CLI protocol. These controls specify the observable
interaction and the division of work between model and harness.
The frontier GPT runs record low reasoning effort; the local direct-output
tests record \texttt{think=false}. Conclusions concern those settings.

\clearpage
\subsection{Integrated-Gradient Numerical Reliability}
\label{sec:app_ig_quality}
The four LLM attribution prompts use the serialized H$_3^+$, depth-2,
seed-97, iteration-0 state with initialization scale 0.12
($\theta_{14}\simeq+0.0366$ rad) and the fixed reference group lists above.
The decision traces use their recorded scale-0.35 state; each analysis
evaluates its own state and score while sharing the task and target IDs.
For the LLM figures, we fix the observed action and teacher-force the JSON
prefix for each emitted ID. The scalar score is the first-token log
probability conditioned on grammar-valid ID prefixes, not the probability
of the entire freely generated mask. Only marked numeric feature-token
embeddings are interpolated. The input is tokenized once: the baseline
replaces these embeddings with repeated embeddings of the token for numeric
zero, preserving sequence length, positions, attention mask, syntax,
instructions, and the fixed output prefix. It does not retokenize a string
of shorter numerical values. This is a synthetic embedding-space baseline,
not a physically valid alternative VQE state. Attribution sums are aggregated
by labeled spans. Absolute-mass normalization summarizes the distribution
of attribution magnitude across input families; its denominator is total
absolute attribution, whereas energy accuracy is evaluated separately.

The Ollama/GGUF generation runtime is not differentiable in this setup;
gradients come from matched Hugging Face model families under the quantization
specified by each figure. Quantization, kernels, and routed experts can
change scores and continuations. Our controller instead uses its native
checkpoint, standardized scalar features, a training-mean baseline, and
1024 quadrature nodes. Therefore cross-model percentages are not calibrated
on a common functional or numerical scale; each panel describes its own
target score and baseline.

For a scored group, let $D=f(x)-f(x_0)$ and
$R=D-\sum_i\mathrm{IG}_i$. We report both absolute and relative error:
small $|R|$ alone is insufficient when $|D|$ is near zero.
The table below is computed from the saved metadata, not from a new experiment.
\begin{table}[ht]
\centering\small
\caption{Completeness diagnostics for the six displayed group scores.
The last column counts scores with $|R|>|D|$; these estimates fail even a
100\% relative completeness check. Residuals accompany the finite-quadrature
estimates for each model.}
\label{tab:ig_numerical_audit}
\begin{tabular}{lrrrr}
\toprule
Model & Nodes & Max $|R|$ & Max $|R|/|D|$ & $|R|>|D|$ \\
\midrule
Qwen3-8B & 32 & 0.1224 & 8.56 & 4/6 \\
Llama-3.1-8B & 32 & 1.1491 & 93.33 & 3/6 \\
DeepSeek-R1-8B & 32 & 0.4373 & 4.63 & 2/6 \\
GPT-OSS-20B & 32 & 1.4140 & 1.40 & 2/6 \\
Our controller & 1024 & 0.0126 & 0.0046 & 0/6 \\
\bottomrule
\end{tabular}
\end{table}
Qwen's first score has a 1.61\% residual relative to its score change;
several later near-saturated continuation scores have substantially larger
relative errors. The Llama, DeepSeek, and GPT-OSS entries likewise show
substantial finite-quadrature residuals. Their plotted family shares are
finite-quadrature estimates under the specified scoring functions.
The native controller uses 1024 nodes and has maximum relative residual
0.46\%; its 12.16\% current-state attribution identifies a resolved
dependence of its group scores on optimization-state inputs. The LLM
panels use different scoring functions and embedding baselines, so the
family percentages describe within-model attribution profiles. Closed-loop
energy and shot cost provide the common cross-model performance measures.

\clearpage
\subsection{Fixed-Snapshot Attribution Stability Audit}
\label{sec:app_ig_stability}
We test whether the feature-family ordering in
Figure~\ref{fig:feature_family_attribution_strip} persists across numerical
integration settings. This completed follow-up evaluates the same four
serialized H$_3^+$ attribution cases described in
Appendix~\ref{sec:app_ig_quality}: depth 2, seed 97, iteration 0,
parameter 14, and initialization scale 0.12. Each model retains its archived
prompt, token positions, reference group IDs, teacher-forced JSON prefix,
score function, and numeric-zero embedding baseline. These are numerical
repeats of fixed cases, not additional physical snapshots or online VQE runs.

\paragraph{Controlled refinement.}
We use 32, 64, 128, and 256 Gauss--Legendre nodes with float64 accumulation.
All four models are evaluated under their original attribution runtime.
For Qwen, Llama, and DeepSeek, a separate FP32-compute diagnostic preserves
the NF4 weight bytes while promoting floating-point layers and quantized
matrix-multiplication compute to FP32. This is a separate arithmetic regime,
not an unquantized model or a bitwise-identical scoring implementation.
GPT-OSS retains its native MXFP4 implementation and hard expert routing.
The audit thus contains 28 model--precision--quadrature configurations,
each scoring six fixed group IDs. No model is trained, no new action is
generated, and the original plotted assets are retained.

\paragraph{Aggregation and ranking.}
For output group $g$, let $I_{gs}$ be the signed IG summed over the tokens
of numeric input span $s$. Its feature-family share is
\begin{equation}
 A_{gF}=100\,\frac{\sum_{s\in F}|I_{gs}|}{\sum_s|I_{gs}|},
 \qquad \overline A_F=\frac{1}{6}\sum_g A_{gF}.
\end{equation}
The family ranking uses $\overline A_F$. Percentages therefore summarize
absolute span attribution, not selection probability, causal effect size,
or energy contribution. Structure includes circuit location and Pauli
support; budget/action covers controller-contract fields. Trajectory
denotes the serialized current-angle and normalized-step features, rather
than a complete history of measured gradients.

\begin{table}[ht]
\centering\small
\setlength{\tabcolsep}{4pt}
\caption{Feature-family ranking and trajectory-share stability in the
fixed-snapshot audit. $S$: circuit/support; $C$: coefficient;
$L$: light cone; $B$: budget/action; $T$: trajectory.
Each model's mean ranking is identical across its tested settings.
Shares are averaged over six scored group IDs; the range spans all tested
node counts and arithmetic regimes, not a confidence interval.
FP32 is a compute diagnostic with unchanged NF4 weights.}
\label{tab:ig_stability_summary}
\begin{tabular}{llrrr}
\toprule
Model & Stable mean ranking & \multicolumn{2}{c}{Trajectory at 256 nodes (\%)} & \shortstack{All-setting\\range (\%)} \\
\cmidrule(lr){3-4}
 & & Original & FP32 & \\
\midrule
Qwen3-8B & $S>C>L>B>T$ & 0.77 & 1.04 & 0.59--1.06 \\
Llama-3.1-8B & $S>C>L>B>T$ & 0.95 & 0.76 & 0.67--0.95 \\
DeepSeek-R1-8B & $S>C>L>B>T$ & 0.99 & 1.01 & 0.96--1.01 \\
GPT-OSS-20B & $S>L>C>B>T$ & 2.49 & -- & 2.49--2.71 \\
\bottomrule
\end{tabular}

\end{table}

\paragraph{Consistent low trajectory attribution.}
Trajectory ranks fifth among the five families for every scored group in
every configuration: 168/168 output--configuration combinations.
The model-specific orderings and trajectory-share ranges are reported in
Table~\ref{tab:ig_stability_summary}. Under the original runtime, increasing from
128 to 256 nodes changes these means by only 0.024, 0.165, 0.007, and
0.037 percentage points. These counts describe numerical robustness within
the fixed cases, rather than independent-sample statistical evidence.

\paragraph{Implication for frozen selection.}
Despite receiving the full input feature set, frozen LLMs do not
automatically make substantial use of trajectory information without
targeted intervention in these audited cases: static physical inputs
dominate their attribution profiles.
This pattern persists across the tested quadrature and precision settings.
It complements the feature-deletion and online evaluations in
Tables~\ref{tab:v80_llm_ablation} and~\ref{tab:v80b_best_prompt_online},
which assess changes in decisions and closed-loop performance.
The consistently low within-model trajectory ranking, rather than an exact
cross-model percentage ratio, is the central result of this refinement.

\clearpage
\paragraph{Numerical checks and their relation to ranking stability.}
The audit predeclares a per-output completeness tolerance
$|R|\leq 0.001+0.05|D|$, with $D$ and $R$ defined in
Appendix~\ref{sec:app_ig_quality}. Between successive node counts, it also
checks maximum family-cell change $\leq1$ percentage point, maximum
trajectory-cell change $\leq0.5$ percentage points, and minimum per-output
family-rank Spearman correlation $\geq0.9$. Refinement can stop after at
least 128 nodes only when two consecutive levels meet all conditions;
otherwise it continues through 256 nodes. All seven model--runtime pairs
reach 256 nodes in this audit.

\begin{table}[ht]
\centering\small
\setlength{\tabcolsep}{4pt}
\caption{Final numerical diagnostics, kept separate from the stable mean
rankings in Table~\ref{tab:ig_stability_summary}. Residual and completeness
pass counts use 256 nodes. Changes and correlations compare 128 with 256
nodes: $\Delta_F$ is the largest absolute family-cell change,
$\Delta_T$ the largest trajectory-cell change (both in percentage points),
and $\rho_{\min}$ the minimum per-output rank correlation. Pass counts
apply the combined absolute/relative completeness tolerance, not a
relative-error-only threshold.}
\label{tab:ig_stability_diagnostics}
\begin{tabular}{llrrrrr}
\toprule
Model & Compute & Max $|R|$ & \shortstack{Complete\\scores} & $\Delta_F$ & $\Delta_T$ & $\rho_{\min}$ \\
\midrule
Qwen3-8B & Original & 1.8269 & 0/6 & 3.296 & 0.205 & 0.90 \\
Qwen3-8B & FP32 & 3.4425 & 2/6 & 2.551 & 0.528 & 0.90 \\
Llama-3.1-8B & Original & 0.9630 & 0/6 & 5.381 & 0.568 & 1.00 \\
Llama-3.1-8B & FP32 & 0.3602 & 2/6 & 2.144 & 0.134 & 1.00 \\
DeepSeek-R1-8B & Original & 0.2188 & 1/6 & 1.276 & 0.047 & 0.90 \\
DeepSeek-R1-8B & FP32 & 0.0222 & 4/6 & 0.812 & 0.013 & 1.00 \\
GPT-OSS-20B & Original & 1.4116 & 0/6 & 1.838 & 0.352 & 1.00 \\
\bottomrule
\end{tabular}

\end{table}

DeepSeek with FP32 compute meets the family-stability conditions at
256 nodes, with a maximum family-cell change of 0.812 percentage points
and identical per-output family ranks. Its completeness tolerance is met
by four of six scores. Across all seven model--runtime pairs, the number
of complete scores ranges from zero to four, so the full conjunction of
numerical criteria is not met. The other pairs retain larger family-cell
changes even when their mean ordering and low trajectory shares persist.
Thus the audit reports ranking robustness and completeness as separate
properties; the percentages remain finite-quadrature estimates. In
particular, normalizing absolute attribution to 100\% is not a correction
for residual error. No post-hoc rescaling is applied.



\end{document}